\documentclass[rmp,aps,lengthcheck,nofootinbib]{revtex4}
\usepackage{longtable}
\usepackage{dcolumn}
\usepackage{epsf}
\usepackage{bm}
\usepackage{amsmath}
\usepackage{graphicx}
\usepackage{xr}
\begin{document}
\newcommand{\greeksym}[1]{{\usefont{U}{psy}{m}{n}#1}}
\newcommand{\rmssmu}{\mbox{\scriptsize{\greeksym{m}}}}
\newcommand{\rmsstau}{\mbox{\scriptsize{\greeksym{t}}}}
\newcommand{\rmssgamma}{\mbox{\scriptsize{\greeksym{g}}}}
\newcommand{\rmmu}{\mbox{\greeksym{m}}}
\newcommand{\rmtau}{\mbox{\greeksym{t}}}
\newcommand{\rmalpha}{\mbox{\greeksym{a}}}
\newcommand{\rmssalpha}{\mbox{\scriptsize{\greeksym{a}}}}
\newcommand{\rmpi}{\mbox{\greeksym{p}}}
\newcommand{\rmsspi}{\mbox{\scriptsize{\greeksym{p}}}}
\newcommand{\rmphi}{\mbox{\greeksym{f}}}
\newcommand\T{\rule{0pt}{2.4ex}}
\newcommand\B{\rule[-1.0ex]{0pt}{0pt}}
\def\lbar{\lambda\hskip-4.5pt\vrule height4.6pt depth-4.3pt width4pt}
\def\moles{$^{\rm o}\hskip-4.85pt\vrule height4.98pt depth-4.89pt width4.4pt\hskip4.85pt$}
\def\fr#1#2{{\textstyle{#1\over#2}}}
\def\iG{\it \Gamma}
\renewcommand{\thesection}{\arabic{section}}
\renewcommand{\thesubsection}{\arabic{subsection}}
\renewcommand{\thesubsubsection}{\arabic{subsubsection}}
\renewcommand{\theparagraph}{\arabic{paragraph}}
\renewcommand{\thetable}{\arabic{table}}
\title{
       CODATA Recommended Values of the Fundamental Physical
       Constants: 2006\footnote{
  This report was prepared by the authors under the auspices of 
  the CODATA Task Group on Fundamental Constants.
  The members of the task group are: \\
  F. Cabiati, Istituto Nazionale di Ricerca Metrologica, Italy \\
  K. Fujii, National Metrology Institute of Japan, Japan \\
  S. G. Karshenboim, D. I. Mendeleyev All-Russian Research Institute for Metrology, Russian Federation \\
  I. Lindgren, Chalmers University of Technology and G\"oteborg University, Sweden \\
  B. A. Mamyrin (deceased), A. F. Ioffe Physical-Technical Institute, Russian Federation \\
  W. Martienssen, Johann Wolfgang Goethe-Universit\"at, Germany \\
  P. J. Mohr, National Institute of Standards and Technology, United States of America \\
  D. B. Newell, National Institute of Standards and Technology, United States of America \\
  F. Nez, Laboratoire Kastler-Brossel, France \\
  B. W. Petley, National Physical Laboratory, United Kingdom \\
  T. J. Quinn, Bureau international des poids et mesures \\
  B. N. Taylor, National Institute of Standards and Technology, United States of America \\
  W. W\"oger, Physikalisch-Technische Bundesanstalt, Germany \\
  B. M. Wood, National Research Council, Canada \\
  Z. Zhang, National Institute of Metrology, China (People's Republic of) \\
  }
  }
\author{
Peter J. Mohr\footnote{Electronic address: mohr@nist.gov},
Barry N. Taylor\footnote{Electronic address: barry.taylor@nist.gov}, and
David B. Newell\footnote{Electronic address: dnewell@nist.gov},}
\affiliation{
    National Institute of Standards and Technology,
    Gaithersburg, Maryland 20899-8420, USA
   }
\date{\today}
\begin{abstract}
This paper gives the 2006 self-consistent set of values of the basic
constants and conversion factors of physics and chemistry recommended by
the Committee on Data for Science and Technology (CODATA) for
international use.  Further, it describes in detail the adjustment of
the values of the constants, including the selection of the final set of
input data based on the results of least-squares analyses.  The 2006
adjustment takes into account the data considered in the 2002 adjustment
as well as the data that became available between 31 December 2002, the
closing date of that adjustment, and 31 December 2006, the closing date
of the new adjustment.  The new data have led to a significant reduction
in the uncertainties of many recommended values.  The 2006 set replaces
the previously recommended 2002 CODATA set and may also be found on the
World Wide Web at physics.nist.gov/constants.
%
\end{abstract}
\maketitle
\tableofcontents
%
%
\section*{Glossary}
\begin{longtable}{lp{7.0cm}}

AMDC & Atomic Mass Data Center, 
Centre de Spectrom\'etrie Nucl\'eaire et de Spectrom\'etrie de Masse 
(CSNSM), Orsay, France \\

AME2003 & 2003 atomic mass evaluation of the AMDC\\

$A_{\rm r}(X)$ & Relative atomic mass of $X$: $A_{\rm r}(X) =
m(X)/m_{\rm u}$ \\

$A_{90}$  & Conventional unit of electric current: \newline
$A_{90} = V_{90}/{\it \Omega}_{90}$  \\

\AA$^\ast$ & \AA ngstr\"om-star: 
${\rm \lambda}({\rm WK}{\rmalpha}_{1}) = 0.209\,010\,0 \ {\rm \AA}^\ast$ 
\\ 

$a_{\rm e}$ & Electron magnetic moment anomaly: \newline $a_{\rm e} =
(|g_{\rm e}|-2$)/2 \\

$a_{\rmssmu}$ & Muon magnetic moment anomaly: \newline $a_{\rmssmu} =
(|g_{\rmssmu}| -2$)/2  \\ 

BIPM & International Bureau of
Weights and Measures, S\`evres, France \\

BNL & Brookhaven National Laboratory, Upton, New York, USA \\

CERN & European Organization for Nuclear Research, Geneva, Switzerland \\ 

CIPM &  International Committee for Weights and Measures \\

CODATA \ \ & Committee on Data for Science and
        Technology of the International Council
        for Science  \\

$CPT$ & Combined charge conjugation, parity inversion, and time reversal \\

$c$ & Speed~of~light~in~vacuum \\

cw & Continuous wave \\

d & Deuteron (nucleus of deuterium D, or $^2$H) \\

$d_{220}$ & $\{220\}$ lattice spacing of an ideal crystal of naturally occurring 
silicon \\

$d_{220}({\scriptstyle X })$ & $\{220\}$ lattice spacing of crystal $X$ of naturally occurring silicon \\

$E_{\rm b}$ & Binding energy \\

e   & Symbol for either member of the electron-positron 
pair; when necessary, e$^-$ or e$^+$ is used to indicate
the electron or positron  \\

$e$ & Elementary charge: absolute value of the charge of the electron \\

$F$ & Faraday constant: $F$ = $N_{\rm A}e$ \\

FCDC & Fundamental Constants Data Center, NIST, USA \\

FSU & Friedrich-Schiller University, Jena, Germany \\

${\cal F}_{90}$ & ${\cal F}_{90} = (F/A_{90})$~A \\

$G$ & Newtonian constant of gravitation \\

$g$ & Local acceleration of free fall \\

$g_{\rm d}$ & Deuteron $g$-factor: $g_{\rm d} = \mu_{\rm d}/\mu_{\rm N}$ \\

$g_{\rm e}$ & Electron $g$-factor: $g_{\rm e} = 2\mu_{\rm e}/\mu_{\rm B}$ \\
 
$g_{\rm p}$ & Proton $g$-factor: $g_{\rm p} = 2\mu_{\rm p}/\mu_{\rm N}$ \\

$g^\prime_{\rm p}$ & Shielded proton $g$-factor: 
$g_{\rm p}^\prime = 2\mu_{\rm p}^\prime/\mu_{\rm N}$ \\

$g_{\rm t}$ & Triton $g$-factor: $g_{\rm t} = 2\mu_{\rm t}/\mu_{\rm N}$ \\

$g_{X}(Y)$ & $g$-factor of particle $X$ in the ground (1S) state of 
hydrogenic atom $Y$ \\

$g_{\rmssmu}$ & Muon $g$-factor: $g_{\rmssmu} = 2\mu_{\rmssmu}/(e\hbar/2m_{\rmssmu})$ \\
 
GSI & Gesellschaft f\"ur Schwerionenforschung, Darmstadt, Germany \\

HD & HD molecule (bound state of hydrogen and deuterium atoms) \\

HT & HT molecule (bound state of hydrogen and tritium atoms) \\

h & Helion (nucleus of $^{3}$He) \\

$h$ & Planck constant; $\hbar = h/2\rmpi$ \\

Harvard;  & Harvard University, Cambridge, Massachusetts, \\
\ HarvU &  USA \\

ILL & Institut Max von Laue-Paul Langevin, Grenoble, France \\

IMGC & Istituto di Metrologia ``T. Colonetti,'' Torino, Italy \\

INRIM  & Istituto Nazionale di Ricerca Metrologica, Torino, Italy \\

IRMM & Institute for Reference Materials and Measurements, Geel, Belgium \\

JINR & Joint Institute for Nuclear Research, Dubna, Russian Federation \\

KRISS & Korea Research Institute of Standards and Science, Taedok Science Town, Republic of Korea \\

KR/VN & KRISS-VNIIM collaboration \\

$K_{\rm J}$ & Josephson constant: $K_{\rm J} = 2e/h$ \\

$K_{\rm J -90}$ & Conventional value of the Josephson constant 
$K_{\rm J}$:
$K_{\rm J -90} = 483\,597.9$~GHz V$^{-1}$ \\

$k$ & Boltzmann constant: $k = R/N_{\rm A}$ \\

LAMPF & Clinton P. Anderson Meson Physics Facility at Los Alamos National Laboratory, Los Alamos, New Mexico, USA \\

LKB & Laboratoire Kastler-Brossel, Paris, France \\

LK/SY & LKB and SYRTE collaboration \\

LNE & Laboratoire national de m\'etrologie et d'essais, Trappes, France \\

MIT & Massachusetts Institute of Technology, Cambridge, Massachusetts, USA \\

MPQ & Max-Planck-Institut f\"ur Quantenoptik, Garching, Germany \\

MSL & Measurement Standards Laboratory, Lower Hutt, New Zealand \\

$M(X)$ & Molar mass of $X$: $M(X) = A_{\rm r}(X) M_{\rm u}$ \\

Mu & Muonium (${\rmmu}^{+} {\rm e}^{-}$ atom) \\

$M_{\rm u}$ & Molar mass constant: $M_{\rm u} = 10^{-3}~{\rm kg~mol^{-1}}$ \\

$m_{\rm u}$ & Unified atomic mass constant: $m_{\rm u} = m(^{12}$C)/12 \\ 

$m_{X}$, $m(X)$ & Mass of $X$ (for the electron e, proton p, and
other elementary particles, the first symbol is used, {\it i.e.,}
$m_{\rm e}$, $m_{\rm p}$, {\it etc.}) \\

$N_{\rm A}$ & Avogadro constant \\

N/P/I & NMIJ-PTB-IRMM combined result \\

NIM & National Institute of Metrology, Beijing, China (People's Republic of) \\

NIST & National Institute of Standards and Technology, Gaithersburg, 
Maryland and Boulder, Colorado, USA  \\

NMI & National Metrology Institute, Lindfield, Australia \\

NMIJ & National Metrology Institute of Japan, Tsukuba, Japan \\

NMR & Nuclear magnetic resonance \\

NPL & National Physical Laboratory, Teddington, UK \\

NRLM & National Research Laboratory of Metrology, Tsukuba, Japan \\

n & Neutron \\

PRC & People's Republic of China \\

PTB & Physikalisch-Technische Bundesanstalt, Braunschweig and Berlin, Germany \\

p & Proton \\

$\overline {\rm p} \,^A$He$^+$ & Antiprotonic helium ($^A$He$^{+}$
+ $\overline {\rm p}$ atom, $A= 3 \mbox{ or } 4$)  \\

QED & Quantum electrodynamics \\

$Q(\chi^2|\nu)$ & Probability that an observed value of chi-square for
$\nu$ degrees of freedom would exceed $\chi^2$ \\

$R$ & Molar gas constant \\

$\overline R$ & Ratio of muon anomaly difference frequency to
free proton NMR frequency \\

$R_{\rm B}$ & Birge ratio: $R_{\rm B} = (\chi^{2}/\nu) ^\frac {1}{2}$ \\ 

$R_{\rm d}$; Rd & Bound-state rms charge radius of the deuteron \\

$R_{\rm K}$ & von Klitzing constant: $R_{\rm K} = h/e^{2}$ \\

$R_{\rm K-90}$ & Conventional value of the von Klitzing constant
$R_{\rm K}$:
$R_{\rm K-90} = 25\,812.807~{\rm \Omega}$ \\

$R_{\rm p}$; Rp& Bound-state rms charge radius of the proton \\

$R_\infty$ & Rydberg constant: $R_\infty = m_{\rm e}c\alpha^{2}
/2h$ \\

$r(x_i,x_j)$ & Correlation coefficient of estimated values $x_i$ and $x_j$:
$r(x_i,x_j) = u(x_i,x_j)/[u(x_i) u(x_j)]$ \\

$r_i$ & Normalized residual of $x_i$: $r_i = (x_i-\hat x_i)/u(x_i)$,
$\hat x_i$ is the adjusted value of $x_i$ \\

rms & Root mean square \\

$S_{\rm c}$ & Self-sensitivity coefficient \\

SI & Syst\`eme international d'unit\'es (International System of Units) \\

Stanford; & Stanford University, Stanford, California, USA \\
\ StanfU & \\

StPtrsb & St. Petersburg, Russian Federation \\

SYRTE & Syst\`emes de r\'ef\'erence Temps Espace, Paris, France \\

$T$ & Thermodynamic temperature \\

t & Triton (nucleus of tritium T, or $^3$H) \\

th & Theory \\

Type A & Uncertainty evaluation by the statistical analysis of series
of observations \\

Type B & Uncertainty evaluation by means other than the statistical analysis of series
of observations \\

$t_{90}$ & Celsius temperature on the International Temperature Scale of 1990 (ITS-90) \\

U. Sussex; & University of Sussex, Sussex, UK \\
\ USus & \\

UK & United Kingdom \\

USA & United States of America \\

UWash & University of Washington, Seattle, Washington, USA \\

u & Unified atomic mass unit (also called the dalton, Da): 1 u = $m_{\rm u}$ = $m(^{12}$C)/12 \\

$u(x_i)$ & Standard uncertainty ({\it i.e.,} estimated standard deviation) of an 
estimated value $x_i$ of a quantity $X_i$ (also simply $u$) \\

$u(x_i,x_j)$ & Covariance of estimated values $x_i$ and $x_j$ \\

$u_{\rm diff}$ & Standard uncertainty of the difference $x_i-x_j$:
$u_{\rm diff}^2 = u^2(x_i)+u^2(x_j)-2\,u(x_i,x_j)$ \\
 
$u_{\rm r}(x_i)$ & Relative standard uncertainty of an 
estimated value $x_i$ of a quantity $X_i$: \newline
$u_{\rm r}(x_i) = u(x_i)/|x_i|, \ x_i \ne 0$ (also simply $u_{\rm r}$) \\

$u_{\rm r}(x_i,x_j)$ & Relative covariance of estimated values $x_i$ and $x_j$:
$u_{\rm r}(x_i,x_j) = u(x_i,x_j)/(x_i x_j)$ \\ 
 
$V_{\rm m}({\rm Si)}$ & Molar volume of naturally occurring silicon \\

VNIIM & D. I. Mendeleyev All-Russian Research Institute for Metrology, St. Petersburg, Russian 
Federation  \\

$V_{90}$ & Conventional unit of voltage based on the Josephson effect and
$K_{\rm J -90}$: $V_{90} = (K_{\rm J -90}/K_{\rm J}$) V \\

WGAC & Working Group on the Avogadro Constant of the CIPM Consultative
Committee for Mass and Related Quantities (CCM) \\

$W_{90}$ & Conventional unit of power:
$W_{90} = V^{2}_{90}/{\it \Omega}_{90}$ \\

XROI & Combined x-ray and optical interferometer \\

xu(CuK${\rmalpha}_1$) & Cu x unit: ${\rm \lambda}$(CuK${\rmalpha}_1$) =
1\,537.400 xu(CuK${\rmalpha}_1$) \\

xu(MoK${\rmalpha}_1$) & Mo x unit: ${\rm \lambda}$(MoK${\rmalpha}_1$) =
707.831 xu(MoK${\rmalpha}_1$) \\

$x(X)$ & Amount-of-substance fraction of $X$ \\

YAG & Yttrium aluminium garnet; Y$_3$Al$_5$O$_{12}$ \\

Yale; YaleU & Yale University, New Haven, Connecticut, USA \\

$\alpha$ & Fine-structure constant: $\alpha = e^2/4\rmpi\epsilon_0\hbar c
\approx 1/137 $ \\

${\rmalpha}$ & Alpha particle (nucleus of $^{4}$He) \\

$\it\Gamma ^{\prime}_{X-{\rm 90}}$(lo) & $\it\Gamma ^{\prime}_{X-{\rm 90}}({\rm lo})
= (\gamma_{X}^\prime \, A_{\rm 90})$~A$^{-1}$, $X$ = p or h \\

$\it\Gamma ^{\prime}_{\rm p-90}$(hi) & $\it\Gamma ^{\prime}_{\rm p-90}({\rm hi})
= (\gamma_{\rm p}^\prime / A_{\rm 90})$~A \\

$\gamma_{\rm p}$ & Proton gyromagnetic ratio:~$\gamma_{\rm p} =
2\mu_{\rm p} / \hbar$ \\

$\gamma_{\rm p}^\prime$ & Shielded proton gyromagnetic ratio:
$\gamma_{\rm p}^\prime = 2\mu^\prime_{\rm p}/\hbar$ \\

$\gamma^{\prime}_{\rm h}$ & Shielded helion gyromagnetic ratio:
$\gamma ^{\prime}_{\rm h} = 2|\mu^{\prime}_{\rm h}|/ \hbar $ \\

$\Delta\nu_{\rm Mu}$ & Muonium ground-state hyperfine splitting \\

$\delta_{\rm e}$ & Additive correction to the theoretical expression 
for the electron magnetic moment
anomaly $a_{\rm e}$ \\

$\delta_{\rm Mu}$ & Additive correction to the theoretical expression for the ground-state
hyperfine splitting of muonium ${\rm \Delta \nu}_{\rm Mu}$ \\

$\delta_{\overline {\rm p} \,{\rm He}}$ & Additive correction to the 
theoretical expression for a particular transition frequency of antiprotonic helium \\

$\delta_{X} (n{\rm L}_{j})$ & Additive correction to the theoretical expression for an energy
level of either hydrogen H or deuterium D with quantum numbers $n$, L, and $j$ \\

$\delta_{\rmssmu}$ & Additive correction to the theoretical expression for the muon magnetic moment anomaly $a_{\rmssmu}$ \\

$\epsilon_{\rm 0}$ & Electric constant: $\epsilon_{\rm 0} = 1/\mu_{\rm 0} c^2$ \\

${\rm \lambda}({X\,{\rm K}}{\rmalpha}_1)$ & Wavelength of K${\rmalpha}_1$ x-ray line of element $X$ \\

${\rm \lambda}_{\rm meas}$ & Measured wavelength of the 2.2 MeV capture 
${\rm \gamma}$-ray emitted in the
reaction ~n + p $\rightarrow$ d + ${\rm \gamma}$ \\  

${\rmmu}$ & Symbol for either member of the muon-antimuon pair; when necessary, ${\rmmu}^{-}$ or ${\rmmu}^{+}$
is used to indicate the negative muon or positive muon \\

$\mu_{\rm B}$ & Bohr magneton: $\mu_{\rm B} = e \hbar/2m_{\rm e}$ \\

$\mu_{\rm N}$ & Nuclear magneton: $\mu_{\rm N} = e\hbar/2m_{\rm p}$ \\

$\mu_{X}(Y)$ & Magnetic moment of particle $X$ in atom or molecule $Y$. \\

$\mu_{\rm 0}$ & Magnetic constant: 
$\mu_{\rm 0} = 4{\rmpi}\times 10^{-7}$~N/A$^{2}$ \\

$\mu_{X}$,~$\mu^\prime_{X}$ & 
Magnetic moment, or shielded magnetic moment, of particle $X$ \\

$\nu$ & Degrees of freedom of a particular adjustment \\

$\nu(f_{\rm p})$ & Difference between muonium hyperfine splitting Zeeman transition frequencies
$\nu_{34}$ and $\nu_{12}$ at a magnetic flux density $B$ corresponding to the free
proton NMR frequency $f_{\rm p}$ \\

$\sigma$ & Stefan-Boltzmann constant: $\sigma = 2{\rmpi}^{5}k^{4}/(15h^{3}c^{2})$ \\

${\rmtau}$ & Symbol for either member of the tau-antitau pair; 
when necessary, ${\rmtau}^{-}$ or
${\rmtau}^{+}$ is used to indicate the negative tau or positive tau \\

${\rm \chi}^{2}$ & The statistic ``chi square'' \\

${\it \Omega}_{90}$ & Conventional unit of resistance based on the quantum Hall
effect and $R_{\rm K -90}: {\it \Omega}_{90} = (R_{\rm K}/R_{\rm K -90})~{\rm \Omega}$ \\

$\doteq$ & Symbol used to relate an input datum to its observational equation 

\end{longtable}
\setcounter{table}{0}
\section{Introduction}
\label{sec:intro}

\subsection{Background}
\label{ssec:bk}

This paper gives the complete 2006 CODATA self-consistent set of recommended
values of the fundamental physical constants and describes in detail the 2006
least-squares adjustment, including the selection of the final set of input
data based on the results of least-squares analyses.  Prepared under the
auspices of the CODATA Task Group on Fundamental Constants, this is the fifth
such report of the Task Group since its establishment in 1969 \footnote{The
Committee on Data for Science and Technology was established in 1966 as an
interdisciplinary committee of the International Council for Science.}  and the
third in the four-year cycle of reports begun in 1998.  The 2006 set of
recommended values replaces its immediate predecessor, the 2002 set.  The
closing date for the availability of the data considered for inclusion in this
adjustment was 31 December 2006.  As a consequence of the new data that became
available in the intervening four years there has been a significant reduction
of the uncertainty of many constants.  The 2006 set of recommended values first
became available on 29 March 2007 at http://physics.nist.gov/constants, a Web
site of the NIST Fundamental Constants Data Center (FCDC).

The 1998 and 2002 reports describing the 1998 and 2002 adjustments
\cite{2000035,2005191}, referred to as CODATA-98 and CODATA-02 throughout this
article, describe in detail much of the currently available data, its analysis,
and the techniques used to obtain a set of best values of the constants using
the standard method of least squares for correlated input data.  This paper
focuses mainly on the new information that has become available since 31
December 2002 and references the discussions in CODATA-98 and CODATA-02 for
earlier work in the interest of brevity.  More specifically, if a potential
input datum is not discussed in detail,  the reader can assume that it (or a
closely related datum) has been reviewed in either CODATA-98 or CODATA-02.

The reader is also referred to these papers for a discussion of the motivation
for and the philosophy behind the periodic adjustment of the values of the
constants and for descriptions of how units, quantity symbols, numerical
values, numerical calculations, and uncertainties are treated, in addition to
how the data are characterized, selected, and evaluated.  Since the
calculations are carried out with more significant figures than are displayed
in the text to avoid rounding errors, data with more digits are available on
the FCDC Web site for possible independent analysis.

However, because of their importance, we  recall in detail the following two
points also discussed in these references. First, although it is generally
agreed that the correctness and over-all consistency of the basic theories and
experimental methods of physics can be tested by comparing values of particular
fundamental constants obtained from widely differing experiments, throughout
this adjustment, as a working principle, we assume the validity of the physical
theory that necessarily underlies it. This includes special relativity, quantum
mechanics, quantum electrodynamics (QED), the Standard Model of particle
physics, including combined charge conjugation, parity inversion, and time
reversal $(CPT)$ invariance, and the theory of the Josephson and quantum Hall
effects, especially the exactness of the relationships between the Josephson
and von Klitzing constants $K_{\rm J}$ and $R_{\rm K}$ and the elementary
charge $e$ and Planck constant $h$. In fact, tests of these relations, $K_{\rm
J}=2e/h$ and $R_{\rm K}=h/e^2$, using the input data of the 2006 adjustment are
discussed in Sec.~\ref{sssec:epstests}.

The second point has to do with the 31 December 2006 closing date for data to
be considered for inclusion in the 2006 adjustment. A datum was considered to
have met this date, even though not yet reported in an archival journal, as
long as a description of the work was available that allowed the Task Group to
assign a valid standard uncertainty $u(x_i)$ to the datum.  Thus, any input
datum labeled with an ``07'' identifier because it was published in 2007 was,
in fact, available by the cutoff date.  Also, some references to results that
became available after the deadline are included, even though they were not
used in the adjustment.

\subsection{Time variation of the constants}
\label{ssec:tvc}

This subject, which was briefly touched upon in CODATA-02, continues to be an
active field of experimental and theoretical research, because of its
importance to our understanding of physics at the most fundamental level.
Indeed, a large number of papers relevant to the topic have appeared in the
last four years; see the FCDC bibliographic database on the fundamental
constants using the keyword ``time variation'' at
http://physics.nist.gov/constantsbib.  For example, see \textcite{2007029,
2007214}.  However, there has been no laboratory observation of time dependence
of any constant that might be relevant to the recommended values.

\subsection{Outline of paper}
\label{ssec:outline}

Section \ref{sec:squ} touches on special quantities and units, that is, those
that have exact values by definition.

Sections~\ref{sec:ram}-\ref{sec:xeq} review all of the available experimental
and theoretical data that might be relevant to the 2006 adjustment of the
values of the constants. As discussed in Appendix E of CODATA-98, in a least
squares analysis of the fundamental constants the numerical data, both
experimental and theoretical, also called \textit{observational data} or
\textit{input data}, are expressed as functions of a set of independent
variables called \textit{adjusted constants}.  The functions that relate the
input data to the adjusted constants are called \textit{observational
equations}, and the least squares procedure provides best estimated values, in
the least squares sense, of the adjusted constants. The focus of the
review-of-data sections is thus the identification and discussion of the input
data and observational equations of interest for the 2006 adjustment.  Although
not all observational equations that we use are explicitly given in the text,
all are summarized in Tables~\ref{tab:pobseqsa}, \ref{tab:pobseqsb1}, and
\ref{tab:pobseqsc} of Sec.~\ref{ssec:mada}.

As part of our discussion of a particular datum, we often deduce from it an
inferred value of a constant, such as the fine-structure constant $\alpha$ or
Planck constant $h$. It should be understood, however, that these inferred
values are for comparison purposes only; the datum from which it is obtained,
not the inferred value, is the input datum in the adjustment.

Although just 4 years separate the 31 December closing dates of the 2002 and
2006 adjustments, there are a number of important new results to consider.
Experimental advances include the 2003 Atomic Mass Evaluation from the Atomic
Mass Data Center (AMDC) that provides new values for the relative atomic masses
$A_{\rm r}(X)$ of a number of relevant atoms; a new value of the electron
magnetic moment anomaly $a_{\rm e}$ from measurements on a single electron in a
cylindrical penning trap that provides a value of the fine-structure constant
$\alpha$; better measurements of the relative atomic masses of $^{2}{\rm H}$,
$^{3}{\rm H}$, and $^{4}{\rm He}$; new measurements of transition frequencies
in antiprotonic helium ($\bar{\rm p}\,^A{\rm He^+}$ atom) that provide a
competitive value of the relative atomic mass of the electron $A_{\rm r}(\rm
e)$; improved measurements of the nuclear magnetic resonance (NMR) frequencies
of the proton and deuteron in the HD molecule and of the proton and triton in
the HT molecule; a highly accurate value of the Planck constant obtained from
an improved measurement of the product $K^2_{\rm J}R_{\rm K}$ using a
moving-coil watt balance; new results using combined x-ray and optical
interferometers for the \{220\} lattice spacing of single crystals of naturally
occurring silicon; and an accurate value of the quotient $h/m(^{87}{\rm Rb})$
obtained by measuring the recoil velocity of rubidium-87 atoms upon absorption
or emission of photons---a result that provides an accurate value of $\alpha$
that is virtually independent of the electron magnetic moment anomaly.

Theoretical advances include improvements in certain aspects of the theory of
the energy levels of hydrogen and deuterium; improvements in the theory of
antiprotonic helium transition frequencies that, together with the new
transition frequency measurements, have led to the aforementioned competitive
value of $A_{\rm r}(\rm e)$; a new theoretical expression for $a_{\rm e}$ that,
together with the new experimental value of $a_{\rm e}$, has led to the
aforementioned value of $\alpha$; improvements in the theory of the $g$-factor
of the bound electron in hydrogenic ions with nuclear spin quantum number $i =
0$ relevant to the determination of $A_{\rm r} (\rm e)$; and improved theory of
the ground state hyperfine splitting of muonium $\Delta\nu_{\rm Mu}$ (the
$\rmmu^+{\rm e}^-$ atom).

Section~\ref{sec:ad} describes the analysis of the data, with the exception of
the Newtonian constant of gravitation which is analyzed in Sec.~\ref{sec:ncg}.
The consistency of the data and potential impact on the determination of the
2006 recommended values were appraised by comparing measured values of the same
quantity, comparing measured values of different quantities through inferred
values of a third quantity such as $\alpha$ or $h$, and finally by using the
method of least squares.  Based on these investigations, the final set of input
data used in the 2006 adjustment was selected.

Section~\ref{sec:2006crv} provides, in several tables, the 2006 CODATA
recommended values of the basic constants and conversion factors of physics and
chemistry, including the covariance matrix of a selected group of constants.

Section~\ref{sec:c} concludes the paper with a comparison of the 2006 and 2002
recommended values of the constants, a survey of implications for physics and
metrology of the 2006 values and adjustment, and suggestions for future work
that can advance our knowledge of the values of the constants.

\section{Special quantities and units}
\label{sec:squ}

Table~\ref{tab:exact} lists those special quantities whose numerical values are
exactly defined.   In the International System of Units (SI) \cite{2006107},
which we use throughout this paper, the definition of the meter fixes the speed
of light in vacuum $c$, the definition of the ampere fixes the magnetic
constant (also called the permeability of vacuum) $\mu_0$, and the definition
of the mole fixes the molar mass of the carbon 12 atom $M(^{12}{\rm C})$ to
have the exact values given in the table.  Since the electric constant (also
called the permittivity of vacuum) is related to $\mu_0$ by $\epsilon_0 =
1/\mu_0c^2$, it too is known exactly.
     
The relative atomic mass $A_{\rm r}(X)$ of an entity $X$ is defined by $A_{\rm
r}(X) = m(X)/ m_{\rm u}$, where $m(X)$ is the mass of $X$ and $m_{\rm u}$ is
the atomic mass constant defined by 
\begin{eqnarray}
m_{\rm u} =  \frac{1}{12}m(^{12}{\rm C}) = 1~{\rm u}
\approx 1.66\times 10^{-27} \mbox{ kg},
\end{eqnarray}
where $m(^{12}{\rm C})$  is the mass of the carbon 12 atom and u is the unified
atomic mass unit (also called the dalton, Da).  Clearly, $A_{\rm r}(X)$ is a
dimensionless quantity and $A_{\rm r}(^{12}{\rm C}) = 12$ exactly.  The molar
mass $M(X)$ of entity $X$, which is the mass of one mole of $X$ with SI unit
kg/mol, is given by 
\begin{eqnarray} 
M(X) = N_{\rm A}m(X) = A_{\rm r}(X)M_{\rm u},
\end{eqnarray} 
where $N_{\rm A} \approx 6.02 \times10^{23}$/mol is the Avogadro constant and
$M_{\rm u} = 10^{-3}$ kg/mol is the molar mass constant.  The numerical value
of $N_{\rm A}$ is the number of entities in one mole, and since the definition
of the mole states that one mole contains the same number of entities as there
are in 0.012 kg of carbon 12, $M(^{12}{\rm C}) = 0.012$ kg/mol exactly.
     
The Josephson and quantum Hall effects have played and continue to play
important roles in adjustments of the values of the constants, because the
Josephson and von Klitzing constants $K_{\rm J}$ and $R_{\rm K}$, which
underlie these two effects, are related to $e$ and $h$ by 
\begin{eqnarray}
K_{\rm J} = {2e\over h}\,;\qquad R_{\rm K} = \frac{h}{e^2} = \frac{\mu_0c}{2\alpha}.
\end{eqnarray}
Although we assume these relations are exact, and no evidence---either
theoretical or experimental---has been put forward that challenges this
assumption, the consequences of relaxing it are explored in
Sec.~\ref{sssec:epstests}.  Some references to recent work related to the
Josephson and quantum Hall effects may be found in the FCDC bibliographic
database (see Sec.~\ref{ssec:tvc}).

\def\vsp#1{\noalign{\vbox to #1 pt {}}}
\def\hsp#1{\hbox to #1 pt {}}
\begin{table*}
\caption{Some exact quantities relevant to the 2006 adjustment.}
\label{tab:exact}
\begin{tabular}{lll}
\toprule
\vbox to 10 pt {}
Quantity & Symbol  \hbox to 2cm{}
& Value \\
\colrule
\vsp{3}
speed of light in vacuum
 &$c$, $c_0$                
 &$299\,792\,458 \ {\rm m \  s^{-1}}$ \\
magnetic constant
 &$\mu_0$                   
 & $4\rmpi\times10^{-7} \  {\rm N \  A^{-2}}$
 $= 12.566\,370\,614... \times 10 ^{-7} \ {\rm N \  A^{-2}}$ \\
electric constant
 &$\epsilon_0$              
 &$(\mu_0c^2)^{-1} $
 $=8.854\,187\,817... \ \times10^{-12} \ {\rm F \  m^{-1}}$ \\
relative atomic mass of $^{12}$C
 &$A_{\rm r}(^{12}{\rm C})$         
 & $12$ \\
molar mass constant
 &$M_{\rm u}$         
 & $10^{-3} \ {\rm kg \  mol^{-1}}$ \\
molar mass of $^{12}$C~ $A_{\rm r}(^{12}{\rm C})\,M_{\rm u}$
 &$M(^{12}{\rm C})$         
 & $12\times10^{-3} \ {\rm kg \  mol^{-1}}$ \\
conventional value of Josephson constant
 &$K_{{\rm J}-90}$          
 & $483\,597.9 \ {\rm GHz \  V^{-1}}$ \\
conventional value of von Klitzing constant  \hbox to 1 cm {}
 &$R_{{\rm K}-90}$          
 & $25\,812.807 \ {\rm \Omega}$ \\
\botrule
\end{tabular}
\end{table*} 
The next-to-last two entries in Table~\ref{tab:exact} are the conventional
values of the Josephson and von Klitzing constants adopted by the International
Committee for Weights and Measures (CIPM) and introduced on 1 January 1990 to
establish worldwide uniformity in the measurement of electrical quantities.  In
this paper, all electrical quantities are expressed in SI units.  However,
those measured in terms of the Josephson and quantum Hall effects with the
assumption that $K_{\rm J}$ and $R_{\rm K}$ have these conventional values are
labeled with a subscript 90.

For high-accuracy experiments involving the force of gravity, such as the
watt-balance, an accurate measurement of the local acceleration of free fall at
the site of the experiment is required.  Fortunately, portable and easy-to-use
commercial absolute gravimeters are available that can provide a local value of
$g$ with a relative standard uncertainty of a few parts in $10^9$.  That these
instruments can achieve such a small uncertainty if properly used is
demonstrated by a periodic international comparison of absolute gravimeters
(ICAG) carried out at the International Bureau of Weights and Measures (BIPM),
S\`evres, France; the seventh and most recent, denoted ICAG-2005, was completed
in September 2005 \cite{tbp05vea}; the next is scheduled for 2009.  In the
future, atom interferometry or Bloch oscillations using ultracold atoms could
provide a competitive or possibly more accurate method for determining a local
value of $g$ \cite{2001070,2002010,2005150}.

\section{Relative atomic masses}
\label{sec:ram}

Included in the set of adjusted constants are the relative atomic masses
$A_{\rm r}(X)$ of a number of particles, atoms, and ions.
Tables~\ref{tab:rmass03}-\ref{tab:rmcovvd} and the following sections summarize
the relevant data.

\subsection{Relative atomic masses of atoms}
\label{ssec:rama}

Most values of the relative atomic masses of neutral atoms used in this
adjustment are taken from the 2003 atomic mass evaluation (AME2003) of the
Atomic Mass Data Center, Centre de Spectrom\'etrie Nucl\'eaire et de
Spectrom\'etrie de Masse (CSNSM), Orsay, France \cite{2003252,2003253,amdc06}.
The results of AME2003 supersede those of both the 1993 atomic mass evaluation
and the 1995 update.  Table~\ref{tab:rmass03} lists the values from AME2003 of
interest here, while Table~\ref{tab:rmcov} gives the covariance for hydrogen
and deuterium \cite{awcovs03}.  Other non-negligible covariances of these
values are discussed in the appropriate sections.

Table~\ref{tab:rmass06} gives six values of $A_{\rm r}(X)$ relevant to the 2006
adjustment reported since the completion and publication of AME2003 in late
2003 that we use in place of the corresponding values in
Table~\ref{tab:rmass03}.

The $^3$H and $^3$He values are those reported by the SMILETRAP group at the
Manne Siegbahn Laboratory (MSL), Stockholm, Sweden \cite{2006032}, using a
Penning trap and a time of flight technique to detect cyclotron resonances.
This new $^3$He result is in good agreement with a more accurate, but still
preliminary, result from the University of Washington group in Seattle, USA
\cite{pcvd2006}.  The AME2003  values for $^3$H and $^3$He were influenced by
an earlier result for $^3$He from the University of Washington group which is
in disagreement with their new result.

The values for $^4$He and $^{16}$O are those reported by the University of
Washington group \cite{2006036} using their improved mass spectrometer; they
are based on a thorough reanalysis of  data that yielded preliminary results
for these atoms which were used in AME2003.  They include an experimentally
determined image-charge correction with a relative standard uncertainty $u_{\rm
r}=7.9 \times 10^{-12}$ in the case of $^4 {\rm He}$ and $u_{\rm r}=4.0 \times
10^{-12}$ in the case of $^{16} {\rm O}$.  The value of $A_{\rm r}(^2 {\rm H})$
is also from this group and is a near-final result based on the analysis of ten
runs carried out over a 4 year period \cite{pcvd2006}. Because the result is
not yet final, the total uncertainty is conservatively assigned; $u_{\rm r}=9.9
\times 10^{-12}$ for the image-charge correction. This value of $A_{\rm r}(^2
{\rm H})$ is consistent with the preliminary value reported by
\textcite{2006036} based on the analysis of only three runs.

The covariance and correlation coefficient of $A_{\rm r}(^3 {\rm H})$ and
$A_{\rm r}(^3 {\rm He})$ given in Table~\ref{tab:rmcovsm} are due to the common
component of uncertainty $u_{\rm r}=1.4 \times 10^{-10}$ of the relative atomic
mass of the ${\rm H}_2 ^{+}$ reference ion used in the SMILETRAP measurements;
the covariances and correlation coefficients of the University of  Washington
values of $A_{\rm r}(^2 {\rm H})$, $A_{\rm r}(^4 {\rm He})$, and $A_{\rm
r}(^{16} {\rm O})$ given in Table~\ref{tab:rmcovvd} are due to the
uncertainties of the image-charge corrections, which are based on the same
experimentally determined relation.

The $^{29}$Si value is that implied by the ratio $A_{\rm
r}(^{29}$Si$^+$)/$A_{\rm r}(^{28}$Si H$^+$)$ =  0.999\,715\,124\,1812(65)$
obtained at the Massachusetts Institute of Technology (MIT), Cambridge, USA,
using a recently developed technique of determining mass ratios by directly
comparing the cyclotron frequencies of two different ions simultaneously
confined in a Penning trap \cite{2005195}.  (The relative atomic mass work of
the MIT group has now been transferred to Florida State University,
Tallahassee, USA.) This approach eliminates many components of uncertainty
arising from systematic effects.  The value for $A_{\rm r}(^{29}$Si) is given
in the Supplementary Information to \textcite{2005195} and has a significantly
smaller uncertainty than the corresponding AME2003 value.

\subsection{Relative atomic masses of ions and nuclei}
\label{ssec:ramnuc}

The relative atomic mass $A_{\rm r}(X)$ of a neutral atom $X$ is given in terms
of the relative atomic mass of an ion of the atom formed by the removal of $n$
electrons by
\begin{eqnarray}
A_{\rm \rm r} (X)
&=&A_{\rm r}(X^{n+}) + n A_{\rm r} ({\rm e})
\nonumber\\ &&
 - \frac{E_{\rm b}(X) 
 - E_{\rm b}(X^{n+})}{m_{\rm u} c^2} \ .
\label{eq:araxn}
\end{eqnarray}
Here $E_{\rm b}(X)/m_{\rm u}c^{2}$ is the relative-atomic-mass equivalent of
the total binding energy of the $Z$ electrons of the atom, where $Z$ is the
atomic number (proton number), and $E_{\rm b}(X^{n+})/m_{\rm u}c^{2}$ is the
relative-atomic-mass-equivalent of the binding energy of the $Z-n$ electrons of
the $X^{n+}$ ion.  For a fully stripped atom, that is, for $n=Z$, $X^{Z+}$ is
$N$, where $N$ represents the nucleus of the atom, and $E_{\rm
b}(X^{Z+})/m_{\rm u}c^{2}$ = 0, which yields the first few equations of
Table~\ref{tab:pobseqsb1} in Sec.~\ref{ssec:mada}.

The binding energies $E_{\rm b}$ used in this work are the same as those used
in the 2002 adjustment; see Table~IV of CODATA-02.  For tritium, which is not
included there, we use the value $1.097\,185\,439\times10^7$ m$^{-1}$
\cite{pcsk2006}.  The uncertainties of the binding energies are negligible for
our application.

\subsection{Cyclotron resonance measurement of the electron relative atomic
mass $\bm{A_{\rm r}({\rm e})}$} 
\label{ssec:ptmare} 

A value of $A_{\rm r}$(e) is available from a Penning-trap measurement carried
out by the University of Washington group \cite{1995160}; it is used as an
input datum in the 2006 adjustment, as it was in the 2002 adjustment: 
\begin{eqnarray}
A_{\rm r}({\rm e}) =  0.000\,548\,579\,9111(12) \qquad [ 2.1\times 10^{-9}] \ .
\label{eq:areexp} 
\end{eqnarray}

\begin{table}[t]
\caption{Values of the relative atomic masses of the neutron and various atoms
as given in the 2003 atomic mass evaluation together with the defined value for
$^{12}$C.}
\label{tab:rmass03}
\begin{tabular}{cD{.}{.}{8.20}l}
\toprule
\vbox to 10pt{}
Atom     &  \multicolumn{1}{c}{Relative atomic}
& Relative standard \\
& \multicolumn{1}{c}{\text{mass $A_{\rm r}({\rm X})$}}  
& \hskip 8pt uncertainty $u_{\rm r}$  \\
\colrule
n \vbox to 10pt{} &   1.008\,664\,915\,74(56) & \hskip 16pt $ 5.6\times 10^{-10}$ \\
$^{1}$H     &   1.007\,825\,032\,07(10) & \hskip 16pt $ 1.0\times 10^{-10}$ \\
$^{2}$H     &    2.014\,101\,777\,85(36)    & \hskip 16pt $ 1.8\times 10^{-10}$    \\
$^{3}$H     &    3.016\,049\,2777(25)   & \hskip 16pt $ 8.2\times 10^{-10}$    \\
$^{3}$He    &    3.016\,029\,3191(26)   & \hskip 16pt $ 8.6\times 10^{-10}$   \\
$^{4}$He    &    4.002\,603\,254\,153(63)   & \hskip 16pt $ 1.6\times 10^{-11}$   \\
$^{12}$C    &   12             & \hskip 16pt    (exact)          \\
$^{16}$O    &    15.994\,914\,619\,56(16)   & \hskip 16pt $ 1.0\times 10^{-11}$   \\
$^{28}$Si   &    27.976\,926\,5325(19)  & \hskip 16pt $ 6.9\times 10^{-11}$  \\
$^{29}$Si   &    28.976\,494\,700(22)  & \hskip 16pt $ 7.6\times 10^{-10}$  \\
$^{30}$Si   &    29.973\,770\,171(32)  & \hskip 16pt $ 1.1\times 10^{-9}$  \\
$^{36}$Ar   &    35.967\,545\,105(28)  & \hskip 16pt $ 7.8\times 10^{-10}$  \\
$^{38}$Ar   &    37.962\,732\,39(36)  & \hskip 16pt $ 9.5\times 10^{-9}$  \\
$^{40}$Ar   &    39.962\,383\,1225(29)  & \hskip 16pt $ 7.2\times 10^{-11}$  \\
$^{87}$Rb   &    86.909\,180\,526(12)  & \hskip 16pt $ 1.4\times 10^{-10}$  \\
$^{107}$Ag  &    106.905\,0968(46) & \hskip 16pt $ 4.3\times 10^{-8}$ \\
$^{109}$Ag  &    108.904\,7523(31) & \hskip 16pt $ 2.9\times 10^{-8}$ \\
$^{133}$Cs  &    132.905\,451\,932(24) & \hskip 16pt $ 1.8\times 10^{-10}$ \\
\botrule
\end{tabular}
\end{table}

\begin{table}
\caption{The variances, covariance, and correlation coefficient of the AME2003
values of the relative atomic masses of hydrogen and deuterium.  The number in
bold above the main diagonal is $10^{18}$ times the numerical value of the
covariance; the numbers in bold on the main diagonal are $10^{18}$ times the
numerical values of the variances; and the number in italics below the main
diagonal is the correlation coefficient.}
\label{tab:rmcov}
\begin{tabular} {l|rr}
\toprule
\vbox to 10 pt {}
& $A_{\rm r}(^1{\rm H})$ \hbox to 0.5 pt{} 
& $A_{\rm r}(^2{\rm H})$ \hbox to 0.5 pt{}  \\
\colrule
\vbox to 10 pt {}
$A_{\rm r}(^1{\rm H})$ & ${\bf  0.0107}$ & ${\bf  0.0027}$ \\
$A_{\rm r}(^2{\rm H})$  & ${\it  0.0735}$ & ${\bf  0.1272}$ \\
\botrule
\end{tabular}
\end{table}

\begin{table}[t]
\caption{Values of the relative atomic masses of various atoms that have become
available since the 2003 atomic mass evaluation.}
\label{tab:rmass06}
\begin{tabular}{cD{.}{.}{8.20}l}
\toprule
\vbox to 10pt{}
Atom     &   \multicolumn{1}{c}{Relative atomic}
& Relative standard \\
& \multicolumn{1}{c}{mass $A_{\rm r}(X)$}  & \hskip 8pt uncertainty $u_{\rm r}$  \\
\colrule
$^{2}$H    &    2.014\,101\,778\,040(80)    & \hskip 16pt $ 4.0\times 10^{-11}$  \vbox to 10pt{} \\
$^{3}$H    &    3.016\,049\,2787(25)    & \hskip 16pt $ 8.3\times 10^{-10}$   \\
$^{3}$He   &    3.016\,029\,3217(26)    & \hskip 16pt $ 8.6\times 10^{-10}$   \\
$^{4}$He   &    4.002\,603\,254\,131(62)    & \hskip 16pt $ 1.5\times 10^{-11}$   \\
$^{16}$O   &   15.994\,914\,619\,57(18)    & \hskip 16pt $ 1.1\times 10^{-11}$   \\
$^{29}$Si  &   28.976\,494\,6625(20)   & \hskip 16pt $ 6.9\times 10^{-11}$   \\
\botrule
\end{tabular}
\end{table}

\begin{table}
\caption{The variances, covariance, and correlation coefficient of the values
of the SMILETRAP relative atomic masses of tritium and helium three.  The
number in bold above the main diagonal is $10^{18}$ times the numerical value
of the covariance; the numbers in bold on the main diagonal are $10^{18}$ times
the numerical values of the variances; and the number in italics below the main
diagonal is the correlation coefficient.}
\label{tab:rmcovsm}
\begin{tabular} {l|rr}
\toprule
\vbox to 10 pt {}
& $A_{\rm r}({\rm ^3H})$ \hbox to 0.5 pt{} 
& $A_{\rm r}({\rm ^3He})$ \hbox to 0.5 pt{}  \\
\colrule
\vbox to 10 pt {}
$A_{\rm r}({\rm ^3H})$ & ${\bf  6.2500}$ & ${\bf  0.1783}$ \\
$A_{\rm r}({\rm ^3He})$  & ${\it  0.0274}$ & ${\bf  6.7600}$ \\
\botrule
\end{tabular}
\end{table}

\begin{table}
\caption{The variances, covariances, and correlation coefficients of the
University of Washington values of the relative atomic masses of deuterium,
helium 4, and oxygen 16.  The numbers in bold above the main diagonal are
$10^{20}$ times the numerical values of the covariances; the numbers in bold on
the main diagonal are $10^{20}$ times the numerical values of the variances;
and the numbers in italics below the main diagonal are the correlation
coefficients.}
\label{tab:rmcovvd}
\begin{tabular} {l|rrr}
\toprule
\vbox to 10 pt {}
& $A_{\rm r}(^2{\rm H})$ \hbox to 0.5pt{} 
& $A_{\rm r}(^4{\rm He})$ \hbox to 0.5pt{} 
& $A_{\rm r}(^{16}{\rm O})$ \hbox to 0.5pt{}  \\
\colrule
\vbox to 10 pt {}
$A_{\rm r}(^2{\rm H})$ & ${\bf  0.6400}$ & ${\bf  0.0631}$ & ${\bf  0.1276}$ \\
$A_{\rm r}(^4{\rm He})$  & ${\it  0.1271}$ & ${\bf  0.3844}$ & ${\bf  0.2023}$ \\
$A_{\rm r}(^{16}{\rm O})$  & ${\it  0.0886}$ & ${\it  0.1813}$ & ${\bf  3.2400}$ \\
\botrule
\end{tabular}
\end{table}

\section{Atomic transition frequencies} 
\label{sec:tf}

Atomic transition frequencies in hydrogen, deuterium, and anti-protonic helium
yield information on the Rydberg constant, the proton and deuteron charge
radii, and the relative atomic mass of the electron.  The hyperfine splitting
in hydrogen and fine-structure splitting in helium do not yield a competitive
value of any constant at the current level of accuracy of the relevant
experiment and/or theory.  All of these topics are discussed in this section.  

\subsection{Hydrogen and deuterium transition frequencies, the Rydberg constant
$\bm{R_\infty}$, and the proton and deuteron charge radii $\bm{R_{\rm p},
R_{\rm d}}$}
\label{ssec:ryd}

The Rydberg constant is related to other constants by the definition
\begin{eqnarray}
R_\infty = \alpha^2{m_{\rm e}c \over 2h} \ .
\end{eqnarray}
It can be accurately determined by comparing measured resonant frequencies of
transitions in hydrogen (H) and deuterium (D) to the theoretical expressions
for the energy level differences in which it is a multiplicative factor.

\subsubsection{Theory relevant to the Rydberg constant}
\label{sssec:rydth}

The theory of the energy levels of hydrogen and deuterium atoms relevant to the
determination of the Rydberg constant $R_\infty$, based on measurements of
transition frequencies, is summarized in this section.  Complete information
necessary to determine the theoretical values of the relevant energy levels is
provided, with an emphasis on results that have become available since the
previous adjustment described in CODATA-02.  For brevity, references to earlier
work, which can be found in \textcite{2001057}, for example, are not included
here.

An important consideration is that the theoretical values of the energy levels
of different states are highly correlated.  For example, for S states, the
uncalculated terms are primarily of the form of an unknown common constant
divided by $n^3$.  This fact is taken into account by calculating covariances
between energy levels in addition to the uncertainties of the individual levels
as discussed in detail in Sec.~\ref{par:teu}.  In order to take these
correlations into account, we distinguish between components of uncertainty
that are proportional to $1/n^3$, denoted by $u_0$, and components of
uncertainty that are essentially random functions of $n$, denoted by $u_n$.

The energy levels of hydrogen-like atoms are determined mainly by the Dirac
eigenvalue, QED effects such as self energy and vacuum polarization, and
nuclear size and motion effects, all of which are summarized in the following
sections.

\paragraph{Dirac eigenvalue}

The binding energy of an electron in a static Coulomb field (the external
electric field of a point nucleus of charge $Ze$ with infinite mass) is
determined predominantly by the Dirac eigenvalue 
\begin{eqnarray} 
E_{\rm D} = f(n,j)\, m_{\rm e}c^2 \ , 
\label{eq:diracen}
\end{eqnarray} 
where
\begin{eqnarray} 
f(n,j) = \left[ 1+{(Z\alpha)^2\over (n-\delta)^2} \right]^{-1/2} \ ,
\label{eq:diracev}
\end{eqnarray} 
$n$ and $j$ are the principal quantum number and total angular momentum of the
state, respectively, and
\begin{eqnarray} 
\delta = j+\fr{1}{2}-\left[(j+\fr{1}{2})^2-(Z\alpha)^2\right]^{1/2} \ . 
\label{eq:delta}
\end{eqnarray} 
Although we are interested only in the case where the nuclear charge is $e$, we
retain the atomic number $Z$ in order to indicate the nature of various terms.

Corrections to the Dirac eigenvalue that approximately take into account the
finite mass of the nucleus $m_{\rm N}$ are included in the more general
expression for atomic energy levels, which replaces Eq.~(\ref{eq:diracen})
\cite{1955001,1990020}:
\begin{eqnarray} 
E_M &=& Mc^2 +[f(n,j)-1]m_{\rm r}c^2 -[f(n,j)-1]^2{m_{\rm r}^2c^2\over 
2M} \nonumber\\ 
&&+\, {1-\delta_{l0}\over \kappa(2l+1)} {(Z\alpha)^4m_{\rm r}^3c^2\over 
2 n^3 m_{\rm N}^2} +\cdots \ , 
\label{eq:relred}
\end{eqnarray} 
where $l$ is the nonrelativistic orbital angular momentum quantum number,
$\kappa$ is the angular-momentum-parity quantum number $\kappa =
(-1)^{j-l+1/2}(j+\fr{1}{2})$, $M = m_{\rm e} + m_{\rm N}$, and $m_{\rm r} =
m_{\rm e}m_{\rm N}/(m_{\rm e}+m_{\rm N})$ is the reduced mass. 

\paragraph{Relativistic recoil}

Relativistic corrections to Eq.~(\ref{eq:relred}) associated with motion of the
nucleus are considered relativistic-recoil corrections. The leading term, to
lowest order in $Z\alpha$ and all orders in $m_{\rm e}/m_{\rm N}$, is
\cite{1977002,1990020}
\begin{eqnarray} 
E_{\rm S} &=& {m_{\rm r}^3\over m_{\rm e}^2 m_{\rm N}}{(Z\alpha)^5\over \rmpi n^3} 
 m_{\rm e} c^2 \nonumber\\ 
&&\times \bigg\{\fr{1}{3}\delta_{l0}\ln (Z\alpha)^{-2} -\fr{8}{3}\ln 
k_0(n,l) -\fr{1}{9}\delta_{l0}-\fr{7}{3}a_n \nonumber\\ 
&&- \, {2\over m_{\rm N}^2-m_{\rm e}^2}\delta_{l0} \left[m_{\rm N}^2\ln 
\Big({m_{\rm e}\over m_{\rm r}}\Big) - m_{\rm e}^2\ln\Big({m_{\rm N}\over m_{\rm r}} 
\Big)\right]\bigg\}, \nonumber\\ 
\label{eq:salp}
\end{eqnarray} 
where
\begin{eqnarray} 
a_n &=& -2\left[\ln\Big({2\over n}\Big) + \sum_{i=1}^n{1\over i} +1-{1 
\over2n}\right]\delta_{l0} \nonumber\\ 
&&+ \, {1-\delta_{l0}\over l(l+1)(2l+1)} \,. 
\end{eqnarray} 

To lowest order in the mass ratio, higher-order corrections in $Z\alpha$ have
been extensively investigated; the contribution of the next two orders in
$Z\alpha$ is
\begin{eqnarray}
E_{\rm R} &=& {m_{\rm e}\over m_{\rm N}}
{(Z\alpha)^6\over n^3}m_{\rm e}c^2 
\nonumber\\ && \times
\left[D_{60}  + D_{72}Z\alpha \ln^2{(Z\alpha)^{-2}} + \cdots \right]
\ ,
\label{eq:err60etc}
\end{eqnarray}
where for $n{\rm S}_{1/2}$ states \cite{1995071,1997084}
\begin{eqnarray}
D_{60} &=& 
4\ln 2 -{7\over 2}  
\label{eq:errd60s}
\end{eqnarray}
and \cite{1999164,1999182}
\begin{eqnarray}
D_{72} &=& -{11\over 60 \rmpi} \ ,
\label{eq:err72}
\end{eqnarray}
and for states with $l \ge 1$ \cite{1995070,1996171,1996109}
\begin{eqnarray}
D_{60} &=& 
\left[3-{l(l+1)\over n^2}\right]
{2\over(4l^2-1)(2l+3)}   \ .
\label{eq:errlg1}
\end{eqnarray}
In Eq.~(\ref{eq:errlg1}) and subsequent discussion, the first subscript on the
coefficient of a term refers to the power of $Z\alpha$ and the second subscript
to the power of $\ln (Z\alpha)^{-2}$.  The relativistic recoil correction used
in the 2006 adjustment is based on Eqs.~(\ref{eq:salp}) to (\ref{eq:errlg1}).
The estimated uncertainty for S states is taken to be 10~\% of
Eq.~(\ref{eq:err60etc}), and for states with $l\ge 1$, it is taken to be 1~\%
of that equation.

Numerical values for the complete contribution of Eq.~(\ref{eq:err60etc}) to
all orders in $Z\alpha$ have been obtained by \cite{1998031}.  Although the
difference between the all-orders calculation and the truncated power series
for S states is about three times their quoted uncertainty, the two results are
consistent within the uncertainty assigned here.  The covariances of the
theoretical values are calculated by assuming that the uncertainties are
predominately due to uncalculated terms proportional to $(m_{\rm e}/m_{\rm
N})/n^3$.

\paragraph{Nuclear polarization}
\label{par:nucpol} 

Interactions between the atomic electron and the nucleus which involve excited
states of the nucleus give rise to nuclear polarization corrections.  For
hydrogen, we use the result \cite{2000054}
\begin{eqnarray}
E_{\rm P}({\rm H}) &=&  -0.070(13) h {\delta_{l0}\over n^3} \ {\rm kHz}
\ .
\label{eq:hnucpol}
\end{eqnarray}
For deuterium, the sum of the proton polarizability, the neutron polarizability
\cite{1998110}, and the dominant nuclear structure polarizability
\cite{1997130}, gives
\begin{eqnarray}
E_{\rm P}({\rm D}) &=&  -21.37(8) h{\delta_{l0}\over n^3} \ {\rm kHz} \ .
\label{eq:dnucpol}
\end{eqnarray}
We assume that this effect is negligible in states of higher $l$.

\paragraph{Self energy}
\label{par:selfen} 

The one-photon electron self energy is given by
\begin{eqnarray}
E_{\rm SE}^{(2)} = {\alpha\over\rmpi}{(Z\alpha)^4\over n^3} F(Z\alpha) \, m_{\rm e} c^2 \ ,
\label{eq:selfen}
\end{eqnarray}
where
\begin{eqnarray}
F(Z\alpha) &=& A_{41}\ln(Z\alpha)^{-2}+A_{40}+A_{50}\,(Z\alpha)
\nonumber\\
&&+A_{62}\,(Z\alpha)^2\ln^2(Z\alpha)^{-2}
+A_{61}\,(Z\alpha)^2\ln(Z\alpha)^{-2}
\nonumber\\
&&+G_{\rm SE}(Z\alpha)\,(Z\alpha)^2 \ .
\label{eq:sepow}
\end{eqnarray}
From \textcite{1965012} and earlier papers cited therein,
\begin{eqnarray}
A_{41} &=& \fr{4}{3} \, \delta_{l0} 
\nonumber\\
A_{40} &=& 
-\fr{4}{3}\ln k_0(n,l)
+\fr{10}{9} \, \delta_{l0}-{1\over 2\kappa(2l+1)}(1-\delta_{l0})
\nonumber\\
A_{50} &=& \left(\fr{139}{32} - 2 \ln 2 \right) \rmpi \, \delta_{l0}
\label{eq:sepows}
\\
A_{62} &=& - \delta_{l0}
\nonumber\\
A_{61} &=& \left[4\left(1 + \frac{1}{2} + \cdots + \frac{1}{n}\right)
+{28\over3}\ln{2}-4\ln{n}\right.
\nonumber\\
&-&\left.{601\over180} - {77\over 45n^2}\right]\delta_{l0}
+\left(1-{1\over n^2}\right)
\left({2\over15}+{1\over3}\delta_{j{1\over2}}\right)\delta_{l1}
\nonumber\\
&+&{96n^2-32l(l+1)\over 3 n^2(2l-1)(2l)(2l+1)(2l+2)(2l+3)}(1-\delta_{l0})\ .
\nonumber
\end{eqnarray}
The Bethe logarithms $\ln k_0(n,l)$ in Eq.~(\ref{eq:sepows}) are given in
Table~\ref{tab:bethe} \cite{1990002}.

\begin{table}
\def\sb{\hbox to 7mm{}} 
\caption{Bethe logarithms $\ln k_0(n,l)$ relevant to the determination of
$R_\infty$.}
\label{tab:bethe} 
\begin{tabular}{c@{\sb}c@{\sb}c@{\sb}c} 
\toprule
\vbox to 10 pt {}
$n$ & S & P & D \\
\colrule
1 & $ 2.984\,128\,556$ & \vbox to 10pt{} & \\
2 & $ 2.811\,769\,893$ & $ -0.030\,016\,709$ & \\
3 & $ 2.767\,663\,612$ &                  &                  \\
4 & $ 2.749\,811\,840$ & $ -0.041\,954\,895$ & $ -0.006\,740\,939$ \\
6 & $ 2.735\,664\,207$ &                  & $ -0.008\,147\,204$ \\
8 & $ 2.730\,267\,261$ &                  & $ -0.008\,785\,043$ \\
12&                    &                  & $ -0.009\,342\,954$ \\
\botrule
\end{tabular} 
\end{table} 

The function $G_{\rm SE}(Z\alpha)$ in Eq.~(\ref{eq:sepow}) is the higher-order
contribution (in $Z\alpha$) to the self energy, and the values for $G_{\rm
SE}(\alpha)$ that we use here are listed in Table~\ref{tab:gse}.  For S and P
states with $n \le 4$ the values in the table are based on direct numerical
evaluations by \textcite{1999001,2001072,2004090,2005127}.  The values of
$G_{\rm SE}(\alpha)$ for the 6S and 8S states are based on the low-$Z$ limit of
this function $G_{\rm SE}(0)=A_{60}$ \cite{2005212} together with
extrapolations of the results of complete numerical calculations of
$F(Z\alpha)$ [see Eq.~(\ref{eq:sepow})] at higher $Z$ \cite{pc06skpm}.  The
values of $G_{\rm SE}(\alpha)$ for D states are from \textcite{hdelse2005}

The dominant effect of the finite mass of the nucleus on the self energy
correction is taken into account by multiplying each term of $F(Z\alpha)$ by
the reduced-mass factor $(m_{\rm r}/m_{\rm e})^3$, except that the magnetic
moment term $-1/[2\kappa(2l+1)]$ in $A_{40}$ is instead multiplied by the
factor $(m_{\rm r}/m_{\rm e})^2$.  In addition, the argument $(Z\alpha)^{-2}$
of the logarithms is replaced by $(m_{\rm e}/m_{\rm r})(Z\alpha)^{-2}$
\cite{1990020}.

The uncertainty of the self energy contribution to a given level arises
entirely from the uncertainty of $G_{\rm SE}(\alpha)$ listed in
Table~\ref{tab:gse} and is taken to be entirely of type $u_n$.

\begin{table*}
\def\sb{\hbox to 5mm{}}
\caption{Values of the function $G_{\rm SE}(\alpha)$.}
\label{tab:gse}
\begin{tabular}{r @{\sb} l @{\sb} l @{\sb} l @{\sb} l @{\sb} l}
\toprule
\vbox to 10 pt {}
$n$& \ \ \ \ S$_{1/2}$& \ \ \ \ P$_{1/2}$& \ \ \ \ P$_{3/2}$ & \ \ \ \ D$_{3/2}$& \ \ \ \ D$_{5/2}$\\
\colrule
\vbox to 10 pt {}
 $ 1  $&$  -30.290\,240(20) $&&&&\\ 
 $ 2  $&$  -31.185\,150(90) $&$  -0.973\,50(20) $&$  -0.486\,50(20) $&&\\
 $ 3  $&$  -31.047\,70(90) $&$                $&$                 $&$                $&$                 $ \\
 $ 4  $&$  -30.9120(40) $&$  -1.1640(20) $&$  -0.6090(20) $&$                $&$  0.031\,63(22) $ \\
 $ 6  $&$  -30.711(47) $&$                $&$                 $&$                $&$  0.034\,17(26) $ \\
 $ 8  $&$  -30.606(47) $&$                $&$                 $&$  0.007\,940(90) $&$  0.034\,84(22) $ \\
 $ 12 $&$                 $&$                $&$                 $&$  0.0080(20) $&$  0.0350(30) $ \\
\botrule
\end{tabular}
\end{table*}

\paragraph{Vacuum polarization}
\label{par:vacpol} 

The second-order vacuum-polarization level shift is 
\begin{eqnarray} 
E_{\rm VP}^{(2)} = {\alpha\over\rmpi}{(Z\alpha)^4\over n^3} H\!(Z\alpha) \, 
m_{\rm e} c^2 \ , 
\label{eq:vacpol} 
\end{eqnarray} 
where the function $H\!(Z\alpha)$ is divided into the part corresponding to the
Uehling potential, denoted here by $H^{(1)}\!(Z\alpha)$, and the higher-order
remainder $H^{({\rm R})}\!(Z\alpha)$, where
\begin{eqnarray} 
H^{(1)}\!(Z\alpha) &=& V_{40}+V_{50}\,(Z\alpha) +V_{61}\,(Z\alpha)^2\ln(Z 
\alpha)^{-2} 
\nonumber\\ 
&&+ \, G_{\rm VP}^{(1)}(Z\alpha)\,(Z\alpha)^2  
\label{eq:uehpow} 
\\ 
H^{({\rm R})}\!(Z\alpha) &=& G_{\rm VP}^{({\rm R})}(Z\alpha)\,(Z\alpha)^2 \ , 
\label{eq:wkvp}
\end{eqnarray} 
with 
\begin{eqnarray} 
V_{40} &=& -\frac{4}{15} \, \delta_{l0} \nonumber\\ 
V_{50} &=& \frac{5}{48}\rmpi \, \delta_{l0} \\ 
V_{61} &=& -\frac{2}{15} \, \delta_{l0}\,. \nonumber 
\end{eqnarray} 
The part $G_{\rm VP}^{(1)}(Z\alpha)$ arises from the Uehling potential with
values given in Table~\ref{tab:gvp1} \cite{1982004,2002148}.  The higher-order
remainder $G_{\rm VP}^{({\rm R})} (Z\alpha)$ has been considered by Wichmann
and Kroll, and the leading terms in powers of $Z\alpha$ are
\cite{1956001,1975028,1983005} 
\begin{eqnarray} 
G_{\rm VP}^{\rm (R)}(Z\alpha) &=& \left(\frac{19}{45} - \frac{\rmpi^2}{27}\right) 
\delta_{l0}\nonumber\\ 
&&+ \left(\frac{1}{16} - \frac{31\rmpi^2}{2880}\right)\rmpi(Z\alpha)
\delta_{l0} + \cdots \ . \qquad
\label{eq:hovp}
\end{eqnarray} 
Higher-order terms omitted from Eq.~(\ref{eq:hovp}) are negligible.

\begin{table*}
\def\sb{\hbox to 5mm{}}
\caption{Values of the function $G_{\rm VP}^{(1)}(\alpha)$.}
\label{tab:gvp1}
\begin{tabular}{r @{\sb} l @{\sb} l @{\sb} l @{\sb} l @{\sb} l}
\toprule
\vbox to 10 pt {}
$n$& \ \ \ \ S$_{1/2}$& \ \ \ \ P$_{1/2}$& \ \ \ \ P$_{3/2}$ & \ \ \ \ D$_{3/2}$& \ \ \ \ D$_{5/2}$\\
\colrule
\vbox to 10 pt {}
 $ 1  $&$  -0.618\,724 $&&&&\\ 
 $ 2  $&$  -0.808\,872 $&$  -0.064\,006 $&$  -0.014\,132 $&&\\
 $ 3  $&$  -0.814\,530 $&$                $&$                 $&$                $&$                 $ \\
 $ 4  $&$  -0.806\,579 $&$  -0.080\,007 $&$  -0.017\,666 $&$                $&$  -0.000\,000 $ \\
 $ 6  $&$  -0.791\,450 $&$                $&$                 $&$                $&$  -0.000\,000 $ \\
 $ 8  $&$  -0.781\,197 $&$                $&$                 $&$  -0.000\,000 $&$  -0.000\,000 $ \\
 $ 12 $&$                 $&$                $&$                 $&$  -0.000\,000 $&$  -0.000\,000 $ \\
\botrule
\end{tabular}
\end{table*}

In a manner similar to that for the self energy, the leading effect of the
finite mass of the nucleus is taken into account by multiplying
Eq.~(\ref{eq:vacpol}) by the factor $(m_{\rm r}/m_{\rm e})^3$ and including a
multiplicative factor of $(m_{\rm e}/m_{\rm r})$ in the argument of the
logarithm in Eq.~(\ref{eq:uehpow}).

There is also a second-order vacuum polarization level shift due to the
creation of virtual particle pairs other than the e$^-$e$^+$ pair.  The
predominant contribution for $n$S states arises from ${\rmmu}^+{\rmmu}^-$, with
the leading term being \cite{1995102,1995005}
\begin{eqnarray} 
E_{{\rmssmu}{\rm VP}}^{(2)} = {\alpha\over\rmpi}{(Z\alpha)^4\over n^3} 
\left(-{4\over15}\right) 
\left({m_{\rm e}\over m_{\rmssmu}}\right)^2
\left({m_{\rm r}\over m_{\rm e}}\right)^3
m_{\rm e} c^2 \ . 
\nonumber\\
\label{eq:vacpolmu} 
\end{eqnarray} 
The next order term in the contribution of muon vacuum polarization to $n$S
states is of relative order $Z\alpha m_{\rm e}/m_{\rm \mu}$ and is therefore
negligible.  The analogous contribution $E_{{\rmsstau}{\rm VP}}^{(2)}$ from
${\rmtau}^+{\rmtau}^-$ ($-18$ Hz for the 1S state) is also negligible at the
level of uncertainty of current interest.  

For the hadronic vacuum polarization contribution, we take the result given by
\textcite{1999042} that utilizes all available e$^+$e$^-$ scattering data:
\begin{eqnarray}
E_{\rm had \, VP}^{(2)} =  0.671(15) E_{{\rmssmu}{\rm VP}}^{(2)} \ ,
\end{eqnarray}
where the uncertainty is of type $u_0$.

The muonic and hadronic vacuum polarization contributions are negligible for P
and D states.

\paragraph{Two-photon corrections}
\label{par:tpc} 

Corrections from two virtual photons have been partially calculated as a power
series in $Z\alpha$:
\begin{eqnarray} 
E^{(4)} &=& \left({\alpha\over\rmpi}\right)^2 {(Z\alpha)^4\over n^3} m_{\rm e}c^2 
F^{(4)}(Z\alpha) \ ,
\label{eq:total4}
\end{eqnarray} 
where
\begin{eqnarray} 
F^{(4)}(Z\alpha) &=& 
B_{40} + B_{50}\,(Z\alpha)
+ B_{63}\,(Z\alpha)^2\ln^3(Z\alpha)^{-2} 
\nonumber\\
&& + B_{62}\,(Z\alpha)^2\ln^2(Z\alpha)^{-2}
\nonumber\\
&& + B_{61}\,(Z\alpha)^2\ln(Z\alpha)^{-2}
+ B_{60}\,(Z\alpha)^2
\nonumber\\ &&+\cdots \ .
\end{eqnarray} 

The leading term $B_{40}$ is well known:
\begin{eqnarray} 
B_{40} &=& \left[\frac{3\rmpi^2}{2}\ln 2-\frac{10\rmpi^2}{27}-\frac{2179}{648}-\frac{9}{4}\zeta(3) 
\right] \delta_{l0} \nonumber\\ 
&&+ \left[\frac{\rmpi^2\ln 2}{2}-\frac{\rmpi^2}{12}-\frac{197}{144}-\frac{3\zeta(3)}{4} 
\right] {1 - \delta_{l0} \over \kappa(2l+1)} \ .
\nonumber\\
\label{eq:b40}
\end{eqnarray}
The second term is \cite{1993025,1997027,1995102,1994092}
\begin{eqnarray} 
B_{50} &=& -21.5561(31)\delta_{l0} \ , 
\label{eq:2phb50}
\end{eqnarray} 
and the next coefficient is \cite{1993019,2000101,2000094,2001059}
\begin{eqnarray}
B_{63} = -{8\over 27}\delta_{l0} \ .
\label{eq:b63}
\end{eqnarray}

For S states the coefficient $B_{62}$ is given by
\begin{eqnarray}
B_{62} &=& {16\over 9} \left[{71\over60}-\ln{2} + \gamma 
+ \psi(n) - \ln n - {1\over n} + {1\over4n^2}
\right] \ , 
\nonumber\\
\label{eq:b62s}
\end{eqnarray}
where $\gamma = 0.577...$ is Euler's constant and $\psi$ is the psi function
\cite{1965020}.  The difference $B_{62}(1)-B_{62}(n)$ was calculated by
\textcite{1996012} and confirmed by \textcite{2001059} who also calculated the
$n$-independent additive constant.  For P states the calculated value is
\cite{1996012}
\begin{eqnarray}
B_{62} = {4\over 27} {n^2 - 1 \over n^2} \ .
\label{eq:b62p}
\end{eqnarray}
This result has been confirmed by \textcite{2002155} who also show that for D
and higher angular momentum states $B_{62}=0$.

Recent work has led to new results for $B_{61}$ and higher-order coefficients.
In \textcite{2005212} an additional state-independent contribution to the
coefficient $B_{61}$ for S states is given, which slightly differs (2~\%) from
the earlier result of \textcite{2001059} quoted in CODATA 2002.  The revised
coefficient for S states is
\begin{eqnarray}
B_{61} &=& {413\,581\over 64\,800} +
{4N(n{\rm S})\over3} + {2027\rmpi^2\over864} - {616\,\ln{2}\over 135}
\nonumber\\&&
- {2\rmpi^2\ln{2}\over 3} 
+ {40\ln^2{2}\over 9}
+ \zeta(3)
+\left(\frac{304}{135}-\frac{32\,\ln{2}}{9}\right)
\nonumber\\ && \times
\left[{3\over4} + \gamma
+ \psi(n) - \ln n - {1\over n} + {1\over4n^2}
\right] \ ,
\label{eq:b61s}
\end{eqnarray}

where $\zeta$ is the Riemann zeta function \cite{1965020}.  The coefficients
$N(n{\rm S})$ are listed in Table~\ref{tab:b61n}.  The state-dependent part
$B_{61}(n{\rm S}) - B_{61}(1{\rm S})$ was confirmed by \textcite{2005212} in
their Eqs.~(4.26) and (6.3).  For higher-$l$ states, $B_{61}$ has been
calculated by \textcite{2005212}; for P states
\begin{eqnarray}
B_{61}(n{\rm P}_{1/2}) &=& \frac{4}{3}\,N(n{\rm P})
+ \frac{n^2-1}{n^2}\left(\frac{166}{405}-\frac{8}{27}\,\ln{2}\right)
,
\qquad
\\
B_{61}(n{\rm P}_{3/2}) &=& \frac{4}{3}\,N(n{\rm P})
+ \frac{n^2-1}{n^2}\left(\frac{31}{405}-\frac{8}{27}\,\ln{2}\right)
,
\qquad
\end{eqnarray}
and for D states
\begin{eqnarray}
B_{61}(n{\rm D}) &=& 0 \ .
\end{eqnarray}
The coefficient $B_{61}$ also vanishes for states with $l>2$.  The necessary
values of $N(n{\rm P})$ are given in Eq.~(17) of \textcite{2003118} and are
listed in Table~\ref{tab:b61n}.

\begin{table}
\caption{Values of $N$ used in the 2006 adjustment}
\label{tab:b61n} 
\begin{tabular}{D{.}{.}{3.3}D{.}{.}{3.12}D{.}{.}{3.12}} 
\toprule
\vbox to 10 pt {}
 n & \multicolumn{1}{c}{$N(n{\rm S})$} & \multicolumn{1}{c}{$N(n{\rm P})$} \\
\colrule
1 &  17.855\,672\,03(1) & \vbox to 10pt{} \\
2 &  12.032\,141\,58(1) &  0.003\,300\,635(1)\\
3 &  10.449\,809(1) & \\
4 &  9.722\,413(1) &  -0.000\,394\,332(1)\\
6 &  9.031\,832(1) & \\
8 &  8.697\,639(1) & \\
\botrule
\end{tabular} 
\end{table} 

The next term is $B_{60}$, and recent work has also been done for this
contribution.  For S states, the state dependence is considered first, and is
given by \textcite{2005093,2005212}
\begin{eqnarray}
B_{60}(n{\rm S}) - B_{60}(1{\rm S}) &=& b_{\rm L}(n{\rm S})
-b_{\rm L}(1{\rm S}) + A(n) \ ,
\qquad
\label{eq:b60diffs}
\end{eqnarray}
where
\begin{eqnarray}
A(n) &=& \left(\frac{38}{45}-\frac{4}{3}\ln{2}\right)
[N(n{\rm S}) - N(1{\rm S})]
\nonumber\\&&
-\frac{337\,043}{129\,600}
-\frac{94\,261}{21\,600\,n} +\frac{902\,609}{129\,600\,n^2}
\nonumber\\&&
+\left(\frac{4}{3}-\frac{16}{9\,n}+\frac{4}{9\,n^2}\right)\ln^2{2}
\nonumber\\&&
+\left(-\frac{76}{45}+\frac{304}{135\,n}-\frac{76}{135\,n^2}\right)\ln{2}
\nonumber\\&&
+\left(-\frac{53}{15}+\frac{35}{2\,n}-\frac{419}{30\,n^2}\right)\zeta(2)\,\ln{2}
\nonumber\\&&
+\left(\frac{28\,003}{10\,800}-\frac{11}{2\,n}+\frac{31\,397}{10\,800\,n^2}\right)\zeta(2)
\nonumber\\&&
+\left(\frac{53}{60}-\frac{35}{8\,n}+\frac{419}{120\,n^2}\right)\zeta(3)
\nonumber\\&&
+\bigg(\frac{37\,793}{10\,800}+\frac{16}{9}\ln^2{2}-\frac{304}{135}\ln{2}
+8\,\zeta(2)\,\ln{2}
\nonumber\\&& \quad
-\frac{13}{3}\zeta(2)-2\,\zeta(3)\bigg)
 \left[\gamma + \psi(n) - \ln{n}\right] \ .
\end{eqnarray}
The term $A(n)$ makes a small contribution in the range 0.3 to 0.4 for the
states under consideration.

The two-loop Bethe logarithms $b_{\rm L}$ in Eq.~(\ref{eq:b60diffs}) are listed
in Table~\ref{tab:b60}.  The values for $n=1$ to 6 are from
\textcite{2003160,2004148}, and the value at $n=8$ is obtained by extrapolation
of the calculated values from $n=4$ to 6 [$b_{\rm L}(5{\rm S}) = -60.6(8)$]
with a function of the form
\begin{eqnarray}
b_{\rm L}(n{\rm S}) = a + \frac{b}{n} + \frac{c}{n(n+1)} \ ,
\end{eqnarray}
which yields
\begin{eqnarray}
b_{\rm L}(n{\rm S}) = -55.8 - \frac{24}{n} \ .
\end{eqnarray}
It happens that the fit gives $c=0$.  An estimate for $B_{60}$ given by
\begin{eqnarray}
B_{60}(n{\rm S}) = b_{\rm L}(n{\rm S}) + \frac{10}{9}N(n{\rm S}) + \cdots  
\label{eq:oldb60}
\end{eqnarray}

was derived by \textcite{2001059}.  The dots represent uncalculated
contributions at the relative level of 15~\% \cite{2003160}.
Equation~(\ref{eq:oldb60}) gives $B_{60}(1{\rm S}) = -61.6(9.2)$.  However,
more recently \textcite{2005346,2005205,2003167,2007142} have calculated the
1S-state two-loop self energy correction for $Z\ge10$.  This is expected to
give the main contribution to the higher-order two-loop correction.  Their
results extrapolated to $Z=1$ yield a value for the contribution of all terms
of order $B_{60}$ or higher of $-127\times(1 \pm 0.3)$, which corresponds to a
value of roughly $B_{60} = -129(39)$, assuming a linear extrapolation from
$Z=1$ to $Z=0$.  This differs by about a factor of two from the result given by
Eq.~(\ref{eq:oldb60}).  In view of this difference between the two
calculations, for the 2006 adjustment, we use the average of the two values
with an uncertainty that is half the difference, which gives
\begin{eqnarray}
    B_{60}(1{\rm S}) =  -95.3( 0.3)( 33.7) 
    \ .
\label{eq:b601s}
\end{eqnarray}
In Eq.~(\ref{eq:b601s}), the first number in parentheses is the state-dependent
uncertainty $u_n(B_{60})$ associated with the two-loop Bethe logarithm, and the
second number in parentheses is the state-independent uncertainty $u_0(B_{60})$
that is common to all S-state values of $B_{60}$.  Values of $B_{60}$ for all
relevant S-states are given in Table~\ref{tab:b60}.  For higher-$l$ states,
$B_{60}$ has not been calculated, so we take it to be zero, with uncertainties
$u_n[B_{60}(n{\rm P})] = 5.0$ and $u_n[B_{60}(n{\rm D})] = 1.0$.  We assume
that these uncertainties account for higher-order P and D state uncertainties
as well.  For S states, higher-order terms have been estimated by
\textcite{2005212} with an effective potential model.  They find that the next
term has a coefficient of $B_{72}$ and is state independent.  We thus assume
that the uncertainty $u_0[B_{60}(n{\rm S})]$ is sufficient to account for the
uncertainty due to omitting such a term and higher-order state-independent
terms.  In addition, they find an estimate for the state dependence of the next
term, given by
\begin{eqnarray}
\Delta B_{71}(n{\rm S}) &=&
B_{71}(n{\rm S}) - B_{71}(1{\rm S}) = 
\rmpi\left(\frac{427}{36} - \frac{16}{3}\,\ln{2}\right)
\nonumber\\
&\times&\left[\frac{3}{4}-\frac{1}{n}+\frac{1}{4n^2}
+\gamma +\psi(n)-\ln{n}\right] 
\end{eqnarray}
with a relative uncertainty of 50\;\%.  We include this additional term, which
is listed in Table~\ref{tab:b60}, along with the estimated uncertainty
$u_n(B_{71}) = B_{71}/2$.

The disagreement of the analytic and numerical calculations results in an
uncertainty of the two-photon contribution that is larger than the estimated
uncertainty used in the 2002 adjustment.  As a result, the uncertainties of the
recommended values of the Rydberg constant and proton and deuteron radii are
slightly larger in the 2006 adjustment, although the 2002 and 2006 recommended
values are consistent with each other.  On the other hand, the uncertainty of
the 2P state fine structure is reduced as a result of the new analytic
calculations.

As in the case of the order $\alpha$ self-energy and vacuum-polarization
contributions, the dominant effect of the finite mass of the nucleus is taken
into account by multiplying each term of the two-photon contribution by the
reduced-mass factor $(m_{\rm r}/m_{\rm e})^3$, except that the magnetic moment
term, the second line of Eq.~(\ref{eq:b40}), is instead multiplied by the
factor $(m_{\rm r}/m_{\rm e})^2$.  In addition, the argument $(Z\alpha)^{-2}$
of the logarithms is replaced by $(m_{\rm e}/m_{\rm r})(Z\alpha)^{-2}$.

\begin{table}
\caption{Values of $b_{\rm L}$, $B_{60}$, and $\Delta B_{71}$ used in the 2006 adjustment}
\label{tab:b60} 
\begin{tabular}{D{.}{.}{3.1}D{.}{.}{5.7}D{.}{.}{4.10} D{.}{.}{10.3}} 
\toprule
\vbox to 10 pt {}
 n & \multicolumn{1}{c}{$b_{\rm L}(n{\rm S})$} & \multicolumn{1}{c}{$B_{60}(n{\rm S})$} 
& \multicolumn{1}{c}{$\Delta B_{71}(n{\rm S})$} \\ 
\colrule
\vbox to 10 pt {}
1 &  -81.4( 0.3) &  -95.3( 0.3)( 33.7) &    \vbox to 10pt{}  \\
2 &  -66.6( 0.3) &  -80.2( 0.3)( 33.7) &  16(8)\phantom{2} \\
3 &  -63.5( 0.6) &  -77.0( 0.6)( 33.7) &  22(11) \\
4 &  -61.8( 0.8) &  -75.3( 0.8)( 33.7) &  25(12) \\
6 &  -59.8( 0.8) &  -73.3( 0.8)( 33.7) &  28(14) \\
8 &  -58.8( 2.0) &  -72.3( 2.0)( 33.7) &  29(15) \\
\botrule
\end{tabular} 
\end{table} 

\paragraph{Three-photon corrections}
\label{par:thpc} 

The leading contribution from three virtual photons is expected to have the
form
\begin{eqnarray} 
E^{(6)} &=& \left({\alpha\over\rmpi}\right)^3 {(Z\alpha)^4\over n^3} m_{\rm e}c^2 
\left[C_{40} + C_{50}(Z\alpha) + \cdots \right] \ ,
\nonumber\\
\label{eq:total6}
\end{eqnarray} 
in analogy with Eq.~(\ref{eq:total4}) for two photons.  The leading term
$C_{40}$ is
\cite{2000019,1996060,1995220,1995158}
\begin{eqnarray}
C_{40} &=& \bigg[
 -{{568\,{\rm a_4}}\over{9}}+{{85\,\zeta(5)}\over{24}}
\nonumber\\&&
-{{121\,\rmpi^{2}\,\zeta(3)}\over{72}}
-{{84\,071\,\zeta(3)}\over{2304}}
-{{71\,\ln ^{4}2}\over{27}}
\nonumber\\&&
-{{239\,\rmpi^{2}\,\ln^{2}2}\over{135}}
+{{4787\,\rmpi^{2}\,\ln 2}\over{108}}
+{{1591\,\rmpi^{4}}\over{3240}}
\nonumber\\&&
-{{252\,251\,\rmpi^{2}}\over{9720}}+{679\,441\over93\,312}
\bigg] \delta_{l0} \nonumber\\
&&+ \bigg[
-{{100\,{\rm a_4}}\over{3}}+{{215\,\zeta(5)}\over{24}}
\nonumber\\&&
-{{83\,\rmpi^{2}\,\zeta(3)}\over{72}}-{{139\,\zeta(3)}\over{18}} 
-{{25\,\ln ^{4}2}\over{18}}
\nonumber\\&&
+{{25\,\rmpi^{2}\,\ln ^{2}2}\over{18}}+{{298\,\rmpi^{2}\,\ln 2}\over{9}}
+{{239\,\rmpi^{4}}\over{2160}}
\nonumber\\&&
-{{17\,101\,\rmpi^{2}}\over{810}}-{28\,259\over5184}
\bigg] {1 - \delta_{l0} \over \kappa(2l+1)} \ ,
\nonumber\\
\label{eq:c40}
\end{eqnarray}
where $a_4 = \sum_{n=1}^\infty 1/(2^n\,n^4) = 0.517\,479\,061\dots$ .
Higher-order terms have not been calculated, although partial results have been
obtained \cite{2007138}.  An uncertainty is assigned by taking $u_0(C_{50}) =
30\delta_{l0}$ and $u_n(C_{63}) = 1$, where $C_{63}$ is defined by the usual
convention.  The dominant effect of the finite mass of the nucleus is taken
into account by multiplying the term proportional to $\delta_{l0}$ by the
reduced-mass factor $(m_{\rm r}/m_{\rm e})^3$ and the term proportional to
$1/[\kappa(2l+1)]$, the magnetic moment term, by the factor $(m_{\rm r}/m_{\rm
e})^2$.  

The contribution from four photons is expected to be of order
\begin{eqnarray}
\left({\alpha\over\rmpi}\right)^4 {(Z\alpha)^4\over n^3} m_{\rm e}c^2 \ ,
\label{eq:fourphoton}
\end{eqnarray}
which is about 10 Hz for the 1S state and is negligible at the level of
uncertainty of current interest.

\paragraph{Finite nuclear size}
\label{par:nucsize} 

At low $Z$, the leading contribution due to the finite size of the nucleus is
\begin{eqnarray} 
E^{(0)}_{\rm NS} = {\cal E}_{\rm NS}
\delta_{l0}  \ , 
\label{eq:ens0}
\end{eqnarray} 
with
\begin{eqnarray} 
{\cal E}_{\rm NS}
 = {2\over3}\left({m_{\rm r}\over m_{\rm e}}\right)^3
{(Z\alpha)^2\over n^3} \ m_{\rm e}c^2 
\left({Z\alpha R_{\rm N}\over\lbar_{\rm C}} 
\right)^2  \ , 
\label{eq:enscoef}
\end{eqnarray} 
where $R_{\rm N}$ is the bound-state root-mean-square (rms) charge radius of
the nucleus and $\lbar_{\rm C}$ is the Compton wavelength of the electron
divided by $2\rmpi$.  The leading higher-order contributions have been examined
by \textcite{1979028,1997168,1997022} [see also \textcite{1979029,1983005}].
The expressions that we employ to evaluate the nuclear size correction are the
same as those discussed in more detail in CODATA-98.

For S states the leading and next-order corrections are given by
\begin{eqnarray}
E_{\rm NS} &=& {\cal E}_{\rm NS}
\Bigg\{1 - C_\eta{ m_{\rm r}\over m_{\rm e}}
{R_{\rm N}\over\lbar_{\rm C}}
Z\alpha
-\bigg[
\ln{\left({m_{\rm r}\over m_{\rm e}}
{R_{\rm N}\over \lbar_{\rm C}}{Z\alpha\over n}\right)}
\nonumber\\
&& + \psi(n) + \gamma - {(5n+9)(n-1)\over4n^2}
-C_\theta
\bigg](Z\alpha)^2
\Bigg\} \ ,
\nonumber\\
\label{eq:ensho}
\end{eqnarray}
where $C_\eta$ and $C_\theta$ are constants that depend on the details of the
assumed charge distribution in the nucleus.  The values used here are $C_\eta =
1.7(1)$ and $C_\theta = 0.47(4)$ for hydrogen or $C_\eta = 2.0(1)$ and
$C_\theta = 0.38(4)$ for deuterium.

For the P$_{1/2}$ states in hydrogen the leading term is
\begin{eqnarray}
E_{\rm NS} = {\cal E}_{\rm NS}
{(Z\alpha)^2(n^2-1)\over 4 n^2} \ .
\label{eq:enpho}
\end{eqnarray}
For P$_{3/2}$ states and D states the nuclear-size contribution is negligible.

\paragraph{Nuclear-size correction to self energy and vacuum polarization}
\label{par:nssevp} 

For the self energy, the additional contribution due to the finite size of the
nucleus is \cite{1993124,1997158,2002194,2003127}
\begin{eqnarray}
E_{\rm NSE} = \left(4\ln{2}-\frac{23}{4}\right)
\alpha(Z\alpha){\cal E}_{\rm NS}\delta_{l0} \ ,
\label{eq:nse}
\end{eqnarray}
and for the vacuum polarization it is \cite{1979031,1979031e,1985047,1997158}
\begin{eqnarray}
E_{\rm NVP} = {3\over4}\alpha(Z\alpha){\cal E}_{\rm NS}\delta_{l0} \ .
\label{eq:nvp}
\end{eqnarray}

For the self-energy term, higher-order size corrections for S states
\cite{2002194} and size corrections for P states have been calculated
\cite{2003118,2003138}, but these corrections are negligible for the current
work, and are not included.  The D-state corrections are assumed to be
negligible.

\paragraph{Radiative-recoil corrections}
\label{par:rrc} 

The dominant effect of nuclear motion on the self energy and vacuum
polarization has been taken into account by including appropriate reduced-mass
factors.  The additional contributions beyond this prescription are termed
radiative-recoil effects with leading terms given by
\begin{eqnarray} 
E_{\rm RR} &=& {m_{\rm r}^3\over m_{\rm e}^2m_{\rm N}}
{\alpha(Z\alpha)^5\over \rmpi^2 \, n^3} m_{\rm e}c^2
\delta_{l0}
\nonumber\\&&\times\bigg[
 6\,\zeta(3) -2\,\rmpi^2\ln{2} + {35\,\rmpi^2\over 36} - {448\over27}
\nonumber\\&& \qquad +
{2\over3}\rmpi(Z\alpha)\,\ln^2{(Z\alpha)^{-2}} + \cdots
\bigg] \ . \qquad
\label{eq:radrec}
\end{eqnarray} 
The constant term in Eq.~(\ref{eq:radrec}) is the sum of the analytic result
for the electron-line contribution \cite{2001060,2001075} and the
vacuum-polarization contribution \cite{1995124,1995100}.  This term agrees with
the numerical value \cite{1995100} used in CODATA-98.  The log-squared term has
been calculated by \textcite{1999164} and by \textcite{1999182}.

For the uncertainty, we take a term of order $(Z\alpha)\ln(Z\alpha)^{-2}$
relative to the square brackets in Eq.~(\ref{eq:radrec}) with numerical
coefficients 10 for $u_0$ and 1 for $u_n$.  These coefficients are roughly what
one would expect for the higher-order uncalculated terms.  For higher-$l$
states in the present evaluation, we assume that the uncertainties of the two-
and three-photon corrections are much larger than the uncertainty of the
radiative-recoil correction.  Thus, we assign no uncertainty for the
radiative-recoil correction for P and D states.

\paragraph{Nucleus self energy}
\label{par:nse} 

An additional contribution due to the self energy of the nucleus has been given
by \textcite{1995100}:
\begin{eqnarray}
E_{\rm SEN} &=& {4Z^2\alpha(Z\alpha)^4 \over 3 \rmpi n^3}
{m_{\rm r}^3\over m_{\rm N}^2}c^2 \nonumber\\
&& \times\left[
\ln{\left({m_{\rm N}\over m_{\rm r}(Z\alpha)^2}\right)}\delta_{l0}
-\ln k_0(n,l) \right] \ . \qquad
\label{eq:nucse}
\end{eqnarray}
This correction has also been examined by \textcite{2001057}, who consider how
it is modified by the effect of structure of the proton.  The structure effect
would lead to an additional model-dependent constant in the square brackets in
Eq.~(\ref{eq:nucse}).  

To evaluate the nucleus self-energy correction, we use Eq.~(\ref{eq:nucse}) and
assign an uncertainty $u_0$ that corresponds to an additive constant of 0.5 in
the square brackets for S states.  For P and D states, the correction is small
and its uncertainty, compared to other uncertainties, is negligible.

\paragraph{Total energy and uncertainty}
\label{par:teu} 

The total energy $E_{n{\rm L}j}^X$ of a particular level (where L = S, P, ...
and $X$ = H, D) is the sum of the various contributions listed above plus an
additive correction $\delta_{n{\rm L}j}^X$ that accounts for the uncertainty in
the theoretical expression for $E_{n{\rm L}j}^X$.  Our theoretical estimate of
the value of $\delta_{n{\rm L}j}^X$ for a particular level is zero with a
standard uncertainty of $u(\delta_{n{\rm L}j}^X)$ equal to the square root of
the sum of the squares of the individual uncertainties of the contributions; as
they are defined above, the contributions to the energy of a given level are
independent.  (Components of uncertainty associated with the fundamental
constants are not included here, because they are determined by the least
squares adjustment itself.)  Thus, we have for the square of the uncertainty,
or variance, of a particular level
\begin{eqnarray}
u^2(\delta_{n{\rm L}j}^X)
= \sum_i {u_{0i}^2(XLj) 
+ u_{ni}^2(XLj) \over n^6 } \ ,
\end{eqnarray}
where the individual values $u_{0i}(XLj)/n^3$ and $u_{ni}(XLj)/n^3$ are the
components of uncertainty from each of the contributions, labeled by $i$,
discussed above.  (The factors of $1/n^3$ are isolated so that $u_{0i}(XLj)$ is
explicitly independent of $n$.)

The covariance of any two $\delta$'s follows from Eq.~(F7) of Appendix~F of
CODATA-98. For a given isotope $X$, we have 
\begin{eqnarray}
u(\delta_{n_1{\rm L}j}^X,
\delta_{n_2{\rm L}j}^X)
= \sum_i{u^2_{0i}(XLj) \over
 (n_1 n_2)^3} \ ,
\end{eqnarray}
which follows from the fact that $u(u_{0i}, u_{ni}) = 0$ and
$u(u_{n_1i},u_{n_2i}) = 0$ for $n_1 \ne n_2$.  We also set
\begin{eqnarray}
u(\delta_{n_1{\rm L_1}j_1}^X,
\delta_{n_2{\rm L_2}j_2}^X)
= 0 \ ,
\end{eqnarray}
if ${\rm L}_1 \ne {\rm L}_2$ or $j_1 \ne j_2$.

For covariances between $\delta$'s for hydrogen and deuterium, we have for
states of the same $n$
\begin{eqnarray}
&&u(\delta_{n{\rm L}j}^{\rm H},
\delta_{n{\rm L}j}^{\rm D})
\nonumber\\
 \ &&= \sum_{i = i_{\rm c}}{ u_{0i}({\rm HL}j)u_{0i}({\rm DL}j)
+ u_{ni}({\rm HL}j)u_{ni}({\rm DL}j) \over n^6} \ , \qquad\quad
\end{eqnarray}
and for $n_1\neq n_2$
\begin{eqnarray}
u(\delta_{n_1{\rm L}j}^{\rm H},
\delta_{n_2{\rm L}j}^{\rm D})
= \sum_{i = i_{\rm c}}{ u_{0i}({\rm HL}j)u_{0i}({\rm DL}j) 
\over (n_1 n_2)^3} \ ,
\end{eqnarray}
where the summation is over the uncertainties common to hydrogen and deuterium.
In most cases, the uncertainties can in fact be viewed as common except for a
known multiplicative factor that contains all of the mass dependence.  We
assume
\begin{eqnarray}
u(\delta_{n_1{\rm L_1}j_1}^{\rm H},
\delta_{n_2{\rm L_2}j_2}^{\rm D})
= 0 \ ,
\end{eqnarray}
if ${\rm L}_1 \ne {\rm L}_2$ or
$j_1 \ne j_2$.

The values of $u(\delta_{n{\rm L}j}^X)$ of interest for the 2006 adjustment are
given in Table~\ref{tab:rdata} of Sec.~\ref{sec:ad}, and the non negligible
covariances of the $\delta$'s are given in the form of correlation coefficients
in Table~\ref{tab:rdcc} of that section.  These coefficients are as large as
0.9999.

Since the transitions between levels are measured in frequency units (Hz), in
order to apply the above equations for the energy level contributions we divide
the theoretical expression for the energy difference $\Delta E$ of the
transition by the Planck constant $h$ to convert it to a frequency.  Further,
since we take the Rydberg constant $R_\infty = \alpha^2m_{\rm e}c/2h$
(expressed in m$^{-1}$) rather than the electron mass $m_{\rm e}$ to be an
adjusted constant, we replace the group of constants $\alpha^2m_{\rm e}c^2/2h$
in $\Delta E /h$ by $cR_\infty$.

\paragraph{Transition frequencies between levels with $\bm{n = 2}$}
\label{par:trfreq} 

As an indication of the consistency of the theory summarized above and the
experimental data, we list below values of the transition frequencies between
levels with $n=2$ in hydrogen.  These results are based on values of the
constants obtained in a variation of the 2006 least squares adjustment in which
the measurements of the directly related transitions (items $A38$, $A39.1$, and
$A39.2$ in Table~\ref{tab:rdata}) are not included, and the weakly coupled
constants $A_{\rm r}$(e), $A_{\rm r}$(p), $A_{\rm r}$(d), and $\alpha$, are
assigned their 2006 adjusted values.  The results are
\begin{eqnarray}
\nu_{\rm H}(2{\rm P}_{1/2} - 2{\rm S}_{1/2}) &=& 
 1\,057\,843.9(2.5) \ {\rm kHz} \quad [ 2.3\times 10^{-6}]  \nonumber \\
\nu_{\rm H}(2{\rm S}_{1/2} - 2{\rm P}_{3/2}) &=& 
 9\,911\,197.6(2.5) \ {\rm kHz} \quad [ 2.5\times 10^{-7}]  \nonumber \\
\nu_{\rm H}(2{\rm P}_{1/2} - 2{\rm P}_{3/2})&&
\nonumber\\&&\hbox to -65pt {} =
 10\,969\,041.475(99) \ {\rm kHz} \quad [ 9.0\times 10^{-9}] \ ,
\label{eq:neq2data}
\end{eqnarray}
which agree well with the relevant experimental results of
Table~\ref{tab:rdata}.  Although the first two values in
Eq.~(\ref{eq:neq2data}) have changed only slightly from the results of the 2002
adjustment, the third value, the fine-structure splitting, has an uncertainty
that is almost an order-of-magnitude smaller than the 2002 value, due mainly to
improvements in the theory of the two-photon correction.

A value of the fine structure constant $\alpha$ can be obtained from the data
on the hydrogen and deuterium transitions.  This is done by running a variation
of the 2006 least-squares adjustment that includes all the transition frequency
data in Table~\ref{tab:rdata} and the 2006 adjusted values of $A_{\rm r}$(e),
$A_{\rm r}$(p), and $A_{\rm r}$(d).  The resulting value is
\begin{eqnarray}
\alpha^{-1} &=&  137.036\,002(48) \qquad [ 3.5\times 10^{-7}] \ ,
\label{eq:alphinvhd}
\end{eqnarray}
which is consistent with the 2006 recommended value, although substantially
less accurate.   This result is included in Table~\ref{tab:alphas}.

\subsubsection{Experiments on hydrogen and deuterium}
\label{sssec:rydex}

Table~\ref{tab:rydfreq} summarizes the transition frequency data relevant to
the determination of ${R_\infty}$. With the exception of the first entry, which
is the most recent result for the 1S$_{1/2}$ -- 2S$_{1/2}$ transition frequency
in hydrogen from the group at the Max-Planck-Institute f\"ur Quantenoptik
(MPQ), Garching, Germany, all of these data are the same as those used in the
2002 adjustment. Since these data are reviewed in CODATA-98 or CODATA-02, they
are not discussed here. For a brief discussion of data not included in
Table~\ref{tab:rydfreq}, see Sec.~II.B.3 of CODATA-02.

\shortcites{2004111,1995159,1998002,1997001,1999072,1999072e,
1996001,1995138,1994090,1986003,1979001}
\def\vsp{\vbox to 10pt{}}
\begin{table*}
\def\sp{\hbox to 2.5mm {}}
\caption{ Summary of measured transition frequencies $\nu$ considered in the
present work for the determination of the Rydberg constant $R_\infty$ (H is
hydrogen and D is deuterium).}
\label{tab:rydfreq}
\begin{tabular}{l@{\sp}l@{\sp}l@{\sp}l@{\sp}l}
\toprule
\noalign{\vbox to 5 pt {}}
 Authors  & Laboratory
  & \hbox to 23 pt {} Frequency interval(s) & Reported value
 & Rel. stand. \\ 
 & & \hbox to 10 pt {} & 
\hbox to 10pt{} $\nu$/kHz & uncert. $u_{\rm r}$ \\
\noalign{\vbox to 5 pt {}}
\colrule
\noalign{\vbox to 5 pt {}}
\cite{2004111} & MPQ
& $\nu_{\rm H}({\rm 1S_{1/2}}-{\rm 2S_{1/2}})$ 
& $ 2\,466\,061\,413\,187.074(34)$ & $ 1.4\times 10^{-14}$ \\ 
\vsp\cite{1995159}  & MPQ
& $\nu_{\rm H}({\rm 2S_{1/2}}-{\rm 4S_{1/2}})
- {1\over4}\nu_{\rm H}({\rm 1S_{1/2}}-{\rm 2S_{1/2}})$
& $ 4\,797\,338(10)$ & $ 2.1\times 10^{-6}$ \\
&    
& $\nu_{\rm H}({\rm 2S_{1/2}}-{\rm 4D_{5/2}})
- {1\over4}\nu_{\rm H}({\rm 1S_{1/2}}-{\rm 2S_{1/2}})$
 & $ 6\,490\,144(24)$ & $ 3.7\times 10^{-6}$ \\
&& $\nu_{\rm D}({\rm 2S_{1/2}}-{\rm 4S_{1/2}})
- {1\over4}\nu_{\rm D}({\rm 1S_{1/2}}-{\rm 2S_{1/2}})$
& $ 4\,801\,693(20)$ & $ 4.2\times 10^{-6}$ \\
&    
& $\nu_{\rm D}({\rm 2S_{1/2}}-{\rm 4D_{5/2}}) 
- {1\over4}\nu_{\rm D}({\rm 1S_{1/2}}-{\rm 2S_{1/2}})$
 & $ 6\,494\,841(41)$ & $ 6.3\times 10^{-6}$ \\
\vsp\cite{1998002} & MPQ
& $\nu_{\rm D}({\rm 1S_{1/2}} -{\rm 2S_{1/2}}) 
- \nu_{\rm H}({\rm 1S_{1/2}} - {\rm 2S_{1/2}})$ 
& $ 670\,994\,334.64(15)$ & $ 2.2\times 10^{-10}$ \\

\vsp\cite{1997001} & LKB/SYRTE
& $\nu_{\rm H}({\rm 2S_{1/2}}-{\rm 8S_{1/2}})$ & 
$ 770\,649\,350\,012.0(8.6)$ & $ 1.1\times 10^{-11}$ \\ 
& & $\nu_{\rm H}({\rm 2S_{1/2}}-{\rm 8D_{3/2}})$ & 
$ 770\,649\,504\,450.0(8.3)$ & $ 1.1\times 10^{-11}$ \\ 
& & $\nu_{\rm H}({\rm 2S_{1/2}}-{\rm 8D_{5/2}})$ & 
$ 770\,649\,561\,584.2(6.4)$ & $ 8.3\times 10^{-12}$ \\ 
& & $\nu_{\rm D}({\rm 2S_{1/2}}-{\rm 8S_{1/2}})$ & 
$ 770\,859\,041\,245.7(6.9)$ & $ 8.9\times 10^{-12}$ \\ 
& & $\nu_{\rm D}({\rm 2S_{1/2}}-{\rm 8D_{3/2}})$ & 
$ 770\,859\,195\,701.8(6.3)$ & $ 8.2\times 10^{-12}$ \\ 
& & $\nu_{\rm D}({\rm 2S_{1/2}}-{\rm 8D_{5/2}})$ & 
$ 770\,859\,252\,849.5(5.9)$ & $ 7.7\times 10^{-12}$ \\ 

\vsp\cite{1999072,1999072e} & LKB/SYRTE
& $\nu_{\rm H}({\rm 2S_{1/2}}-{\rm 12D_{3/2}})$ & 
$ 799\,191\,710\,472.7(9.4)$ & $ 1.2\times 10^{-11}$ \\ 
& & $\nu_{\rm H}({\rm 2S_{1/2}}-{\rm 12D_{5/2}})$ & 
$ 799\,191\,727\,403.7(7.0)$ & $ 8.7\times 10^{-12}$ \\ 
& & $\nu_{\rm D}({\rm 2S_{1/2}}-{\rm 12D_{3/2}})$ & 
$ 799\,409\,168\,038.0(8.6)$ & $ 1.1\times 10^{-11}$ \\ 
& & $\nu_{\rm D}({\rm 2S_{1/2}}-{\rm 12D_{5/2}})$ & 
$ 799\,409\,184\,966.8(6.8)$ & $ 8.5\times 10^{-12}$ \\ 

\vsp \cite{1996001} & LKB
& $\nu_{\rm H}({\rm 2S_{1/2}}-{\rm 6S_{1/2}})
- {1\over4}\nu_{\rm H}({\rm 1S_{1/2}}-{\rm 3S_{1/2}})$
& $ 4\,197\,604(21)$ & $ 4.9\times 10^{-6}$ \\ 
&& $\nu_{\rm H}({\rm 2S_{1/2}}-{\rm 6D_{5/2}})
- {1\over4}\nu_{\rm H}({\rm 1S_{1/2}}-{\rm 3S_{1/2}})$
& $ 4\,699\,099(10)$ & $ 2.2\times 10^{-6}$ \\ 

\vsp\cite{1995138} & Yale 
& $\nu_{\rm H}({\rm 2S_{1/2}}-{\rm 4P_{1/2}})
- {1\over4}\nu_{\rm H}({\rm 1S_{1/2}}-{\rm 2S_{1/2}})$
& $ 4\,664\,269(15)$ & $ 3.2\times 10^{-6}$ \\ 
&& $\nu_{\rm H}({\rm 2S_{1/2}}-{\rm 4P_{3/2}})
- {1\over4}\nu_{\rm H}({\rm 1S_{1/2}}-{\rm 2S_{1/2}})$
& $ 6\,035\,373(10)$ & $ 1.7\times 10^{-6}$ \\ 

\vsp\cite{1994090} & Harvard
& $\nu_{\rm H}({\rm 2S_{1/2}}-{\rm 2P_{3/2}})$
& $ 9\,911\,200(12)$ & $ 1.2\times 10^{-6}$ \\ 

\vsp\cite{1986003} & Harvard
& $\nu_{\rm H}({\rm 2P_{1/2}}-{\rm 2S_{1/2}})$
& $ 1\,057\,845.0(9.0)$ & $ 8.5\times 10^{-6}$ \\ 

\vsp\cite{1979001} & U. Sussex
& $\nu_{\rm H}({\rm 2P_{1/2}}-{\rm 2S_{1/2}})$
& $ 1\,057\,862(20)$ & $ 1.9\times 10^{-5}$ \\
\botrule
\end{tabular}
\end{table*}

The new MPQ result,
\begin{eqnarray}
\nu_{\rm H}({\rm 1S_{1/2}}-{\rm 2S_{1/2}})
&=&  2\,466\,061\,413\,187.074(34)~{\rm kHz}
\nonumber\\&&\qquad\qquad
[ 1.4\times 10^{-14}] ,
\label{eq:g12h04}
\end{eqnarray}
was obtained in the course of an experiment to search for a temporal variation
of the fine-structure constant $\alpha$ \cite{2004111, 2004049, 2005082,
pc06tu}. It is consistent with, but has a somewhat smaller uncertainty than,
the previous result from the MPQ group, $\nu_{\rm H}({\rm 1S_{1/2}}-{\rm
2S_{1/2}})=2.466\,061\,413\,187.103(46)~\rm kHz~[1.9\times10^{-14}]$
\cite{2000034}, which was the value used in the 2002 adjustment. The
improvements that led to the reduction in uncertainty include a more stable
external reference cavity for locking the 486\,\,nm cw dye laser, thereby
reducing its linewidth; an upgraded vacuum system that lowered the background
gas pressure in the interaction region, thereby reducing the background gas
pressure shift and its associated uncertainty; and a significantly reduced
within-day Type A ({\it i.e.,} statistical) uncertainty due to the narrower laser
linewidth and better signal-to-noise ratio.

The MPQ result in Eq.~(\ref{eq:g12h04}) and Table~\ref{tab:rydfreq} for
$\nu_{\rm H}({\rm 1S_{1/2}}-{\rm 2S_{1/2}})$ was provided by \textcite{pc06tu}
of the MPQ group. It follows from the measured value $\nu_{\rm H}({\rm
1S_{1/2}}-{\rm 2S_{1/2}})= 2.466\,061\,102\,474.851(34)~\rm kHz~[1.4 \times
10^{-14}]$ obtained for the $(1{\rm S},\, F=1,\, m_F =\pm
1)\longrightarrow(2{\rm S},\, F^{\prime}=1,\, m_F^{\prime} =\pm 1)$ transition
frequency \cite{2004111, 2004049, 2005082} by using the well known 1S and 2S
hyperfine splittings \cite{1990021, 2004006} to convert it to the frequency
corresponding to the hyperfine centroid.

\subsubsection{Nuclear radii}
\label{sssec:nucrad}

The theoretical expressions for the finite nuclear size correction to the
energy levels of hydrogen H and deuterium D (see Sec.~\ref{par:nucsize}) are
functions of the bound-state nuclear rms charge radius for the proton, $R_{\rm
p}$, and for the deuteron, $R_{\rm d}$.  These values are treated as variables
in the adjustment, so the transition frequency data, together with theory,
determine values for the radii.  The radii are also determined by elastic
electron-proton scattering data in the case of $R_{\rm p}$ and from elastic
electron-deuteron scattering data in the case of $R_{\rm d}$.  These
independently determined values are used as additional information on the
radii.  There have been no new results during the last 4 years and thus we take
as input data for these two radii the values used in the 2002 adjustment:
\begin{eqnarray}
R_{\rm p} &=&  0.895(18) \ {\rm fm} 
\label{eq:rp}\\
R_{\rm d} &=&  2.130(10) \ {\rm fm}. 
\label{eq:rd}
\end{eqnarray}
The result for $R_{\rm p}$ is due to \textcite{2003222} [see also
\textcite{2007137}]. The result for $R_{\rm d}$ is that given in Sec.~III.B.7
of CODATA-98 based on the analysis of \textcite{1998074}.

An experiment currently underway to measure the Lamb shift in muonic hydrogen
may eventually provide a significantly improved value of $R_{\rm p}$ and hence
an improved value of $R_\infty$ \cite{2007129}.

\subsection{Antiprotonic helium transition frequencies and $\bm{A_{\rm r}({\rm
e})}$}
\label{ssec:aph}

The antiprotonic helium atom is a three-body system consisting of a $^4$He or
$^3$He nucleus, an antiproton, and an electron, denoted by ${\rm \bar p}\,\rm
He^+$.  Even though the Bohr radius for the antiproton in the field of the
nucleus is about 1836 times smaller than the electron Bohr radius, in the
highly-excited states studied experimentally, the average orbital radius of the
antiproton is comparable to the electron Bohr radius, giving rise to relatively
long-lived states.  Also, for the high-$l$ states studied, because of the
vanishingly small overlap of the antiproton wavefunction with the helium
nucleus, strong interactions between the antiproton and the nucleus are
negligible.  

One of the goals of the experiments is to measure the antiproton-electron mass
ratio.  However, since we assume that $CPT$ is a valid symmetry, for the
purpose of the least squares adjustment we take the masses of the antiproton
and proton to be equal and use the data to determine the proton-electron mass
ratio.  Since the proton mass is known more accurately than the electron mass
from other experiments, the mass ratio yields information primarily on the
electron mass.  Other experiments have demonstrated the equality of the
charge-to-mass ratio of p and $\bar{\rm p}$ to within 9 parts in $10^{11}$; see
\textcite{2006037}.

\subsubsection{Theory relevant to antiprotonic helium}
\label{sssec:aphelth}

Calculations of transition frequencies of antiprotonic helium have been done by
\textcite{2003333} and by \textcite{2003324,2005330}.  The uncertainties of
calculations by \textcite{2005330} are of the order of 1 MHz to 2 MHz, while
the uncertainties and scatter relative to the experimental values of the
results of \textcite{2003333} are substantially larger, so we use the results
\textcite{2005330} in the 2006 adjustment.  [See also the remarks in
\textcite{2007139} concerning the theory.]

The dominant contribution to the energy levels is just the non-relativistic
solution of the Schr\"odinger equation for the three-body system together with
relativistic and radiative corrections treated as perturbations.  The
nonrelativistic levels are resonances, because the states can decay by the
Auger effect in which the electron is ejected.  \textcite{2005330} calculates
the nonrelativistic energy by using one of two formalisms, depending on whether
the Auger rate is small or large.  In the case where the rate is small, the
Feshbach formalism is used with an optical potential.  The optical potential is
omitted in the calculation of higher-order relativistic and radiative
corrections.  For broad resonances with a higher Auger rate, the
nonrelativistic energies are calculated with the Complex Coordinate rotation
method.  In checking the convergence of the nonrelativistic levels, attention
was paid to the convergence of the expectation value of the the delta function
operators used in the evaluation of the relativistic and radiative corrections.

\textcite{2005330} evaluated the relativistic and radiative corrections as
perturbations to the nonrelativistic levels, including relativistic corrections
of order $\alpha^2R_\infty$, anomalous magnetic moment corrections of order
$\alpha^3R_\infty$ and higher, one-loop self-energy and vacuum-polarization
corrections of order $\alpha^3R_\infty$, higher-order one-loop and leading
two-loop corrections of order $\alpha^4R_\infty$.  Higher-order relativistic
corrections of order $\alpha^4R_\infty$ and radiative corrections of order
$\alpha^5R_\infty$ were estimated with effective operators.  The uncertainty
estimates account for uncalculated terms of order
$\alpha^5\ln{\alpha}\,R_\infty$.

Transition frequencies obtained by \textcite{2005330,pc06vk} using the
CODATA-02 values of the relevant constants are listed in Table~\ref{tab:aphe}
under the column header ``Calculated Value.''  We denote these values of the
frequencies by $\nu_{\bar{\rm p}\,{\rm
He}}^{(0)}(n,l:n^\prime,l^\prime)$, where He is either $^3$He$^+$ or
$^4$He$^+$.  Also calculated are the leading-order changes in the theoretical
values of the transition frequencies as a function of the relative changes in
the mass ratios $A_{\rm r}({\rm \bar p})/A_{\rm r}({\rm e})$ and $A_{\rm
r}(N)/A_{\rm r}({\rm \bar p})$; here $N$ is either $^3$He$^{2+}$ or
$^4$He$^{2+}$.  If we denote the transition frequencies as functions of these
mass ratios by $\nu_{\bar{\rm p}\,{\rm He}}(n,l:n^\prime,l^\prime)$, then
the changes can be written as
\begin{eqnarray}
 a_{\rm \bar p\,{\rm He}}(n,l:n^\prime,l^\prime) 
&=& 
\left(\frac{A_{\rm r}({\rm \bar p})}{A_{\rm r}({\rm e})}\right)^{(0)}
\frac{\partial \nu_{\rm \bar p\,{\rm He}}(n,l:n^\prime,l^\prime)}
{\partial \left(\frac{A_{\rm r}({\rm \bar p})}{A_{\rm r}({\rm e})}\right)}
\nonumber \\ \\
 b_{\rm \bar p\,{\rm He}}(n,l:n^\prime,l^\prime) 
&=&\left(\frac{A_{\rm r}({He})}{A_{\rm r}({\rm \bar p})}\right)^{(0)}
\frac{\partial \nu_{\rm \bar p\,{\rm He}}(n,l:n^\prime,l^\prime)}
{\partial \left(\frac{A_{\rm r}({N})}{A_{\rm r}({\rm \bar p})}\right)} \ . 
\nonumber \\ 
\end{eqnarray}
Values of these derivatives, in units of $2cR_\infty$, are listed in
Table~\ref{tab:aphe} in the columns with the headers ``$a$'' and ``$b$,''
respectively.  The zero-order frequencies and the derivatives are used in the
expression
\begin{eqnarray}
&& \hskip -10 pt  \nu_{\rm \bar p\,{\rm He}}\,(n,l:n^\prime,l^\prime)
= \nu_{\rm \bar p\,{\rm He}}^{(0)}(n,l:n^\prime,l^\prime)
\nonumber\\ &&
+ a_{\rm \bar p\,{\rm He}}(n,l:n^\prime,l^\prime)
\left[\left(\frac{A_{\rm r}({\rm e})}{A_{\rm r}({\rm \bar p)}}\right)^{\!(0)} \!\!
\left(\frac{A_{\rm r}({\rm \bar p}\,)}{A_{\rm r}({\rm e})}\right)-1 \right]
\label{eq:apheth}
\\ &&
+ b_{\bar{\rm p}\,{\rm He}}(n,l:n^\prime,l^\prime)
\left[\left(\frac{A_{\rm r}({\rm \bar p})}{A_{\rm r}({N)}}\right)^{\!(0)} \!\!
\left(\frac{A_{\rm r}({N})}{A_{\rm r}({\rm \bar p})}\right)-1 \right] + \dots \, ,
\nonumber
\end{eqnarray}
which provides a first-order approximation to the transition frequencies as a
function of changes to the mass ratios.  This expression is used to incorporate
the experimental data and the calculations for the antiprotonic system as a
function of the mass ratios into the least-squares adjustment.  It should be
noted that even though the mass ratios are the independent variables in
Eq.~(\ref{eq:apheth}) and the atomic relative masses $A_{\rm r}({\rm e})$,
$A_{\rm r}({\rm p}\,)$, and $A_{\rm r}({N)}$ are the adjusted constants in the
2006 least-squares adjustment, the primary effect of including this data in the
adjustment is on the electron relative atomic mass, because independent data in
the adjustment constrains the proton and helium nuclei relative atomic masses
with smaller uncertainties.

The uncertainties in the theoretical expressions for the transition frequencies
are included in the adjustment as additive constants $\delta_{\rm \bar
p\,He}(n,l:n^\prime,l^\prime)$.  Values for the theoretical uncertainties and
covariances used in the adjustment are given in Sec.~\ref{sec:ad},
Tables~\ref{tab:cdata} and \ref{tab:cdcc}, respectively \cite{pc06vk}.

\begin{table*}
\newcommand\s[1]{\hbox to #1 pt {}}
\caption{Summary of data related to 
the determination of $A_{\rm r}({\rm e})$ from measurements on antiprotonic
helium}
\label{tab:aphe}
\begin{tabular}{c@{\hbox to 40pt {}}D{.}{.}{13.10}D{.}{.}{13.10}D{.}{.}{6.5}D{.}{.}{6.5}}
\toprule
\vbox to 10 pt {}
Transition   & \text{Experimental} \s{-17}& \text{Calculated}\s{-10} 
& a\s{-10}  & b\s{-10} \\
$(n,l)\rightarrow(n^\prime,l^\prime)$& \text{Value (MHz)}\s{-15} &  
\text{Value (MHz)}\s{-15} & (2cR_\infty)\s{-25} & (2cR_\infty)\s{-25}\\
\colrule
\vbox to 10 pt {}
$\bar{\rm p}^4$He$^+$: $(32,31) \rightarrow (31,30)$ &  1\,132\,609\,209(15)\s{-16} &  1\,132\,609\,223.50(82) &  0.2179 &  0.0437  \\
$\bar{\rm p}^4$He$^+$: $(35,33) \rightarrow (34,32)$ &  804\,633\,059.0(8.2)        &  804\,633\,058.0(1.0) &  0.1792 &  0.0360  \\
$\bar{\rm p}^4$He$^+$: $(36,34) \rightarrow (35,33)$ &  717\,474\,004(10)\s{-16} &  717\,474\,001.1(1.2) &  0.1691 &  0.0340  \\
$\bar{\rm p}^4$He$^+$: $(37,34) \rightarrow (36,33)$ &  636\,878\,139.4(7.7)        &  636\,878\,151.7(1.1) &  0.1581 &  0.0317  \\
$\bar{\rm p}^4$He$^+$: $(39,35) \rightarrow (38,34)$ &  501\,948\,751.6(4.4)        &  501\,948\,755.4(1.2) &  0.1376 &  0.0276  \\
$\bar{\rm p}^4$He$^+$: $(40,35) \rightarrow (39,34)$ &  445\,608\,557.6(6.3)        &  445\,608\,569.3(1.3) &  0.1261 &  0.0253  \\
$\bar{\rm p}^4$He$^+$: $(37,35) \rightarrow (38,34)$ &  412\,885\,132.2(3.9)        &  412\,885\,132.8(1.8) &  -0.1640 &  -0.0329  \\
\colrule
\vbox to 10 pt {}
$\bar{\rm p}^3$He$^+$: $(32,31) \rightarrow (31,30)$ &  1\,043\,128\,608(13)\s{-16} &  1\,043\,128\,579.70(91) &  0.2098 &  0.0524  \\
$\bar{\rm p}^3$He$^+$: $(34,32) \rightarrow (33,31)$ &  822\,809\,190(12)\s{-16} &  822\,809\,170.9(1.1) &  0.1841 &  0.0460  \\
$\bar{\rm p}^3$He$^+$: $(36,33) \rightarrow (35,32)$ &  646\,180\,434(12)\s{-16} &  646\,180\,408.2(1.2) &  0.1618 &  0.0405  \\
$\bar{\rm p}^3$He$^+$: $(38,34) \rightarrow (37,33)$ &  505\,222\,295.7(8.2)        &  505\,222\,280.9(1.1) &  0.1398 &  0.0350  \\
$\bar{\rm p}^3$He$^+$: $(36,34) \rightarrow (37,33)$ &  414\,147\,507.8(4.0)        &  414\,147\,509.3(1.8) &  -0.1664 &  -0.0416  \\
\botrule
\end{tabular}
\end{table*}

\subsubsection{Experiments on antiprotonic helium}
\label{sssec:aphelex}

Experimental work on antiprotonic helium began in the early 1990s and it
continues to be an active field of research; a comprehensive review through
2000 is given by \textcite{2002276} and a very concise review through 2006 by
\textcite{2007139}. The first measurements of $\bar{\rm p}\,\rm He^+$
transition frequencies at CERN with $u_{\rm r}<10^{-6}$ were reported in 2001
\cite{2001090}, improved results were reported in 2003 \cite{2003194}, and
transition frequencies with uncertainties sufficiently small that they can,
together with the theory of the transitions, provide a competitive value of
$A_{\rm r}$(e), were reported in 2006 \cite{2006056}. 

The 12 transition frequencies---seven for $^4$He and five for $^3$He given by
\textcite{2006056}---which we take as input data in the 2006 adjustment are
listed in column 2 of Table~\ref{tab:aphe} with the corresponding transitions
indicated in column 1.  To reduce rounding errors, an additional digit for both
the frequencies and their uncertainties as provided by \textcite{pc06mh} have
been included. All twelve frequencies are correlated; their correlation
coefficients, based on detailed uncertainty budgets for each, also provided by
\textcite{pc06mh}, are given in Table~\ref{tab:cdcc} in Sec~\ref{sec:ad}.

In the current version of the experiment, 5.3~MeV antiprotons from the CERN
Antiproton Decelerator (AD) are decelerated using a radio-frequency quadrupole
decelerator (RFQD) to energies in the range 10~keV to 120~keV controlled by a
dc potential bias on the RFQD's electrodes. The decelerated antiprotons, about
$30~\%$ of the antiprotons entering the RFQD, are then diverted to a low
pressure cryogenic helium gas target at 10~K by an achromatic momentum
analyzer, the purpose of which is to eliminate the large background that the
remaining $70~\%$ of undecelerated antiprotons would have produced.

About $3~\%$ of the $\bar{\rm p}$ stopped in the target form $\bar{\rm p}\,\rm
He^+$, in which a $\bar{\rm p}$ with large principle quantum number
($n\approx{38}$) and angular momentum quantum number ($l\approx {n}$)
circulates in a localized, nearly circular orbit around the He$^{2+}$ nucleus
while the electron occupies the distributed 1S state. These $\bar{\rm p}$
energy levels are metastable with lifetimes of several microseconds and
de-excite radiatively. There are also short lived $\bar{\rm p}$ states with
similar values of $n$ and $l$ but with lifetimes on the order of 10\, ns and
which de-excite by Auger transitions to form $\bar{\rm p}$\,He$^{2+}$
hydrogen-like ions. These undergo Stark collisions, which cause the rapid
annihilation of the $\bar{\rm p}$ in the helium nucleus. The annihilation rate
vs. time elapsed since $\bar{\rm p}\,\rm He^+$ formation, or delayed
annihilation time spectrum (DATS), is measured using Cherenkov counters. 

With the exception of the $(36, 34)\rightarrow(35, 33)$ transition frequency,
all of the frequencies given in Table~\ref{tab:aphe} were obtained by
stimulating transitions from the $\bar{\rm p}\,\rm He^+$ metastable states with
values of $n$ and $l$ indicated in column one on the left-hand side of the
arrow to the short lived, Auger-decaying states with values of $n$ and $l$
indicated on the right-hand side of the arrow. 

The megawatt-scale light intensities needed to induce the ${\rm \bar p}\,\rm
He^+$ transitions, which cover the wavelength range 265~nm to 726~nm, can only
be provided by a pulsed laser. Frequency and linewidth fluctuations and frequency
calibration problems associated with such lasers were overcome by starting with
a cw ``seed'' laser beam of frequency $\nu_{\rm cw}$, known with $u_{\rm
r}<4\times10^{-10}$ through its stabilization by an optical frequency comb, and
then amplifying the intensity of the laser beam by a factor of $10{^6}$ in a cw
pulse amplifier consisting of three dye cells pumped by a pulsed Nd:YAG laser.
The 1~W seed laser beam with wavelength in the range 574~nm to 673~nm was
obtained from a pumped cw dye laser, and the 1~W seed laser beam with
wavelength in the range 723~nm to 941~nm was obtained from a pumped cw
Ti:sapphire laser. The shorter wavelengths (265~nm to 471~nm) for inducing
transitions were obtained by frequency doubling the amplifier output at 575~nm
and 729~nm to 941~nm or by frequency tripling its 794~nm output. The frequency
of the seed laser beam $\nu_{\rm cw}$, and thus the frequency $\nu_{\rm pl}$ of
the pulse amplified beam, was scanned over a range of $\pm4~\rm GHz$ around the
$\bar{\rm p}\,\rm He^+$ transition frequency by changing the repetition
frequency $f_{\rm rep}$ of the frequency comb. 

The resonance curve for a transition was obtained by plotting the area under
the resulting DATS peak vs. $\nu_{\rm pl}$. Because of the approximate 400~MHz
Doppler broadening of the resonance due to the 10~K thermal motion of the
$\bar{\rm p}\,\rm He^+$ atoms, a rather sophisticated theoretical line shape
that takes into account many factors must be used to obtained the desired
transition frequency.

Two other effects of major importance are the so-called chirp effect and linear
shifts in the transition frequencies due to collisions between the $\bar{\rm
p}\,\rm He^+$ and background helium atoms. The frequency $\nu_{\rm pl}$ can
deviate from $\nu_{\rm cw}$ due to sudden changes in the index of refraction of
the dye in the cells of the amplifier. This chirp, which can be expressed as
$\Delta\nu_{\rm c}(t)= \nu_{\rm pl}(t)-\nu_{\rm cw}$, can shift the measured
$\bar{\rm p}\,\rm He^+$ frequencies from their actual values.
\textcite{2006056} eliminated this effect by measuring $\Delta\nu_{\rm c}(t)$
in real time and applying a frequency shift to the seed laser, thereby
canceling the dye-cell chirp. This effect is the predominant contributor to the
correlations among the 12 transitions \cite{pc06mh}.  The collisional shift was
eliminated by measuring the frequencies of ten transitions in helium gas
targets with helium atom densities $\rho$ in the range $2\times10^{18}/\rm
cm{^3}$ to $3\times10^{21}/\rm cm{^3}$ to determine $d\nu/d\rho$.  The $in$
$vacuo$ ($\rho=0$) values were obtained by applying a suitable correction in
the range $-14~\rm MHz$ to $1~\rm MHz$ to the initially measured frequencies
obtained at $\rho \approx2\times10^{18}/\rm cm^3$.

In contrast to the other 11 transition frequencies in Table~\ref{tab:aphe},
which were obtained by inducing a transition from a long-lived, metastable
state to a short-lived, Auger-decaying state, the $(36, 34)\rightarrow(35, 33)$
transition frequency was obtained by inducing a transition from the (36, 34)
metastable state to the (35, 33) metastable state using three different lasers.
This was done by first depopulating at time $t_1$ the (35, 33) metastable state
by inducing the $(35, 33)\rightarrow(34, 32)$ metastable-state to
short-lived-state transition, then at time $t_2$ inducing the $(36,
34)\rightarrow(35, 33)$ transition using the cw pulse-amplified laser, and then
at time $t_3$ again inducing the $(35, 33)\rightarrow(34, 32)$ transition. The
resonance curve for the $(36, 34)\rightarrow(35, 33)$ transition was obtained
from the DATS peak resulting from this last induced transition.

The 4~MHz to 15~MHz standard uncertainties of the transition frequencies in
Table~\ref{tab:aphe} arise from the resonance line shape fit (3~MHz to 13~MHz,
statistical or Type A), not completely eliminating the chirp effect (2~MHz to
4~MHz, nonstatistical or Type B), collisional shifts (0.1~MHz to 2~MHz, Type
B), and frequency doubling or tripling (1~MHz to 2~MHz, Type B). 

\subsubsection{Values of $A_{\rm r}({\rm e})$ inferred from antiprotonic
helium}
\label{sssec:apheare}

From the theory of the 12 antiprotonic transition frequencies discussed in
Sec~\ref{sssec:aphelth}, the 2006 recommended values of the relative atomic
masses of the proton, alpha particle (nucleus of the $^4$He atom), and the
helion (nucleus of the $^3$He atom), $A_{\rm r}({\rm p})$, $A_{\rm r}({\rm
alpha})$, and $A_{\rm r}({\rm h})$, respectively, together with the 12
experimental values for these frequencies given in Table~\ref{tab:aphe}, we
find the following three values for $A_{\rm r}({\rm e})$ from the seven
$\bar{\rm p}\,^4\rm He^+$ frequencies alone, from the five $\bar{\rm p}\,^3\rm
He^+$ frequencies alone, and from the 12 frequencies together:
\begin{eqnarray}
A_{\rm r}({\rm e}) &=&   0.000\,548\,579\,9103(12) \quad [ 2.1\times 10^{-9}] 
\label{eq:areaphe4}
\\
A_{\rm r}({\rm e}) &=&   0.000\,548\,579\,9053(15) \quad [ 2.7\times 10^{-9}]
\label{eq:areaphe3}
\\
A_{\rm r}({\rm e}) &=&   0.000\,548\,579\,908\,81(91) \quad [ 1.7\times 10^{-9}] \, . \qquad
\label{eq:areaphe}
\end{eqnarray}
The separate inferred values from the $\bar{\rm p}\,^4\rm He^+$ and $\bar{\rm
p}\,^3\rm He^+$ frequencies differ somewhat, but the value from all 12
frequencies not only agrees with the three other available results for $A_{\rm
r}({\rm e})$ (see Table~\ref{tab:ares}, Sec~\ref{ssec:cod}), but has a
competitive level of uncertainty as well.

\subsection{Hyperfine structure and fine structure}
\label{ssec:hfsfs}

\subsubsection{Hyperfine structure}
\label{sssec:hfs}

Because the ground-state hyperfine transition frequencies $\Delta\nu_{\rm H}$,
$\Delta\nu_{\rm Mu}$, and $\Delta\nu_{\rm Ps}$ of the comparatively simple
atoms hydrogen, muonium, and positronium, respectively, are proportional to
$\alpha^{2}R_{\infty}c$, in principle a value of $\alpha$ can be obtained by
equating an experimental value of one of these transition frequencies to its
presumed readily calculable theoretical expression. However, currently only
measurements of $\Delta\nu_{\rm Mu}$ and the theory of the muonium hyperfine
structure have sufficiently small uncertainties to provide a useful result for
the 2006 adjustment, and even in this case the result is not a competitive
value of $\alpha$, but rather the most accurate value of the electron-muon mass
ratio $m_{\rm e}/m_{\rmssmu}$. Indeed, we discuss the relevant experiments and
theory in Sec.\ref{ssec:muhfs}.

Although the ground-state hyperfine transition frequency of hydrogen has long
been of interest as a potential source of an accurate value of $\alpha$ because
it is experimentally known with $u_{\rm r} \approx 10^{-12}$ \cite{1990021},
the relative uncertainty of the theory is still of the order of $10^{-6}$.
Thus, $\Delta\nu_{\rm H}$ cannot yet provide a competitive value of the
fine-structure constant. At present, the main sources of uncertainty in the
theory arise from the internal structure of the proton, namely (i) the electric
charge and magnetization densities of the proton, which are taken into account
by calculating the proton's so-called Zemach radius; and (ii) the
polarizability of the proton (that is, protonic excited states).  For details
of the progress made over the last four years in reducing the uncertainties
from both sources, see \cite{2007136, 2007183, pc07is} and the references cited
therein. Because the muon is a structureless point-like particle, the theory of
$\Delta\nu_{\rm Mu}$ is free of such uncertainties. 


It is also not yet possible to obtain a useful value of $\alpha$ from
$\Delta\nu_{\rm Ps}$ since the most accurate experimental result has $u_{\rm r}
= 3.6\times 10^{-6}$ \cite{1984031}.  The uncertainty of the theory of
$\Delta\nu_{\rm Ps}$ is not significantly smaller and may in fact be larger
\cite{2004246, 2002086}.

\subsubsection{Fine structure}
\label{sssec:fs}

As in the case of hyperfine splittings, fine-structure transition frequencies
are proportional to $\alpha^2R_{\infty}c$ and could be used to deduce a value
of $\alpha$.  Some data related to the fine structure of hydrogen and deuterium
are discussed in Sec.~\ref{sssec:rydex} in connection with the Rydberg
constant.  They are included in the adjustment because of their influence on
the adjusted value of $R_\infty$.  However, the value of $\alpha$ that can be
derived from these data is not competitive; see Eq.~(\ref{eq:alphinvhd}). See
also Sec.~III.B.3 of CODATA-02 for a discussion of why earlier fine
structure-related results in H and D are not considered.

Because the transition frequencies corresponding to the differences in energy
of the three 2$^3$P levels of $^4$He can be both measured and calculated with
reasonable accuracy, the fine structure of $^4$He has long been viewed as a
potential source of a reliable value of $\alpha$.  The three frequencies of
interest are $\nu_{01} \approx 29.6$~GHz, $\nu_{12}\approx2.29$~GHz, and
$\nu_{02} \approx 31.9$~GHz, which correspond to the intervals $2^{3}{\rm
P}_{1} - 2^{3}{\rm P}_{0}, 2^{3}{\rm P}_{2} - 2^{3}{\rm P}_{1}$, and $2^{3}{\rm
P}_{2} - 2^{3}{\rm P}_{0}$, respectively. The value with the smallest
uncertainty for any of these frequencies was obtained at Harvard
\cite{2005131}:
\begin {eqnarray}
\nu_{01} = 29\, 616\, 951.66(70)\ {\rm kHz}\qquad [2.4\times 10^{-8}]\ .
\label{eq:nu01hav05}
\end{eqnarray}
It is consistent with the value of $\nu_{01}$ reported by \textcite{2001254}
with $u_{\rm r}=3.0\times 10^{-8}$, and that reported by \textcite{2005145}
with $u_{\rm r}=3.4\times 10^{-8}$.  If the theoretical expression for
$\nu_{01}$ were exactly known, the weighted mean of the three results would
yield a value of $\alpha$ with $u_{\rm r} \approx 8\times 10^{-9}$.

However, as discussed in CODATA-02, the theory of the $2^{3}{\rm P}_{J}$
transition frequencies is far from satisfactory. First, different calculations
disagree, and because of the considerable complexity of the calculations and
the history of their evolution, there is general agreement that results that
have not been confirmed by independent evaluation should be taken as tentative.
Second, there are significant disagreements between theory and experiment.
Recently, \textcite{2006053} has advanced the theory by calculating the
complete contribution to the $2^{3}{\rm P}_{J}$ fine-structure levels of order
$m\alpha^7$ (or $\alpha^5$ Ryd), with the final theoretical result for
$\nu_{01}$ being 
\begin {eqnarray}
\nu_{01} = 29\, 616\, 943.01(17)\ {\rm kHz}\qquad [5.7\times 10^{-9}]\ .
\end{eqnarray}
This value disagrees with the experimental value given in
Eq~(\ref{eq:nu01hav05}) as well as with the theoretical value $\nu_{01} = 29\,
616\, 946.42(18)\ {\rm kHz}$ $[6.1\times 10^{-9}]$ given by \textcite{2002166},
which also disagrees with the experimental value. These disagreements suggest
that there is a problem with theory and/or experiment which must be resolved
before a meaningful value of $\alpha$ can be obtained from the helium fine
structure \cite{2006053}. Therefore, as in the 2002 adjustment, we do not
include $^4$He fine-structure data in the 2006 adjustment.

\section{Magnetic moment anomalies and $\bm g$-factors}
\label{sec:mmagf}

In this section, the theory and experiment for the magnetic moment anomalies of
the free electron and muon and the bound-state $g$-factor of the electron in
hydrogenic carbon ($^{12}$C$^{5+}$) and in hydrogenic oxygen ($^{16}$O$^{7+}$)
are reviewed.  

\def\m{\phantom{-}}
\begin{table*}
\caption{Summary of data related to magnetic moments of the electron and muon
and inferred values of the fine structure constant.  (The source data and not
the inferred values given here are used in the adjustment.)}
\label{tab:gfree}
\begin{tabular}{llc@{\quad}l@{\quad}l}
\toprule
\vbox to 10 pt {}
Quantity     &  \hbox to 12pt {} Value
& Relative standard
& Identification & Sect. and Eq. \\
&& uncertainty $u_{\rm r}$&& \\
\colrule
$a_{\rm e}$ \vbox to 12pt {}& $\phantom{-} 1.159\,652\,1883(42)\times 10^{-3}$ & $ 3.7\times 10^{-9}$ & 
    UWash-87 & 
\ref{par:aeuw} (\ref{eq:aeuwash}) \\
\ \ $\alpha^{-1}(a_{\rm e})$ &\ \ $\phantom{-} 137.035\,998\,83(50)$&$ 3.7\times 10^{-9}$& &
    \ref{sssec:alphaae} (\ref{eq:alphinvuwash87}) \\
&&&&\\

$a_{\rm e}$ & $\phantom{-} 1.159\,652\,180\,85(76)\times 10^{-3}$ & $ 6.6\times 10^{-10}$ & 
    HarvU-06 & 
\ref{par:harvard} (\ref{eq:aeharv06}) \\
\ \ $\alpha^{-1}(a_{\rm e})$ &\ \ $\phantom{-} 137.035\,999\,711(96)$&$ 7.0\times 10^{-10}$& &
    \ref{sssec:alphaae} (\ref{eq:alphinvharvu06}) \\
&&&&\\

$\overline{R}$ &
   $\phantom{-} 0.003\,707\,2064(20)$ & $ 5.4\times 10^{-7}$ & 
    BNL-06 &
    \ref{sssec:amb} (\ref{eq:rbar06}) \\
\ \ $a_{\rmssmu}$ & \ \ $\phantom{-} 1.165\,920\,93(63)\times 10^{-3}$ & $ 5.4\times 10^{-7}$ & & 
\ref{sssec:amb} (\ref{eq:amu06}) \\
\ \ $\alpha^{-1}(\overline{R})$ &\ \ $\phantom{-} 137.035\,67(26)$&$ 1.9\times 10^{-6}$& &
\ref{par:tamual} (\ref{eq:alphiam06}) \\

\botrule
\end{tabular}
\end{table*}

The magnetic moment of any of the three charged leptons $\ell={\rm
e},\,\rmmu,\,\rmtau$ is written as
\begin{eqnarray}
\bm\mu_\ell = g_\ell { e \over 2m_\ell}\bm s \ ,
\label{eq:lgdef}
\end{eqnarray}
where $g_\ell$ is the $g$-factor of the particle, $m_\ell$ is its mass, and
$\bm s$ is its spin.  In Eq.~(\ref{eq:lgdef}), $e$ is the elementary charge and
is positive.  For the negatively charged leptons $\ell^{\,-}$, $g_\ell$ is
negative, and for the corresponding antiparticles $\ell^{\,+}$, $g_\ell$ is
positive.  $CPT$ invariance implies that the masses and absolute values of the
$g$-factors are the same for each particle-antiparticle pair.  These leptons
have eigenvalues of spin projection $s_z = \pm \hbar/2$, and it is conventional
to write, based on Eq.~(\ref{eq:lgdef}),
\begin{eqnarray}
\mu_{\rm \ell} = {g_{\ell}\over2} \,\frac{e\hbar}{2m_\ell} \ ,
\label{eq:gedef}
\end{eqnarray}
where in the case of the electron, $\mu_{\rm B} = e\hbar/2m_{\rm e}$ is the
Bohr magneton.

The free lepton magnetic moment anomaly $a_\ell$ is defined as
\begin{eqnarray}
|g_\ell| &=& 2(1+a_\ell) \ ,
\label{eq:adef}
\end{eqnarray}
where $g_{\rm D} = -2$ is the value predicted by the free-electron Dirac
equation.  The theoretical expression for $a_\ell$ may be written as
\begin{eqnarray}
a_\ell({\rm th}) = a_\ell({\rm QED}) + a_\ell({\rm weak})
 + a_\ell({\rm had}) \ ,
\label{eq:aeth}
\end{eqnarray}
where the terms denoted by QED, weak, and had account for the purely quantum
electrodynamic, predominantly electroweak, and predominantly hadronic (that is,
strong interaction) contributions to $a_\ell$, respectively.  

The QED contribution may be written as \cite{1990005}
\begin{eqnarray}
a_\ell({\rm QED})&=& A_1 + A_2(m_\ell/m_{\ell^\prime}) 
+ A_2(m_\ell/m_{\ell^{\prime\prime}})
\nonumber\\
&&+ A_3(m_\ell/m_{\ell^\prime},m_\ell/m_{\ell^{\prime\prime}}) \ ,
\label{eq:aqed}
\end{eqnarray}
where for the electron, $(\ell,\,\ell^\prime,\,\ell^{\prime\prime}) =({\rm
e},\,{\rmmu},\,{\rmtau})$, and for the muon,
$(\ell,\,\ell^\prime,\,\ell^{\prime\prime}) =({\rmmu},\,{\rm e},\,{\rmtau})$.
The anomaly for the $\rmtau$, which is poorly known experimentally
\cite{2006110}, is not considered here.  For recent work on the theory of
$a_{\rmsstau}$, see \textcite{2007181}.  In Eq.~(\ref{eq:aqed}), the term $A_1$
is mass independent, and the mass dependence of $A_2$ and $A_3$ arises from
vacuum polarization loops with lepton $\ell^\prime$, $\ell^{\prime\prime}$, or
both.  Each of the four terms on the right-hand side of Eq.~(\ref{eq:aqed}) can
be expressed as a power series in the fine-structure constant $\alpha$:
\begin{eqnarray}
A_i &=& A_i^{(2)}\left({\alpha\over\rmpi}\right)
+A_i^{(4)}\left({\alpha\over\rmpi}\right)^2
+A_i^{(6)}\left({\alpha\over\rmpi}\right)^3
\nonumber\\
&&+A_i^{(8)}\left({\alpha\over\rmpi}\right)^4
+A_i^{(10)}\left({\alpha\over\rmpi}\right)^5
+\cdots \ ,
\label{eq:aeqedsal}
\end{eqnarray}
where $A_2^{(2)}= A_3^{(2)} = A_3^{(4)}=0$.  Coefficients proportional to
$(\alpha/\rmpi)^n$ are of order $e^{2n}$ and are referred to as $2n$th-order
coefficients.

The second-order coefficient is known exactly, and the fourth- and sixth-order
coefficients are known analytically in terms of readily evaluated functions:
\begin{eqnarray}
A_1^{(2)} &=& \fr{1}{2}
\label{eq:a12}
\\
A_1^{(4)} &=&  -0.328\,478\,965\,579\ldots
\label{eq:a14}
\\
A_1^{(6)} &=&  1.181\,241\,456\ldots \ .
\label{eq:a16}
\end{eqnarray}

A total of 891 Feynman diagrams give rise to the mass-independent eighth-order
coefficient $A_1^{(8)}$, and only a few of these are known analytically.
However, in an effort that has its origins in the 1960s, Kinoshita and
collaborators have calculated all of $A_1^{(8)}$ numerically, with the result
of this ongoing project that was used in the 2006 adjustment being
\cite{2006003, 2006080, 2006080e}
\begin{eqnarray}
A_1^{(8)} =  -1.7283(35) \ . 
\label{eq:a18}
\end{eqnarray}

Work was done in the evaluation and checking of this coefficient in an effort
to obtain a reliable quantitative result.  A subset of 373 diagrams containing
closed electron loops was verified by more than one independent formulation.
The remaining 518 diagrams with no closed electron loops were formulated in
only one way.  As a check on this set, extensive cross checking was performed
on the renormalization terms both among themselves and with lower-order
diagrams that are known exactly \cite{2006003} [see also \textcite{2006080,
2006080e}].  For the final numerical integrations, an adaptive-iterative Monte
Carlo routine was used.  A time-consuming part of the work was checking for
round-off error in the integration.

The $ 0.0035$ standard uncertainty of $A_1^{(8)}$ contributes a standard
uncertainty to $a_{\rm e}({\rm th})$ of $ 0.88\times 10^{-10}\,a_{\rm e}$,
which is smaller than the uncertainty due to uncalculated higher-order
contributions.  Independent work is in progress on analytic calculations of
eighth-order integrals.  See, for example, \textcite{2001339,2001106,2004210}.

Little is known about the tenth-order coefficient $A_1^{(10)}$ and higher-order
coefficients, although \textcite{2006374} are starting the numerical evaluation
of the 12\,672 Feynman diagrams for this coefficient.  To evaluate the
contribution to the uncertainty of $a_{\rm e}({\rm th})$ due to lack of
knowledge of $A_1^{(10)}$, we follow CODATA-98 to obtain $A_1^{(10)} = 0.0(
3.7)$.  The $ 3.7$ standard uncertainty of $A_1^{(10)}$ contributes a standard
uncertainty component to $a_{\rm e}({\rm th})$ of $ 2.2\times 10^{-10} \,
a_{\rm e}$; the uncertainty contributions to $a_{\rm e}({\rm th})$ from all
other higher-order coefficients, which should be significantly smaller, are
assumed to be negligible.

The 2006 least-squares adjustment was carried out using the theoretical results
given above, including the value of $A_1^{(8)}$ given in Eq.~(\ref{eq:a18}).
Well after the deadline for new data and the recommended values from the
adjustment were made public \cite{web}, it was discovered by \textcite{2007167}
that 2 of the 47 integrals representing 518 QED diagrams that had not
previously been confirmed independently required a corrected treatment of
infrared divergences.  The revised value they give is
\begin{eqnarray}
A_1^{(8)} &=& -1.9144(35) \ ,
\end{eqnarray}
although the new calculation is still tentative \cite{2007167}.  This result
would lead to the value 
\begin{eqnarray}
\alpha^{-1} &=& 137.035\,999\,070(98) \qquad[7.1\times10^{-10}]
\end{eqnarray}
for the inverse fine-structure constant derived from the electron anomaly using
the Harvard measurement result for $a_{\rm e}$ \cite{2006080, 2006080e}.  This
number is shifted down from the previous result by $641\times10^{-9}$ and its
uncertainty is increased from (96) to (98) (see Sec.~\ref{sssec:alphaae}), but
it is still consistent with the values obtained from recoil experiments (see
Table~\ref{tab:dhom}).  If this result for $A_1^{(8)}$ had been used in the
2006 adjustment, the recommended value of the inverse fine-structure constant
would differ by a similar, although slightly smaller, change.  The effect on
the muon anomaly theory is completely negligible.

The mass independent term $A_1$ contributes equally to the free electron and
muon anomalies and the bound-electron $g$-factors.  The mass-dependent terms
are different for the electron and muon and are considered separately in the
following.  For the bound-electron $g$-factor, there are bound-state
corrections in addition to the free-electron value of the $g$-factor, as
discussed below.

\subsection{Electron magnetic moment anomaly $\bm{a_{\rm e}}$ and the
fine-structure constant $\bm{\alpha$}}
\label{ssec:emma}

The combination of theory and experiment for the electron magnetic moment
anomaly yields the value for the fine-structure constant $\alpha$ with the
smallest estimated uncertainty (see Table~\ref{tab:gfree} for the values
corresponding to the 2006 adjustment).

\subsubsection{Theory of $a_{\rm e}$}
\label{sssec:ath}

The mass-dependent coefficients of interest and corresponding contributions to
the theoretical value of the anomaly $a_{\rm e}({\rm th})$, based on the 2006
recommended values of the mass ratios, are

\begin{eqnarray}
A_2^{(4)}\!(m_{\rm e}/m_{\rmssmu}) &=&  5.197\,386\,78(26)\times 10^{-7}
\nonumber\\
&&\quad\rightarrow 
 24.182\times 10^{-10}a_{\rm e}
\\
\nonumber\\
A_2^{(4)}\!(m_{\rm e}/m_{\rmsstau}) &=&  1.837\,63(60)\times 10^{-9}
\nonumber\\
&&\quad\rightarrow  0.085\times 10^{-10}a_{\rm e}
\\ 
\nonumber\\ 
A^{(6)}_2\!(m_{\rm e}/m_{\rmssmu}) &=&  -7.373\,941\,72(27)\times 10^{-6}
\nonumber\\
&&\quad\rightarrow  -0.797\times 10^{-10}a_{\rm e}
\label{eq:a26s}
\\ 
\nonumber\\
A^{(6)}_2\!(m_{\rm e}/m_{\rmsstau}) &=&  -6.5819(19)\times 10^{-8} 
\nonumber\\
&&\quad\rightarrow  -0.007\times 10^{-10}a_{\rm e} \ ,
\end{eqnarray}
where the standard uncertainties of the coefficients are due to the
uncertainties of the mass ratios, which are negligible.  The contributions from
$A_3^{(6)}\!(m_{\rm e}/m_{\rmssmu},m_{\rm e}/m_{\rmsstau})$ and all
higher-order mass-dependent terms are negligible as well.

The value for $A^{(6)}_2\!(m_{\rm e}/m_{\rmssmu})$ in Eq.~(\ref{eq:a26s}) has
been updated from the value in CODATA-02 and is in agreement with the result of
\textcite{2007014} based on a calculation to all orders in the mass ratio.  The
change is given by the term
\begin{eqnarray}
 && {17\,x^6\,\zeta(3)\over{36}}-{{4381\,x^6\,\ln ^2x}\over{30240}}+{{
 24761\,x^6\,\ln x}\over{158760}}
\nonumber \\ &&
-{{13\,\pi^2\,x^6}\over{1344}}-{{
 1840256147\,x^6}\over{3556224000}} \ ,
\end{eqnarray}
where $x = m_{\rm e}/m_{\rmssmu}$, which was not included in CODATA-02.  The
earlier result was based on Eq.~(4) of \textcite{1993015}, which only included
terms to order $x^4$.  The additional term was kindly provided by
\textcite{lar2006}.

For the electroweak contribution we have
\begin{eqnarray}
a_{\rm e}({\rm weak})
&=&  0.029\,73(52)\times 10^{-12}
\nonumber\\
&=&  0.2564(45)\times 10^{-10}a_{\rm e} \ ,
\label{eq:aeweak}
\end{eqnarray}
as calculated in CODATA-98 but with the current values of $G_{\rm F}$ and ${\rm
sin}^2\theta_{\rm W}$ (see Sec.~\ref{ssec:lisq}).

The hadronic contribution is 
\begin{eqnarray}
a_{\rm e}({\rm had}) &=&  1.682(20)\times 10^{-12}
\nonumber\\
                     &=&  1.450(17)\times 10^{-9}a_{\rm e} \ .
\label{eq:aehad}
\end{eqnarray}
It is the sum of the following three contributions:
$a_{\rm e}^{(4)}({\rm had}) =  1.875(18)\times 10^{-12}$ 
obtained by \textcite{1998089};
$a_{\rm e}^{(6a)}({\rm had}) =  -0.225(5)\times 10^{-12}$ 
given by \textcite{1997005}; and
$a_{\rm e}^{({\rmssgamma}{\rmssgamma})}({\rm had}) =  0.0318(58)\times 10^{-12}$
calculated by multiplying the corresponding result for the muon given in
Sec.~\ref{sssec:amuth} by the factor $(m_{\rm e}/m_{\rmssmu})^2$, since $a_{\rm
e}^{({\rmssgamma}{\rmssgamma})}({\rm had})$ is assumed to vary approximately as
the square of the mass.

Since the dependence on $\alpha$ of any contribution other than $a_{\rm e}({\rm
QED})$ is negligible, the anomaly as a function of $\alpha$ is given by
combining terms that have like powers of $\alpha/\rmpi$ to yield
\begin{eqnarray}
a_{\rm e}({\rm th}) = a_{\rm e}({\rm QED}) + a_{\rm e}({\rm weak})
 + a_{\rm e}({\rm had}) \ ,
\label{eq:appaeth}
\end{eqnarray}
where
\begin{eqnarray}
a_{\rm e}({\rm QED}) 
&=& C_{\rm e}^{(2)}\left({\alpha\over\rmpi}\right) 
+ C_{\rm e}^{(4)}\left({\alpha\over\rmpi}\right)^2
+ C_{\rm e}^{(6)}\left({\alpha\over\rmpi}\right)^3
\nonumber\\
&&+ C_{\rm e}^{(8)}\left({\alpha\over\rmpi}\right)^4
+ C_{\rm e}^{(10)}\left({\alpha\over\rmpi}\right)^5
+ \cdots,
\label{eq:appaeqed}
\end{eqnarray}
with
\begin{eqnarray}
C_{\rm e}^{(2)} &=& 
 0.5
\nonumber\\
C_{\rm e}^{(4)} &=& 
 -0.328\,478\,444\,00
\nonumber\\
C_{\rm e}^{(6)} &=& 
 1.181\,234\,017
\nonumber\\
C_{\rm e}^{(8)} &=& 
 -1.7283(35)
\nonumber\\
C_{\rm e}^{(10)} &=& 
 0.0(3.7) \ ,
\label{eq:appaecs}
\end{eqnarray}
and where $a_{\rm e}({\rm weak})$ and $a_{\rm e}({\rm had})$ are given in
Eqs.~(\ref{eq:aeweak}) and (\ref{eq:aehad}).

The standard uncertainty of $a_{\rm e}({\rm th})$ from the uncertainties of the
terms listed above, other than that due to $\alpha$, is
\begin{eqnarray}
u[a_{\rm e}({\rm th)}] =  0.27\times 10^{-12} =  2.4\times 10^{-10}\, a_{\rm e},
\label{eq:uncaeth}
\end{eqnarray}
and is dominated by the uncertainty of the coefficient $C_{\rm e}^{(10)}$.

For the purpose of the least-squares calculations carried out in
Sec.~\ref{ssec:mada}, we define an additive correction $\delta_{\rm e}$ to
$a_{\rm e}({\rm th)}$ to account for the lack of exact knowledge of $a_{\rm
e}({\rm th)}$, and hence the complete theoretical expression for the electron
anomaly is
\begin{eqnarray}
a_{\rm e}(\alpha,\delta_{\rm e}) = a_{\rm e}({\rm th)} + \delta_{\rm e} \ .
\label{eq:aefth}
\end{eqnarray}
Our theoretical estimate of $\delta_{\rm e}$ is zero and its standard
uncertainty is $u[a_{\rm e}({\rm th})]$:
\begin{eqnarray}
\delta_{\rm e} =  0.00(27)\times 10^{-12} \ .
\label{eq:caeth}
\end{eqnarray}

\subsubsection{Measurements of $a_{\rm e}$}
\label{sssec:aemeas}

\paragraph{Measurement of $a_{\rm e}$: University of Washington.}
\label{par:aeuw}

The classic series of measurements of the electron and positron anomalies
carried out at the University of Washington by \textcite{1987003} yield the
value
\begin{eqnarray}
a_{\rm e} =  1.159\,652\,1883(42)\times 10^{-3} ~~ [ 3.7\times 10^{-9}] \ ,
\label{eq:aeuwash}
\end{eqnarray}
as discussed in CODATA-98.  This result assumes that $CPT$ invariance holds
for the electron-positron system.

\paragraph{Measurement of $a_{\rm e}$: Harvard University.}
\label{par:harvard}

A new determination of the electron anomaly using a cylindrical Penning trap
has been carried out by \textcite{2006081} at Harvard University, yielding the
value 
\begin{eqnarray}
a_{\rm e} =  1.159\,652\,180\,85(76)\times 10^{-3} \quad [ 6.6\times 10^{-10}] \ ,
\label{eq:aeharv06}
\end{eqnarray}
which has an uncertainty that is nearly six times smaller than that of the
University of Washington result. 

As in the University of Washington experiment, the anomaly is obtained in
essence from the relation $a_{\rm e} =f_{\rm a}/f_{\rm c}$ by determining, in
the same magnetic flux density $B$ (about 5~T), the anomaly difference
frequency $f_{\rm a} = f_{\rm s} - f_{\rm c}$ and cyclotron frequency $f_{\rm
c} = eB/2{\pi}m_{\rm e}$, where $f_{\rm s} = g_{\rm e}\mu_{\rm B}B/h$ is the
electron spin-flip (often called precession) frequency.  The marked improvement
achieved by the Harvard group, the culmination of a 20 year effort, is due in
large part to the use of a cylindrical Penning trap with a resonant cavity that
interacts with the trapped electron in a readily calculable way, and through
its high $Q$ resonances, significantly increases the lifetime of the electron
in its lowest few energy states by inhibiting the decay of these states through
spontaneous emission. Further, cooling the trap and its vacuum enclosure to
100~mK by means of a dilution refrigerator eliminates blackbody radiation that
could excite the electron from these states. 

The frequencies $f_{\rm a}$ and $f_{\rm c}$ are determined by applying
quantum-jump spectroscopy (QJS) to transitions between the lowest spin ($m_{\rm
s}=\pm 1/2$) and cyclotron ($n = 0, 1, 2$) quantum states of the electron in
the trap. (In QJS, the quantum jumps per attempt to drive them are measured as
a function of drive frequency.) The transitions are induced by applying a
signal of frequency $\approx f_{\rm a}$ to trap electrodes or by transmitting
microwaves of frequency $\approx f_{\rm c}$ into the trap cavity. A change in
the cyclotron or spin state of the electron is reflected in a shift in
$\overline{\nu}_{z}$, the self excited axial oscillation of the electron. (The
trap axis and $B$ are in the $z$ direction.)  This oscillation induces a signal
in a resonant circuit that is amplified and fed back to the trap to drive the
oscillation. Saturated nickel rings surrounding the trap produce a small
magnetic bottle that provides quantum nondemolition couplings of the spin and
cyclotron energies to $\overline{\nu}_{z}$.  Failure to resolve the cyclotron
energy levels would result in an increase of uncertainty due to the leading
relativistic correction $\delta/f_{\rm c}\equiv hf_{\rm c}/mc^2 \approx
10^{-9}$. 

Another unique feature of the Harvard experiment is that the effect of the trap
cavity modes on $f_{\rm c}$, and hence on the measured value of $a_{\rm e}$,
are directly observed for the first time.  The modes are quantitatively
identified as the familiar transverse electric (TE) and transverse magnetic
(TM) modes by observing the response of a cloud of electrons to an axial
parametric drive, and, based on the work of \textcite{1986027}, the range of
possible shifts of $f_{\rm c}$ for a cylindrical cavity with a $Q > 500$ as
used in the Harvard experiment can be readily calculated. Two measurements of
$a_{\rm e}$ were made: one, which resulted in the value of $a_{\rm e}$ given in
Eq.~(\ref{eq:aeharv06}), was at a value of $B$ for which $f_{\rm c} = 149$~GHz,
far from modes that couple to the cyclotron motion; the other was at 146.8 GHz,
close to mode TE$_{127}$. Within the calibration and identification
uncertainties for the mode frequencies, very good agreement was found between
the measured and predicted difference in the two values. Indeed, their weighted
mean gives a value of $a_{\rm e}$ that is larger than the value in
Eq.~(\ref{eq:aeharv06}) by only the fractional amount $0.5\times 10^{-10}$,
with $u_{\rm r}$ slightly reduced to $6.5\times 10^{-10}$. 

The largest component of uncertainty, $5.2 \times 10^{-10}$, in the $6.6 \times
10^{-10}$ $u_{\rm r}$ of the Harvard result for $a_{\rm e}$ arises from fitting
the resonance line shapes for $f_{\rm a}$ and $f_{\rm c}$ obtained from the
quantum jump spectroscopy data. It is based on the consistency of three
different methods of extracting these frequencies from the line shapes.  The
method that yielded the best fits and which was used to obtain the reported
value of $a_{\rm e}$ weights each drive frequency, spin flip or cyclotron, by
the number of quantum jumps it produces, and then uses the weighted average of
the resulting spin flip and cyclotron frequencies in the final calculation of
$a_{\rm e}$. Although the cavity shifts are well characterized, they account
for the second largest fractional uncertainty component, $3.4 \times 10^{-10}$.
The statistical (Type A) component, which is the next largest, is only $1.5
\times 10^{-10}$. 

\subsubsection{Values of $\alpha$ inferred from $a_{\rm e}$}
\label{sssec:alphaae}

Equating the theoretical expression with the two experimental values of $a_{\rm
e}$ given in Eqs.~(\ref{eq:aeuwash}) and (\ref{eq:aeharv06}) yields 
\begin{eqnarray}
\alpha^{-1}(a_{\rm e}) =  137.035\,998\,83(50) 
~~ [ 3.7\times 10^{-9}]
\label{eq:alphinvuwash87}
\end{eqnarray}
from the University of Washington result and 
\begin{eqnarray}
\alpha^{-1}(a_{\rm e}) =  137.035\,999\,711(96) 
~~ [ 7.0\times 10^{-10}]
\label{eq:alphinvharvu06}
\end{eqnarray}
from the Harvard University result.  The contribution of the uncertainty in
$a_{\rm e}({\rm th})$ to the relative uncertainty of either of these results is
$ 2.4\times 10^{-10}$.  The value in Eq.~(\ref{eq:alphinvharvu06}) has the
smallest uncertainty of any value of alpha currently available.  Both values
are included in Table~\ref{tab:gfree}. 

\subsection{Muon magnetic moment anomaly $a_{\rmssmu}$}
\label{ssec:mmma}

Comparison of theory and experiment for the muon magnetic moment anomaly gives
a test of the theory of the hadronic contributions, with the possibility of
revealing physics beyond the Standard Model. 

\subsubsection{Theory of $a_{\rmssmu}$}
\label{sssec:amuth}

The current theory of $a_{\rmssmu}$ has been throughly reviewed in a number of
recent publications by different authors, including a book devoted solely to the
subject; see, for example, \textcite{2005258, 2006173, 2006196, 2007093,
2007247}. 

The relevant mass-dependent terms and the corresponding contributions to
$a_{\rmssmu}$(th), based on the 2006 recommended values of the mass ratios, are 
\begin{eqnarray}
A_2^{(4)}\!(m_{\rmssmu}/m_{\rm e}) &=&
 1.094\,258\,3088(82)
\\
 &\rightarrow&  506\,386.4561(38)\times 10^{-8} a_{\rmssmu} \ ,
\nonumber \\ \nonumber \\
A_2^{(4)}\!(m_{\rmssmu}/m_{\rmsstau}) &=&
 0.000\,078\,064(25)
\\
&\rightarrow&  36.126(12)\times 10^{-8} a_{\rmssmu} \ ,
\nonumber \\ \nonumber \\
A^{(6)}_2\!(m_{\rmssmu}/m_{\rm e}) &=&  22.868\,379\,97(19)
\label{eq:a26lmume}
\\ &\rightarrow&  24\,581.766\,16(20)\times 10^{-8} a_{\rmssmu} \ ,
\nonumber \\ \nonumber\\
A^{(6)}_2\!(m_{\rmssmu}/m_{\rmsstau}) &=&  0.000\,360\,51(21)
\label{eq:a26smumtau}
\\ &\rightarrow&  0.387\,52(22)\times 10^{-8} a_{\rmssmu} \ ,
\nonumber \\ \nonumber\\
A_2^{(8)}\!(m_{\rmssmu}/m_{\rm e}) &=&  132.6823(72)
\label{eq:a28mume}
\\ &\rightarrow&  331.288(18)\times 10^{-8} a_{\rmssmu} \ ,
\nonumber \\ \nonumber\\
A_2^{(10)}\!(m_{\rmssmu}/m_{\rm e}) &=&  663(20)
\label{eq:a210mume}
\\ &\rightarrow&  3.85(12)\times 10^{-8} a_{\rmssmu} \ ,
\end{eqnarray}
\begin{eqnarray}
A_3^{(6)}\!(m_{\rmssmu}/m_{\rm e},m_{\rmssmu}/m_{\rmsstau}) &=&  0.000\,527\,66(17)
\label{eq:a36mumemumtau}  
\\ &\rightarrow&  0.567\,20(18)\times 10^{-8} a_{\rmssmu \ ,}
\nonumber \\ \nonumber\\
A_3^{(8)}\!(m_{\rmssmu}/m_{\rm e},m_{\rmssmu}/m_{\rmsstau}) &=&  0.037\,594(83)
\label{eq:a38mumemumtau}
\\ &\rightarrow&  0.093\,87(21)\times 10^{-8} a_{\rmssmu} \ .
\nonumber 
\end{eqnarray}
These contributions and their uncertainties, as well as the values (including
their uncertainties) of $a_{\rmssmu}({\rm weak})$ and $a_{\rmssmu}({\rm had})$
given below, should be compared with the $54\times 10^{-8} a_{\rmssmu}$
standard uncertainty of the experimental value of $a_{\rmssmu}$ from Brookhaven
National Laboratory (BNL) (see next section).

Some of the above terms reflect the results of recent calculations.  The value
of $A_2^{(6)}\!(m_{\rmssmu}/m_{\rmsstau})$ in Eq.~(\ref{eq:a26smumtau})
includes an additional contribution as discussed in connection with
Eq.~(\ref{eq:a26s}).  The terms $A_2^{(8)}\!(m_{\rmssmu}/m_{\rm e})$ and
$A_3^{(8)}\!(m_{\rmssmu}/m_{\rm e},m_{\rmssmu}/m_{\rmsstau})$ have been updated
by \textcite{2004131}, with the resulting value for
$A_2^{(8)}\!(m_{\rmssmu}/m_{\rm e})$ in Eq.~(\ref{eq:a28mume}) differing from
the previous value of $127.50(41)$ due to the elimination of various problems
with the earlier calculations, and the resulting value for
$A_3^{(8)}\!(m_{\rmssmu}/m_{\rm e},m_{\rmssmu}/m_{\rmsstau})$ in
Eq.~(\ref{eq:a38mumemumtau}) differing from the previous value of $0.079(3)$,
because diagrams that were thought to be negligible do in fact contribute to
the result. Further, the value for $A_2^{(10)}\!(m_{\rmssmu}/m_{\rm e})$ in
Eq.~(\ref{eq:a210mume}) from \textcite{2006016} replaces the previous value,
930(170). These authors believe that their result, obtained from the numerical
evaluation of all of the integrals from 17 key subsets of Feynman diagrams,
accounts for the leading contributions to $A_2^{(10)}\!(m_{\rmssmu}/m_{\rm
e})$, and the work of \textcite{2006239}, based on the so-called
renormalization group-inspired scheme-invariant approach, strongly supports
this view. 

The electroweak contribution to $a_{\rmssmu}({\rm th})$ is taken to be
\begin{eqnarray}
a_{\rmssmu}({\rm weak}) =  154(2)\times 10^{-11} \ , 
\label{eq:amw}
\end{eqnarray}
as given by \textcite{2003052,2003052e}.  This value was used in the 2002
adjustment and is discussed in CODATA-02.

The hadronic contribution to $a_{\rmssmu}({\rm th})$ may be written as 
\begin{eqnarray}
a_{\rmssmu}({\rm had}) &=& a_{\rmssmu}^{(4)}({\rm had}) + a_{\rmssmu}^{(6a)}({\rm had})
+ a_{\rmssmu}^{({\rmssgamma}{\rmssgamma})}({\rm had}) +\cdots \ ,
\nonumber\\
\label{eq:hadcontr}
\end{eqnarray}
where $a_{\rmssmu}^{(4)}({\rm had})$ and $a_{\rmssmu}^{(6a)}({\rm had})$ arise
from hadronic vacuum polarization and are of order $(\alpha/\rmpi)^2$ and
$(\alpha/\rmpi)^3$, respectively; and
$a_{\rmssmu}^{({\rmssgamma}{\rmssgamma})}({\rm had})$, which arises from
hadronic light-by-light vacuum polarization, is also of order
$(\alpha/\rmpi)^3$.

Values of $a_{\rmssmu}^{(4)}({\rm had})$ are obtained from calculations that
evaluate dispersion integrals over measured cross sections for the scattering
of e$^+$e$^-$ into hadrons.  In addition, in some such calculations, data on
decays of the $\rmtau$ into hadrons is used to replace the e$^+$e$^-$ data in
certain parts of the calculation. In the 2002 adjustment, results from both
types of calculation were averaged to obtain a value that would be
representative of both approaches. 

There have been improvements in the calculations that use only e$^+$e$^-$ data
with the addition of new data from the detectors CMD-2 at Novosibirsk, KLOE at
Frascati, BaBar at the Stanford Linear Accelerator Center, and corrected data
from the detector SND at Novosibirsk \cite{2007147, 2007203, 2007247}.
However, there is a persistent disagreement between the results that include
the $\rmtau$ decay data and those that use only e$^+$e$^-$ data. In view of the
improvements in the results based solely on e$^+$e$^-$ data and the unresolved
questions concerning the assumptions required to incorporate the $\rmtau$ data
into the analysis \cite{2006173, 2006196, 2007203}, we use in the 2006
adjustment results based solely on e$^+$e$^-$ data.  The value employed is 
\begin{eqnarray} 
a_{\rmssmu}^{(4)}({\rm had}) &=&  690(21)\times 10^{-10} \ , 
\label{eq:amh4}
\end{eqnarray}
which is the unweighted mean of the values $a_{\rmssmu}^{(4)}({\rm had}) =
689.4(4.6)\times 10^{-10}$ \cite{2007147} and $a_{\rmssmu}^{(4)}({\rm had}) =
690.9(4.4)\times 10^{-10}$ \cite{2007203}. The uncertainty assigned the value
of $a_{\rmssmu}^{(4)}({\rm had})$, as expressed in Eq~(\ref{eq:amh4}), is
essentially the difference between the values that include $\rmtau$ data and
those that do not.  In particular, the result that includes $\rmtau$ data that
we use to estimate the uncertainty is $711.0(5.8)\times 10^{-11}$ from
\textcite{2003223}; the value of $a_{\rmssmu}^{(4)}({\rm had})$ used in the
2002 adjustment was based in part on this result. Although there is the smaller
value $701.8(5.8)\times 10^{-11}$ from \textcite{2005180}, we use only the
larger value in order to obtain an uncertainty that covers the possibility of
physics beyond the Standard Model not included in the calculation of
$a_{\rmssmu}$(th). Other, mostly older results for $a_{\rmssmu}^{(4)}({\rm
had})$, but which in general agree with the two values we have averaged, are
summarized in Table~III of \textcite{2007247}.

For the second term in Eq.~(\ref{eq:hadcontr}), we employ the value
\begin{eqnarray}
a_{\rmssmu}^{(6a)}({\rm had}) =  -97.90(95)\times 10^{-11} \ 
\label{eq:amh6a}
\end{eqnarray}
calculated by \textcite{2004126}, which was also used in the 2002 adjustment.

The light-by-light contribution in Eq.~(\ref{eq:hadcontr}) has been calculated
by \textcite{2004130, 2006196}, who obtain the value
\begin{eqnarray}
a_{\rmssmu}^{({\rmssgamma}{\rmssgamma})}({\rm had})
 =  136(25)\times 10^{-11} \ .
\label{eq:amhgg}
\end{eqnarray}
It is somewhat larger than earlier results, because it includes short distance
constraints imposed by quantum chromodynamics (QCD) that were not included in
the previous calculations. It is consistent with the 95~\% confidence limit
upper bound of $159\times 10^{-11}$ for
$a_{\rmssmu}^{({\rmssgamma}{\rmssgamma})}({\rm had})$ obtained by
\textcite{2006240}, the value $110(40)\times 10^{-11}$ proposed by
\textcite{2007151}, and the value $125(35)\times 10^{-11}$ suggested by
\textcite{2004223}.

The total hadronic contribution is 
\begin{eqnarray}
a_{\rmssmu}({\rm had}) &=&  694(21)\times 10^{-10}
\nonumber\\
                 &=& 595(18)\times 10^{-7}a_{\rmssmu}  \ .
\label{eq:muhadtot}
\end{eqnarray}

Combining terms in $a_{\rmssmu}({\rm QED})$ that have like powers of
$\alpha/\rmpi$, we summarize the theory of $a_{\rmssmu}$ as follows:
\begin{eqnarray}
a_{\rmssmu}({\rm th}) = a_{\rmssmu}({\rm QED}) + a_{\rmssmu}({\rm weak})
 + a_{\rmssmu}({\rm had}) \ ,
\label{eq:appamth}
\end{eqnarray}
where
\begin{eqnarray}
a_{\rmssmu}({\rm QED}) 
&=& C_{\rmssmu}^{(2)}\left({\alpha\over\rmpi}\right) 
+ C_{\rmssmu}^{(4)}\left({\alpha\over\rmpi}\right)^2
+ C_{\rmssmu}^{(6)}\left({\alpha\over\rmpi}\right)^3
\nonumber\\
&&+ C_{\rmssmu}^{(8)}\left({\alpha\over\rmpi}\right)^4
+ C_{\rmssmu}^{(10)}\left({\alpha\over\rmpi}\right)^5
+ \cdots,
\label{eq:appamqed}
\end{eqnarray}
with
\begin{eqnarray}
C_{\rmssmu}^{(2)} &=& 
 0.5
\nonumber\\
C_{\rmssmu}^{(4)} &=& 
 0.765\,857\,408(27)
\nonumber\\
C_{\rmssmu}^{(6)} &=& 
 24.050\,509\,59(42)
\nonumber\\
C_{\rmssmu}^{(8)} &=& 
 130.9916(80)
\nonumber\\
C_{\rmssmu}^{(10)} &=& 
 663(20) \ ,
\label{eq:appamcs}
\end{eqnarray}
and where $a_{\rmssmu}({\rm weak})$ and $a_{\rmssmu}({\rm had})$ are as given
in Eqs.~(\ref{eq:amw}) and (\ref{eq:muhadtot}).  The standard uncertainty of
$a_{\rmssmu}({\rm th})$ from the uncertainties of the terms listed above, other
than that due to $\alpha$, is
\begin{eqnarray}
u[a_{\rmssmu}({\rm th)}] =  2.1\times 10^{-9} =  1.8\times 10^{-6}\, a_{\rmssmu},
\label{eq:uncamth}
\end{eqnarray}
and is primarily due to the uncertainty of $a_{\rmssmu}({\rm had})$.

For the purpose of the least-squares calculations carried out in
Sec.~\ref{ssec:mada}, we define an additive correction $\delta_{\rmssmu}$ to
$a_{\rmssmu}({\rm th)}$ to account for the lack of exact knowledge of
$a_{\rmssmu}({\rm th)}$, and hence the complete theoretical expression for the
muon anomaly is
\begin{eqnarray}
a_{\rmssmu}(\alpha,\delta_{\rmssmu}) = 
a_{\rmssmu}({\rm th)} + \delta_{\rmssmu} \ .
\label{eq:amfth}
\end{eqnarray}
Our theoretical estimate of $\delta_{\rmssmu}$ is zero and its standard
uncertainty is $u[a_{\rmssmu}({\rm th})]$:
\begin{eqnarray}
\delta_{\rmssmu} =  0.0(2.1)\times 10^{-9} \ .
\label{eq:camth}
\end{eqnarray}
Although $a_{\rmssmu}({\rm th)}$ and $a_{\rm e}({\rm th})$ have
some common components of uncertainty, the covariance
of $\delta_{\rmssmu}$ and $\delta_{\rm e}$ is negligible.

\subsubsection{Measurement of $a_{\rm \rmssmu}$: Brookhaven.}
\label{sssec:amb}

Experiment E821 at Brookhaven National Laboratory (BNL), Upton, New York, was
initiated by the Muon $g-2$ Collaboration in the early-1980s with the goal of
measuring $a_{\rmssmu}$ with a significantly smaller uncertainty than $u_{\rm
r}=7.2 \times 10^{-6}$. This is the uncertainty achieved in the third $g-2$
experiment carried out at the European Organization for Nuclear Research
(CERN), Geneva, Switzerland, in the mid-1970s using both positive and negative
muons and which was the culmination of nearly 20 years of effort
\cite{1979010}.

The basic principle of the experimental determination of $a_{\rmssmu}$ is
similar to that used to determine $a_{\rm e}$ and involves measuring the
anomaly difference frequency $f_{\rm a} = f_{\rm s} - f_{\rm c}$, where $f_{\rm
s} = |g_{\rmssmu}| (e\hbar/2m_{\rmssmu})B/h$ is the muon spin-flip (often
called precession) frequency in the applied magnetic flux density $B$ and where
$f_{\rm c} = eB/2\rmpi m_{\rmssmu}$ is the corresponding muon cyclotron
frequency. However, instead of eliminating $B$ by measuring $f_{\rm c}$ as is
done for the electron, $B$ is determined from proton nuclear magnetic resonance
(NMR) measurements. As a consequence, the value of $\mu_{\rmssmu }/\mu_{\rm p}$
is required to deduce the value of $a_{\rmssmu}$ from the data. The relevant
equation is
\begin{eqnarray}
a_{\rmssmu} = \frac{\overline{R}}{|\mu_{\rmssmu}/\mu_{\rm p}| - \overline{R}} \; ,
\label{eq:amurbar}
\end{eqnarray} 
where $\overline{R} = f_{\rm a}/\overline{f}_{\rm p}$, and $\overline{f}_{\rm
p}$ is the free proton NMR frequency corresponding to the average flux density
seen by the muons in their orbits in the muon storage ring used in the
experiment. (Of course, in the corresponding experiment for the electron, a
Penning trap is employed rather than a storage ring.)

The BNL $a_{\rmssmu}$ experiment was discussed in both CODATA-98 and CODATA-02.
In the 1998 adjustment, the CERN final result for $\overline{R}$ with $u_{\rm
r}=7.2 \times10^{-6}$, and the first BNL result for $\overline{R}$, obtained
from the 1997 engineering run using positive muons and with $u_{\rm r}=13
\times10^{-6}$, were taken as input data.  By the time of the 2002 adjustment,
the BNL experiment had progressed to the point where the CERN result was no
longer competitive, and the input datum used was the BNL mean value of
$\overline{R}$ with $u_{\rm r}=6.7 \times10^{-7}$ obtained from the 1998, 1999,
and 2000 runs using ${\rmmu^+}$. The final run of the BNL E821 experiment was
carried out in 2001 with ${\rmmu^-}$ and achieved an uncertainty for
$\overline{R}$ of $u_{\rm r}=7.0 \times10^{-7}$, but the result only became
available in early 2004, well after the closing date of the 2002 adjustment. 

Based on the data obtained in all five runs and assuming $CPT$ invariance, an
assumption justified by the consistency of the values of $\overline{R}$
obtained from either ${\rmmu^+}$ or ${\rmmu^-}$, the final report on the E821
experiment gives as the final value of $\overline{R}$ \cite{2006132} [see also
\cite{2007093}]
\begin{eqnarray}
\overline{R} &=&  0.003\,707\,2064(20)\quad [ 5.4\times 10^{-7}] \ ,
\label{eq:rbar06}
\end{eqnarray}
which we take as an input datum in the 2006 adjustment.  A new BNL experiment
to obtain a value of $\overline{R}$ with a smaller uncertainty is under discussion
\cite{2007208}.

The experimental value of $a_{\rmssmu}$ implied by this value of $\overline{R}$
is, from Eq.~(\ref{eq:amurbar}) and the 2006 recommended value of $\mu_{\rmssmu
}/\mu_{\rm p}$, the uncertainty of which is inconsequential in this
application,
\begin{eqnarray}
a_{\rmssmu}(\rm exp) =  1.165\,920\,93(63)\times 10^{-3}\quad[ 5.4\times 10^{-7}] \ . \qquad
\label{eq:amu06}
\end{eqnarray}
Further, with the aid of Eq.~(\ref{eq:mumemump}) in Sec.~\ref{ssec:muhfs},
Eq.~(\ref{eq:amurbar}) can be written as
\begin{eqnarray}
\overline{R} = -\frac{a_{\rmssmu}(\alpha,\delta_{\rmssmu })}
{1 + a_{\rm e}(\alpha,\delta_{\rm e})}\frac{m_{\rm e}
}{m_{\rmssmu }}\frac{\mu_{{\rm e}^-}}{\mu_{\rm p}} \ ,
\label{eq:rbarobs}
\end{eqnarray}
where we have used the relations $g_{\rm e} = -2(1 + a_{\rm e})$ and
$g_{\rmssmu} = -2(1 + a_{\rmssmu})$ and replaced $a_{\rm e}$ and $a_{\rmssmu}$
with their complete theoretical expressions $a_{\rm e}(\alpha,\delta_{\rm e})$
and $a_{\rmssmu}(\alpha,\delta_{\rmssmu})$, which are discussed in
Sec.~\ref{sssec:ath} and Sec.~\ref{sssec:amuth}, respectively.
Equation~(\ref{eq:rbarobs}) is, in fact, the observational equation for the
input datum $\overline{R}$.

\paragraph{Theoretical value of $a_{\rmssmu}$ and inferred value of $\alpha$}
\label{par:tamual}

Evaluation of the theoretical expression for $a_{\rmssmu}$ in
Eq.~(\ref{eq:appamth}) with the 2006 recommended value of $\alpha$, the
uncertainty of which is negligible in this context, yields
\begin{eqnarray}
a_{\rmssmu}(\rm th) =  1.165\,9181(21)\times 10^{-3} \quad [ 1.8\times 10^{-6}] \ , \qquad
\label{eq:amuthnum}
\end{eqnarray}
which may be compared to the value in Eq.~(\ref{eq:amu06}) deduced from the BNL
result for $\overline{R}$ given in Eq.~(\ref{eq:rbar06}).  The experimental
value exceeds the theoretical value by $ 1.3\,u_{\rm diff}$, where $u_{\rm
diff}$ is the standard uncertainty of the difference.  It should be recognized,
however, that this reasonable agreement is a consequence of the comparatively
large uncertainty we have assigned to $a_{\rmssmu}^{(4)}(\rm had)$ [see
Eq.~(\ref{eq:uncamth})].  If the result for $a_{\rmssmu}^{(4)}(\rm had)$ that
includes $\rm tau$ data were ignored and the uncertainty of
$a_{\rmssmu}^{(4)}(\rm had)$ were based on the estimated uncertainties of the
calculated values using only $\rm e^{+}\rm e^{-}$ data, then the experimental
value would exceed the theoretical value by $ 3.5\,u_{\rm diff}$.  This
inconsistency is well known to the high-energy physics community and is of
considerable interest because it may be an indication of ``New Physics'' beyond
the Standard Model, such as supersymmetry \cite{2007032}.

One might ask, why include the theoretical value for $a_{\rmssmu}$ in the 2006
adjustment given its current problems?  By retaining the theoretical expression
with an increased uncertainty, we ensure that the 2006 recommended value of
$a_{\rmssmu}$ reflects, even though with a comparatively small weight, the
existence of the theoretical value.

The consistency between theory and experiment may also be examined by
considering the value of $\alpha$ obtained by equating the theoretical
expression for $a_{\rmssmu}$ with the BNL experimental value, as was done for
$a_{\rm e}$ in Sec.~\ref{sssec:alphaae}.  The result is \begin{eqnarray}
\alpha^{-1} =  137.035\,67(26) \quad [ 1.9\times 10^{-6}] \ ,
\label{eq:alphiam06}
\end{eqnarray}
which is the value included in Table~\ref{tab:gfree}.

\subsection{Bound electron $\bm{g}$-factor in $\bm{^{12}{\rm C}^{5+}}$ and in
$\bm{^{16}{\rm O}^{7+}}$ and $\bm{A_{\rm r}({\rm e})}$}
\label{ssec:ehco}

Precise measurements and theoretical calculations of the $g$-factor of the
electron in hydrogenic $^{12}$C and in hydrogenic $^{16}$O lead to values of
$A_{\rm r}$(e) that contribute to the determination of the 2006 recommended
value of this important constant.

For a ground-state hydrogenic ion $^{A}X^{(Z-1)+}$ with mass number $A$, atomic
number (proton number) $Z$, nuclear spin quantum number $i$ = 0, and $g$-factor
$g_{\rm e^-} (^{A}{X}^{(Z-1)+})$ in an applied magnetic flux density $B$, the
ratio of the electron's spin-flip (often called precession) frequency $f_{\rm
s}=|g_{{\rm e}^-} (^{A}{X}^{(Z-1)+})| (e\hbar/2m_{\rm e})B/h$ to the cyclotron
frequency of the ion $f_{\rm c} = (Z-1) eB/2\rmpi m(^{A}{X}^{(Z-1)+})$ in the
same magnetic flux density is
\begin{eqnarray}
\frac{f_{\rm s} (^A{X}^{(Z-1)+})}{f_{\rm c} (
^A{X}^{(Z-1)+})}&=&-\frac{g_{{\rm e}^-}(
^A{X}^{(Z-1)+})}{2(Z-1)}\frac{A_{\rm r}(
^A{X}^{(Z-1)+})}{A_{\rm r} ({\rm e})} \ , 
\nonumber\\
\label{eq:fsfcgx}
\end{eqnarray}
where as usual, $A_{\rm r}(X)$ is the relative atomic mass of particle $X$. If
the frequency ratio $f_{\rm s}/f_{\rm c}$ is determined experimentally with
high accuracy, and $A_{\rm r}(^A{X}^{(Z-1)+})$ of the ion is also accurately
known, then this expression can be used to determine an accurate value of
$A_{\rm r}$(e), assuming the bound-state electron $g$-factor can be calculated
from QED theory with sufficient accuracy; or the $g$-factor can be determined
if $A_{\rm r}$(e) is accurately known from another experiment.  In fact, a
broad program involving workers from a number of European laboratories has been
underway since the mid-1990s to measure the frequency ratio and calculate the
$g$-factor for different ions, most notably (to date) $^{12}{\rm C}^{5+}$ and
$^{16}{\rm O}^{7+}$.  The measurements themselves are being performed at the
Gesellschaft f\"{u}r Schwerionenforschung, Darmstadt, Germany (GSI) by GSI and
University of Mainz researchers, and we discuss the experimental determinations
of $f_{\rm s}/f_{\rm c}$ for $^{12}{\rm C}^{5+}$ and $^{16}{\rm O}^{7+}$ at GSI
in Secs.~\ref{par:gec} and \ref{par:geo}. The theoretical expressions for the
bound-electron $g$-factors of these two ions are reviewed in the next section.

\subsubsection{Theory of the bound electron $g$-factor}
\label{sssec:thbegf}

In this section, we consider an electron in the 1S state of hydrogen like carbon
12 or oxygen 16 within the framework of bound-state QED.  The measured quantity
is the transition frequency between the two Zeeman levels of the atom in an
externally applied magnetic field.

The energy of a free electron with spin projection $s_z$ in a magnetic flux
density $B$ in the $z$ direction is
\begin{eqnarray}
E &=& -\bm \mu \cdot \bm B = - g_{\rm e^-} {e\over2m_{\rm e}}s_z B\ ,
\label{eq:eemf}
\end{eqnarray}
and hence the spin-flip energy difference is
\begin{eqnarray}
\Delta E = - g_{\rm e^-} \mu_{\rm B} B \ .
\label{eq:deemf}
\end{eqnarray}
(In keeping with the definition of the $g$-factor in Sec.~\ref{sec:mmagf}, the
quantity $g_{\rm e^-}$ is negative.) The analogous expression for ions with no
nuclear spin is
\begin{eqnarray}
\Delta E_{\rm b}(X) &=& - g_{\rm e^-}(X) \mu_{\rm B} B \ ,
\label{eq:deebmf}
\end{eqnarray}
which defines the bound-state electron $g$-factor, and where $X$ is either
$^{12}{\rm C}^{5+}$ or $^{16}${\rm O}$^{7+}$.

The theoretical expression for $g_{\rm e^-}(X)$ is written as
\begin{eqnarray}
g_{\rm e^-}({\rm X}) 
= g_{\rm D} + \Delta g_{\rm rad} + \Delta g_{\rm rec} 
+ \Delta g_{\rm ns} + \cdots
\ ,
\label{eq:gsumdef}
\end{eqnarray}
where the individual terms are the Dirac value, the radiative corrections, the
recoil corrections, and the nuclear size corrections, respectively.  These
theoretical contributions are discussed in the following paragraphs; numerical
results based on the 2006 recommended values are summarized in
Tables~\ref{tab:gfactthc} and \ref{tab:gfacttho}.  In the 2006 adjustment
$\alpha$ in the expression for $g_{\rm D}$ is treated as a variable, but the
constants in the rest of the calculation of the $g$-factors are taken as fixed
quantities.

\cite{1928001} obtained the exact value
\begin{eqnarray}
g_{\rm D} &=& - \frac{2}{3}\left[1+2\sqrt{1-(Z\alpha)^2}\right]
\nonumber\\
&=& - 2\left[1-\frac{1}{3}(Z\alpha)^2 -\frac{1}{12}(Z\alpha)^4 
-\frac{1}{24}(Z\alpha)^6
 + \cdots\right]
\nonumber\\
\label{eq:diracg}
\end{eqnarray}
from the Dirac equation for an electron in the field of a fixed point charge of
magnitude $Ze$, where the only uncertainty is that due to the uncertainty in
$\alpha$.

The radiative corrections may be written as
\begin{eqnarray}
\Delta g_{\rm rad} &=& 
- 2\left[C_{\rm e}^{(2)}(Z\alpha)\left({\alpha\over\rmpi}\right)
+ C_{\rm e}^{(4)}(Z\alpha)\left({\alpha\over\rmpi}\right)^2
+ \cdots\right] \ ,
\nonumber\\
\end{eqnarray}
where the coefficients $C_{\rm e}^{(2n)}(Z\alpha)$, corresponding to $n$
virtual photons, are slowly varying functions of $Z\alpha$.  These coefficients
are defined in direct analogy with the corresponding coefficients for the free
electron $C_{\rm e}^{(2n)}$ given in Eq.~(\ref{eq:appaecs}) so that
\begin{eqnarray}
\lim_{Z\alpha\rightarrow 0}C_{\rm e}^{(2n)}(Z\alpha)
= C_{\rm e}^{(2n)} \ .
\end{eqnarray}
\def\shrink{\hbox to -1pt{}}

The first two terms of the coefficient $C_{\rm e}^{(2)}\shrink(Z\alpha)$ have
been known for some time \cite{1970020,1970019,1971019}.  Recently,
\textcite{2004078, 2004078e, 2005079} have calculated additional terms with the
result
\begin{eqnarray}
C_{\rm e,SE}^{(2)}\shrink(Z\alpha) 
&=& \frac{1}{2}\bigg\{1 + \frac{(Z\alpha)^2}{6}
+(Z\alpha)^4\left[
\frac{32}{9}\,\ln{(Z\alpha)^{-2}} 
\right.
\nonumber\\ &&\left.
\quad + \frac{247}{216}
-\frac{8}{9}\,\ln{k_0}
- \frac{8}{3}\, \ln{k_3}
\right] 
\nonumber\\ && \quad +
 (Z\alpha)^5\,R_{\rm SE}(Z\alpha) \bigg\} \ ,
\label{eq:pachetal}
\end{eqnarray}
where
\begin{eqnarray}
\ln{k_0} &=&  2.984\,128\,556
\label{eq:lnk0}
\\
\ln{k_3} &=&  3.272\,806\,545
\label{eq:lnk3}
\\
R_{\rm SE}(6\alpha) &=&  22.160(10)
\\
R_{\rm SE}(8\alpha) &=&  21.859(4) \ .
\end{eqnarray}
The quantity $\ln{k_0}$ is the Bethe logarithm for the 1S state (see
Table~\ref{tab:bethe}) and $\ln{k_3}$ is a generalization of the Bethe
logarithm relevant to the $g$-factor calculation.  The remainder function
$R_{\rm SE}(Z\alpha)$ was obtained by \textcite{2004078, 2004078e} by
extrapolation of the results of numerical calculations of the self energy for
$Z>8$ by \textcite{2002126} using Eq.~(\ref{eq:pachetal}) to remove the
lower-order terms.  For $Z=6$ and $Z=8$ this yields
\begin{eqnarray}
C_{\rm e,SE}^{(2)}\shrink(6\alpha) &=&  0.500\,183\,606\,65(80)
\nonumber\\
C_{\rm e,SE}^{(2)}\shrink(8\alpha) &=&  0.500\,349\,2887(14)
 \ .
\label{eq:yerokgco}
\end{eqnarray}

The lowest-order vacuum-polarization correction consists of a wave-function
correction and a potential correction.  The wave-function correction has been
calculated numerically by \textcite{2000111}, with the result (in our notation)
\begin{eqnarray}
C_{\rm e,VPwf}^{(2)}\shrink(6\alpha) &=&  -0.000\,001\,840\,3431(43) \ .
\nonumber\\
C_{\rm e,VPwf}^{(2)}\shrink(8\alpha) &=&  -0.000\,005\,712\,028(26) \ .
\label{eq:gvpwfco}
\end{eqnarray}
Each of these values is the sum of the Uehling potential contribution and the
higher-order Wichmann-Kroll contribution, which were calculated separately with
the uncertainties added linearly, as done by \textcite{2000111}.  The values in
Eq.~(\ref{eq:gvpwfco}) are consistent with the result of an evaluation of the
correction in powers of $Z\alpha$ \cite{2000037,2001061,2001218}.  For the
potential correction, \textcite{2000111} found that the Uehling potential
contribution is zero and calculated the Wichmann-Kroll contribution numerically
over a wide range of $Z$ \cite{2000153}.  An extrapolation of the numerical
values from higher-$Z$, taken together with the analytic result of
\textcite{2002139},
\begin{eqnarray}
C_{\rm e,VPp}^{(2)}\shrink(Z\alpha) &=& {7\rmpi\over432}(Z\alpha)^5
+\cdots \ ,
\label{eq:kmvp}
\end{eqnarray}
for the lowest-order Wichmann-Kroll contribution, yields
\begin{eqnarray}
C_{\rm e,VPp}^{(2)}\shrink(6\alpha) &=&  0.000\,000\,007\,9595(69)
\nonumber\\
C_{\rm e,VPp}^{(2)}\shrink(8\alpha) &=&  0.000\,000\,033\,235(29) \ .
\label{eq:gvppcobeier}
\end{eqnarray}
More recently, \textcite{2005090} have obtained the result
\begin{eqnarray}
C_{\rm e,VPp}^{(2)}\shrink(6\alpha) &=&  0.000\,000\,008\,201(11)
\nonumber\\
C_{\rm e,VPp}^{(2)}\shrink(8\alpha) &=&  0.000\,000\,034\,23(11) \ .
\label{eq:gvppcokarsh}
\end{eqnarray}
The values in Eq.~(\ref{eq:gvppcobeier}) and Eq.~(\ref{eq:gvppcokarsh})
disagree somewhat, so in the present analysis, we use a value that is an
unweighted average of the two, with half the difference for the uncertainty.
These average values are
\begin{eqnarray}
C_{\rm e,VPp}^{(2)}\shrink(6\alpha) &=&  0.000\,000\,008\,08(12)
\nonumber\\
C_{\rm e,VPp}^{(2)}\shrink(8\alpha) &=&  0.000\,000\,033\,73(50) \ .
\label{eq:gvppco}
\end{eqnarray}
The total one-photon vacuum polarization coefficients are given
by the sum of Eqs.~(\ref{eq:gvpwfco}) and (\ref{eq:gvppco}):
\begin{eqnarray}
C_{\rm e,VP}^{(2)}\shrink(6\alpha) &=&
C_{\rm e,VPwf}^{(2)}\shrink(6\alpha)+
C_{\rm e,VPp}^{(2)}\shrink(6\alpha) \nonumber\\
&=& -0.000\,001\,832\,26(12)
\nonumber\\
C_{\rm e,VP}^{(2)}\shrink(8\alpha) &=&
C_{\rm e,VPwf}^{(2)}\shrink(8\alpha)+
C_{\rm e,VPp}^{(2)}\shrink(8\alpha) \nonumber\\
&=& -0.000\,005\,678\,30(50) \ .
\label{eq:cvp2co}
\end{eqnarray}

The total for the one-photon coefficient $C_{\rm e}^{(2)}\shrink(Z\alpha)$,
given by the sum of Eqs.~(\ref{eq:yerokgco}) and (\ref{eq:cvp2co}), is
\begin{eqnarray}
C_{\rm e}^{(2)}\shrink(6\alpha) &=&
C_{\rm e,SE}^{(2)}\shrink(6\alpha)+
C_{\rm e,VP}^{(2)}\shrink(6\alpha) \nonumber\\
&=&  0.500\,181\,774\,38(81)
\nonumber\\
C_{\rm e}^{(2)}\shrink(8\alpha) &=&
C_{\rm e,SE}^{(2)}\shrink(8\alpha)+
C_{\rm e,VP}^{(2)}\shrink(8\alpha) \nonumber\\
&=&  0.500\,343\,6104(14) \ ,
\label{eq:c2co}
\end{eqnarray}
and the total one-photon contribution $\Delta g^{(2)}$ to the $g$-factor is
thus
\begin{eqnarray}
\Delta g^{(2)} &=& 
- 2\,C_{\rm e}^{(2)}(Z\alpha)\left({\alpha\over\rmpi}\right)
\nonumber\\
&=&  -0.002\,323\,663\,914(4) \quad {\rm for}~Z = 6
\nonumber\\
&=&  -0.002\,324\,415\,746(7) \quad {\rm for}~Z = 8 \ .
\nonumber\\
\end{eqnarray}
The separate one-photon self energy and vacuum polarization contributions to
the $g$-factor are given in Tables~\ref{tab:gfactthc} and \ref{tab:gfacttho}.

Calculations by \textcite{1997162} using the Bargmann-Michel-Telegdi equation
and by \textcite{2001004} using an effective potential approach yield
\begin{eqnarray}
C_{\rm e}^{(2n)}\shrink(Z\alpha) = C_{\rm e}^{(2n)}
\left(1 + {(Z\alpha)^2\over 6} + \cdots \right)
\label{eq:egbinding}
\end{eqnarray}
as the leading binding correction to the free electron coefficients $C_{\rm
e}^{(2n)}$ for any order $n$.  For $C_{\rm e}^{(2)}(Z\alpha)$, this correction
was known for some time.  For higher-order terms, it provides the leading
binding effect.

The two-loop contribution of relative order $(Z\alpha)^4$ has recently been
calculated by \textcite{2005079,2006041} for any S state.  Their result for the
ground-state correction is
\begin{eqnarray}
C_{\rm e}^{(4)}\shrink(Z\alpha) &=& C_{\rm e}^{(4)}
\left(1 + {(Z\alpha)^2\over 6} \right)
\nonumber\\ && \hbox to -2cm {}+
(Z\alpha)^4\,\bigg[
\frac{14}{9}\,\ln{(Z\alpha)^{-2}} +
\frac{991343}{155520} - \frac{2}{9}\,\ln{k_0} - \frac{4}{3}\,\ln{k_3}
\nonumber\\ && \hbox to -2cm {}+
\frac{679\,\rmpi^2}{12960} - \frac{1441\,\rmpi^2}{720}\,\ln{2} +
\frac{1441}{480}\,\zeta({3})
\bigg]
 + {\cal O}(Z\alpha)^5
\nonumber\\
\nonumber\\ &=&
 -0.328\,5778(23) \quad {\rm for}~Z = 6
\nonumber\\ &=&
 -0.328\,6578(97) \quad {\rm for}~Z = 8,
\label{eq:c4co}
\end{eqnarray}
where $\ln{k_0}$ and $\ln{k_3}$ are given in Eqs.~(\ref{eq:lnk0}) and
(\ref{eq:lnk3}).  The uncertainty due to uncalculated terms is estimated by
assuming that the unknown higher-order terms, of order $(Z\alpha)^5$ or higher
for two loops, are comparable to the higher-order one-loop terms scaled by the
free-electron coefficients in each case, with an extra factor of 2 included
\cite{2005079}:
\begin{eqnarray}
u\left[C_{\rm e}^{(4)}\shrink(Z\alpha)\right] &=& 
2\,\left|
(Z\alpha)^5 \,
C_{\rm e}^{(4)}\,
R_{\rm SE}(Z\alpha)\right| \ .
\end{eqnarray}

The three- and four-photon terms are calculated with the leading binding
correction included:
\begin{eqnarray}
C_{\rm e}^{(6)}\shrink(Z\alpha) &=& C_{\rm e}^{(6)}
\left(1 + {(Z\alpha)^2\over 6} + \cdots \right)
\nonumber\\ &=&
 1.181\,611\dots \quad {\rm for}~Z = 6
\nonumber\\ &=&
 1.181\,905\dots \quad {\rm for}~Z = 8 \ ,
\label{eq:c6co}
\end{eqnarray}
where $C_{\rm e}^{(6)} = 1.181\,234\dots$~,
and
\begin{eqnarray}
C_{\rm e}^{(8)}\shrink(Z\alpha) &=& C_{\rm e}^{(8)}
\left(1 + {(Z\alpha)^2\over 6} + \cdots \right)
\nonumber\\ &=&
 -1.7289(35)\dots \quad {\rm for}~Z = 6
\nonumber\\ &=&
 -1.7293(35)\dots \quad {\rm for}~Z = 8 \ ,
\label{eq:c8co}
\end{eqnarray}
where $C_{\rm e}^{(8)} =  -1.7283(35)$ \cite{2006003}.  This value would shift
somewhat if the more recent tentative value $C_{\rm e}^{(8)} = -1.9144(35)$
\cite{2007167} were used (see Sec.~\ref{sec:mmagf}).  An uncertainty estimate
\begin{eqnarray}
C_{\rm e}^{(10)}\shrink(Z\alpha) &\approx&  C_{\rm e}^{(10)} =  0.0(3.7)
\label{eq:c10co}
\end{eqnarray}
is included for the five-loop correction.

The recoil correction to the bound-state $g$-factor associated with the finite
mass of the nucleus is denoted by $\Delta g_{\rm rec}$, which we write here as
the sum $\Delta g_{\rm rec}^{(0)} + \Delta g_{\rm rec}^{(2)}$ corresponding to
terms that are zero- and first-order in $\alpha/\rmpi$, respectively.  For
$\Delta g_{\rm rec}^{(0)}$, we have
\begin{eqnarray}
\Delta g_{\rm rec}^{(0)}  &=&
\bigg\{- (Z\alpha)^2 + {(Z\alpha)^4\over
3[1+\sqrt{1-(Z\alpha)^2}]^2}
\nonumber\\
&& -(Z\alpha)^5\,P(Z\alpha)\bigg\}{m_{\rm e}\over m_{\rm N}}
+{\cal O}\left({m_{\rm e}\over m_{\rm N}}\right)^{\!2}
\nonumber\\ \nonumber \\
&=&  -0.000\,000\,087\,71(1) \dots ~ {\rm for}~Z=6 
\nonumber \\
&=&  -0.000\,000\,117\,11(1) \dots ~ {\rm for}~Z=8 \ ,
\label{eq:grr0}
\end{eqnarray}

where $m_{\rm N}$ is the mass of the nucleus.  The mass ratios, obtained from
the 2006 adjustment, are
${m_{\rm e}/ m(^{12}{\rm C}^{6+})} =  0.000\,045\,727\,5\ldots$ and
${m_{\rm e}/ m(^{16}{\rm O}^{8+})} =  0.000\,034\,306\,5\ldots$.
The recoil terms are the same as in CODATA-02 and references to the original
calculations are given there.  An additional term of the order of the mass
ratio squared is included as
\begin{eqnarray}
S_Z (Z\alpha)^2\left({m_{\rm e}\over m_{\rm N}}\right)^{\!2}
\ ,
\label{eq:msqgfcorr}
\end{eqnarray}
where $S_Z$ is taken to be the average of the disagreeing values $1+Z$,
obtained by \textcite{1997162,pc02me}, and $Z/3$ obtained by
\textcite{2002033,2001312} for this term.  The uncertainty in $S_Z$ is taken to
be half the difference of the two values.

For $\Delta g_{\rm rec}^{(2)}$, we have
\begin{eqnarray}
\Delta g_{\rm rec}^{(2)}  &=&
{\alpha\over\rmpi}{(Z\alpha)^2\over3}
{m_{\rm e}\over m_{\rm N}} +\cdots
\nonumber\\
&=&  0.000\,000\,000\,06\ldots ~ {\rm for}~Z=6
\nonumber\\
&=&  0.000\,000\,000\,09\ldots ~ {\rm for}~Z=8 \ .
\label{eq:grr2}
\end{eqnarray}

There is a small correction to the bound-state $g$-factor due to the finite
size of the nucleus, of order
\begin{eqnarray}
\Delta g_{\rm ns} = - {8\over3}(Z\alpha)^4
\left({R_{\rm N}\over \lbar_{\rm C}}\right)^2 + \cdots \ ,
\label{eq:gnsg}
\end{eqnarray}
where $R_{\rm N}$ is the bound-state nuclear rms charge radius and $\lbar_{\rm
C}$ is the Compton wavelength of the electron divided by $2\rmpi$.  This term
is calculated as in CODATA-02 \cite{2002058} with updated values for the
nuclear radii $R_{\rm N} =  2.4703(22)$ fm and $R_{\rm N} =  2.7013(55)$ from
the compilation of \textcite{2004198} for $^{12}$C and $^{16}$O, respectively.
This yields the correction
\begin{eqnarray}
\Delta g_{\rm ns} &=&  -0.000\,000\,000\,408(1) \quad {\rm for}~^{12}{\rm C}
\nonumber\\
\Delta g_{\rm ns} &=&  -0.000\,000\,001\,56(1) \quad {\rm for}~^{16}{\rm O} \ .
\label{eq:gns}
\end{eqnarray}

The theoretical value for the $g$-factor of the electron in hydrogenic carbon
12 or oxygen 16 is the sum of the individual contributions discussed above and
summarized in Tables~\ref{tab:gfactthc} and \ref{tab:gfacttho}:
\begin{eqnarray}
g_{\rm e^-}(^{12}{\rm C}^{5+})
&=&  -2.001\,041\,590\,203(28)
\nonumber\\
g_{\rm e^-}(^{16}{\rm O}^{7+})
&=&  -2.000\,047\,020\,38(11) \ .
\nonumber\\
\label{eq:gco}
\end{eqnarray}

For the purpose of the least-squares calculations carried out in
Sec.~\ref{ssec:mada}, we define $g_{\rm C}({\rm th})$ to be the sum of $g_{\rm
D}$ as given in Eq.~(\ref{eq:diracg}), the term $-2(\alpha/\rmpi)C_{\rm
e}^{(2)}$, and the numerical values of the remaining terms in
Eq.~(\ref{eq:gsumdef}) as given in Table~\ref{tab:gfactthc}, where the standard
uncertainty of these latter terms is
\begin{eqnarray}
u[g_{\rm C}({\rm th})] &=&  0.3\times 10^{-10} =
 1.4\times 10^{-11}|g_{\rm C}({\rm th})| \ .
\nonumber\\
\end{eqnarray}

The uncertainty in $g_{\rm C}({\rm th})$ due to the uncertainty in $\alpha$
enters the adjustment primarily through the functional dependence of $g_{\rm
D}$ and the term $-2(\alpha/\rmpi)C_{\rm e}^{(2)}$ on $\alpha$.  Therefore this
particular component of uncertainty is not explicitly included in $u[g_{\rm
C}({\rm th})]$.  To take the uncertainty $u[g_{\rm C}({\rm th})]$ into account
we employ as the theoretical expression for the $g$-factor
\begin{eqnarray}
g_{\rm C}(\alpha,\delta_{\rm C}) &=& g_{\rm C}({\rm th})
+ \delta_{\rm C} \ ,
\end{eqnarray}
where the input value of the additive correction $\delta_{\rm C}$ is taken to
be zero and its standard uncertainty is $u[g_{\rm C}({\rm th})]$:
\begin{eqnarray}
\delta_{\rm C} =  0.00(27)\times 10^{-10} \ .
\label{eq:cgcth}
\end{eqnarray}
Analogous considerations apply for the $g$-factor in oxygen:
\begin{eqnarray}
u[g_{\rm O}({\rm th})] &=&  1.1\times 10^{-10} =
 5.3\times 10^{-11}|g_{\rm O}({\rm th})| \nonumber\\
\\
g_{\rm O}(\alpha,\delta_{\rm O}) &=& g_{\rm O}({\rm th})
+ \delta_{\rm O}
\\
\delta_{\rm O} &=&  0.0(1.1)\times 10^{-10} \ .
\label{eq:cgoth}
\end{eqnarray}

Since the uncertainties of the theoretical values of the carbon and oxygen
$g$-factors arise primarily from the same sources, the quantities $\delta_{\rm
C}$ and $\delta_{\rm O}$ are highly correlated.  Their covariance is
\begin{eqnarray}
u(\delta_{\rm C},\delta_{\rm O}) =  27\times 10^{-22} \ ,
\end{eqnarray}
which corresponds to a correlation coefficient of $r(\delta_{\rm C},\delta_{\rm
O})= 0.92$.

The theoretical value of the ratio of the two $g$-factors, which is relevant to
the comparison to experiment in Sec.~\ref{par:geco}, is
\begin{eqnarray}
{g_{\rm e^-}(^{12}{\rm C}^{5+}) \over
g_{\rm e^-}(^{16}{\rm O}^{7+})}
&=&  1.000\,497\,273\,218(41) \ ,
\label{eq:gcogo}
\end{eqnarray}
where the covariance, including the contribution from the uncertainty in
$\alpha$ for this case, is taken into account.

\begin{table}[t]
\def\m{\phantom{-}}
\caption{Theoretical contributions and total for the $g$-factor of the electron
in hydrogenic carbon 12 based on the 2006 recommended values of the constants.}
\label{tab:gfactthc}
\def\sp{\hbox to 5mm {}}
\begin{center}
\begin{tabular}{c@{\sp}l@{\sp}c}
\toprule
\vbox to 10 pt {}
Contribution   &  \hbox to .8 cm{} Value & Source \\
\colrule
\vbox to 10 pt {}
Dirac $g_{\rm D}$                 & $   -1.998\,721\,354\,402(2)  $ & Eq.~(\ref{eq:diracg}) \\
 $\Delta g^{(2)}_{\rm SE}       $ & $   -0.002\,323\,672\,426(4)$ & Eq.~(\ref{eq:yerokgco}) \\
 $\Delta g^{(2)}_{\rm VP}       $ & $\m 0.000\,000\,008\,512(1)$ & Eq.~(\ref{eq:cvp2co})  \\
 $\Delta g^{(4)}                $ & $\m 0.000\,003\,545\,677(25)  $ & Eq.~(\ref{eq:c4co})  \\
 $\Delta g^{(6)}                $ & $   -0.000\,000\,029\,618  $ & Eq.~(\ref{eq:c6co})  \\
 $\Delta g^{(8)}                $ & $\m 0.000\,000\,000\,101  $ & Eq.~(\ref{eq:c8co})  \\
 $\Delta g^{(10)}                $ & $\m 0.000\,000\,000\,000(1)  $ & Eq.~(\ref{eq:c10co})  \\
 $\Delta g_{\rm rec}            $ & $   -0.000\,000\,087\,639(10) $ & Eqs.~(\ref{eq:grr0})-(\ref{eq:grr2}) \\
 $\Delta g_{\rm ns}             $ & $   -0.000\,000\,000\,408(1)  $ & Eq.~(\ref{eq:gns})  \\
\B $g_{\rm e^-}(^{12}{\rm C}^{5+})$ & $   -2.001\,041\,590\,203(28) \vbox to .4 cm {}  $ & Eq.~(\ref{eq:gco}) \\
\botrule
\end{tabular}
\end{center}
\end{table}

\begin{table}[t]
\def\m{\phantom{-}}
\caption{Theoretical contributions and total for the $g$-factor of the electron
in hydrogenic oxygen 16 based on the 2006 recommended values of the constants.}
\label{tab:gfacttho}
\begin{center}
\def\sp{\hbox to 6mm {}}
\begin{tabular}{c@{\sp}l@{\sp}c}
\toprule
\vbox to 10 pt {}
Contribution   &  \hbox to .8 cm{} Value & Source \\
\colrule
\vbox to 10 pt {}
Dirac $g_{\rm D}$                 & $   -1.997\,726\,003\,08  $ & Eq.~(\ref{eq:diracg}) \\
 $\Delta g^{(2)}_{\rm SE}       $ & $   -0.002\,324\,442\,12(1)$ & Eq.~(\ref{eq:yerokgco}) \\
 $\Delta g^{(2)}_{\rm VP}       $ & $\m 0.000\,000\,026\,38$ & Eq.~(\ref{eq:cvp2co})  \\
 $\Delta g^{(4)}                $ & $\m 0.000\,003\,546\,54(11)  $ & Eq.~(\ref{eq:c4co})  \\
 $\Delta g^{(6)}                $ & $   -0.000\,000\,029\,63  $ & Eq.~(\ref{eq:c6co})  \\
 $\Delta g^{(8)}                $ & $\m 0.000\,000\,000\,10  $ & Eq.~(\ref{eq:c8co})  \\
 $\Delta g^{(10)}                $ & $\m 0.000\,000\,000\,00  $ & Eq.~(\ref{eq:c10co})  \\
 $\Delta g_{\rm rec}            $ & $   -0.000\,000\,117\,02(1) $ & Eqs.~(\ref{eq:grr0})-(\ref{eq:grr2}) \\
 $\Delta g_{\rm ns}             $ & $   -0.000\,000\,001\,56(1)  $ & Eq.~(\ref{eq:gns})  \\
\B $g_{\rm e^-}(^{16}{\rm O}^{7+})$ & $   -2.000\,047\,020\,38(11) \vbox to .4 cm {}  $ & Eq.~(\ref{eq:gco}) \\
\botrule
\end{tabular}
\end{center}
\end{table}

\subsubsection{Measurements of $g_{\rm e}({\rm ^{12}C^{5+}})$ and $g_{\rm
e}({\rm ^{16}O^{7+}})$.}
\label{sssec:bsgfexps}

The experimental data on the electron bound-state $g$-factor in hydrogenic
carbon and oxygen and the inferred values of $A_{\rm r}({\rm e})$ are
summarized in Table~\ref{tab:gfbs}.

\def\m{\phantom{-}}
\def\fixh{\vbox to 9pt {}}
\begin{table*}
\caption{Summary of experimental data on the electron bound-state $g$-factor in
hydrogenic carbon and oxygen and inferred values of the relative atomic mass of
the electron.}
\label{tab:gfbs}
\begin{tabular}{ll@{\qquad}l@{\qquad}c@{\qquad}l}
\toprule
\vbox to 10 pt {}
Input datum  &  \hbox to 15pt {} Value
& \hbox to -20pt {} Relative standard
& Identification & Sec. and Eq. \\
 & & \hbox to -5pt {}uncertainty
$u_{\rm r}$ &  \\
\colrule
\fixh $f_{\rm s}({\rm ^{12}C^{5+}})/f_{\rm c}({\rm ^{12}C^{5+}})$ & $\phantom{-} 4376.210\,4989(23)$ & $ 5.2\times 10^{-10}$ & GSI-02 & \ref{par:gec} (\ref{eq:rfsfcc02}) \\
\ \ $A_{\rm r}({\rm e})$ & $\phantom{-} 0.000\,548\,579\,909\,32(29)$ & $ 5.2\times 10^{-10}$ & & \ref{par:gec} (\ref{eq:arec02}) \\
\fixh $f_{\rm s}({\rm ^{16}O^{7+}})/f_{\rm c}({\rm ^{16}O^{7+}})$ & $\phantom{-} 4164.376\,1837(32)$ & $ 7.6\times 10^{-10}$ & GSI-02 & \ref{par:geo} (\ref{eq:rfsfco02}) \\
\ \ $A_{\rm r}({\rm e})$ & $\phantom{-} 0.000\,548\,579\,909\,58(42)$ & $ 7.6\times 10^{-10}$ & & \ref{par:geo} (\ref{eq:areo02}) \\
\botrule
\end{tabular}
\end{table*}

\paragraph{Experiment on $g_{\rm e}({\rm ^{12}C^{5+}})$.}
\label{par:gec}

The accurate determination of the frequency ratio $f_{\rm
s}(^{12}$C$^{5+})/f_{\rm c}(^{12}$C$^{5+})$ at GSI based on the double
Penning-trap technique was discussed in CODATA-02. [See also the recent concise
review by \textcite{2006035}.]  Since the result used as an input datum in the
2002 adjustment is unchanged, we take it as an input datum in the 2006
adjustment as well \cite{2002007,2003022,pc03gw}:
\begin{eqnarray}
{f_{\rm s}\left(^{12}{\rm C}^{5+}\right)\over
 f_{\rm c}\left(^{12}{\rm C}^{5+}\right)} =
 4376.210\,4989(23) \ .
\label{eq:rfsfcc02}
\end{eqnarray}

From Eq.~(\ref{eq:fsfcgx}) and Eq.~(\ref{eq:araxn}) we have
\begin{eqnarray}
{f_{\rm s}\left(^{12}{\rm C}^{5+}\right)\over
 f_{\rm c}\left(^{12}{\rm C}^{5+}\right)} &=&
-{g_{\rm e^-}\left(^{12}{\rm C}^{5+}\right)\over
10 A_{\rm r}({\rm e})}
\nonumber\\&&\hbox to -2 cm {} \times
\left[12-5A_{\rm r}({\rm e}) + {E_{\rm b}\left(^{12}{\rm C}\right)
-E_{\rm b}\left(^{12}{\rm C}^{5+}\right)\over m_{\rm u}c^2}\right] \ ,
\qquad
\label{eq:rfsfccoe}
\end{eqnarray}
which is the basis for the observational equation for the $^{12}$C$^{5+}$
frequency-ratio input datum.

Evaluation of this expression using the result for $f_{\rm
s}(^{12}$C$^{5+})/f_{\rm c}(^{12}$C$^{5+})$ in Eq.~(\ref{eq:rfsfcc02}), the
theoretical result for $g_{{\rm e}^-}(^{12}$C$^{5+})$ in
Table~\ref{tab:gfactthc}, and the relevant binding energies in Table~IV of
CODATA-02, yields
\begin{eqnarray}
A_{\rm r}({\rm e}) =  0.000\,548\,579\,909\,32(29) \quad [ 5.2\times 10^{-10}] \ . \quad
\label{eq:arec02}
\end{eqnarray}
This value is consistent with that from antiprotonic helium given in
Eq.~(\ref{eq:areaphe}) and that from the University of Washington given in
Eq.~(\ref{eq:areexp}), but has about a factor of three to four smaller
uncertainty.\\

\paragraph{Experiment on $g_{\rm e}({\rm ^{16}O^{7+}})$.}
\label{par:geo}

The double Penning-trap determination of the frequency ratio $f_{\rm
s}(^{16}$O$^{7+})/f_{\rm c}(^{16}$O$^{7+})$ at GSI was also discussed in
CODATA-02, but the value used as an input datum was not quite final
\cite{2002180,pc03gw,2003008}. A slightly different value for the ratio was
given in the final report of the measurement \cite{2002008}, which is the value
we take as the input datum in the 2006 adjustment but modified slightly as
follows based on information provided by \textcite{pc06jv}: (i) an unrounded
instead of a rounded value for the correction due to extrapolating the axial
temperature $T_{z}$ to 0\,K was added to the uncorrected ratio
($-0.000\,004\,7$ in place of $-0.000\,005$); and (ii) a more detailed
uncertainty budget was employed to evaluate the uncertainty of the ratio. The
resulting value is
\begin{eqnarray}
{f_{\rm s}\left(^{16}{\rm O}^{7+}\right)\over
 f_{\rm c}\left(^{16}{\rm O}^{7+}\right)} =
 4164.376\,1837(32)  \ .
\label{eq:rfsfco02}
\end{eqnarray}

In analogy with what was done above with the ratio $f_{\rm
s}(^{12}$C$^{5+})/f_{\rm c}(^{12}$C$^{5+})$, from Eq.~(\ref{eq:fsfcgx}) and
Eq.~(\ref{eq:araxn}) we have
\begin{eqnarray}
&&{f_{\rm s}\left(^{16}{\rm O}^{7+}\right)\over
 f_{\rm c}\left(^{16}{\rm O}^{7+}\right)} =
-{g_{\rm e^-}\left(^{16}{\rm O}^{7+}\right)\over
14 A_{\rm r}({\rm e})} \,
A_{\rm r}\!\left(^{16}{\rm O}^{7+}\right)
\label{eq:rfsfcooe}
\end{eqnarray}
with
\begin{eqnarray}
A_{\rm r}\left(^{16}{\rm O}\right) &=&
A_{\rm r}\left(^{16}{\rm O}^{7+}\right)
+7A_{\rm r}({\rm e})
\nonumber\\ &&\qquad - {
E_{\rm b}\left(^{16}{\rm O}\right) -
E_{\rm b}\left(^{16}{\rm O}^{7+}\right)
\over m_{\rm u}c^2} \ ,
\label{eq:aroobseq}
\end{eqnarray}
which are the basis for the observational equations for the oxygen frequency
ratio and $A_{\rm r}(^{16}$O), respectively. The first expression, evaluated
using the result for $f_{\rm s}(^{16}$O$^{7+})/f_{\rm c}(^{16}$O$^{7+})$ in
Eq.~(\ref{eq:rfsfco02}) and the theoretical result for $g_{{\rm e}^-}
(^{16}$O$^{7+})$ in Table~\ref{tab:gfacttho}, in combination with the second
expression, evaluated using the value of $A_{\rm r}(^{16}$O) in
Table~\ref{tab:rmass06} and the relevant binding energies in Table~IV of
CODATA-02, yields
\begin{eqnarray}
A_{\rm r}({\rm e}) =  0.000\,548\,579\,909\,58(42) \quad [ 7.6\times 10^{-10}] \ . \quad
\label{eq:areo02}
\end{eqnarray}
It is consistent with both the University of Washington value in
Eq.~(\ref{eq:areexp}) and the value in Eq.~(\ref{eq:arec02}) obtained from
$f_{\rm s}(^{12}$C$^{5+})/f_{\rm c}(^{12}$C$^{5+})$.

\paragraph{Relations between $g_{\rm e}({\rm ^{12}C^{5+}})$ and $g_{\rm e}({\rm
^{16}O^{7+}})$.}
\label{par:geco}

It should be noted that the GSI frequency ratios for $^{12}$C$^{5+}$ and
$^{16}$O$^{7+}$ are correlated.  Based on the detailed uncertainty budgets of
the two results \cite{pc03gw, pc06jv},  we find the correlation coefficient to
be
\begin{eqnarray}
r\left[
{f_{\rm s}\left(^{12}{\rm C}^{5+}\right)\over
 f_{\rm c}\left(^{12}{\rm C}^{5+}\right)}
,
{f_{\rm s}\left(^{16}{\rm O}^{7+}\right)\over
 f_{\rm c}\left(^{16}{\rm O}^{7+}\right)}
\right] =  0.082 \ .
\label{eq:rfcfo}
\end{eqnarray}

Finally, as a consistency test, it is of interest to compare the experimental
and theoretical values of the ratio of $g_{{\rm e}^-}(^{12}$C$^{5+})$ to
$g_{{\rm e}^-}(^{16}$O$^{7+})$ \cite{2002168}. The main reason is that the
experimental value of the ratio is only weakly dependent on the value of
$A_{\rm r}({\rm e})$.  The theoretical value of the ratio is given in
Eq.~(\ref{eq:gcogo}) and takes into account the covariance of the two
theoretical values.  The experimental value of the ratio can be obtained by
combining Eqs.~(\ref{eq:rfsfcc02}), (\ref{eq:rfsfccoe}), (\ref{eq:rfsfco02}) to
(\ref{eq:aroobseq}) and (\ref{eq:rfcfo}), and using the 2006 recommended value
for $A_{\rm r}({\rm e})$.  Because of the weak dependence of the experimental
ratio on $A_{\rm r}$(e), the value used is not at all critical.  The result is
\begin{eqnarray}
\frac{g_{{\rm e}^-}(^{12}{\rm C}^{5+})}{g_{{\rm e}^-}(^{16}{\rm O}^{7+})}
&=&  1.000\,497\,273\,68(89) ~ [ 8.9\times 10^{-10}] \ , \qquad
\label{eq:gcogoex}
\end{eqnarray}
in agreement with the theoretical value.

\section{Magnetic moment ratios and the muon-electron mass ratio}
\label{sec:mmrmemr}

Magnetic moment ratios and the muon-electron mass ratio are determined by
experiments on bound states of the relevant particles.  The free electron and
muon magnetic moments are discussed in Sec.~\ref{sec:mmagf} and the theory of
the $g$-factor of an electron bound in an atom with no nuclear spin is
considered in Sec.~\ref{sssec:thbegf}.

For nucleons or nuclei with spin $\bm I$, the magnetic moment can be written as
\begin{eqnarray}
\bm\mu = g { e \over 2m_{\rm p}}\bm I \ ,
\label{eq:ngdef}
\end{eqnarray}
or
\begin{eqnarray}
\mu = g \mu_{\rm N} i \ .
\label{eq:ngdeff}
\end{eqnarray}
In Eq.~(\ref{eq:ngdeff}), $\mu_{\rm N} = e\hbar/2m_{\rm p}$ is the nuclear
magneton, defined in analogy with the Bohr magneton, and $i$ is the spin
quantum number of the nucleus defined by $\bm I^2 = i(i+1)\hbar^2$ and $I_z =
-i\hbar, ... , (i-1)\hbar, i\hbar$, where $I_z$ is the spin projection.
However, in some publications, moments of nucleons are expressed in terms of
the Bohr magneton with a corresponding change in the definition of the
$g$-factor.

For atoms with a nonzero nuclear spin, bound state $g$-factors are defined by
considering the contribution to the Hamiltonian from the interaction of the
atom with an applied magnetic flux density $B$.  For example, for hydrogen, in
the framework of the Pauli approximation, we have  
\begin{eqnarray}
{\cal H} &=& \beta({\rm H})\bm\mu_{\rm e^-}\cdot\bm\mu_{\rm p}
 -\bm\mu_{\rm e^-}({\rm H})\cdot \bm B
  -\bm\mu_{\rm p}({\rm H})\cdot \bm B
\nonumber\\
    &=& {2\rmpi\over\hbar}\Delta\nu_{\rm H}\bm s \cdot \bm I
     - g_{\rm e^-}({\rm H})\,{\mu_{\rm B}\over\hbar}\ \bm s \cdot \bm B
    - g_{\rm p}({\rm H})\,{\mu_{\rm N}\over\hbar}\ \bm I \cdot \bm B \ ,
\nonumber\\
\label{eq:gdefs}
\end{eqnarray}
where $\beta({\rm H})$ characterizes the strength of the hyperfine interaction,
$\Delta\nu_{\rm H}$ is the ground-state hyperfine frequency, $\bm s$ is the
spin of the electron, and $\bm I$ is the spin of the nucleus, that is, the
proton.  Equation~(\ref{eq:gdefs}), or its analog for other combinations of
particles, serves to define the corresponding bound-state $g$-factors, which
are $g_{\rm e^-}({\rm H})$ and $g_{\rm p}({\rm H})$ in this case.

\subsection{Magnetic moment ratios}
\label{ssec:mmr}

A number of magnetic moment ratios are of interest for the 2006 adjustment.
The results of measurements and the inferred values of various quantities are
summarized in Sec.~\ref{sssec:exps}, and the measurement results themselves are
also summarized in Table~\ref{tab:momrats}.

The inferred moment ratios depend on the relevant theoretical binding
corrections that relate the $g$-factor measured in the bound state to the
corresponding free-particle $g$-factor.  To use the results of these
experiments in the 2006 adjustment, we employ theoretical expressions that give
predictions for the moments and $g$-factors of the bound particles in terms of
free-particle moments and $g$-factors as well as adjusted constants; this is
discussed in the following section.  However, in a number of cases, the
differences between the bound-state and free-state values are sufficiently
small that the adjusted constants can be taken as exactly known.

\subsubsection{Theoretical ratios of atomic bound-particle to free-particle
$g$-factors}
\label{sssec:thbfrats}

Theoretical $g$-factor-related quantities used in the 2006 adjustment are the
ratio of the $g$-factor of the electron in the ground state of hydrogen to that
of the free electron $g_{\rm e^-}({\rm H})/g_{\rm e^-}$; the ratio of the
$g$-factor of the proton in hydrogen to that of the free proton $g_{\rm p}({\rm
H})/g_{\rm p}$; the analogous ratios for the electron and deuteron in
deuterium, $g_{\rm e^-}({\rm D})/g_{\rm e^-}$ and $g_{\rm d}({\rm D})/g_{\rm
d}$, respectively; and the analogous ratios for the electron and positive muon
in muonium, $g_{\rm e^-}({\rm Mu})/g_{\rm e^-}$ and $g_{\rm \mu^+}({\rm
Mu})/g_{\rm \mu^+}$, respectively.

These ratios and the references for the relevant calculations are discussed in
CODATA-98 and CODATA-02; only a summary of the results is included here.

For the electron in hydrogen, we have
\begin{eqnarray}
{g_{\rm e^-}({\rm H}) \over g_{\rm e^-}}
&=& 1 -\fr{1}{3}(Z\alpha)^2 - \fr{1}{12}(Z\alpha)^4 
+ \fr{1}{4}(Z\alpha)^2\left({\alpha\over\rmpi}\right)
\nonumber\\ &&
+ \fr{1}{2}(Z\alpha)^2{m_{\rm e}\over m_{\rm p}}
+ \fr{1}{2}\left(A_1^{(4)}-\fr{1}{4}\right)(Z\alpha)^2
\left(\alpha\over\rmpi\right)^2 
\nonumber \\ &&
-\fr{5}{12}(Z\alpha)^2\left({\alpha\over\rmpi}\right)
{m_{\rm e}\over m_{\rm p}}
+ \cdots \ ,
\label{eq:ehgrat}
\end{eqnarray}
where $A_1^{(4)}$ is given in Eq.~(\ref{eq:a14}).  
For the proton in hydrogen, we have
\begin{eqnarray}
{g_{\rm p}({\rm H}) \over g_{\rm p}} &=& 1 - \fr{1}{3}\alpha(Z\alpha)
- \fr{97}{108}\alpha(Z\alpha)^3
\nonumber\\
&&+\fr{1}{6} \alpha(Z\alpha) {m_{\rm e} \over m_{\rm p}} 
{3+4a_{\rm p}\over 1+a_{\rm p}} + \cdots \ ,
\label{eq:phgrat}
\end{eqnarray}
where the proton magnetic moment anomaly $a_{\rm p}$ is defined by
\begin{eqnarray}
a_{\rm p} &=& {\mu_{\rm p} \over \left(e\hbar/2m_{\rm p}\right)}-1
\approx 1.793
\ .
\label{eq:apdef}
\end{eqnarray}

For deuterium, similar expressions apply for the electron
\begin{eqnarray}
{g_{\rm e^-}({\rm D}) \over g_{\rm e^-}}
&=& 1 -\fr{1}{3}(Z\alpha)^2 - \fr{1}{12}(Z\alpha)^4 
+ \fr{1}{4}(Z\alpha)^2\left({\alpha\over\rmpi}\right)
\nonumber \\ &&
+ \fr{1}{2}(Z\alpha)^2{m_{\rm e}\over m_{\rm d}}
+ \fr{1}{2}\left(A_1^{(4)}-\fr{1}{4}\right)(Z\alpha)^2
\left(\alpha\over\rmpi\right)^2 
\nonumber \\ &&
-\fr{5}{12}(Z\alpha)^2\left({\alpha\over\rmpi}\right)
{m_{\rm e}\over m_{\rm d}}
+ \cdots \ ,
\label{eq:edgrat}
\end{eqnarray}
and deuteron
\begin{eqnarray}
{g_{\rm d}({\rm D}) \over g_{\rm d}} &=& 1 - \fr{1}{3}\alpha(Z\alpha)
- \fr{97}{108}\alpha(Z\alpha)^3
\nonumber\\
&&+\fr{1}{6} \alpha(Z\alpha) {m_{\rm e} \over m_{\rm d}} 
{3+4a_{\rm d}\over 1+a_{\rm d}} + \cdots \ ,
\label{eq:ddgrat}
\end{eqnarray}
where the deuteron magnetic moment anomaly $a_{\rm d}$ is defined by
\begin{eqnarray}
a_{\rm d} = {\mu_{\rm d}\over
 \left(e\hbar/ m_{\rm d}\right)} - 1 \approx -0.143 \ .
\end{eqnarray}

In the case of muonium Mu, some additional higher-order terms are included
because of the larger mass ratio.  For the electron in muonium, we have
\begin{eqnarray}
{g_{\rm e^-}({\rm Mu}) \over g_{\rm e^-}}
&=& 1 -\fr{1}{3}(Z\alpha)^2 - \fr{1}{12}(Z\alpha)^4 
+ \fr{1}{4}(Z\alpha)^2\left({\alpha\over\rmpi}\right)
\nonumber \\ &&
+ \fr{1}{2}(Z\alpha)^2{m_{\rm e}\over m_{\rmssmu}}
+ \fr{1}{2}\left(A_1^{(4)}-\fr{1}{4}\right)(Z\alpha)^2
\left(\alpha\over\rmpi\right)^2 
\nonumber \\ && \hbox to -0.5 cm {}
-\fr{5}{12}(Z\alpha)^2\left({\alpha\over\rmpi}\right)
{m_{\rm e}\over m_{\rmssmu}}
- \fr{1}{2}(1+Z)(Z\alpha)^2\left({m_{\rm e}\over m_{\rmssmu}}\right)^{\!2}
\nonumber \\ && \hbox to -0.5 cm {}
+ \cdots \ ,
\label{eq:emugrat}
\end{eqnarray}
and for the muon in muonium, the ratio is
\begin{eqnarray}
{g_{{\rmssmu}^+}({\rm Mu}) \over g_{{\rmssmu}^+}} &=& 
1 - \fr{1}{3}\alpha(Z\alpha) - \fr{97}{108}\alpha(Z\alpha)^3
\nonumber\\ &&
+ \fr{1}{2} \alpha(Z\alpha) {m_{\rm e} \over m_{\rmssmu}}
+ \fr{1}{12} \alpha(Z\alpha) 
\left({\alpha\over\rmpi}\right)
{m_{\rm e} \over m_{\rmssmu}}
\nonumber\\ &&
- \fr{1}{2}(1+Z)\alpha(Z\alpha) \left({m_{\rm e} \over m_{\rmssmu}}\right)^2
 + \cdots \ . 
\nonumber\\
\label{eq:mumugrat}
\end{eqnarray}

The numerical values of the corrections in Eqs.~(\ref{eq:ehgrat}) to
(\ref{eq:mumugrat}), based on the 2006 adjusted values of the relevant
constants, are listed in Table \ref{tab:gfactrat}.  Uncertainties are
negligible at the level of uncertainty of the relevant experiments.

\begin{table}[t]
\caption{Theoretical values for various bound-particle to free-particle
$g$-factor ratios relevant to the 2006 adjustment based on the 2006 recommended
values of the constants.}
\label{tab:gfactrat}
\begin{center}
\begin{tabular}{c@{\qquad}l}
\toprule
Ratio   &   \qquad Value \\
\colrule
 $g_{\rm e^-}({\rm H})/g_{\rm e^-}$         & $  1 -17.7054\times 10^{-6} $ \T \\
 $g_{\rm p}({\rm H})/g_{\rm p}$             & $  1 -17.7354\times 10^{-6} $ \\
 $g_{\rm e^-}({\rm D})/g_{\rm e^-}$         & $  1 -17.7126\times 10^{-6} $ \\
 $g_{\rm d}({\rm D})/g_{\rm d}$             & $  1 -17.7461\times 10^{-6} $ \\
 $g_{\rm e^-}({\rm Mu})/g_{\rm e^-}$        & $  1 -17.5926\times 10^{-6} $ \\
 $g_{\rm \mu^+}({\rm Mu})/g_{\rm \mu^+}$ \B & $  1 -17.6254\times 10^{-6} $ \B \\
\botrule
\end{tabular}
\end{center}
\end{table}

\subsubsection{Ratio measurements}
\label{sssec:exps}

\paragraph{Electron to proton magnetic moment ratio ${\mu_{\rm e}/\mu_{\rm
p}}$.}
\label{par:epmmr}

The ratio $\mu_{\rm e}/\mu_{\rm p}$ is obtained from measurements of the ratio
of the magnetic moment of the electron to the magnetic moment of the proton in
the 1S state of hydrogen $\mu_{\rm e^-}({\rm H})/\mu_{\rm p}({\rm H})$. We use
the value obtained by \textcite{1972028} at MIT:
\begin{eqnarray}
{\mu_{\rm e^-}({\rm H})\over\mu_{\rm p}({\rm H})} =  -658.210\,7058(66)
\quad [ 1.0\times 10^{-8}] \ ,
\label{eq:wkmw}
\end{eqnarray}
where a minor typographical error in the original publication has been
corrected \cite{pc97dk}.  The free-particle ratio $\mu_{\rm e}/\mu_{\rm p}$
follows from the bound-particle ratio and the relation
\begin{eqnarray}
{\mu_{\rm e^-}\over\mu_{\rm p}} &=&
{g_{\rm p}({\rm H}) \over g_{\rm p}}\left(
{g_{\rm e^-}({\rm H}) \over g_{\rm e^-}}\right)^{-1}
{\mu_{\rm e^-}({\rm H})\over\mu_{\rm p}({\rm H})}
\nonumber\\
&=&  -658.210\,6860(66) \quad [ 1.0\times 10^{-8}] \ ,
\label{eq:muemup}
\end{eqnarray}
where the bound-state $g$-factor ratios are given in Table \ref{tab:gfactrat}. 

\begin{table*}
\def\m{\phantom{-}}
\caption{Summary of data for magnetic moment ratios of various bound
particles.}
\label{tab:momrats}
\begin{tabular}{l@{\qquad}lcll}
\toprule
\vbox to 10 pt {}
Quantity     &  \hbox to 12pt {} Value
& Relative standard
& Identification & Sect. and Eq. \\
&& uncertainty $u_{\rm r}$&& \\
\colrule
$\mu_{\rm e^-}({\rm H})/\mu_{\rm p}({\rm H})$ & $ -658.210\,7058(66)$ 
    & $ 1.0\times 10^{-8}$ & 
    MIT-72 \vbox to 12 pt {} &
    \ref{par:epmmr} (\ref{eq:wkmw}) \\
$\mu_{\rm d}({\rm D})/\mu_{\rm e^-}({\rm D})$ & $ -4.664\,345\,392(50)\times 10^{-4}$ & $ 1.1\times 10^{-8}$ & 
    MIT-84 &
    \ref{par:demmr} (\ref{eq:mddmedmit78}) \\
$\mu_{\rm p}({\rm HD})/\mu_{\rm d}({\rm HD})$ & $\phantom{-}  3.257\,199\,531(29)$ & $ 8.9\times 10^{-9}$ & StPtrsb-03 & \ref{par:dpmmr} (\ref{eq:muphd}) \\
$\sigma_{\rm dp}$ & $\phantom{-}  15(2)\times 10^{-9}$ & & StPtrsb-03 & \ref{par:dpmmr} (\ref{eq:sdp03}) \\
$\mu_{\rm t}({\rm HT})/\mu_{\rm p}({\rm HT})$ & $\phantom{-}  1.066\,639\,887(10)$ & $ 9.4\times 10^{-9}$ & StPtrsb-03 & \ref{par:dpmmr} (\ref{eq:mutht}) \\
$\sigma_{\rm tp}$ & $\phantom{-}  20(3)\times 10^{-9}$ & & StPtrsb-03 & \ref{par:dpmmr} (\ref{eq:stp03}) \\
$\mu_{\rm e^-}({\rm H})/\mu_{\rm p}^\prime$ & 
   $ -658.215\,9430(72)$ & $ 1.1\times 10^{-8}$ & 
    MIT-77 &
    \ref{par:espmm} (\ref{eq:muehmuppmit77}) \\
$\mu_{\rm h}^\prime/\mu_{\rm p}^\prime$ &
   $ -0.761\,786\,1313(33)$ & $ 4.3\times 10^{-9}$ & 
    NPL-93 &
    \ref{par:hspmmr} (\ref{eq:muhe3muppnpl93}) \\
$\mu_{\rm n}/\mu_{\rm p}^\prime$  \B &
   $ -0.684\,996\,94(16)$ & $ 2.4\times 10^{-7}$ & 
    ILL-79 &
    \ref{par:nsprmr} (\ref{eq:munmupphar79}) \\
\botrule
\end{tabular}
\end{table*}

\paragraph{Deuteron to electron magnetic moment ratio ${\mu_{\rm d}/\mu_{\rm
e}}$.}
\label{par:demmr}

From measurements of the ratio $\mu_{\rm d}({\rm D})/\mu_{\rm e^-}({\rm D})$ in
the 1S state of deuterium, \textcite{pc84pkw} at MIT obtained
\begin{eqnarray}
{\mu_{\rm d}({\rm D})\over\mu_{\rm e^-}({\rm D})} = -4.664\,345\,392(50)\times 10^{-4}
\quad [ 1.1\times 10^{-8}] \ .
\nonumber\\
\label{eq:mddmedmit78}
\end{eqnarray}
Although this result has not been published, as in the 1998 and 2002
adjustments, we include it as an input datum, because the method is described
in detail by \textcite{1972028} in connection with their measurement of
$\mu_{\rm e^-}({\rm H})/\mu_{\rm p}({\rm H})$.  The free-particle ratio is
given by
\begin{eqnarray}
{\mu_{\rm d}\over\mu_{\rm e^-}} &=&
{g_{\rm e^-}({\rm D}) \over g_{\rm e^-}}
\left({g_{\rm d}({\rm D}) \over g_{\rm d}}\right)^{-1}
{\mu_{\rm d}({\rm D})\over\mu_{\rm e^-}({\rm D})}
\nonumber\\
&=&  -4.664\,345\,548(50)\times 10^{-4} \quad [ 1.1\times 10^{-8}] \ ,
\nonumber\\&&
\label{eq:rmudmue}
\end{eqnarray}
with the bound-state $g$-factor ratios given in Table \ref{tab:gfactrat}.

\paragraph{Proton to deuteron and triton to proton magnetic moment ratios 
${\mu_{\rm p}}/\mu_{\rm d}$ and ${\mu_{\rm t}}/\mu_{\rm p}$}
\label{par:dpmmr}

The ratios ${\mu_{\rm p}}/\mu_{\rm d}$ and ${\mu_{\rm t}}/\mu_{\rm p}$ can be
determined by nuclear magnetic resonance (NMR) measurements on the HD molecule
(bound state of hydrogen and deuterium) and the HT molecule (bound state of
hydrogen and tritium, $^3{\rm H}$), respectively.  The relevant expressions are
(see CODATA-98)
\begin{eqnarray}
\frac{\mu_{\rm p}({\rm HD})}{\mu_{\rm d}({\rm HD})} &=&
\left[1 + \sigma_{\rm d}({\rm HD}) - \sigma_{\rm p}({\rm HD})
\right]
\frac{\mu_{\rm p}}{\mu_{\rm d}}
\label{eq:mupmudhd}
\\
\frac{\mu_{\rm t}({\rm HT})}{\mu_{\rm p}({\rm HT})} &=&
\left[1 - \sigma_{\rm t}({\rm HT}) + \sigma_{\rm p}({\rm HT})
\right]
\frac{\mu_{\rm t}}{\mu_{\rm p}} \ ,
\label{eq:mutmupht}
\end{eqnarray}
where $\mu_{\rm p}({\rm HD})$ and $\mu_{\rm d}({\rm HD})$ are the proton and
deuteron magnetic moments in HD, respectively, and $\sigma_{\rm p}({\rm HD})$
and $\sigma_{\rm d}({\rm HD})$ are the corresponding nuclear magnetic shielding
corrections.  Similarly, $\mu_{\rm t}({\rm HT})$ and $\mu_{\rm p}({\rm HT})$
are the triton (nucleus of tritium) and proton magnetic moments in HT,
respectively, and $\sigma_{\rm t}({\rm HT})$ and $\sigma_{\rm p}({\rm HT})$ are
the corresponding nuclear magnetic shielding corrections.  [Note that $\mu(\rm
bound)=(1-\sigma)\mu(\rm free)$ and the nuclear magnetic shielding corrections
are small.]

The determination of ${\mu_{\rm d}}/{\mu_{\rm p}}$ from NMR measurements on HD
by \textcite{1953009} and by a Russian group working in St. Petersburg
\cite{1975024, 1989006} was discussed in CODATA-98. However, for reasons given
there, mainly the lack of sufficient information to assign a reliable
uncertainty to the reported values of $\mu_{\rm d}({\rm HD})/{\mu_{\rm p}({\rm
HD})}$ and also to the nuclear magnetic shielding correction difference
$\sigma_{\rm d}(\rm HD) -\sigma_{\rm p}(\rm HD)$, the results were not used in
the 1998 or 2002 adjustments. Further, since neither of these adjustments
addressed quantities related to the triton, the determination of ${\mu_{\rm
t}}/{\mu_{\rm p}}$ from measurements on HT by the Russian group \cite{1977015}
was not considered in either of these adjustments.  It may be recalled that
a systematic error related to the use of separate inductance coils for the
proton and deuteron NMR resonances in the measurements of \textcite{1975024}
was eliminated in the HT measurements of \textcite{1977015} as well as in the
HD measurements of \textcite{1989006}.

Recently, a member of the earlier St. Petersburg group together with one or
more other Russian colleagues in St. Petersburg published the following results
based in part on new measurements and re-examination of relevant theory
\cite{2003262, 2005103}:
\begin{eqnarray}
\frac{\mu_{\rm p}({\rm HD})}{\mu_{\rm d}({\rm HD})}
&=&  3.257\,199\,531(29) \qquad [ 8.9\times 10^{-9}]
\label{eq:muphd}
\quad
\\
\frac{\mu_{\rm t}({\rm HT})}{\mu_{\rm p}({\rm HT})}
&=&  1.066\,639\,887(10) \qquad [ 9.4\times 10^{-9}]
\label{eq:mutht}
\quad
\end{eqnarray}
\begin{eqnarray}
\sigma_{\rm dp} &\equiv&
\sigma_{\rm d}({\rm HD}) - \sigma_{\rm p}({\rm HD})
=  15(2)\times 10^{-9}
\label{eq:sdp03}
\\
\sigma_{\rm tp} &\equiv&
\sigma_{\rm t}({\rm HT}) - \sigma_{\rm p}({\rm HT})
=  20(3)\times 10^{-9} \ ,
\label{eq:stp03}
\end{eqnarray}
which together with Eqs.~(\ref{eq:mupmudhd}) and (\ref{eq:mutmupht})
yield
\begin{eqnarray}
\frac{\mu_{\rm p}}{\mu_{\rm d}}
&=&  3.257\,199\,482(30) \qquad [ 9.1\times 10^{-9}]
\label{eq:mupsmudstp03}
\quad
\\
\frac{\mu_{\rm t}}{\mu_{\rm p}}
&=&  1.066\,639\,908(10) \qquad [ 9.8\times 10^{-9}] \ .
\label{eq:mutsmupstp03}
\quad
\end{eqnarray}

The purpose of the new work \cite{2003262, 2005103} was (i) to check whether
rotating the NMR sample and using a high-pressure gas as the sample (60 to 130
atmospheres), which was the case in most of the older Russian experiments,
influenced the results and to report a value of $\mu_{\rm p}({\rm HD})/\mu_{\rm
d}({\rm HD})$ with a reliable uncertainty; and (ii) to re-examine the
theoretical values of the nuclear magnetic shielding correction differences
$\sigma_{\rm dp}$ and $\sigma_{\rm tp}$ and their uncertainties. It was also
anticipated that based on this new work, a value of ${\mu_{\rm t}({\rm
HT})}/{\mu_{\rm p}({\rm HT})}$ with a reliable uncertainty could be obtained
from the highly precise measurements of \textcite{1977015}.  However,
\textcite{1989006}, as part of their experiment to determine $\mu_{\rm
d}/\mu_{\rm p}$, compared the result from a 100 atmosphere HD rotating sample
with a 100 atmosphere HD non-rotating sample and found no statistically
significant difference.

To test the effect of sample rotation and sample pressure, \textcite{2003262}
performed measurements using a commercial NMR spectrometer operating at a
magnetic flux density of about 7~T and a non-rotating 10 atmosphere HD gas
sample. Because of the relatively low pressure, the NMR signals were
comparatively weak and a measurement time of 1~h was required.  To simplify
the measurements, the frequency of the proton NMR signal from HD was determined
relative to the frequency of the more easily measured proton NMR signal from
acetone, $(\rm CH_{3})_{2}$CO. Similarly, the frequency of the deuteron NMR
signal from HD was determined relative to the frequency of the more easily
measured deuteron NMR signal from deuterated acetone, $(\rm CD_{3})_{2}$CO. A
number of tests involving the measurement of the hyperfine interaction constant
in the case of the proton triplet NMR spectrum, and the isotopic shift in the
case of the deuteron, where the deuteron HD doublet NMR spectrum was compared
with the singlet spectrum of $\rm D_{2}$, were carried out to investigate the
reliability of the new data. The results of the tests were in good agreement
with the older results obtained with sample rotation and high gas pressure. 

The more recent result for $\mu_{\rm p}({\rm HD})/ \mu_{\rm d}({\rm HD})$
reported by \textcite{2005103}, which was obtained with the same NMR
spectrometer employed by \textcite{2003262} but with a 20 atmosphere
non-rotating gas sample, agrees with the 10 atmosphere non-rotating sample result
of the latter researchers and is interpreted by \textcite{2005103} as
confirming the 2003 result.  Although the values of ${\mu_{\rm p}({\rm
HD})}/{\mu_{\rm d}({\rm HD})}$ reported by the Russian researchers in 2005,
2003, and 1989 agree, the 2003 result as given in Eq.~(\ref{eq:muphd}) and
Table~\ref{tab:momrats}, the uncertainty of which is dominated by the proton
NMR line fitting procedure, is taken as the input datum in the 2006 adjustment
because of the attention paid to possible systematic effects, both experimental
and theoretical.

Based on their HD measurements and related analysis, especially the fact that
sample pressure and rotation do not appear to be a problem at the current level
of uncertainty, \textcite{2003262} conclude that the result for ${\mu_{\rm
t}({\rm HT})}/{\mu_{\rm p}({\rm HT})}$ reported by \textcite{1977015} is
reliable but that it should be assigned about the same relative uncertainty as
their result for ${\mu_{\rm p}({\rm HD})}/{\mu_{\rm d}({\rm HD})}$.  We
therefore include as an input datum in the 2006 adjustment the result for
${\mu_{\rm t}({\rm HT})}/{\mu_{\rm p}({\rm HT})}$ given in Eq.~(\ref{eq:mutht})
and Table~\ref{tab:momrats}.

Without reliable theoretically calculated values for the shielding correction
differences $\sigma_{\rm dp}$ and $\sigma_{\rm tp}$, reliable experimental
values for the ratios ${\mu_{\rm p}({\rm HD})}/{\mu_{\rm d}({\rm HD})}$ and
${\mu_{\rm t}({\rm HT})}/{\mu_{\rm p}({\rm HT})}$ are of little use.  Although
\textcite{1977015} give theoretical estimates of these quantities based on
their own calculations, they do not discuss the uncertainties of their
estimates. To address this issue, \textcite{2003262} carefully examined the
calculations and concluded that a reasonable estimate of the relative
uncertainty is 15~\%. This leads to the values for $\sigma_{\rm dp}$ and
$\sigma_{\rm tp}$ in Eqs.~(\ref{eq:sdp03}) and (\ref{eq:stp03}) and
Table~\ref{tab:momrats}, which we also take as input data for the 2006
adjustment. [For simplicity, we use StPtrsb-03 as the identifier in
Table~\ref{tab:momrats} for ${\mu_{\rm p}({\rm HD})}/{\mu_{\rm d}({\rm HD})}$,
${\mu_{\rm t}({\rm HT})}/{\mu_{\rm p}({\rm HT})}$, $\sigma_{\rm dp}$, and
$\sigma_{\rm tp}$, because they are directly or indirectly a consequence of the
work of \textcite{2003262}.]

The equations for the measured moment ratios 
${\mu_{\rm p}({\rm HD})}/{\mu_{\rm d}({\rm HD})}$ and 
${\mu_{\rm t}({\rm HT})}/{\mu_{\rm p}({\rm HT})}$
in terms of the adjusted constants 
${\mu_{\rm e^-}}/{\mu_{\rm p}}$, 
${\mu_{\rm d}}/{\mu_{\rm e^-}}$, 
${\mu_{\rm t}}/{\mu_{\rm p}}$, 
$\sigma_{\rm dp}$, and 
$\sigma_{\rm tp}$ 
are, from Eqs.~({\ref{eq:mupmudhd}) and ({\ref{eq:mutmupht}),
\begin{eqnarray}
\frac{\mu_{\rm p}({\rm HD})}{\mu_{\rm d}({\rm HD})} &=&
\left[1 + \sigma_{\rm dp} \right]
\left(\frac{\mu_{\rm e^-}}{\mu_{\rm p}}\right)^{-1}
\left(\frac{\mu_{\rm d}}{\mu_{\rm e^-}}\right)^{-1}
\label{eq:obsmupmudhd}
\\
\frac{\mu_{\rm t}({\rm HT})}{\mu_{\rm p}({\rm HT})} &=&
\left[1 - \sigma_{\rm tp} \right]
\frac{\mu_{\rm t}}{\mu_{\rm p}} \ .
\label{eq:obsmutmupht}
\end{eqnarray}

\paragraph{Electron to shielded proton magnetic moment ratio ${\mu_{\rm
e}/\mu_{\rm p}^\prime}$.}
\label{par:espmm}

Based on the measurement of the ratio of the electron moment in the 1S state of
hydrogen to the shielded proton moment at 34.7 $^\circ$C by \textcite{1977016}
at MIT, and temperature-dependence measurements of the shielded proton moment
by \textcite{1984035} at the National Physical Laboratory (NPL), Teddington,
UK, we have
\begin{eqnarray}
{\mu_{\rm e^-}({\rm H})\over \mu_{\rm p}^\prime}
 &=&  -658.215\,9430(72) \quad [ 1.1\times 10^{-8}] \ ,
\label{eq:muehmuppmit77}
\end{eqnarray}
where the prime indicates that the protons are in a spherical sample of pure
H$_2$O at 25 $^\circ$C surrounded by vacuum.  Hence
\begin{eqnarray}
{\mu_{\rm e^-}\over \mu_{\rm p}^\prime} &=&
\left({g_{\rm e^-}({\rm H})\over g_{\rm e^-} }\right)^{-1}
{\mu_{\rm e^-}({\rm H})\over \mu_p^\prime}
\nonumber\\
&=& -658.227\,5971(72) \quad [ 1.1\times 10^{-8}] \ ,
\label{eq:muemuppmit77}
\end{eqnarray}
where the bound-state $g$-factor ratio is given in Table \ref{tab:gfactrat}.
Support for the MIT result in Eq.~(\ref{eq:muemuppmit77}) from measurements at
NPL on the helion (see the following section) is discussed in CODATA-02.

\paragraph{Shielded helion to shielded proton magnetic moment ratio
${\mu^\prime_{\rm h}/\mu_{\rm p}^\prime}$.}
\label{par:hspmmr}

The ratio of the magnetic moment of the helion h, the nucleus of the $^3$He
atom, to the magnetic moment of the proton in H$_2$O was determined in a
high-accuracy experiment at NPL \cite{1993113} with the result
\begin{eqnarray}
{\mu^\prime_{\rm h}\over \mu_{\rm p}^\prime} =
 -0.761\,786\,1313(33) \quad [ 4.3\times 10^{-9}] \ .
\label{eq:muhe3muppnpl93}
\end{eqnarray}
The prime on the symbol for the helion moment indicates that the helion is not
free, but is bound in a helium atom. Although the exact shape and temperature
of the gaseous $^3$He sample is unimportant, we assume that it is spherical, at
25~$^\circ$C, and surrounded by vacuum.

\paragraph{Neutron to shielded proton magnetic moment ratio ${\mu_{\rm
n}/\mu_{\rm p}^\prime}$.}
\label{par:nsprmr}

Based on a measurement carried out at the Institut Max von Laue-Paul Langevin
(ILL) in Grenoble, France \cite{1979014, 1977020}, we have 
\begin{eqnarray}
{\mu_{\rm n}\over\mu_{\rm p}^\prime} =
 -0.684\,996\,94(16) \quad [ 2.4\times 10^{-7}] \ .
\label{eq:munmupphar79}
\end{eqnarray}

The observational equations for the measured values of 
${\mu^\prime_{\rm h}/\mu_{\rm p}^\prime}$ and 
${\mu_{\rm n}/\mu_{\rm p}^\prime}$ are simply 
\begin{eqnarray}
{\mu^\prime_{\rm h}}/{\mu_{\rm p}^\prime} &=&
{\mu^\prime_{\rm h}}/{\mu_{\rm p}^\prime} 
\end{eqnarray}
and 
\begin{eqnarray}
\mu_{\rm n}/{\mu_{\rm p}^\prime} &=&
\mu_{\rm n}/{\mu_{\rm p}^\prime}, 
\end{eqnarray}
while 
the observational equations for the measured values of 
$\mu_{\rm e^-}({\rm H})/\mu_{\rm p}({\rm H})$, 
$\mu_{\rm d}({\rm D})/\mu_{\rm e^-}({\rm D})$, and 
$\mu_{\rm e^-}({\rm H})/\mu_{\rm p}^{\prime}$ follow directly 
from Eqs.~(\ref{eq:muemup}), (\ref{eq:rmudmue}), and 
(\ref{eq:muemuppmit77}), respectively.

\subsection{Muonium transition frequencies, the muon-proton magnetic moment
ratio ${\mu_{\rmssmu}/\mu_{\rm p}}$, and muon-electron mass ratio
$m_{\rmssmu}/m_{\rm e}$}
\label{ssec:muhfs}

Measurements of transition frequencies between Zeeman energy levels in muonium
(the $\rmmu^+{\rm e}^-$ atom) yield values of $\mu_{\rmssmu}/\mu_{\rm p}$ and
the muonium ground-state hyperfine splitting $\Delta \nu_{\rm Mu}$ that depend
weakly on theory.  The relevant expression for the magnetic moment ratio is
\begin{eqnarray}
{\mu_{{\rmssmu}^+} \over \mu_{\rm p}} &=&
{\Delta \nu_{\rm Mu}^2 - \nu^2(f_{\rm p}) + 2 s_{\rm e}f_{\rm p}\,\nu(f_{\rm p})
\over 4 s_{\rm e} f_{\rm p}^2 - 2 f_{\rm p}\,\nu(f_{\rm p})}
\left({g_{{\rmssmu}^+}(\rm Mu})\over g_{{\rmssmu}^+}\right)^{-1} \ ,
\nonumber\\
\label{eq:murat}
\end{eqnarray}
where $\Delta \nu_{\rm Mu}$ and $\nu(f_{\rm p})$ are the sum and difference of
two measured transition frequencies, $f_{\rm p}$ is the free proton NMR
reference frequency corresponding to the magnetic flux density used in the
experiment, $g_{{\rmssmu}^+}({\rm Mu})/ g_{{\rmssmu}^+}$ is the bound-state
correction for the muon in muonium given in Table \ref{tab:gfactrat}, and
\begin{eqnarray}
s_{\rm e} = {\mu_{\rm e^-}\over\mu_{\rm p}}
{g_{\rm e^-}({\rm Mu})\over g_{\rm e^-}} \ ,
\label{eq:se}
\end{eqnarray}
where $g_{\rm e^-}({\rm Mu}) / g_{\rm e^-}$ is the bound-state correction for
the electron in muonium given in the same table.

The muon to electron mass ratio $m_{\rmssmu}/m_{\rm e}$ and the muon to proton
magnetic moment ratio $\mu_{\rmssmu}/\mu_{\rm p}$ are related by
\begin{eqnarray}
{m_{\rmssmu}\over m_{\rm e}} =
\left({\mu_{\rm e}\over \mu_{\rm p}}\right)
\left({\mu_{\rmssmu}\over \mu_{\rm p}}\right)^{-1}
\left({g_{\rmssmu}\over g_{\rm e}}\right) \ .
\label{eq:mumemump}
\end{eqnarray}

The theoretical expression for the hyperfine splitting $\Delta\nu_{\rm Mu}({\rm
th})$ is discussed in the following section and may be written as 
\begin{eqnarray}
\Delta\nu_{\rm Mu}({\rm th}) &=& {16\over3} c R_\infty\alpha^2
{m_{\rm e}\over m_{\rmssmu}}\left(1+{m_{\rm e}\over m_{\rmssmu}}
\right)^{-3} {\cal F}
\hbox to -2pt {}\left(\alpha, {m_{\rm e}/ m_{\rmssmu}}\right)
\nonumber\\
&=&\Delta\nu_{\rm F}
       {\cal F}\hbox to -2pt {}
\left(\alpha, {m_{\rm e}/ m_{\rmssmu}}\right) \ ,
\label{eq:hfsth}
\end{eqnarray}
where the function ${\cal F}$ depends weakly on $\alpha$ and $m_{\rm e}/
m_{\rmssmu}$.  By equating this expression to an experimental value of
$\Delta\nu_{\rm Mu}$, one can calculate a value of $\alpha$ from a given value
of $m_{\rmssmu}/ m_{\rm e}$ or one can calculate a value of $m_{\rmssmu}/
m_{\rm e}$ from a given value of $\alpha$.

\subsubsection{Theory of the muonium ground-state hyperfine splitting}
\label{sssec:muhfs}

This section gives a brief summary of the present theory of $\Delta \nu_{\rm
Mu}$, the ground-state hyperfine splitting of muonium (${\rmmu}^+{\rm e}^-$
atom).  There has been essentially no change in the theory since the 2002
adjustment.  Although complete results of the relevant calculations are given
here, references to the original literature included in CODATA-98 or CODATA-02
are generally not repeated.

The hyperfine splitting is given mainly by the Fermi formula:
\begin{eqnarray} 
\Delta\nu_{\rm F} 
= {16\over 3} c R_\infty Z^3\alpha^2 {m_{\rm e}\over m_{\rmssmu}} 
\left[ 1+{m_{\rm e}\over m_{\rmssmu}} \right]^{-3}\ .
\label{eq:dfermi}
\end{eqnarray} 
Some of the theoretical expressions correspond to a muon with charge $Ze$
rather than $e$ in order to identify the source of the terms.  The theoretical
value of the hyperfine splitting is given by
\begin{eqnarray} 
\Delta \nu_{\rm Mu} (\hbox{th}) 
&=& \Delta \nu_{\rm D} + \Delta \nu_{\rm rad} + \Delta \nu_{\rm rec} 
\nonumber\\
&&+ \Delta \nu_{\rm r\hbox{-}r} 
+ \Delta \nu_{\rm weak} + \Delta \nu_{\rm had}\ , 
\label{eq:dmuth}
\end{eqnarray} 
where the terms labeled D, rad, rec, r-r, weak, and had account for the Dirac
(relativistic), radiative, recoil, radiative-recoil, electroweak, and hadronic
(strong interaction) contributions to the hyperfine splitting, respectively. 

The contribution $\Delta \nu_{\rm D}$, given by the Dirac equation, is
\begin{eqnarray}
\Delta \nu_{\rm D} &=& 
\Delta \nu_{\rm F}(1 + a_{\rmssmu})
\left[1+\fr{3}{2}(Z\alpha)^2 + \fr{17}{8}(Z\alpha)^4
+ \cdots \ \right] ,
\nonumber\\
\label{eq:dbreit}
\end{eqnarray}
where $a_{\rmssmu}$ is the muon magnetic moment anomaly.

The radiative corrections are written as
\begin{eqnarray}
\Delta \nu_{\rm rad} &=& 
\Delta \nu_{\rm F}(1 + a_{\rmssmu})
\Big[
D^{(2)}\!(Z\alpha) \left({\alpha\over\rmpi}\right)
\nonumber \\
&& + D^{(4)}\!(Z\alpha) \left({\alpha\over\rmpi}\right)^2 +
 D^{(6)}\!(Z\alpha) \left({\alpha\over\rmpi}\right)^3 +
\cdots \Big] \ ,
\nonumber\\ &&
\label{eq:drad}
\end{eqnarray}
where the functions $D^{(2n)}\!(Z\alpha)$ are contributions associated with $n$
virtual photons.  The leading term is
\begin{eqnarray}
D^{(2)}\!(Z\alpha) &=&
A_{1}^{(2)} + \left(\ln 2 - \fr{5}{2}\right)\rmpi Z\alpha
\nonumber\\
&& + \Big[-\fr{2}{3}\ln^2 (Z\alpha)^{-2} 
 + \left(\fr{281}{360}-\fr{8}{3}\ln 2\right)\ln(Z\alpha)^{-2} 
\nonumber\\
&& + 16.9037\ldots\Big](Z\alpha)^2
\nonumber\\
&&+ \Big[\left(\fr{5}{2}\ln 2 - \fr{547}{96}\right)
\ln(Z\alpha)^{-2}\Big]\rmpi(Z\alpha)^3 
\nonumber\\
&& + G(Z\alpha)(Z\alpha)^3 \ ,
\label{eq:mud2}
\end{eqnarray}
where $A_{1}^{(2)} = \fr{1}{2}$, as in Eq.~(\ref{eq:a12}).  The function
$G(Z\alpha)$ accounts for all higher-order contributions in powers of
$Z\alpha$, and can be divided into parts that correspond to the self-energy or
vacuum polarization, $G(Z\alpha) = G_{\rm SE}(Z\alpha) + G_{\rm VP}(Z\alpha)$.
We adopt the value
\begin{eqnarray}
G_{\rm SE}(\alpha) = -14(2)\ ,
\label{eq:gse}
\end{eqnarray}
which is the simple mean and standard deviation of the three values: 
$G_{\rm SE}(\alpha) = -12.0(2.0)$
from \textcite{1997086};
$G_{\rm SE}(0) = -15.9(1.6)$ from
\textcite{2001353,pc02nio}; and
$G_{\rm SE}(\alpha) = -14.3(1.1)$ from \textcite{2001252}.
The vacuum polarization part $G_{\rm VP}(Z\alpha)$ has been calculated to
several orders of $Z\alpha$ by \textcite{2000031,1999173}.  Their expression
yields
\begin{eqnarray}
G_{\rm VP}(\alpha) = 7.227(9)\ . 
\end{eqnarray}
 
For $D^{(4)}\!(Z\alpha)$, as in CODATA-02, we have
\begin{eqnarray}
D^{(4)}\!(Z\alpha) &=&
A_{1}^{(4)} + 0.7717(4) \rmpi Z\alpha
+ \Big[-\fr{1}{3}\ln^2 (Z\alpha)^{-2} 
\nonumber\\ && 
-0.6390 \ldots\times\ln (Z\alpha)^{-2}
+10(2.5) \Big](Z\alpha)^2
\nonumber\\ && 
+ \cdots \ ,
\label{eq:mud4}
\end{eqnarray}
where $A_1^{(4)}$ is given in Eq.~(\ref{eq:a14}).

Finally,
\begin{eqnarray}
D^{(6)}\!(Z\alpha) =
A_{1}^{(6)} 
+ \cdots \ ,
\label{eq:mud6}
\end{eqnarray}
where only the leading contribution $A_{1}^{(6)}$ as given in
Eq.~(\ref{eq:a16}) is known.  Higher-order functions $D^{(2n)}\!(Z\alpha)$ with
$n>3$ are expected to be negligible.

The recoil contribution is given by
\begin{eqnarray} 
\label{eq:hfsrec} 
\Delta \nu_{\rm rec} &=& 
\Delta\nu_{\rm F}{m_{\rm e}\over m_{\rmssmu}}
\Bigg(-{3\over1-\left(m_{\rm e}/ m_{\rmssmu}\right)^2}
\ln \Big({m_{\rmssmu}\over m_{\rm e}}\Big){Z\alpha\over\rmpi} \nonumber\\ 
&& +\, {1\over\left(1+m_{\rm e}/m_{\rmssmu}\right)^2} 
\bigg\{\ln {(Z\alpha)^{-2}}-8\ln 
2 + {65\over 18}
\nonumber\\&&
+\bigg[{9\over 2\rmpi^2}
\ln^2\left({m_{\rmssmu}\over m_{\rm e}}\right)
+\left({27\over 2\rmpi^2}-1\right)
\ln\left({m_{\rmssmu}\over m_{\rm e}}\right)
\nonumber\\&&
+{93\over4\rmpi^2} + {33\zeta(3)\over\rmpi^2}
-{13\over12}-12\ln2 \bigg] 
{m_{\rm e}\over m_{\rmssmu}}
\bigg\}(Z\alpha)^2 \nonumber\\ 
&& +\, \bigg\{-{3\over 2} \ln{\Big({m_ 
{\rmssmu}\over m_{\rm e}}\Big)} \ln(Z\alpha)^{-2} 
-{1\over6}\ln^2 {(Z\alpha)^{-2}} 
\nonumber\\ && 
+\left({101\over18} - 10 \ln 2\right)\ln(Z\alpha)^{-2}
\nonumber\\&&\qquad\qquad
+40(10)\bigg\}
{(Z\alpha)^3\over\rmpi}
\Bigg)
 + \cdots \,, 
\nonumber\\&&
\end{eqnarray} 
\label{eq:drec}
as discussed in CODATA-02

The radiative-recoil contribution is
\begin{eqnarray} 
\label{eq:hfsradrec} 
\Delta \nu_{\rm r\hbox{-}r} &=& \nu_{\rm F} 
\left({\alpha\over \rmpi}\right)^2 {m_{\rm e}\over m_{\rmssmu}} 
\bigg\{\bigg[ -2\ln^2\Big({m_{\rmssmu}\over m_{\rm e}}\Big) +{13\over
12}\ln{\Big({m_{\rmssmu}\over m_{\rm e}}\Big)} \nonumber\\ 
&& +\, {21\over2}\zeta(3)+{\rmpi^2\over6}+{35\over9}\bigg] 
+\bigg[\, {4\over3} \ln^2\alpha^{-2}
\nonumber\\
&&+\left({16\over3} \ln 2 - {341\over180}\right) \ln\alpha^{-2}
-40(10) \bigg]\rmpi\alpha \nonumber\\ 
&&+\bigg[-{4\over3}\ln^3 \Big({m_{\rmssmu}\over 
m_{\rm e}}\Big) 
+{4\over3}\ln^2 \Big({m_{\rmssmu}\over m_{\rm e}}\Big)
\bigg] {\alpha\over \rmpi}\bigg\} 
\nonumber\\ 
&& - \nu_{\rm F}\alpha^2\!\left(m_{\rm e}\over m_{\rmssmu}\right)^{\!2}
\left(6\ln{2} + {13\over6}\right) + \cdots \ ,
\label{eq:dradrec}
\end{eqnarray} 
where, for simplicity, the explicit dependence on $Z$ is not shown.

The electroweak contribution due to the exchange of a Z$^0$ boson is
\cite{1996059}
\begin{eqnarray} 
\Delta \nu_{\rm weak} &=& -65 \ {\rm Hz} \ .
\label{eq:dweak}
\end{eqnarray} 

For the hadronic vacuum polarization contribution we use the result of
\textcite{2002136},
\begin{eqnarray}
\Delta \nu_{\rm had} &=& 236(4) \ {\rm Hz}\ ,
\label{eq:dhad}
\end{eqnarray} 
which takes into account experimental data on the cross section for ${\rm
e}^-{\rm e}^+ \rightarrow {\rmpi}^+{\rmpi}^-$ and on the ${\rmphi}$ meson
leptonic width.  The leading hadronic contribution is 231.2(2.9) and the next
order term is 5(2) giving a total of 236(4).  The pion and kaon contributions
to the hadronic correction have been considered within a chiral unitary
approach and found to be in general agreement with (but have a three times
larger uncertainty) the corresponding contributions given in earlier studies
using data from e$^+$-e$^-$ scattering \cite{2003214}.

The standard uncertainty of $\Delta\nu_{\rm Mu}(\rm th)$ was fully discussed in
Appendix~E of CODATA-02. The four principle sources of uncertainty are the
terms $\Delta\nu_{\rm rad}$, $\Delta\nu_{\rm rec}$, $\Delta\nu_{\rm r-r}$, and
$\Delta\nu_{\rm had}$ in Eq.~(\ref{eq:dmuth}).  Included in the 67~Hz
uncertainty of $\Delta\nu_{\rm r-r}$ is a 41~Hz component, based on the partial
calculations of \textcite{2002004,2003146,1993001}, to account for a possible
uncalculated radiative-recoil contribution of order $\Delta\nu_{\rm F}(m_{\rm
e}/(m_{\rmssmu}) (\alpha/\rmpi)^3\ln(m_{\rmssmu}/m_{\rm e})$ and
non-logarithmic terms.  Since the completion of the 2002 adjustment, the
results of additional partial calculations have been published that, if taken
at face value, would lead to a small reduction in the 41~Hz estimate
\cite{2003243, 2004241,2004242,2005012,2007138}. However, because the
calculations are not yet complete and the decrease of the 101~Hz total
uncertainty assigned to $\Delta\nu_{\rm Mu}(\rm th)$ for the 2002 adjustment
would only be a few percent, the Task Group decided to retain the 101~Hz
uncertainty for the 2006 adjustment.

We thus have for the standard uncertainty of the theoretical expression for the
muonium hyperfine splitting
$\Delta \nu_{\rm Mu} \rm (th)$
\begin{eqnarray}
u[\Delta \nu_{\rm Mu} {\rm(th)}] =  101 \ {\rm Hz} 
\quad [ 2.3\times 10^{-8}] \ .
\label{eq:runcmuhfs}
\end{eqnarray}
For the least-squares calculations, we use as the theoretical expression for
the hyperfine splitting
\begin{eqnarray}
\Delta \nu_{\rm Mu}\!\!\left(R_\infty,\alpha,{m_{\rm e}\over m_{\rmssmu}},
\delta_{\rmssmu},\delta_{\rm Mu}\right) = 
\Delta \nu_{\rm Mu} \rm (th) + \delta_{\rm Mu} \ ,
\nonumber\\
\end{eqnarray}
where $\delta_{\rm Mu}$ is assigned, a priori, the value
\begin{eqnarray}
\delta_{\rm Mu} =  0(101) \ {\rm Hz}
\label{eq:cmu}
\end{eqnarray}
in order to account for the uncertainty of the theoretical expression. 

The theory summarized above predicts
\begin{eqnarray}
\Delta \nu_{\rm Mu} =  4\,463\,302\,881(272) \ {\rm Hz}
\quad [ 6.1\times 10^{-8}] \ ,
\label{eq:dnmup}
\end{eqnarray}
based on values of the constants obtained from a variation of the 2006
least-squares adjustment that omits as input data the two LAMPF measured values
of $\Delta\nu_{\rm Mu}$ discussed in the following section.

The main source of uncertainty in this value is the mass ratio $m_{\rm
e}/m_{\rmssmu}$ that appears in the theoretical expression as an overall
factor.  See the text following Eq.~(D14) of Appendix D of CODATA-98 for an
explanation of why the relative uncertainty of the predicted value of $\Delta
\nu_{\rm Mu}$ in Eq.~(\ref{eq:dnmup}) is smaller than the relative uncertainty
of the electron-muon mass ratio as given in Eq.~(\ref{eq:mesmmuiL}) of
Sec.~\ref{par:clampf}.

\subsubsection{Measurements of muonium transition frequencies and values of
$\mu_{\rmssmu}/\mu_{\rm p}$ and $m_{\rmssmu}/m_{\rm e}$}
\label{sssec:mufreqs}

The two most precise determinations of muonium Zeeman transition frequencies
were carried out at the Clinton P. Anderson Meson Physics Facility at Los
Alamos (LAMPF), USA, and were reviewed in detail in CODATA-98.  The following
three sections and Table~\ref{tab:muhfs} give the key results. 

\begin{table*}
\def\m{\phantom{-}}
\caption{ Summary of data related to the hyperfine splitting in muonium and
inferred values of $\mu_{\rmssmu}/\mu_{\rm p}$, $m_{\rmssmu}/m_{\rm e}$, and
$\alpha$ from the combined 1982 and 1999 LAMPF data.}
\label{tab:muhfs}
\def\sp{\hbox to 10 pt{}}
\begin{tabular}{l@{\sp}l@{\sp}c@{\sp}l@{\sp}l}
\toprule
\vbox to 10 pt {}
Quantity     &  \hbox to 12pt {} Value
& Relative standard
& Identification & Sect. and Eq. \\
&& uncertainty $u_{\rm r}$&& \\
\colrule
\vbox to 11 pt {}

$\Delta\nu_{\rm Mu}$ &
    $\phantom{-} 4\,463\,302.88(16)$ kHz& $ 3.6\times 10^{-8}$ & 
    LAMPF-82 &
    \ref{par:lampf1982} (\ref{eq:nupL82}) \\
$\nu(f_{\rm p})$ &
   $\phantom{-} 627\,994.77(14)$ kHz & $ 2.2\times 10^{-7}$ & 
    LAMPF-82 &
    \ref{par:lampf1982} (\ref{eq:numL82}) \\
&&&&\\

$\Delta\nu_{\rm Mu}$ &
    $\phantom{-} 4\,463\,302\,765(53)$ Hz& $ 1.2\times 10^{-8}$ & 
    LAMPF-99 &
    \ref{par:lampf1999} (\ref{eq:nupL99}) \\
$\nu(f_{\rm p})$ &
   $\phantom{-} 668\,223\,166(57)$ Hz & $ 8.6\times 10^{-8}$ & 
    LAMPF-99 &
    \ref{par:lampf1999} (\ref{eq:numL99}) \\
&&&&\\

\ \ $\mu_{\rmssmu}/\mu_{\rm p}$ 
     & $\phantom{-} 3.183\,345\,24(37)$ & $ 1.2\times 10^{-7}$ & 
     LAMPF & \ref{par:clampf} (\ref{eq:mumupsmupL}) \\

\ \ $m_{\rmssmu}/m_{\rm e}$ 
     & $\phantom{-} 206.768\,276(24)$ & $ 1.2\times 10^{-7}$ &
     LAMPF & \ref{par:clampf} (\ref{eq:mesmmuiL})   \\

\ \ $\alpha^{-1}$  
     & $\phantom{-} 137.036\,0017(80)$ & $ 5.8\times 10^{-8}$ &
     LAMPF & \ref{par:clampf} (\ref{eq:alphiL})   \\

\botrule
\end{tabular}
\end{table*}

\paragraph{LAMPF 1982}
\label{par:lampf1982}

The results obtained by \textcite{1982003, pc81mar}, which we take as input data
in the current adjustment as in the two previous adjustments, may be summarized
as follows:
\begin{eqnarray}
\Delta \nu_{\rm Mu} =  4\,463\,302.88(16) \ {\rm kHz} \quad [ 3.6\times 10^{-8}]
\label{eq:nupL82}
\end{eqnarray}
\begin{eqnarray}
\nu(f_{\rm p}) =  627\,994.77(14) \ {\rm kHz} \quad [ 2.2\times 10^{-7}]
\label{eq:numL82}
\end{eqnarray}
\begin{eqnarray}
r[\Delta \nu_{\rm Mu},\nu(f_{\rm p})] =  0.23 \ ,
\label{eq:rnupm82}
\end{eqnarray}
where
$f_{\rm p}$ is very nearly $57.972\,993$ MHz, corresponding to the flux density
of about $1.3616$ T used in the experiment, and $r[\Delta \nu_{\rm
Mu},\nu(f_{\rm p})]$ is the correlation coefficient of $\Delta \nu_{\rm Mu}$
and $\nu(f_{\rm p})$.

\paragraph{LAMPF 1999}
\label{par:lampf1999}

The results obtained by \textcite{1999002}, which we also take as input data in
the current adjustment as in the 1998 and 2002 adjustments, may be summarized
as follows:
\begin{eqnarray}
\Delta \nu_{\rm Mu} =  4\,463\,302\,765(53) \ {\rm Hz} \quad [ 1.2\times 10^{-8}]
\label{eq:nupL99}
\end{eqnarray}
\begin{eqnarray}
\nu(f_{\rm p}) =  668\,223\,166(57) \ {\rm Hz} \quad [ 8.6\times 10^{-8}]
\label{eq:numL99}
\end{eqnarray}
\begin{eqnarray}
r[\Delta \nu_{\rm Mu},\nu(f_{\rm p})] =  0.19 \ ,
\label{eq:rnupm99}
\end{eqnarray}
where $f_{\rm p}$ is exactly $72.320\,000$ MHz, corresponding to the flux
density of approximately $1.7$ T used in the experiment, and $r[\Delta \nu_{\rm
Mu},\nu(f_{\rm p})]$ is the correlation coefficient of $\Delta \nu_{\rm Mu}$
and $\nu(f_{\rm p})$.

\paragraph{Combined LAMPF results}
\label{par:clampf}

By carrying out a least-squares adjustment using only the LAMPF 1982 and LAMPF
1999 data, the 2006 recommended values of the quantities $R_{\infty}$,
$\mu_{\rm e}/\mu_{\rm p}$, $g_{\rm e}$, and $g_{\rmssmu}$, together with
Eqs.~(\ref{eq:murat}) to (\ref{eq:hfsth}), we obtain
\begin{eqnarray}
{\mu_{{\rmssmu}^+}\over \mu_{\rm p}} &=&   3.183\,345\,24(37) \quad [ 1.2\times 10^{-7}] 
\label{eq:mumupsmupL}
\\
{m_{\rmssmu}\over m_{\rm e}} &=&   206.768\,276(24) \quad [ 1.2\times 10^{-7}]
\label{eq:mesmmuiL}
\\
\alpha^{-1} &=&   137.036\,0017(80) \quad [ 5.8\times 10^{-8}] \ ,
\label{eq:alphiL}
\end{eqnarray}
where this value of $\alpha$ may be called the muonium value of the
fine-structure constant and denoted as $\alpha^{-1}(\Delta\nu_{\rm Mu})$.

It is noteworthy that the uncertainty of the value of the mass ratio
$m_{\rmssmu}/m_{\rm e}$ given in Eq.~(\ref{eq:mesmmuiL}) is about four times
the uncertainty of the 2006 recommended value.  The reason is that taken
together, the experimental value of and theoretical expression for the
hyperfine splitting essentially determine only the value of the product
$\alpha^2 m_{\rmssmu}/m_{\rm e}$, as is evident from Eq.~(\ref{eq:hfsth}). In
the full adjustment the value of $\alpha$ is determined by other data with an
uncertainty significantly smaller than that of the value in
Eq.~(\ref{eq:alphiL}), which in turn determines the value of
$m_{\rmssmu}/m_{\rm e}$ with a smaller uncertainty than that of
Eq.~(\ref{eq:mesmmuiL}).

\section{Electrical measurements}
\label{sec:elmeas}

This section is devoted to the discussion of quantities that require electrical
measurements of the most basic kind for their determination: the gyromagnetic
ratios of the shielded proton and helion, the von Klitzing constant $R_{\rm
K}$, the Josephson constant $K_{\rm J}$, the product $K_{\rm J}^2R_{\rm K}$,
and the Faraday constant. However, some of the results we discuss were taken as
input data for the 2002 adjustment but were not included in the final
least-squares adjustment from which the 2002 recommended values were obtained,
mainly because of their comparatively large uncertainties and hence low weight.
Nevertheless, we take them as input data in the 2006 adjustment because they
provide information on the overall consistency of the available data and tests
of the exactness of the relations $K_{\rm J}=2e/h$ and $R_{\rm K}=h/e^2$.  The
lone exception is the low-field measurement of the gyromagnetic ratio of the
helion reported by \textcite{1989078}.  Because of its large uncertainty and
strong disagreement with many other data, we no longer consider it---see
CODATA-02.

\subsection{Shielded gyromagnetic ratios $\bm{\gamma^{\,\prime}}$, the
fine-structure constant $\bm{\alpha}$, and the Planck constant $\bm{h}$}
\label{ssec:sgyror}

The gyromagnetic ratio $\gamma$ of a bound particle of spin quantum number $i$
and magnetic moment $\mu$ is given by
\begin{eqnarray}
\gamma = { {2\rmpi f }\over{B}} = {\omega\over B} = { {|\mu|}\over {i \hbar} } \ ,
\label{eq:gammadef2}
\end{eqnarray}
where $f$ is the precession (that is, spin-flip) frequency and $\omega$ is the
angular precession frequency of the particle in the magnetic flux density $B$.
The SI unit of $\gamma$ is ${\rm s}^{-1}~{\rm T}^{-1}$ = ${\rm C~kg}^{-1}$ =
${\rm A~s~kg}^{-1}$.  In this section we summarize measurements of the
gyromagnetic ratio of the shielded proton
\begin{eqnarray}
\gamma_{\rm p}^{\,\prime} = { {2\mu_{\rm p}^{\,\prime}}\over {\hbar} } \ ,
\label{eq:gampp}
\end{eqnarray}
and of the shielded helion
\begin{eqnarray}
\gamma_{\rm h}^{\,\prime} = 
{ {2|\mu_{\rm h}^{\,\prime}|}\over {\hbar} } \ ,
\label{eq:gamhe}
\end{eqnarray}
where, as in previous sections that dealt with magnetic-moment ratios involving
these particles, the protons are those in a spherical sample of pure H$_2$O at
25~$^\circ$C surrounded by vacuum; and the helions are those in a spherical
sample of low-pressure, pure $^3$He gas at 25~$^\circ$C surrounded by vacuum.  

As discussed in detail in CODATA-98, two methods are used to determine the
shielded gyromagnetic ratio $\gamma^{\,\prime}$ of a particle: the low-field
method and the high-field method.  In either case the measured current $I$ in
the experiment can be expressed in terms of the product $K_{\rm J}R_{\rm K}$,
but $B$ depends on $I$ differently in the two cases. In essence, the low-field
experiments determine $\gamma^\prime/ K_{\rm J}  R_{\rm K}$ and the high-field
experiments determine $\gamma^\prime K_{\rm J}  R_{\rm K}$. This leads to the
relations
\begin{eqnarray}
\gamma^{\,\prime} &=& 
{\it \Gamma}_{90}^{\,\prime}({\rm lo})
{K_{\rm J} \, R_{\rm K}\over K_{{\rm J}-90} \, R_{{\rm K}-90}}
\label{eq:sgammaslo90}
\\ \nonumber\\
\gamma^{\,\prime} &=& 
{\it \Gamma}_{90}^{\,\prime}({\rm hi})
{K_{{\rm J}-90} \, R_{{\rm K}-90} \over
K_{\rm J} \, R_{\rm K}} \ ,
\label{eq:sgammashi90}
\end{eqnarray}
where ${\it \Gamma}_{90}^{\,\prime}({\rm lo})$ and ${\it
\Gamma}_{90}^{\,\prime}({\rm hi})$ are the experimental values of
$\gamma^\prime$ in SI units that would result from the low- and high-field
experiments if $K_{\rm J}$ and $R_{\rm K}$ had the exactly known conventional
values of $K_{{\rm J}-90}$ and $R_{{\rm K}-90}$, respectively.  The quantities
${\it \Gamma}_{90}^{\,\prime}({\rm lo})$ and ${\it \Gamma}_{90}^{\,\prime}({\rm
hi})$ are the input data used in the adjustment, but the observational
equations take into account the fact that $K_{{\rm J}-90} \ne K_{\rm J}$ and
$R_{{\rm K}-90} \ne R_{\rm K}$.

Accurate values of ${\it \Gamma}_{90}^{\,\prime}({\rm lo})$ and ${\it
\Gamma}_{90}^{\,\prime}({\rm hi})$ for the proton and helion are of potential
importance because they provide information on the values of $\alpha$ and $h$.
Assuming the validity of the relations $K_{\rm J}=2e/h$ and $R_{\rm K}=h/e^2$,
the following expressions apply to the four available proton results and one
available helion result:
\begin{eqnarray}
{\it\Gamma}_{\rm p-90}^{\,\prime}({\rm lo}) =
{K_{\rm J-90} \, R_{\rm K-90} \, g_{\rm e^-} \over
4\mu_0 R_\infty } \, 
{\mu_{\rm p}^\prime\over{\mu_{\rm e^-}}} \,
\alpha^3 \ ,
\label{eq:gpplo90alph}
\end{eqnarray}
\begin{eqnarray}
{\it\Gamma}_{\rm h-90}^{\,\prime}({\rm lo}) =
-{K_{\rm J-90} \, R_{\rm K-90} \, g_{\rm e^-} \over
4\mu_0 R_\infty } \, 
{\mu_{\rm h}^\prime\over{\mu_{\rm e^-}}} \,
\alpha^3 \ ,
\label{eq:ghlow90alph}
\end{eqnarray}
\begin{eqnarray}
{\it\Gamma}_{\rm p-90}^{\,\prime}({\rm hi}) &=&
{c \, \alpha^2 g_{\rm e^-} \over 2 K_{\rm J-90}\,  R_{\rm K-90}\,R_\infty} \,
{\mu^{\,\prime}_{\rm p}\over \mu_{\rm e^-}} \,
{1\over h} \ .
\label{eq:gpphi90h}
\end{eqnarray}
Since the five experiments, including necessary corrections, were discussed
fully in CODATA-98, only a brief summary is given in the following sections.
The five results, together with the value of $\alpha$ inferred from each
low-field measurement and the value of $h$ inferred from each high-field
measurement, are collected in Table~\ref{tab:sgamma}.

\begin{table*}
\def\m{\phantom{-}}
\caption{Summary of data related to shielded gyromagnetic ratios of the proton
and helion, and inferred values of $\alpha$ and~$h$.}
\label{tab:sgamma}
\def\sp{\hbox to 15pt {}}
\begin{tabular}{l@{\sp}l@{\sp}l@{\sp}l@{\sp}l}
\toprule
\vbox to 12 pt {}
Quantity     &  \hbox to 15pt {} Value
& \hbox to -20pt {} Relative standard
& Identification & Sect. and Eq. \\
& &  \hbox to -5pt {}uncertainty $u_{\rm r}$ &  \\
\colrule
${\it\Gamma}_{\rm p-90}^{\,\prime}({\rm lo})$ \vbox to 15pt{}& 
$ 2.675\,154\,05(30)\times 10^{8}$ s$^{-1}$ T$^{-1}$ 
& $ 1.1\times 10^{-7}$ & NIST-89 & \ref{par:nistlf} (\ref{eq:gppnist89}) \\

\ \ $\alpha^{-1}$ &$\phantom{-} 137.035\,9879(51)$&$ 3.7\times 10^{-8}$& 
    & \ref{par:nistlf} (\ref{eq:alphinistg89}) \\

${\it\Gamma}_{\rm p-90}^{\,\prime}({\rm lo})$ \vbox to 15pt{}& 
$ 2.675\,1530(18)\times 10^{8}$ s$^{-1}$ T$^{-1}$ 
& $ 6.6\times 10^{-7}$ & NIM-95 & \ref{par:nimlo} (\ref{eq:gpplnim95}) \\

\ \ $\alpha^{-1}$ &$\phantom{-} 137.036\,006(30)$&$ 2.2\times 10^{-7}$& 
    & \ref{par:nimlo} (\ref{eq:alignim95}) \\

${\it\Gamma}_{\rm h-90}^{\,\prime}({\rm lo})$ \vbox to 15pt{}& 
$ 2.037\,895\,37(37)\times 10^{8}$ s$^{-1}$ T$^{-1}$ 
& $ 1.8\times 10^{-7}$ & KR/VN-98 & \ref{par:kvlf} (\ref{eq:ghkv97}) \\

\ \ $\alpha^{-1}$ &$\phantom{-} 137.035\,9852(82)$&$ 6.0\times 10^{-8}$& 
    & \ref{par:kvlf} (\ref{eq:alphinvkv97}) \\

${\it\Gamma}_{\rm p-90}^{\,\prime}({\rm hi})$ \vbox to 15pt{}& 
$ 2.675\,1525(43)\times 10^{8}$ s$^{-1}$ T$^{-1}$ 
& $ 1.6\times 10^{-6}$ & NIM-95 & \ref{par:nimhi} (\ref{eq:gpphnim95}) \\

\ \ $h$ &$\phantom{-} 6.626\,071(11)\times 10^{-34}$ J s &$ 1.6\times 10^{-6}$& 
    & \ref{par:nimhi} (\ref{eq:hnimg95}) \\

${\it\Gamma}_{\rm p-90}^{\,\prime}({\rm hi})$ \vbox to 15pt{}& 
$ 2.675\,1518(27)\times 10^{8}$ s$^{-1}$ T$^{-1}$ 
& $ 1.0\times 10^{-6}$ & NPL-79 & \ref{par:nplhf} (\ref{eq:gppnpl79}) \\

\ \ \B $h$ &$\phantom{-} 6.626\,0729(67)\times 10^{-34}$ J s &$ 1.0\times 10^{-6}$& 
    & \ref{par:nplhf} (\ref{eq:hnplg79}) \\

\botrule
\end{tabular}
\end{table*}

\subsubsection{Low-field measurements}
\label{sssec:logmr}

A number of national metrology institutes have long histories of measuring the
gyromagnetic ratio of the shielded proton, motivated, in part, by their need to
monitor the stability of their practical unit of current based on groups of
standard cells and standard resistors.  This was prior to the development of
the Josephson and quantum hall effects for the realization of practical
electric units.

\paragraph{NIST: Low field}
\label{par:nistlf}

The most recent National Institute of Standards and Technology (NIST),
Gaithersburg, USA, low-field measurement was reported by \textcite{1989008}.
Their result is
\begin{eqnarray}
{\it\Gamma}_{\rm p-90}^{\,\prime}({\rm lo}) &=&  2.675\,154\,05(30)\times 10^{8} 
\ {\rm s}^{-1} \ {\rm T}^{-1}
\nonumber
\\&& \quad [ 1.1\times 10^{-7}] \ ,
\label{eq:gppnist89}
\end{eqnarray}
where ${\it\Gamma}_{\rm p-90}^{\,\prime}(\rm{\rm lo})$ is related to
$\gamma_{\rm p}^{\,\prime}$ by Eq.~(\ref{eq:sgammaslo90}).

The value of $\alpha$ that may be inferred from this result follows from
Eq.~(\ref{eq:gpplo90alph}). Using the 2006 recommended values for the other
relevant quantities, the uncertainties of which are significantly smaller than
the uncertainty of the NIST  result (statements that also apply to the
following four similar calculations), we obtain
\begin{eqnarray}
\alpha^{-1} =  137.035\,9879(51)\quad[ 3.7\times 10^{-8}] \ ,
\label{eq:alphinistg89}
\end{eqnarray}
where the relative uncertainty is about one-third the relative uncertainty of
the NIST value of ${\it\Gamma}_{\rm p-90}^{\,\prime}({\rm lo})$ because of the
cube-root dependence of $\alpha$ on ${\it\Gamma}_{\rm p-90}^{\,\prime}({\rm
lo})$.

\paragraph{NIM: Low field}
\label{par:nimlo}

The latest low-field proton gyromagnetic ratio experiment carried out by
researchers at the National Institute of Metrology (NIM), Beijing, PRC, yielded
\cite{1995233}
\begin{eqnarray}
{\it\Gamma}_{\rm p-90}^\prime({\rm lo}) &=&  2.675\,1530(18)\times 10^{8}
\ {\rm s}^{-1} \ {\rm T}^{-1}
\nonumber
\\&& \quad [ 6.6\times 10^{-7}] \ .
\label{eq:gpplnim95}
\end{eqnarray}

Based on Eq.~(\ref{eq:gpplo90alph}), the inferred value of $\alpha$ from the
NIM result is
\begin{eqnarray}
\alpha^{-1} =  137.036\,006(30) \quad  [ 2.2\times 10^{-7}] \ .
\label{eq:alignim95}
\end{eqnarray}

\paragraph{KRISS/VNIIM: Low field}
\label{par:kvlf}

The determination of $\gamma_{\rm h}^{\,\prime}$ at the Korea Research
Institute of Standards and Science (KRISS), Taedok Science Town, Republic of
Korea, was carried out in a collaborative effort with researchers from the
Mendeleyev All-Russian Research Institute for Metrology (VNIIM), St.
Petersburg, Russian Federation \cite{1999153,1999083,1998117,1998132,1995040}.
The result of this work can be expressed as
\begin{eqnarray}
{\it\Gamma}_{\rm h-90}^{\prime}({\rm lo})
&=& 2.037\,895\,37(37)\times 10^{8} \ {\rm s}^{-1} \ {\rm T}^{-1}
\nonumber\\&& \quad [ 1.8\times 10^{-7}] \ ,
\label{eq:ghkv97}
\end{eqnarray}
and the value of $\alpha$ that may be inferred from it through
Eq.~(\ref{eq:ghlow90alph}) is 
\begin{eqnarray}
\alpha^{-1} =  137.035\,9852(82)
\quad[ 6.0\times 10^{-8}] \ .
\label{eq:alphinvkv97}
\end{eqnarray}

\subsubsection{High-field measurements}
\label{sssec:higmr}

\paragraph{NIM:high field}
\label{par:nimhi}

The latest high-field proton gyromagnetic ratio experiment at NIM yielded
\cite{1995233}
\begin{eqnarray}
{\it\Gamma}_{\rm p-90}^\prime({\rm hi}) &=&  2.675\,1525(43)\times 10^{8}
\ {\rm s}^{-1} \ {\rm T}^{-1}
\nonumber
\\&& \quad [ 1.6\times 10^{-6}] \ ,
\label{eq:gpphnim95}
\end{eqnarray}
where ${\it\Gamma}_{\rm p-90}^{\,\prime}(\rm{\rm hi})$ is related to
$\gamma_{\rm p}^{\,\prime}$ by Eq.~(\ref{eq:sgammashi90}). Its correlation
coefficient with the NIM low-field result in Eq.~(\ref{eq:gpplnim95}) is
\begin{eqnarray}
r({\rm lo,hi}) =  -0.014 \ .
\end{eqnarray}

Based on Eq.~(\ref{eq:gpphi90h}), the value of $h$ that may be inferred from
the NIM high-field result is 
\begin{eqnarray}
h =  6.626\,071(11)\times 10^{-34} \ {\rm J \ s}\quad [ 1.6\times 10^{-6}] \ .
\label{eq:hnimg95}
\end{eqnarray}

\paragraph{NPL: High field}
\label{par:nplhf}

The most accurate high-field $\gamma_{\rm p}^{\,\prime}$ experiment was carried
out at NPL by \textcite{1979012}, with the result
\begin{eqnarray}
{\it\Gamma}_{\rm p-90}^\prime({\rm hi}) &=&  2.675\,1518(27)\times 10^{8}
\ {\rm s}^{-1} \ {\rm T}^{-1}
\nonumber
\\&& \quad [ 1.0\times 10^{-6}] \ .
\label{eq:gppnpl79}
\end{eqnarray}
This leads to the inferred value
\begin{eqnarray}
h =  6.626\,0729(67)\times 10^{-34} \ {\rm J \ s} \quad [ 1.0\times 10^{-6}] \ ,
\label{eq:hnplg79}
\end{eqnarray}
based on Eq.~(\ref{eq:gpphi90h}).

\subsection{von Klitzing constant $\bm{R_{\rm K}}$ and $\bm{\alpha}$}
\label{ssec:vkc}

Since the the quantum Hall effect, the von Klitzing constant $R_{\rm K}$
associated with it, and the available determinations of $R_{\rm K}$ are fully
discussed in CODATA-98 and CODATA-02, we only outline the main points here.

The quantity $R_{\rm K}$ is measured by comparing a quantized Hall resistance
$R_{\rm H}(i) = R_{\rm K}/i$, where $i$ is an integer, to a resistance $R$
whose value is known in terms of the SI unit of resistance ${\rm \Omega}$.  In
practice, the latter quantity, the ratio $R/{\rm \Omega}$, is determined by
means of a calculable cross capacitor, a device based on a theorem in
electrostatics discovered in the 1950s \cite{1956004,1957015}.  The theorem
allows one to construct a cylindrical capacitor, generally called a
Thompson-Lampard calculable capacitor \cite{1959004}, whose capacitance, to
high accuracy, depends only on its length.

As indicated in Sec.~\ref{sec:squ}, if one assumes the validity of the relation
$R_{\rm K} = h/e^2$, then $R_{\rm K}$ and the fine-structure constant $\alpha$
are related by
\begin{eqnarray}
\alpha = \mu_0 c/2 R_{\rm K} \ .
\label{eq:alrk}
\end{eqnarray}
Hence, the relative uncertainty of the value of $\alpha$ that may be inferred
from a particular experimental value of $R_{\rm K}$ is the same as the relative
uncertainty of that value. 

The values of $R_{\rm K}$ we take as input data in the 2006 adjustment
and the corresponding inferred values values of $\alpha$ are given in
the following sections and are summarized in Table~\ref{tab:rk}.

\begin{table*}
\def\m{\phantom{-}}
\caption{Summary of data related to the 
von Klitzing constant $R_{\rm K}$ and inferred values of $\alpha$.}
\label{tab:rk}
\def\sp{\hbox to 15 pt {}}
\begin{tabular}{l@{\sp}l@{\sp}c@{\sp}l@{\sp}l}
\toprule
Quantity \T  &  \hbox to 12pt {} Value
& Relative standard
& Identification & Sect. and Eq. \\
&& uncertainty $u_{\rm r}$ &&\\
\colrule
$R_{\rm K}$ \vbox to 12 pt {}& $ 25\,812.808\,31(62) \ {\rm \Omega}$ & $ 2.4\times 10^{-8}$ 
& NIST-97 & \ref{sssec:nistvkc} (\ref{eq:rknist97}) \\
\ \ $\alpha^{-1}$ &\ \ $ 137.036\,0037(33)$&$ 2.4\times 10^{-8}$& &
    \ref{sssec:nistvkc} (\ref{eq:airknist97}) \\
&&&&\\
$R_{\rm K}$ & $ 25\,812.8071(11) \ {\rm \Omega}$ & $ 4.4\times 10^{-8}$ 
& NMI-97 & \ref{sssec:nmlvkc} (\ref{eq:rknml97}) \\
\ \ $\alpha^{-1}$ &\ \ $ 137.035\,9973(61)$&$ 4.4\times 10^{-8}$& &
    \ref{sssec:nmlvkc} (\ref{eq:airknml97}) \\
&&&&\\
$R_{\rm K}$ & $ 25\,812.8092(14) \ {\rm \Omega}$ & $ 5.4\times 10^{-8}$ 
& NPL-88 & \ref{sssec:nplvkc} (\ref{eq:rknpl88}) \\
\ \ $\alpha^{-1}$ &\ \ $ 137.036\,0083(73)$&$ 5.4\times 10^{-8}$& &
    \ref{sssec:nplvkc} (\ref{eq:airknpl88}) \\
&&&&\\
$R_{\rm K}$ & $ 25\,812.8084(34) \ {\rm \Omega}$ & $ 1.3\times 10^{-7}$ 
& NIM-95 & \ref{sssec:nimvkc} (\ref{eq:rknim95}) \\
\ \ $\alpha^{-1}$ &\ \ $ 137.036\,004(18)$&$ 1.3\times 10^{-7}$& &
    \ref{sssec:nimvkc} (\ref{eq:airknim95}) \\
&&&&\\
$R_{\rm K}$ & $ 25\,812.8081(14) \ {\rm \Omega}$ & $ 5.3\times 10^{-8}$ 
& LNE-01 & \ref{sssec:lcievkc} (\ref{eq:rklcie01}) \\
\ \ \B $\alpha^{-1}$ &\ \ $ 137.036\,0023(73)$&$ 5.3\times 10^{-8}$& &
    \ref{sssec:lcievkc} (\ref{eq:airklcie01}) \\
\botrule
\end{tabular}
\end{table*}

\subsubsection{NIST: Calculable capacitor}
\label{sssec:nistvkc}

The result obtained at NIST is \cite{1997033} [see also \textcite{1998049}] 
\begin{eqnarray}
R_{\rm K} &=& 25\,812.8 \, [1 +  0.322(24)\times 10^{-6}] \ {\rm \Omega}
\nonumber\\
&=&  25\,812.808\,31(62) \ {\rm \Omega} \quad [ 2.4\times 10^{-8}] \ ,
\label{eq:rknist97}
\end{eqnarray}
and is viewed as superseding the NIST result reported in 1989 by
\textcite{1989002}.  Work by \textcite{1999105} provides additional support for
the uncertainty budget of the NIST calculable capacitor.

The value of $\alpha$ that may be inferred from the NIST value of $R_{\rm K}$
is, from Eq.~(\ref{eq:alrk}),
\begin{eqnarray}
\alpha^{-1} =  137.036\,0037(33) \quad [ 2.4\times 10^{-8}] \ .
\label{eq:airknist97}
\end{eqnarray}

\subsubsection{NMI: Calculable capacitor}
\label{sssec:nmlvkc}

Based on measurements carried out at the National Metrology Institute (NMI),
Lindfield, Australia, from December 1994 to April 1995 and a complete
reassessment of uncertainties associated with their calculable capacitor and
associated apparatus, \textcite{1997146} reported the result 
\begin{eqnarray}
R_{\rm K} &=& R_{\rm K-90} \, [1 +  0.4(4.4)\times 10^{-8}]
\nonumber\\
&=&  25\,812.8071(11) \ {\rm \Omega} \quad [ 4.4\times 10^{-8}] \ .
\label{eq:rknml97}
\end{eqnarray}
The value of $\alpha$ it implies is
\begin{eqnarray}
\alpha^{-1} =  137.035\,9973(61) \quad [ 4.4\times 10^{-8}] \ .
\label{eq:airknml97}
\end{eqnarray}

Because of problems associated with the 1989 NMI value of $R_{\rm K}$, only the
result reported in 1997 is used in the 2006 adjustment, as was the case in the
1998 and 2002 adjustments.

\subsubsection{NPL: Calculable capacitor}
\label{sssec:nplvkc}

The NPL calculable capacitor is similar in design to those of NIST and NMI.
The result for $R_{\rm K}$ reported by \textcite{ccerknpl88} is
\begin{eqnarray}
R_{\rm K} &=& 25\,812.8 \, [1 +  0.356(54)\times 10^{-6}] \ {\rm \Omega}
\nonumber\\
&=&  25\,812.8092(14) \ {\rm \Omega} \quad [ 5.4\times 10^{-8}] \ ,
\label{eq:rknpl88}
\end{eqnarray}
and the value of $\alpha$ that one may infer from it is
\begin{eqnarray}
\alpha^{-1} =  137.036\,0083(73) \quad [ 5.4\times 10^{-8}] \ .
\label{eq:airknpl88}
\end{eqnarray}

\subsubsection{NIM: Calculable capacitor}
\label{sssec:nimvkc}

The NIM calculable cross capacitor differs markedly from the version used at
NIST, NMI, and NPL.  The four bars (electrodes) that comprise the capacitor are
horizontal rather than vertical and the length that determines its known
capacitance is fixed rather than variable.  The NIM result for $R_{\rm K}$, as
reported by \textcite{1995237}, is
\begin{eqnarray}
R_{\rm K} =  25\,812.8084(34) \ {\rm \Omega} \quad [ 1.3\times 10^{-7}] \ ,
\label{eq:rknim95}
\end{eqnarray}
which implies 
\begin{eqnarray}
\alpha^{-1} =  137.036\,004(18) \quad [ 1.3\times 10^{-7}] \ .
\label{eq:airknim95}
\end{eqnarray}

\subsubsection{LNE: Calculable capacitor}
\label{sssec:lcievkc}

The value of $R_{\rm K}$ obtained at the Laboratoire National d'Essais (LNE),
Trappes, France, is \cite{2001026,2003191} 
\begin{eqnarray}
R_{\rm K} =  25\,812.8081(14) \ {\rm \Omega} \quad [ 5.3\times 10^{-8}] \ ,
\label{eq:rklcie01}
\end{eqnarray}
which implies 
\begin{eqnarray}
\alpha^{-1} =  137.036\,0023(73) \quad [ 5.3\times 10^{-8}] \ .
\label{eq:airklcie01}
\end{eqnarray}
The LNE Thompson-Lampard calculable capacitor is unique among all calculable
capacitors in that it consists of five horizontal bars arranged at the corners
of a regular pentagon.

\subsection{Josephson constant $\bm{K_{\rm J}}$ and $\bm{h}$}
\label{ssec:jc}

Again, since the Josephson effect, the Josephson constant $K_{\rm J}$
associated with it, and the available determinations of $K_{\rm J}$ are fully
discussed in CODATA-98 and CODATA-02, we only outline the main points here.

The quantity $K_{\rm J}$ is measured by comparing a Josephson voltage $U_{\rm
J}(n) = nf/K_{\rm J}$ to a high voltage $U$ whose value is known in terms of
the SI unit of voltage V. Here, $n$ is an integer and $f$ is the frequency of
the microwave radiation applied to the Josephson device.  In practice, the
latter quantity, the ratio $U$/V, is determined by counterbalancing an
electrostatic force arising from the voltage $U$ with a known gravitational
force. 

A measurement of $K_{\rm J}$ can also provide a value of $h$.  If, as discussed
in Sec.~\ref{sec:squ}, we assume the validity of the relation $K_{\rm J} =
2e/h$ and recall that $\alpha = e^2/4\rmpi\epsilon_0\hbar = \mu_0ce^2/2h$, we
have
\begin{eqnarray}
h = {8\alpha\over\mu_0cK^2_{\rm J}} \ .
\label{eq:halkj2}
\end{eqnarray}
Since $u_{\rm r}$ of the fine-structure constant is significantly smaller than
$u_{\rm r}$ of the measured values of $K_{\rm J}$, the $u_{\rm r}$ of $h$
derived from Eq.~(\ref{eq:halkj2}) will be essentially twice the $u_{\rm r}$ of
$K_{\rm J}$.

The values of $K_{\rm J}$ we take as input data in the 2006 adjustment, and the
corresponding inferred values of $h$, are given in the following two sections
and are summarized in Table~\ref{tab:hdata}. Also summarized in that table are
the measured values of the product $K_{\rm J}^2R_{\rm K}$ and the quantity
${\cal F}_{90}$ related to the Faraday constant $F$, together with their
corresponding inferred values of $h$.  These results are discussed below in
Secs.~\ref{ssec:kj2rk} and \ref{ssec:f}. 

\begin{table*}
\def\m{\phantom{-}}
\caption{Summary of data related to the Josephson constant
$K_{\rm J}$, the product $K_{\rm J}^2R_{\rm K}$, and the Faraday constant $F$, 
and inferred values of $h$.}
\label{tab:hdata}
\def\sp{\hbox to 15 pt {}}
\begin{tabular}{l@{\sp}l@{\sp}c@{\sp}l@{\sp}l}
\toprule
Quantity \T    &  \hbox to 12pt {} Value
& Relative standard
& Identification & Sect. and Eq. \\
&& uncertainty $u_{\rm r}$ &&\\
\colrule
$K_{\rm J}$ \vbox to 12pt{} & $ 483\,597.91(13) \ {\rm GHz \ V^{-1}}$ & $ 2.7\times 10^{-7}$ 
& NMI-89 & \ref{sssec:nmljc} (\ref{eq:kjnml89}) \\
\ \ $h$ &\ \ $ 6.626\,0684(36)\times 10^{-34}\ {\rm J \ s}$&$ 5.4\times 10^{-7}$& &
    \ref{sssec:nmljc} (\ref{eq:hnml89}) \\
&&&&\\
$K_{\rm J}$ & $ 483\,597.96(15) \ {\rm GHz \ V^{-1}}$ & $ 3.1\times 10^{-7}$ 
& PTB-91 & \ref{sssec:ptbjc} (\ref{eq:kjptb91}) \\
\ \ $h$ &\ \ $ 6.626\,0670(42)\times 10^{-34}\ {\rm J \ s}$&$ 6.3\times 10^{-7}$& &
    \ref{sssec:ptbjc} (\ref{eq:hptb91}) \\
&&&&\\
$K_{\rm J}^2R_{\rm K}$ & $ 6.036\,7625(12)\times 10^{33}\ {\rm J^{-1} \ s^{-1}}$ & $ 2.0\times 10^{-7}$ 
& NPL-90 & \ref{sssec:kj2rknpl} (\ref{eq:kj2rknpl90}) \\
\ \ $h$ &\ \ $ 6.626\,0682(13)\times 10^{-34}\ {\rm J \ s}$&$ 2.0\times 10^{-7}$& &
    \ref{sssec:kj2rknpl} (\ref{eq:hwbnpl90}) \\
&&&&\\
$K_{\rm J}^2R_{\rm K}$ & $ 6.036\,761\,85(53)\times 10^{33}\ {\rm J^{-1} \ s^{-1}}$ & $ 8.7\times 10^{-8}$ 
& NIST-98 & \ref{par:kj2rk98} (\ref{eq:kj2rknist98}) \\
\ \ $h$ &\ \ $ 6.626\,068\,91(58)\times 10^{-34}\ {\rm J \ s}$&$ 8.7\times 10^{-8}$& &
    \ref{par:kj2rk98} (\ref{eq:hwbnist98}) \\
&&&&\\
$K_{\rm J}^2R_{\rm K}$ & $ 6.036\,761\,85(22)\times 10^{33}\ {\rm J^{-1} \ s^{-1}}$ & $ 3.6\times 10^{-8}$ 
& NIST-07 & \ref{par:kj2rk07} (\ref{eq:kj2rknist07}) \\
\ \ $h$ &\ \ $ 6.626\,068\,91(24)\times 10^{-34}\ {\rm J \ s}$&$ 3.6\times 10^{-8}$& &
    \ref{par:kj2rk07} (\ref{eq:hwbnist07}) \\
&&&&\\
${\cal F}_{90}$ & $ 96\,485.39(13)\ {\rm C \ mol^{-1}}$ & $ 1.3\times 10^{-6}$ 
& NIST-80 & \ref{sssec:fcnist} (\ref{eq:f90nist80}) \\
\ \ $h$ &\ \ $ 6.626\,0657(88)\times 10^{-34}\ {\rm J \ s}$&$ 1.3\times 10^{-6}$& &
    \ref{sssec:fcnist} (\ref{eq:hfnist80}) \\
\botrule
\end{tabular}
\end{table*}

\subsubsection{NMI: Hg electrometer}
\label{sssec:nmljc}

The determination of $K_{\rm J}$ at NMI, carried out using an apparatus called
a liquid-mercury electrometer, yielded the result \cite{1989050}
\begin{eqnarray} 
K_{\rm J} &=& 483\,594 \left[1 +  8.087(269)\times
10^{-6}\right] \ {\rm GHz/V} \nonumber\\ &=&  483\,597.91(13) \ {\rm GHz/V}
\quad [ 2.7\times 10^{-7}] \ .  
\label{eq:kjnml89} 
\end{eqnarray}
Equation~(\ref{eq:halkj2}), the NMI value of $K_{\rm J}$, and the 2006
recommended value of $\alpha$, which has a much smaller $u_{\rm r}$, yields an
inferred value for the Planck constant of 
\begin{eqnarray} h =
6.626\,0684(36)\times 10^{-34} \ {\rm J \ s} \quad [ 5.4\times 10^{-7}] \ .
\label{eq:hnml89} 
\end{eqnarray}

\subsubsection{PTB: Capacitor voltage balance}
\label{sssec:ptbjc}

The determination of $K_{\rm J}$ at PTB was carried out by using a voltage
balance consisting of two coaxial cylindrical electrodes
\cite{1991025,1986043,1985042}.  Taking into account the correction associated
with the reference capacitor used in the PTB experiment as described in
CODATA-98, the result of the PTB determination is

\begin{eqnarray}
K_{\rm J} =  483\,597.96(15) \ {\rm GHz/V} \quad [ 3.1\times 10^{-7}] \ ,
\label{eq:kjptb91}
\end{eqnarray}
from which we infer, using Eq.~(\ref{eq:halkj2}),
\begin{eqnarray}
h =  6.626\,0670(42)\times 10^{-34} \ {\rm J \ s}\quad [ 6.3\times 10^{-7}] \ .
\label{eq:hptb91}
\end{eqnarray}

\subsection{Product $\bm{K^{2}_{\rm J}R_{\rm K}}$ and $\bm{h}$}
\label{ssec:kj2rk}

A value of the product $K_{\rm J}^2R_{\rm K}$ is of importance to the
determination of the Planck constant $h$, because if one assumes that the
relations $K_{\rm J} = 2e/h$ and $R_{\rm K} = h/e^2$ are valid, then
\begin{eqnarray} 
h = {4\over K_{\rm J}^2R_{\rm K}} \ .
\label{eq:hkjrk} 
\end{eqnarray} 
The product $K_{\rm J}^2R_{\rm K}$ is determined by comparing electrical power
known in terms of a Josephson voltage and quantized Hall resistance to the
equivalent mechanical power known in the SI unit W = m$^2$~kg~s$^{-3}$.  The
comparison is carried out using an apparatus known as a moving-coil watt
balance first proposed by \textcite{1975027} at NPL. To date two laboratories,
NPL and NIST, have determined $K_{\rm J}^2R_{\rm K}$ using this method.

\subsubsection{NPL: Watt balance}
\label{sssec:kj2rknpl}

Shortly after Kibble's original proposal in 1975, \textcite{1977025} carried
out a feasibility study of the idea based on experience with the NPL apparatus
that was used to determine $\gamma^{\,\prime}_{\rm p}$ by the high-field method
\cite{1979012}. The work continued and led to the publication in 1990 by
\textcite{1990057} of a result with an uncertainty of about 2 parts in
$10^{7}$. This result, discussed in detail in CODATA-98 and which was taken as
an input datum in the 1998 and 2002 adjustments, and which we also take as an
input datum in the 2006 adjustment, may be expressed as 
\begin{eqnarray}
K_{\rm J}^2R_{\rm K} &=& K_{\rm J-NPL}^2R_{\rm K-NPL}
[1 +  16.14(20)\times 10^{-6} ]
\nonumber\\
&=&  6.036\,7625(12)\times 10^{33} \ {\rm J}^{-1} \ {\rm s}^{-1} 
\nonumber\\ && \quad [ 2.0\times 10^{-7}] \ ,
\label{eq:kj2rknpl90}
\end{eqnarray}
where $K_{\rm J-NPL} =  483\,594 $ GHz/V and $R_{\rm K-NPL} =  25\,812.809\,2 $
${\rm \Omega}$.  The value of $h$ that may be inferred from the NPL result is,
according to Eq.~(\ref{eq:hkjrk}),
\begin{eqnarray}
h =  6.626\,0682(13)\times 10^{-34} \ {\rm J \ s} \quad [ 2.0\times 10^{-7}] \ .
\label{eq:hwbnpl90}
\end{eqnarray}

Based on the experience gained in this experiment, NPL researchers designed and
constructed what is essentially a completely new apparatus, called the NPL
Mark~II watt balance, that could possibly achieve a result for $K_{\rm J}^2
R_{\rm K}$ with an uncertainty of a few parts in $10^{8}$ \cite{1997052,
2003204}. Although the balance itself employs the same balance beam as the
previous NPL watt balance, little else from that experiment is retained in the
new experiment.

Over 1000 measurements in vacuum were carried out with the MK~II between
January 2000 and November 2001. Many were made in an effort to identify the
cause of an observed fractional change in the value of $K_{\rm J}^2 R_{\rm K}$
of about $3\times 10^{-7}$ that occurred in mid-April 2000 \cite{2002232}. A
change in the alignment of the apparatus was suspected of contributing to the
shift. 

Significant improvements were subsequently made in the experiment and very
recently, based on measurements carried out from October 2006 to March 2007,
the initial result from MK~II, $h = 6.626\,070\,95(44)~{\rm J~s}~[6.6
\times{10}^{-8}]$, was reported by \textcite{2007209} assuming the validity of
Eq.~(\ref{eq:hkjrk}). Although this result became available much too late to be
considered for the 2006 adjustment, we do note that it lies between the value
of $h$ inferred from the 2007 NIST result for $K_{\rm J}^2R_{\rm K}$ discussed
in Sec.~\ref{par:kj2rk07}, and that inferred from the measurement of the molar
volume of silicon $V_{\rm m}$(Si) discussed in Sec.~\ref{ssec:mvsi}. The NPL
work is continuing and a result with a smaller uncertainty is anticipated
\cite{2007209}.

\subsubsection{NIST: Watt balance}
\label{sssec:kj2rknist}
\paragraph{1998 measurement}
\label{par:kj2rk98}
Work on a moving-coil watt balance at NIST began shortly after Kibble made his
1975 proposal. A first result with $u_{\rm r}=1.3\times10^{-6}$ was reported by
NIST researchers in 1989 \cite{1989002}. Significant improvements were then
made to the apparatus and the final result from this phase of the NIST effort
was reported in 1998 by \textcite{1998071}:
\begin{eqnarray}
K_{\rm J}^2R_{\rm K} &=& K_{\rm J-90}^2R_{\rm K-90} 
[1  -8(87)\times 10^{-9} ]
\nonumber\\
&=&  6.036\,761\,85(53)\times 10^{33} \  {\rm J}^{-1} \ {\rm s}^{-1} 
\nonumber\\ && \quad  [ 8.7\times 10^{-8}] \ .
\label{eq:kj2rknist98}
\end{eqnarray}
A lengthy paper giving the details of the NIST 1998 watt balance experiment was
published in 2005 by \textcite{2005053}.  This was the NIST result taken as an
input datum in the 1998 and 2002 adjustments; although the 1989 result was
consistent with that of 1998, its uncertainty was about 15 times larger.  The
value of $h$ implied by the 1998 NIST result for $K_{\rm J}^2 R_{\rm K}$ is
\begin{eqnarray} h =  6.626\,068\,91(58)\times 10^{-34} \ {\rm J \ s} \quad [
8.7\times 10^{-8}] \ .  \label{eq:hwbnist98} \end{eqnarray}

\paragraph{2007 measurement}
\label{par:kj2rk07}

Based on the lessons learned in the decade-long effort with a watt balance
operating in air that led to their 1998 result for $K_{\rm J}^2 R_{\rm K}$, the
NIST watt-balance researchers initiated a new program with the goal of
measuring $K_{\rm J}^2 R_{\rm K}$ with $u_{\rm r} \approx 10^{-8}$. The
experiment was completely disassembled and renovations to the research facility
were made to improve vibration isolation, reduce electromagnetic interference,
and incorporate a multilayer temperature control system. A new watt balance
with major changes and improvements was constructed with little remaining of
the earlier apparatus except the superconducting magnet used to generate the
required radial magnetic flux density and the wheel used as the balance. 
\vspace{0pt}

The most notable change in the experiment is that in the new apparatus, the
entire balance mechanism and moving coil are in vacuum, which eliminates the
uncertainties of the corrections in the previous experiment for the index of
refraction of air in the laser position measurements ($u_{\rm
r}=43\times10^{-9}$) and for the buoyancy force exerted on the mass
standard ($u_{\rm r}=23\times10^{-9}$). Alignment uncertainties were reduced by
over a factor of four by (i) incorporating a more comprehensive understanding
of all degrees of freedom involving the moving coil; and (ii) the application of
precise alignment techniques for all degrees of freedom involving the moving
coil, the superconducting magnet, and the velocity measuring interferometers.
Hysteresis effects were reduced by a factor of four by using a diamond-like
carbon coated knife edge and flat \cite{2001232}, employing a hysteresis
erasure procedure, and reducing the balance deflections during mass exchanges
with improved control systems. A programmable Josephson array voltage standard
\cite{1997152} was connected directly to the experiment, eliminating two
voltage transfers required in the old experiment and reducing the voltage
traceability uncertainty by a factor of 15.

A total of 6023 individual values of $W_{90}$/W were obtained over the two year
period from March 2003 to February 2005 as part of the effort to develop and
improve the new experiment.  The results are converted to the notation used
here by the relation $W_{90}/{\rm W} = K_{{\rm J}-90}^2R_{{\rm K}-90}/K_{\rm
J}^2R_{\rm K}$ discussed in CODATA-98.  The initial result from that work was
reported in 2005 by \textcite{2005016}:
\begin{eqnarray}
K_{\rm J}^2R_{\rm K} &=& K_{\rm J-90}^2R_{\rm K-90} 
[1  -24(52)\times 10^{-9} ]
\nonumber\\
&=&  6.036\,761\,75(31)\times 10^{33} \  {\rm J}^{-1} \ {\rm s}^{-1} 
\nonumber\\ && \quad  [ 5.2\times 10^{-8}] \ .
\label{eq:kj2rknist05}
\end{eqnarray}
This yields a value for the Planck constant of
\begin{eqnarray}
h =  6.626\,069\,01(34)\times 10^{-34} \ {\rm J \ s} \quad [ 5.2\times 10^{-8}] \ .
\label{eq:hwbnist05}
\end{eqnarray}

This result for $K_{\rm J}^2R_{\rm K}$ was obtained from data spanning the
final 7 months of the 2 year period. It is based on the weighted mean of 48
$W_{90}$/W measurement sets using a Au mass standard and 174 sets using a PtIr
mass standard, where a typical measurement set consists of 12 to 15 individual
values of $W_{90}$/W. The 2005 NIST result is consistent with the 1998 NIST
result but its uncertainty has been reduced by a factor of 1.7.

Following this initial effort with the new apparatus, further improvements were
made to it in order to reduce the uncertainties from various systematic
effects, the most notable reductions being in the determination of the local
acceleration due to gravity $g$ (a factor of 2.5), the effect of balance wheel
surface roughness (a factor of 10), and the effect of the magnetic
susceptibility of the mass standard (a factor of 1.6). An improved result was
then obtained based on 2183 values of $W_{90}$/W recorded in 134 measurement
sets from January 2006 to June 2006. Due to a wear problem with the gold mass
standard, only a PtIr mass standard was used in these measurements. The result,
first reported at a conference in 2006 and subsequently published in the
proceedings of the conference in 2007 by \textcite{2007085}, is
\begin{eqnarray} K_{\rm J}^2R_{\rm K} &=& K_{\rm J-90}^2R_{\rm K-90} [1
-8(36)\times 10^{-9} ] \nonumber\\ &=&  6.036\,761\,85(22)\times 10^{33} \
{\rm J}^{-1} \ {\rm s}^{-1} \nonumber\\ && \quad  [ 3.6\times 10^{-8}] \ .
\label{eq:kj2rknist07} \end{eqnarray} The value of $h$ that may be inferred
from this value of $K_{\rm J}^2R_{\rm K}$ is \begin{eqnarray} h =
6.626\,068\,91(24)\times 10^{-34} \ {\rm J \ s} \quad [ 3.6\times 10^{-8}] \ .
\label{eq:hwbnist07} \end{eqnarray}

The 2007 NIST result for $K_{\rm J}^2R_{\rm K}$ is consistent with and has an
uncertainty smaller by a factor of 1.4 than the uncertainty of the 2005 NIST
result. However, because the two results are from essentially the same
experiment and hence are highly correlated, we take only the 2007 result as an
input datum in the 2006 adjustment.

On the other hand, the experiment on which the NIST 2007 result is based is
only slightly dependent on the experiment on which the NIST 1998 result is
based, as can be seen from the above discussions. Thus, in keeping with our
practice in similar cases, most notably the 1982 and 1999 LAMPF measurements of
muonium Zeeman transition frequencies (see Sec.~\ref{sssec:mufreqs}), we also
take the NIST 1998 result in Eq.~(\ref{eq:kj2rknist98}) as an input datum in
the 2006 adjustment. But to ensure that we do not give undue weight to the NIST
work, an analysis of the uncertainty budgets of the 1998 and 2007 NIST results
was performed to determine the level of correlation. Of the relative
uncertainty components listed in Table II of \textcite{1998071} and in Table 2
of \textcite{2005016} but as updated in Table 1 of \textcite{2007085}, the
largest common relative uncertainty components were from the magnetic flux
profile fit due to the use of the same analysis routine ($16\times10^{-9}$);
leakage resistance and electrical grounding since the same current supply was
used in both experiments ($10\times10^{-9}$); and the determination of the
local gravitational acceleration $g$ due to the use of the same absolute
gravimeter ($7\times10^{-9}$). The correlation coefficient was thus determined
to be \begin{eqnarray} r(K_{\rm J}^2\,R_{\rm k}\mbox{-}98,K_{\rm J}^2\,R_{\rm
k}\mbox{-}07) =  0.14 \ , \end{eqnarray} which we take into account in our
calculations as appropriate.

\subsubsection{Other values}
\label{sssec:kj2rkother}

Although there is no competitive published value of $K_{\rm J}^2R_{\rm K}$
other than those from NPL and NIST discussed above, it is worth noting that at
least three additional laboratories have watt-balance experiments in progress:
the Swiss Federal Office of Metrology and Accreditation (METAS), Bern-Wabern,
Switzerland, the LNE, and the BIPM.  Descriptions of these efforts may be found
in the papers by \textcite{2003041}, \textcite{2005037}, and
\textcite{2007083}, respectively.

\subsubsection{Inferred value of $K_{\rm J}$}
\label{sssec:kjfromkj2rk}

It is of interest to note that a value of $K_{\rm J}$ with an uncertainty
significantly smaller than those of the directly measured values discussed in
Sec.~\ref{ssec:jc} can be obtained from the directly measured watt-balance
values of $K^{2}_{\rm J}R_{\rm K}$, together with the directly measured
calculable-capacitor values of $R_{\rm K}$, without assuming the validity of
the relations $K_{\rm J} = 2e/h$ and $R_{\rm K} = h/e^{2}$. The relevant
expression is simply $K_{\rm J} = [(K^{2}_{\rm J}R_{\rm K})_{\rm W}/(R_{\rm
K})_{\rm C}]^{1/2}$, where $(K^{2}_{\rm J}R_{\rm K})_{\rm W}$ is from the
watt-balance, and $(R_{\rm K})_{\rm C}$ is from the calculable capacitor. 

Using the weighted mean of the three watt-balance results for $K^{2}_{\rm
J}R_{\rm K}$ discussed in this section and the weighted mean of the five
calculable-capacitor results for $R_{\rm K}$ discussed in Sec~\ref{ssec:vkc},
we have
\begin{eqnarray}
K_{\rm J} &=& K_{\rm J-90} 
[1   -2.8(1.9)\times 10^{-8} ]
\nonumber\\
&=&  483\,597.8865(94) \  {\rm GHz/V} 
 \quad  [ 1.9\times 10^{-8}] \ , \qquad
\label{eq:kjwbcc}
\end{eqnarray}
which is consistent with the directly measured values but has an uncertainty
that is smaller by more than an order of magnitude.  This result is implicitly
included in the least-squares adjustment, even though the explicit value for
$K_{\rm J}$ obtained here is not used as an input datum.

\subsection{Faraday constant $\bm{F}$ and $\bm{h}$}
\label{ssec:f}

The Faraday constant $F$ is equal to the Avogadro constant $N_{\rm A}$ times
the elementary charge $e$, $F = N_{\rm A}e$; its SI unit is coulomb per mol, C
mol$^{-1}$ = A s mol$^{-1}$.  It determines the amount of substance $n(X)$ of
an entity $X$ that is deposited or dissolved during electrolysis by the passage
of a quantity of electricity, or charge, $Q = It$, due to the flow of a current
$I$ in a time $t$.  In particular, the Faraday constant $F$ is related to the
molar mass $M(X)$ and valence $z$ of entity $X$ by
\begin{eqnarray}
F = ItM(X) \over z {m_{\rm d}(X)} \ ,
\label{eq:fmze}
\end{eqnarray}
where $m_{\rm d}(X)$ is the mass of entity $X$ dissolved as the result of
transfer of charge $Q=It$ during the electrolysis.  It follows from the
relations $F = N_{\rm A}e$, $e^2 = 2\alpha h/\mu_{\rm 0} c$, $m_{\rm e} =
2R_{\infty}h/c\alpha^2$, and $N_{\rm A} = A_{\rm r}({\rm e})M_{\rm u}/m_{\rm
e}$, where $M_{\rm u} = 10^{-3}$~kg~mol$^{-1}$, that 
\begin{eqnarray}
F = {A_{\rm r}({\rm e})M_{\rm u} \over R_{\infty} } 
\left({c\over2\mu_{\rm 0}}
{\alpha^5\over h}\right)^{\!\! 1/2}   \ .
\label{eq:fconst}
\end{eqnarray}
Since, according to Eq.~(\ref{eq:fmze}), $F$ is proportional to the current
$I$, and $I$ is inversely proportional to the product $K_{\rm J}R_{\rm K}$ if
the current is determined in terms of the Josephson and quantum Hall effects,
we may write
\begin{eqnarray}
{\cal F}_{90} = 
{K_{\rm J}R_{\rm K}\over
K_{\rm J-90}R_{\rm K-90}}
{A_{\rm r}({\rm e})M_{\rm u} \over R_{\infty} }
\left({c\over2\mu_{\rm 0}}
{\alpha^5\over h}\right)^{\!\! 1/2}   \ ,
\label{eq:f90}
\end{eqnarray}
where ${\cal F}_{90}$ is the experimental value of $F$ in SI units that would
result from the Faraday experiment if $K_{\rm J}=K_{\rm J-90}$ and $R_{\rm
K}=R_{\rm K-90}$.  The quantity ${\cal F}_{90}$ is the input datum used in the
adjustment, but the observational equation accounts for the fact that $K_{\rm
J-90} \ne K_{\rm J}$ and $R_{\rm K-90} \ne R_{\rm K}$.  If one assumes the
validity of the expressions $K_{\rm J} = 2e/h$ and $R_{\rm K} = h/e^2$, then in
terms of adjusted constants, Eq.~(\ref{eq:f90}) can be written as
\begin{eqnarray}
{\cal F}_{90} = {c M_{\rm u} \over 
K_{\rm J-90}R_{\rm K-90}}
{A_{\rm r}({\rm e})\alpha^2\over R_{\infty}h} \ .
\label{eq:f90const}
\end{eqnarray}

\subsubsection{NIST: Ag coulometer}
\label{sssec:fcnist}

There is one high-accuracy experimental value of ${\cal F}_{90}$ available,
that from NIST \cite{1980029}.  The NIST experiment used a silver dissolution
coulometer based on the anodic dissolution by electrolysis of silver, which is
monovalent, into a solution of perchloric acid containing a small amount of
silver perchlorate. The basic chemical reaction is Ag $\rightarrow$ Ag$^+$ +
e$^-$ and occurs at the anode, which in the NIST work was a highly purified
silver bar.

As discussed in detail in CODATA-98, the NIST experiment leads to 
\begin{eqnarray}
{\cal F}_{90} =  96\,485.39(13) \ {\rm C \ mol}^{-1} 
\quad [ 1.3\times 10^{-6}] \ .
\label{eq:f90nist80}
\end{eqnarray}
[Note that the new AME2003 values of $A_{\rm r}(^{107}{\rm Ag})$ and $A_{\rm
r}(^{109}{\rm Ag})$ in Table~\ref{tab:rmass03} have no effect on this result.]

The value of $h$ that may be inferred from the NIST result,
Eq.~(\ref{eq:f90const}), and the 2006 recommended values for the other
quantities is
\begin{eqnarray}
h =  6.626\,0657(88)\times 10^{-34} \ {\rm J \ s} \quad [ 1.3\times 10^{-6}] \ ,
\label{eq:hfnist80}
\end{eqnarray}
where the uncertainties of the other quantities are negligible compared to the
uncertainty of ${\cal F}_{90}$.

\section{Measurements involving silicon crystals}
\label{sec:msc}

Here we discuss experiments relevant to the 2006 adjustment that use highly
pure, nearly crystallographically perfect, single crystals of silicon.
However, because one such experiment determines the quotient $h/m_{\rm n}$,
where $m_{\rm n}$ is the mass of the neutron, for convenience and because any
experiment that determines the ratio of the Planck constant to the mass of a
fundamental particle or atom provides a value of the fine-structure constant
$\alpha$, we also discuss in this section two silicon-independent experiments:
the 2002 Stanford University, Stanford, USA, measurement of $h/m(^{133} \rm
Cs)$ and the 2006 Laboratoire Kastler-Brossel or LKB measurement of $h/m(^{87}
\rm Rb)$.

In this section, W4.2a, NR3, W04 and NR4 are shortened forms of the full
crystal designations WASO~4.2a, NRLM3, WASO~04, and NRLM4, respectively, for
use in quantity symbols.  No distinction is made between different crystals
taken from the same ingot.  As we use the current laboratory name to identify a
result rather than the laboratory name at the time the measurement was carried
out, we have replaced IMGC and NRLM with INRIM and NMIJ---see the glossary in
CODATA-98. 

\subsection{\{220\} lattice spacing of silicon $\bm{d_{220}}$}
\label{ssec:lss}

A value of the \{220\} lattice spacing of a silicon crystal in meters is
relevant to the 2006 adjustment not only because of its role in determining
$\alpha$ from $h/m_{\rm n}$ (see Sec.~\ref{sssec:pcnmr}), but also because of
its role in determining the relative atomic mass of the neutron $A_{\rm r}(\rm
n)$ (see~Sec.\ref{ssec:arn}).  Further, together with the measured value of the
molar volume of silicon $V_{\rm m}$(Si), it can provide a competitive value of
$h$ (see Sec.~\ref{ssec:mvsi}).

Various aspects of silicon and its crystal plane spacings of interest here are
reviewed in CODATA-98 and CODATA-02.  [See also the reviews of
\textcite{2003227}, \textcite{2001322}, and \textcite{2001268}].  Some points
worth noting are that silicon is a cubic crystal with $n = 8$ atoms per
face-centered cubic unit cell of edge length (or lattice parameter) $a =
543$~pm with $d_{220}=a/\sqrt{8}$. The three naturally occurring isotopes of Si
are $^{28}$Si, $^{29}$Si, and $^{30}$Si, and the amount-of-substance fractions
$x(^{28}$Si), $x(^{29}$Si), and $x(^{30}$Si) of natural silicon are
approximately 0.92, 0.05, and 0.03, respectively.

Although the \{220\} lattice spacing of Si is not a fundamental constant in the
usual sense, for practical purposes one can consider $a$, and hence $d_{220}$,
of an impurity-free, crystallographically perfect or ``ideal'' silicon crystal
under specified conditions, principally of temperature, pressure, and isotopic
composition, to be an invariant of nature. The reference temperature and
pressure currently adopted are $t_{90} = 22.5~^\circ$C and $p$ = 0 (that is,
vacuum), where $t_{90}$ is Celsius temperature on the International Temperature
Scale of 1990 (ITS-90) \cite{1990064,1990064e}.  However, no reference values
for $x(^{A}$Si) have yet been adopted, because the variation of $a$ due to the
variation of the isotopic composition of the crystals used in high-accuracy
experiments is taken to be negligible at the current level of experimental
uncertainty in $a$. A much larger effect on $a$ is the impurities that the
silicon crystal contains---mainly carbon (C), oxygen (O), and nitrogen
(N)---and corrections must be applied to convert the \{220\} lattice spacing
$d_{220}({\scriptstyle X})$ of a real crystal $X$ to the \{220\} lattice
spacing $d_{220}$ of an ``ideal'' crystal. 

Nevertheless, we account for the possible variation in the lattice spacing of
different samples taken from the same ingot by including an additional
component (or components) of relative standard uncertainty in the uncertainty
of any measurement result involving a silicon lattice spacing (or spacings).
This additional component is typically $\sqrt{2}\times10^{-8}$ for each
crystal, but it can be larger, for example, $(3/2)\sqrt{2} \times10^{-8}$ in
the case of crystal $\rm MO^{*}$ discussed below, because it is known to
contain a comparatively large amount of carbon; see Secs. III.A.c and III.I of
CODATA-98 for details. For simplicity, we do not explicitly mention our
inclusion of such components in the following discussion. 

Further, because of this component and the use of the same samples in different
experiments, and because of the existence of other common components of
uncertainty in the uncertainty budgets of different experimental results
involving silicon crystals, many of the input data discussed in the following
sections are correlated. In most cases we do not explicity give the relevant
correlation coefficients in the text; instead Table~\ref{tab:pdcc} in
Sec.~\ref{sec:ad} provides all the non-negligible correlation coefficients of
the input data listed in Table~\ref{tab:pdata}.

\subsubsection{X-ray/optical interferometer measurements of
$d_{220}({\scriptstyle X})$}
\label{sssec:xroidx}

High accuracy measurements of $d_{220}({\scriptstyle X})$, where $X$ denotes
any one of various crystals, are carried out using a combined x-ray and optical
interferometer (XROI) fabricated from a single crystal of silicon taken from
one of several well-characterized single crystal ingots or boules. As discussed
in CODATA-98, an XROI is a device that enables x-ray fringes of unknown period
$d_{220}({\scriptstyle X})$ to be compared with optical fringes of known period
by moving one of the crystal plates of the XROI, called the analyzer. Also
discussed there are the XROI measurements of $d_{220}({\rm {\scriptstyle
W4.2a}})$, $d_{220}({\rm {\scriptstyle MO^*}})$, and $d_{220}({\rm
{\scriptstyle NR3}})$, which were carried out at the PTB in Germany
\cite{1981017}, the Istituto Nazionale di Ricerca Metrologica, Torino, Italy
(INRIM) \cite{1994140}, and the National Metrology Institute of Japan (NMIJ),
Tsukuba, Japan \cite{1997071}, respectively. 

For the reasons discussed in CODATA-02 and subsequently documented by
\textcite{2004001,2004001e}, only the NMIJ 1997 result was taken as an input
datum in the 2002 adjustment. However, further work, published in the Erratum
to that paper, showed that the results obtained at INRIM given in the paper
were in error. After the error was discovered, additional work was carried out
at INRIM to fully understand and correct it. New results were then reported at
a conference in 2006 and published in the conference proceedings
\cite{2007061}.  Thus, as summarized in Table~\ref{tab:d220abs} and compared in
Fig.~\ref{fig:silicon}, we take as input data the four absolute \{220\} lattice
spacing values determined in three different laboratories, as discussed in the
following three sections.  The last value in the table, which is not an XROI
result, is discussed in Sec.~ \ref{sssec:pcnmr}. 

\begin{figure}
\rotatebox{-90}{\resizebox{!}{4.2in}{
\includegraphics[clip,trim=10 40 40 10]{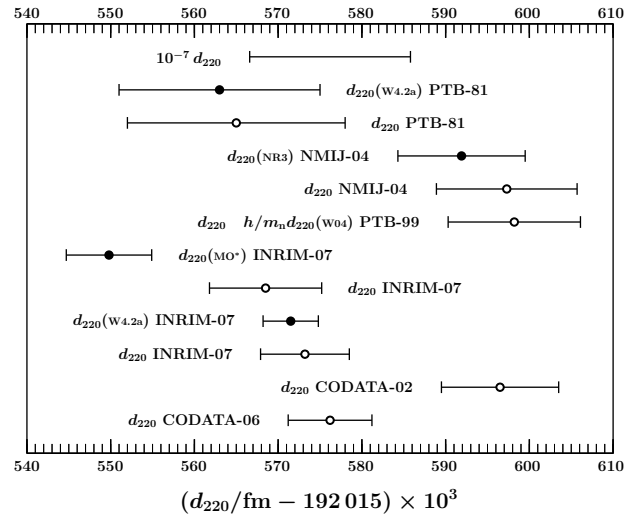}
}}
\caption{ Inferred values (open circles) of $d_{220}$ from various measurements
(solid circles) of $d_{220}(X)$.  For comparison, the 2002 and 2006 CODATA
recommended values of $d_{220}$ are also shown.}
\label{fig:silicon}
\end{figure}

\begin{table*}
\def\m{\phantom{-}}
\caption{Summary of measurements of the absolute \{220\} lattice spacing of various silicon
crystals and inferred values of $d_{220}$.}
\def\vsp{\vbox to 13 pt {}}
\def\svsp{\vbox to 8 pt {}}
\label{tab:d220abs}
\def\sp{\hbox to 10 pt {}}
\begin{tabular}{l@{\sp}l@{\sp}c@{\sp}l@{\sp}l}
\toprule
Quantity \T    &  \hbox to 12pt {} Value
& Relative standard
& Identification & Sect. and Eq. \\
&& uncertainty $u_{\rm r}$&& \\
\colrule
$d_{220}({\rm {\scriptstyle{W4.2a}}})$\vsp & $\phantom{-}  192\,015.563(12)$ fm & $ 6.2\times 10^{-8}$ & PTB-81 & \ref{par:ptblat} (\ref{eq:dw42ptb81}) \\
\ \ $d_{220}$\svsp & $\phantom{-}  192\,015.565(13)$ fm & $ 6.5\times 10^{-8}$ &  & \\
$d_{220}({\rm {\scriptstyle{NR3}}})$ \vsp & $\phantom{-}  192\,015.5919(76)$ fm & $ 4.0\times 10^{-8}$ & NMIJ-04 & \ref{par:nmijlat} (\ref{eq:dnr3nmij04}) \\
\ \ $d_{220}$\svsp & $\phantom{-}  192\,015.5973(84)$ fm & $ 4.4\times 10^{-8}$ &  & \\
$d_{220}({\rm {\scriptstyle{W4.2a}}})$ \vsp & $\phantom{-}  192\,015.5715(33)$ fm & $ 1.7\times 10^{-8}$ & INRIM-07 & \ref{par:inrimlat} (\ref{eq:dw42ainrimptb07}) \\
\ \ $d_{220}$\svsp & $\phantom{-}  192\,015.5732(53)$ fm & $ 2.8\times 10^{-8}$ &  & \\
$d_{220}({\rm {\scriptstyle{MO^*}}})$ \vsp & $\phantom{-}  192\,015.5498(51)$ fm & $ 2.6\times 10^{-8}$ & INRIM-07 & \ref{par:inrimlat} (\ref{eq:dmo4inrim07}) \\
\ \ $d_{220}$\svsp & $\phantom{-}  192\,015.5685(67)$ fm & $ 3.5\times 10^{-8}$ &  & \\
$h/m_{\rm n}d_{220}({\rm {\scriptstyle{W04}}})$ \vsp {}& $\phantom{-}  2060.267\,004(84)$ m s$^{-1}$
& $ 4.1\times 10^{-8}$ & PTB-99 & \ref{sssec:pcnmr} (\ref{eq:homndw04}) \\
\ \ $d_{220}$\svsp & $\phantom{-}  192\,015.5982(79)$ fm & $ 4.1\times 10^{-8}$ & 
& \ref{sssec:pcnmr} (\ref{eq:d220homn}) \\
\botrule
\end{tabular}
\end{table*}

We point out that not only do we take the \{220\} lattice spacings of the
crystals WASO~4.2a, NRLM3, and $\rm MO^{*}$ as adjusted constants, but also the
\{220\} lattice spacings of the crystals N, WASO~17, ILL, WASO~04, and NRLM4,
because they too were involved in various experiments, including the $d_{220}$
lattice spacing fractional difference measurements discussed in
Sec~\ref{sssec:d220diff}. 

\paragraph{PTB measurement of $d_{220}({\rm \scriptstyle{W4.2a}})$}
\label{par:ptblat}

The following value, identified as PTB-81 in Table~\ref{tab:d220abs} and
Fig.~\ref{fig:silicon}, is the original result obtained at PTB as reported by
\textcite{1981017} and discussed in CODATA-98: 
\begin{eqnarray}
d_{220}({\scriptstyle{\rm W4.2a}}) =  192\,015.563(12) \ {\rm fm}
\quad [ 6.2\times 10^{-8}] \ . \qquad
\label{eq:dw42ptb81}
\end {eqnarray}

\paragraph{NMIJ measurement of $d_{220}({\rm \scriptstyle{NR3}})$}
\label{par:nmijlat}

The following value, identified as NMIJ-04 in Table~\ref{tab:d220abs} and
Fig.~\ref{fig:silicon}, reflects the NMIJ efforts in the early and mid-1990s as
well as the work carried out in the early 2000s:
\begin{eqnarray}
d_{220}({\scriptstyle{\rm NR3}}) =  192\,015.5919(76) \ {\rm fm}
\quad [ 4.0\times 10^{-8}] \ . \qquad
\label{eq:dnr3nmij04}
\end {eqnarray}
This value, reported by \textcite{2004001,2004001e}, is the weighted mean of
the 1997 NIMJ result of \textcite{1997071} discussed in CODATA-98 and
CODATA-02, and the result from a new series of measurements performed at NMIJ
from December 2002 to February 2003 with nearly the same apparatus. One of the
principle differences from the earlier experiment was the much improved
temperature control system of the room in which the NMIJ XROI was located; the
new system provided a temperature stability of about 1~mK/d and allowed the
temperature of the XROI to be set to within 20~mK of $22.5~^\circ\rm C$.

The result for $d_{220}({\scriptstyle{\rm NR3}})$ from the 2002-2003
measurements is based on 61 raw data. In each measurement, the phases of the
x-ray and optical fringes (optical orders) were compared at the 0th, 100th, and
201st optical orders, and then with the analyzer moving in the reverse
direction, at the 201st, 100th, and 0th orders. The $n/m$ ratio was calculated
from the phase of the x-ray fringe at the 0th and 201st orders, where $n$ is
the number of x-ray fringes in $m$ optical fringes (optical orders) of period
$\lambda/2$, where $\lambda$ is the wavelength of the laser beam used in the
optical interferometer and $d_{220}({\scriptstyle{\rm
NR3}})=(\lambda/2)/(n/m)$. 

In the new work, the fractional corrections to $d_{220}({\scriptstyle{\rm
NR3}})$, in parts in $10^{9}$, total 181(35), the largest by far being the
correction 173(33) for laser beam diffraction. The next largest is 5.0(7.1) for
laser beam alignment. The statistical uncertainty is 33 (Type A).

Before calculating the weighted mean of the new and 1997 results for
$d_{220}({\scriptstyle{\rm NR3}})$, \textcite{2004001,2004001e} revised the
1997 value based on a reanalysis of the old experiment, taking into account
what was learned in the new experiment. Not only did the reanalysis result in a
reduction of the statistical uncertainty from (again, in parts in $10^{9}$) 50
to 1.8 due to a better understanding of the undulation of $n/m$ values as a
function of time, but also in more reliable estimates of the corrections for
laser beam diffraction and laser beam alignment. Indeed, the  fractional
corrections for the revised 1997 NMIJ value of $d_{220}({\scriptstyle{\rm
NR3}})$ total 190(38) compared to the original total of 173(14), and the final
uncertainty of the revised 1997 value is $u_{\rm r} = 3.8\times{10^{-9}}$
compared to $u_{\rm r} = 4.8\times{10^{-9}}$ of the new value.

For completeness, we note that two possible corrections to the NMIJ result have
been discussed in the literature. In the Erratum to \textcite{2004001,
2004001e}, it is estimated that a fractional correction to the value of
$d_{220}({\scriptstyle{\rm NR3}})$ in Eq.~(\ref{eq:dnr3nmij04}) of
$-1.3\times10^{-8}$ may be required to account for the contamination of the
NMIJ laser by a parasitic component of laser radiation as in the case of the
INRIM laser discussed in the next section. However, it is not applied, because
of its comparatively small size and the fact that no measurements of
$d_{220}({\scriptstyle{\rm NR3}})$ have yet been made at NMIJ (or INRIM) with a
problem-free laser that confirm the correction, as has been done at INRIM for
the crystals WASO~4.2a and $\rm MO^{*}$. 

In \textcite{2007070}, it is estimated, based on a Monte Carlo simulation, that
the fractional correction to $d_{220}({\scriptstyle{\rm NR3}})$ labeled
``Fresnel diffraction'' in Table I of \textcite{1997071} and equal to
$16.0(8)\times10^{-8}$ should be $10(3)\times10^{-8}$. The change arises from
taking into account the misalignment of the interfering beams in the laser
interferometer. Because this additional diffraction effect was present in both
the 1997 and 2002-2003 measurements but was not considered in the reanalysis of
the 1997 result nor in the analysis of the 2002-2003 data, it implies that the
weighted mean value for $d_{220}({\scriptstyle{\rm NR3}})$ in
Eq.~(\ref{eq:dnr3nmij04}) should be reduced by this amount and its $u_{\rm r}$
increased from $4.0\times10^{-8}$ to $5.0\times10^{-8}$. However, because the
data required for the calculation were not precisely known (they were not
logged in the laboratory notebooks because the experimenters were unaware of
their importance), the correction is viewed as somewhat conjectural and thus
that applying it would not be justified \cite{pc06fmn}.

\paragraph{INRIM measurement of $d_{220}({\rm \scriptstyle{W4.2a}})$ and
$d_{220}({\scriptstyle{\rm MO^*}})$}  
\label{par:inrimlat}

The following two new INRIM values, with identifier INRIM-07, were reported by
\textcite{2007061}: 
\begin{eqnarray}
d_{220}({\scriptstyle{\rm W4.2a}}) =  192\,015.5715(33) \ {\rm fm}
\quad [ 1.7\times 10^{-8}]
\label{eq:dw42ainrimptb07} \qquad
\end {eqnarray}
\begin{eqnarray}
d_{220}({\scriptstyle{\rm MO^*}}) =  192\,015.5498(51) \ {\rm fm}
\quad [ 2.6\times 10^{-8}] \ . \qquad
\label{eq:dmo4inrim07}
\end {eqnarray}
The correlation coefficient of these values is 0.057, based on the detailed
uncertainty budget for $d_{220}({\scriptstyle{\rm MO^*}})$ in
\textcite{2004001, 2004001e} and the similar uncertainty budget for
$d_{220}({\scriptstyle{\rm W4.2a }})$ provided by \textcite{pc06mm}.  Although
the 2007 result for $d_{220}({\scriptstyle{\rm MO^*}})$ of \textcite{2007061}
in Eq.~(\ref{eq:dmo4inrim07}) agrees with the 1994 INRIM result of
\textcite{1994140}, which was used as an input datum in the 1998 adjustment,
because of the many advances incorporated in the new work, we no longer
consider the old result.

In addition to the determination, described in the previous section, of the
\{220\} lattice spacing of crystal NRLM3 carried out at NMIJ in 2002-2003 using
the NMIJ NRLM3 x-ray interferometer and associated NMIJ apparatus,
\textcite{2004001, 2004001e} reported the results of measurements carried out
at INRIM of the \{220\} lattice spacings of crystals $\rm MO^{*}$ and NRLM3,
where in the latter case it was an INRIM-NMIJ joint effort that used the NIMJ
NRLM3 x-ray interferometer but the INRIM associated apparatus. But as indicated
above, both results were subsequently found to be in error: the optical laser
beam used to measure the displacement of the x-ray interferometer's analyzer
crystal was contaminated by a parasitic component with a frequency that
differed by about 1.1~GHz from the frequency assigned the laser beam. 

After eliminating the error by replacing the problem laser with a 633~nm He-Ne
external-cavity diode laser locked to a $^{127}{\rm I}_{2}$ stabilized laser,
the INRIM researchers repeated the measurements they had previously carried out
with the INRIM $\rm MO^{*}$ x-ray interferometer and with the refurbished PTB
WASO~4.2a x-ray interferometer originally used in the PTB experiment that led
to the 1981 value of $d_{220}({\rm \scriptstyle{W4.2a}})$ in
Eq.~(\ref{eq:dw42ptb81}).  The PTB WASO~4.2a x-ray interferometer was
refurbished at PTB through remachining, but the result for $d_{220}({\rm
\scriptstyle{W4.2a}})$ obtained at INRIM with the contaminated laser was not
included in \textcite{2004001, 2004001e}.  The values of $d_{220}({\rm
\scriptstyle{W4.2a}})$ and $d_{220}({\rm \scriptstyle{\rm MO^{*}}})$ in
Eqs.~\ref{eq:dw42ainrimptb07} and \ref{eq:dmo4inrim07} resulted from the
repeated measurements \cite{2007061}.

In principle, based on the experimentally observed shifts in the measured
values of $d_{220}({\rm \scriptstyle{W4.2a}})$ and $d_{220}({\rm
\scriptstyle{\rm MO^{*}}})$ obtained with the malfunctioning laser and the
properly functioning laser, the value of $d_{220}({\scriptstyle{\rm NR3}})$
obtained in the INRIM-NMIJ joint effort using the malfunctioning laser
mentioned above, and the value of $d_{220}({\scriptstyle{\rm WS5C}})$ also
obtained with this laser, could be corrected and taken as input data.  WS5C is
an XROI manufactured by INRIM from a WASO~04 sample, but the value of
$d_{220}({\scriptstyle{\rm WS5C}})$ obtained using the contaminated laser was
also not included in \textcite{2004001, 2004001e}.  However, because of the
somewhat erratic history of silicon lattice spacing measurements, the Task
Group decided to use only data obtained with a laser known to be functioning
properly.

The improvements in the INRIM XROI apparatus since the 1994 $d_{220}({\rm
\scriptstyle{\rm MO^{*}}})$ measurement of  \textcite{1994140} include (i) a
new two-axis ``tip-tilt'' platform for the XROI that is electronically
controlled to compensate for parasitic rotations and straightness error of the
guiding system that moves the platform; (ii) imaging the x-ray interference
pattern formed by the x-ray beam transmitted through the moving analyzer in
such a way that detailed information concerning lattice distortion and analyzer
pitch can be extracted on line from the analysis of the phases of the x-ray
fringes; and (iii) an upgraded computer-aided system for combined
interferometer displacement and control, x-ray and optical fringe scanning,
signal digitization and sampling, environmental monitoring, and data analysis.

The values of $d_{220}({\scriptstyle{\rm W4.2a }})$ and
$d_{220}({\scriptstyle{\rm MO^*}})$ in Eqs.~(\ref{eq:dw42ainrimptb07}) and
(\ref{eq:dmo4inrim07}) are the means of tens of individual values, with each
value being the average of about ten data points collected in 1~h measurement
cycles during which the analyzer was translated back and forth by 300 optical
orders. For the two crystals, respectively, the statistical uncertainties in
parts in $10^{9}$ are 3.5 and 11.6, and the various corrections and their
uncertainties are laser beam wavelength, $-0.8(4)$, $-0.8(4)$; laser beam
diffraction, 12.0(2.2), 12.0(2.2); laser beam alignment, 2.5(3.5), 2.5(3.5);
Abbe error, 0.0(2.8), 0.0(3.7); trajectory error, 0.0(1.4), 0.0(3.6); analyzer
temperature, 1.0(5.2), 1.0(7.9); and abberations, 0.0(5.0), 0.0(2.0). The total
uncertainties are 9.6 and 15.7.

\subsubsection{$d_{220}$ difference measurements}
\label{sssec:d220diff}

To relate the lattice spacings of crystals used in various experiments, highly
accurate measurements are made of the fractional difference
$\left[d_{220}({\scriptstyle X}) - d_{220}({\rm ref})\right] /d_{220}({\rm
ref})$ of the \{220\} lattice spacing of a sample of a single crystal ingot $X$
and that of a reference crystal ``ref''. Both NIST and PTB have carried out
such measurements, and the fractional differences from these two laboratories
that we take as input data in the 2006 adjustment are given in the following
two sections and are summarized in Table~\ref{tab:d220rel}. For details
concerning these measurements, see CODATA-98 and CODATA-02. 

\paragraph{NIST difference measurements}
\label{par:nistdiffs}

The following fractional difference involving a crystal denoted simply as ``N''
was obtained as part of the NIST effort to measure the wavelengths in meters of
the ${\rm K}\alpha_{1}$ x-ray lines of Cu, Mo, and W; see Sec.~\ref{ssec:xru}.
\begin{eqnarray}
{d_{220}({\rm {\scriptstyle W17}})-
d_{220}({\rm {\scriptstyle N}})\over
d_{220}({\rm {\scriptstyle W17}})} &=&
 7(22)\times 10^{-9} \ .
\label{eq:drw17n}
\end{eqnarray}

The following three fractional differences involving crystals from the four
crystals denoted ILL, WASO~17, MO*, and NRLM3 were obtained as part of the NIST
effort, discussed in Sec.~\ref{ssec:arn}, to determine the relative atomic mass
of the neutron $A_{\rm r}(\rm n)$ \cite{1999052}:
\begin{eqnarray}
{d_{220}({\rm {\scriptstyle ILL}})-
d_{220}({\rm {\scriptstyle W17}})\over
d_{220}({\rm {\scriptstyle ILL}})} &=&
 -8(22)\times 10^{-9} 
\label{eq:drillw17} \\
{d_{220}({\rm {\scriptstyle ILL}})-
d_{220}({\rm {\scriptstyle MO^*}})\over
d_{220}({\rm {\scriptstyle ILL}})} &=&
 86(27)\times 10^{-9} 
\label{eq:drillmo4} \\
{d_{220}({\rm {\scriptstyle ILL}})-
d_{220}({\rm {\scriptstyle NR3}})\over
d_{220}({\rm {\scriptstyle ILL}})} &=&
 34(22)\times 10^{-9} \ .
\label{eq:drillnr3}
\end{eqnarray}

The following more recent NIST difference measurements, which we also take as
input data in the 2006 adjustment, were provided by \textcite{pc06ek} of NIST
and are updates of the results reported by \textcite{2005010}:
\begin{eqnarray}
{d_{220}({\rm {\scriptstyle NR3}})-
d_{220}({\rm {\scriptstyle W04}})\over
d_{220}({\rm {\scriptstyle W04}})} &=&
 -11(21)\times 10^{-9} 
\label{eq:drnr3w04hk} \\
{d_{220}({\rm {\scriptstyle NR4}})-
d_{220}({\rm {\scriptstyle W04}})\over
d_{220}({\rm {\scriptstyle W04}})} &=&
 25(21)\times 10^{-9} 
\label{eq:drnr4w04hk} \\
{d_{220}({\rm {\scriptstyle W17}})-
d_{220}({\rm {\scriptstyle W04}})\over
d_{220}({\rm {\scriptstyle W04}})} &=&
 11(21)\times 10^{-9} \ .
\label{eq:drw17w04hk}
\end{eqnarray}
The full designations of the two new crystals involved in these comparisons are
WASO 04 and NRLM4. The measurements benefited significantly from the relocation
of the NIST lattice comparator to a new laboratory where the temperature varied
by only about 5~mK in several weeks compared to the previous laboratory where
the temperature varied by about 40~mK in one day \cite{2005010}. 

\paragraph{PTB difference measurements}
\label{par:ptbdiffs}

Results for the \{220\} lattice-spacing fractional differences of various
crystals that we also take as input data in the 2006 adjustment have been
obtained at the PTB \cite{1998127}:
\begin{eqnarray}
{d_{220}({\rm {\scriptstyle W4.2a}})-
d_{220}({\rm {\scriptstyle W04}})\over
d_{220}({\rm {\scriptstyle W04}})} &=&
 -1(21)\times 10^{-9}
\label{eq:drw42w04} \\
{d_{220}({\rm {\scriptstyle W17}})-
d_{220}({\rm {\scriptstyle W04}})\over
d_{220}({\rm {\scriptstyle W04}})} &=&
 22(22)\times 10^{-9}
\label{eq:drw17w04} \\
{d_{220}({\rm {\scriptstyle MO^*}})-
d_{220}({\rm {\scriptstyle W04}})\over
d_{220}({\rm {\scriptstyle W04}})} &=&
 -103(28)\times 10^{-9}
\label{eq:drmo4w04} \\
{d_{220}({\rm {\scriptstyle NR3}})-
d_{220}({\rm {\scriptstyle W04}})\over
d_{220}({\rm {\scriptstyle W04}})} &=&
 -23(21)\times 10^{-9} \ .
\label{eq:drnr3w04}
\end{eqnarray}

To relate $d_{220}({\rm {\scriptstyle W04}})$ to the \{220\} lattice spacing
$d_{220}$ of an ``ideal'' silicon crystal, we take as an input datum
\begin{eqnarray}
{d_{220} - d_{220}({\rm {\scriptstyle W04}})\over
d_{220}({\rm {\scriptstyle W04}})} &=&
 10(11)\times 10^{-9} 
\label{eq:ddw04}
\end{eqnarray}
given by \textcite{2003137}, who obtained it by taking into account the known
carbon, oxygen, and nitrogen impurities in WASO 04.  However, following what
was done in the 1998 and 2002 adjustments, we have included an additional
component of uncertainty of $1\times 10^{-8}$ to account for the possibility
that, even after correction for C, O, and N impurities, the crystal WASO~04,
although very well characterized as to its purity and crystallographic
perfection, does not meet all of the criteria for an ideal crystal.  Indeed, in
general, we prefer to use experimentally measured fractional lattice spacing
differences rather than differences implied by the C, O, and N impurity content
of the crystals in order to avoid the need to assume that all crystals of
interest meet these criteria.

In order to include this fractional difference in the 2002 adjustment, the
quantity $d_{220}$ is also taken as an adjusted constant.

\begin{table*}
\def\m{\phantom{-}}
\caption{Summary of measurements of the relative \{220\} lattice spacings of
silicon crystals.}
\def\vsp{\vbox to 12 pt {}}
\label{tab:d220rel}
\def\sp{\hbox to 15 pt {}}
\begin{tabular}{l@{\sp}l@{\sp}c@{\sp}l}
\toprule
Quantity \T  &  \hbox to 12pt {} Value
& Identification & Sect. and Eq. \\
\colrule
$1-d_{220}({\rm {\scriptstyle{W17}}})/d_{220}({\rm {\scriptstyle ILL}})$ \vsp & $  -8(22)\times 10^{-9}$ & NIST-99 & \ref{par:nistdiffs} (\ref{eq:drillw17}) \\
$1-d_{220}({\rm {\scriptstyle{MO^*}}})/d_{220}({\rm {\scriptstyle ILL}})$ \vsp & $\phantom{-}  86(27)\times 10^{-9}$ & NIST-99 & \ref{par:nistdiffs} (\ref{eq:drillmo4}) \\
$1-d_{220}({\rm {\scriptstyle{NR3}}})/d_{220}({\rm {\scriptstyle ILL}})$ \vsp & $\phantom{-}  34(22)\times 10^{-9}$ & NIST-99 & \ref{par:nistdiffs} (\ref{eq:drillnr3}) \\
$1-d_{220}({\rm {\scriptstyle{N}}})/d_{220}({\rm {\scriptstyle W17}})$ \vsp & $ \phantom{-}  7(22)\times 10^{-9}$ & NIST-97 & \ref{par:nistdiffs} (\ref{eq:drw17n}) \\
$d_{220}({\rm {\scriptstyle{NR3}}})/d_{220}({\rm {\scriptstyle W04}})-1$ \vsp & $  -11(21)\times 10^{-9}$ & NIST-06 & \ref{par:nistdiffs} (\ref{eq:drnr3w04hk}) \\
$d_{220}({\rm {\scriptstyle{NR4}}})/d_{220}({\rm {\scriptstyle W04}})-1$ \vsp & $ \phantom{-}  25(21)\times 10^{-9}$ & NIST-06 & \ref{par:nistdiffs} (\ref{eq:drnr4w04hk}) \\
$d_{220}({\rm {\scriptstyle{W17}}})/d_{220}({\rm {\scriptstyle W04}})-1$ \vsp & $\phantom{-}  11(21)\times 10^{-9}$ & NIST-06 & \ref{par:nistdiffs} (\ref{eq:drw17w04hk}) \\
$d_{220}({\rm {\scriptstyle{W4.2a}}})/d_{220}({\rm {\scriptstyle W04}})-1$ \vsp & $  -1(21)\times 10^{-9}$ & PTB-98 & \ref{par:ptbdiffs} (\ref{eq:drw42w04}) \\
$d_{220}({\rm {\scriptstyle{W17}}})/d_{220}({\rm {\scriptstyle W04}})-1$ \vsp & $\phantom{-}  22(22)\times 10^{-9}$ & PTB-98 & \ref{par:ptbdiffs} (\ref{eq:drw17w04}) \\
$d_{220}({\rm {\scriptstyle{MO^*}}})/d_{220}({\rm {\scriptstyle W04}})-1$ \vsp & $  -103(28)\times 10^{-9}$ & PTB-98 & \ref{par:ptbdiffs} (\ref{eq:drmo4w04}) \\
$d_{220}({\rm {\scriptstyle{NR3}}})/d_{220}({\rm {\scriptstyle W04}})-1$ \vsp & $  -23(21)\times 10^{-9}$ & PTB-98 & \ref{par:ptbdiffs} (\ref{eq:drnr3w04}) \\
$d_{220}/d_{220}({\rm {\scriptstyle W04}})-1$ \B & $\phantom{-}  10(11)\times 10^{-9}$ \vsp & PTB-03 & \ref{par:ptbdiffs} (\ref{eq:ddw04}) \\
\botrule
\end{tabular}
\end{table*}

\subsection{Molar volume of silicon $\bm{V_{\rm m}({\rm Si})}$ and the Avogadro
constant $\bm{N_{\rm A}}$}
\label{ssec:mvsi}

The definition of the molar volume of silicon $V_{\rm m}$(Si) and its
relationship to the Avogadro constant $N_{\rm A}$ and Planck constant $h$ as
well as other constants is discussed in CODATA-98 and summarized in CODATA-02.
In brief we have 
\begin {eqnarray}
m{\rm (Si)} &=& \rho{\rm (Si)}{a^3\over n}\ , 
\label{eq:msiarho}
\\
V_{\rm m}{\rm (Si)} &=& {M{\rm (Si)} \over \rho{\rm (Si)}} 
= {A_{\rm r}({\rm Si})M_{\rm u}\over \rho({\rm Si})} \ ,
\label{eq:vmsi}
\\
N_{\rm A} &=& {V_{\rm m}{\rm (Si)}\over a^3/n} 
= {A_{\rm r}({\rm Si})M_{\rm u}\over\sqrt{8}\,d_{220}^{\,3}\,\rho{\rm (Si)}} \ ,
\label{eq:navmrho}
\\
V_{\rm m}({\rm Si)} &=& {\sqrt{2}\,cM_{\rm u}A_{\rm r}({\rm e})\alpha^{2}d^{\,3}_{220}
\over R_{\rm \infty}h} \ ,
\label{eq:vmsith}
\end{eqnarray}
which are to be understood in the context of an impurity free,
crystallographically perfect, ``ideal'' silicon crystal at the reference
conditions $t_{90} = 22.5~^\circ$C and $p = 0$, and of isotopic composition in
the range normally observed for crystals used in high-accuracy experiments.
Thus $m$(Si), $V_{\rm m}$(Si), $M$(Si), and $A_{\rm r}$(Si) are the mean mass,
mean molar volume, mean molar mass, and mean relative atomic mass of the
silicon atoms in such a crystal, respectively, and $\rho$(Si) is the crystal's
macroscopic mass density.  Equation~(\ref{eq:vmsith}) is the observational
equation for a measured value of $V_{\rm m}$(Si).

It follows from Eq.~(\ref{eq:vmsi}) that the experimental determination of
$V_{\rm m}$(Si) requires (i) measurement of the amount-of-substance ratios
$n(^{29}\rm Si)$/$n(^{28}\rm Si)$ and $n(^{30}\rm Si)$/$n(^{28}\rm Si)$ of a
nearly perfect silicon crystal---and hence amount of substance fractions
$x(^{A}\rm Si)$---and then calculation of $A_{\rm r}(\rm Si)$ from the
well-known values of $A_{\rm r}(^{A}\rm Si)$; and (ii) measurement of the
macroscopic mass density $\rho$(Si) of the crystal.  Determining $N_{\rm A}$
from Eq.~(\ref{eq:navmrho}) by measuring $V_{\rm m}$(Si) in this way and
$d_{220}$ using x rays is called the x-ray-crystal-density (XRCD) method.

An extensive international effort has been under way since the early 1990s to
determine $N_{\rm A}$ using this technique with the smallest possible
uncertainty. The effort is being coordinated by the Working Group on the
Avogadro Constant (WGAC) of the Consultative Committee for Mass and Related
Quantities (CCM) of the CIPM. The WGAC, which has representatives from all
major research groups working in areas relevant to the determination of $N_{\rm
A}$, is currently chaired by P. Becker of PTB.

As discussed at length in CODATA-02, the value of $V_{\rm m}$(Si) used as an
input datum in the 2002 adjustment was provided to the CODATA Task Group by the
WGAC and was a consensus value based on independent measurements of $\rho \rm
(Si)$ at NMIJ and PTB using a number of different silicon crystals, and
measurements of their molar masses $M(\rm Si)$ using isotopic mass spectrometry
at the Institute for Reference Materials and Measurements (IRMM), European
Commission, Geel, Belgium. This value, identified as N/P/I-03 in recognition of
the work done by researchers at NMIJ, PTB, and IRMM, is $V_{\rm m}(\rm
Si)=12.058\,8257(36) \times10^{-6}~{\rm m^{3}}~{\rm mol}^{-1}$~[$3.0
\times{10}^{-7}$]. Since then, the data used to obtain it were reanalyzed by
the WGAC, resulting in the slightly revised value \cite{2005032}
\begin{eqnarray} V_{\rm m}({\rm Si}) &=&  12.058\,8254(34)\times 10^{-6} ~{\rm
m^3~mol^{-1}} \nonumber\\ &&\hbox to 3 cm {} [ 2.8\times 10^{-7}] \ ,
\label{eq:vmsilwgac05} \end{eqnarray} which we take as an input datum in the
2006 adjustment and identify as N/P/I-05. The slight shift in value and
reduction in uncertainty is due to the fact that the effect of nitrogen
impurities in the silicon crystals used in the NMIJ measurements was taken into
account in the reanalysis \cite{2005032}.  Note that the new value of $A_{\rm
r}(^{29}{\rm Si})$ in Table~\ref{tab:rmass06} has no effect on this result. 

Based on Eq.~(\ref{eq:vmsith}) and the 2006 recommended values of $A_{\rm r}({\rm
e})$, $\alpha$, $d_{220}$, and $R_{\rm \infty}$, the value of $h$ implied by
this result is 

\begin{eqnarray}
h =  6.626\,0745(19)\times 10^{-34}~{\rm J~s}\quad[ 2.9\times 10^{-7}] \ .
\label{eq:hvmsilwgac05}
\end{eqnarray}
A comparison of this value of $h$ with those in Tables~\ref{tab:sgamma} and
\ref{tab:hdata} shows that it is generally not in good agreement with the most
accurate of the other values. 

In this regard, two relatively recent publications, the first describing work
performed in China \cite{2005237} and the second describing work performed in
Switzerland \cite{2006115}, reported results which, if taken at face value,
seem to call into question the uncertainty with which the molar mass of
naturally occurring silicon is currently known. [See also \textcite{2005354}.]
These results highlight the importance of the current WGAC project to measure
$V_{\rm m}(\rm Si)$ using highly enriched silicon crystals with $x(^{28}\rm
Si)> 0.99985$ \cite{2006075}, which should simplify the determination of the
molar mass of such crystals.

\subsection{Gamma-ray determination of the neutron relative atomic mass
$\bm{A_{\rm r}({\rm n})}$}
\label{ssec:arn}

Although the value of $A_{\rm r}({\rm n})$ listed in Table~\ref{tab:rmass03} is
a result of AME2003, it is not used in the 2006 adjustment. Instead, $A_{\rm
r}({\rm n})$ is obtained as discussed in this section in order to ensure that
its recommended value is consistent with the best current information on the
\{220\} lattice spacing of silicon.

The value of $A_{\rm r}$(n) can be obtained by measuring the wavelength of the
2.2~MeV $\gamma$ ray in the reaction n + p $\rightarrow$ d + $\gamma$ in terms
of the $d_{220}$ lattice spacing of a particular silicon crystal corrected to
the commonly used reference conditions $t_{90} = 22.5~^\circ$C and $p$ = 0. The
result for the wavelength-to-lattice spacing ratio, obtained from Bragg-angle
measurements carried out in 1995 and 1998 using a flat crystal spectrometer of
the GAMS4 diffraction facility at the high-flux reactor of the Institut Max von
Laue-Paul Langevin (ILL), Grenoble, France, in a NIST and ILL collaboration, is
\cite{1999052}
\begin{eqnarray}
\frac{\lambda_{\rm meas}}{ d_{220}({\rm {\scriptstyle ILL}})}
=  0.002\,904\,302\,46(50) \qquad [ 1.7\times 10^{-7}] \ , \quad
\label{eq:ladill99}
\end{eqnarray}
where $d_{220}({\rm {\scriptstyle ILL}})$ is the \{220\} lattice spacing of the
silicon crystals of the ILL GAMS4 spectrometer at $t_{90} = 22.5~^\circ$C and
$p$ = 0.  Relativistic kinematics of the reaction yields the equation
\begin{eqnarray}
{\lambda_{\rm meas}\over d_{220}({\rm {\scriptstyle ILL}})}
&=& {\alpha^2 A_{\rm r}({\rm e})
\over R_\infty d_{220}({\rm {\scriptstyle ILL}})}
{A_{\rm r}({\rm n}) + A_{\rm r}({\rm p})
\over \left[A_{\rm r}({\rm n}) + A_{\rm r}({\rm p})\right]^2
- A_{\rm r}^2({\rm d})} \ ,
\nonumber\\
\label{eq:relkingarm}
\end{eqnarray}
where all seven quantities on the right-hand side are adjusted constants.

Recently, \textcite{2005195, 2006033} reported determinations of the
wavelengths of the gamma rays emitted in the cascade from the neutron capture
state to the ground state in the reactions 
n + $^{28}$Si $\rightarrow$ $^{29}$Si + $2\gamma$, 
n + $^{32}$S $\rightarrow$ $^{33}$Si + $3\gamma$, and
n + $^{35}$Cl $\rightarrow$ $^{36}$Cl + $2\gamma$.
The gamma-ray energies are 3.5 MeV and 4.9 MeV for the Si reaction, 5.4 MeV,
2.4 MeV, and 0.8 MeV for the S reaction, and 6.1 MeV, 0.5 MeV, and 2.0 MeV for
the Cl reaction.  While these data together with the relevant relative atomic
masses are potentially an additional source of information on the neutron
relative atomic mass, the uncertainties are too large for this purpose; the
inferred value of $A_{\rm r}$(n) has an uncertainty nearly an order of
magnitude larger than that obtained from Eq.~(\ref{eq:relkingarm}). Instead,
this work is viewed as the most accurate test of $E=mc^2$ to date
\cite{2005195}.

\subsection{Quotient of Planck constant and particle mass $\bm{h/m(X)}$ and
$\bm{\alpha}$}
\label{ssec:pcpmq}

The relation $R_{\infty} = \alpha^2m_{\rm e}c/2h$ leads to
\begin{eqnarray}
\alpha = \left[{2R_{\infty}\over c}{A_{\rm r}(X)\over
A_{\rm r}({\rm e})} {h\over m(X)}\right]^{1/2} \ ,
\label{eq:alhmx}
\end{eqnarray}
where $A_{\rm r}(X)$ is the relative atomic mass of particle $X$ with mass
$m(X)$ and $A_{\rm r}({\rm e})$ is the relative atomic mass of the electron.
Because $c$ is exactly known, $u_{\rm r}$ of $R_\infty$ and $A_{\rm r}({\rm
e})$ are less than $7\times10^{-12}$ and $5\times10^{-10}$, respectively, and
$u_{\rm r}$ of $A_{\rm r}(X)$ for many particles and atoms is less than that of
$A_{\rm r}$(e), Eq.~(\ref{eq:alhmx}) can provide a value of $\alpha$ with a
competitive uncertainty if $h/m(X)$ is determined with a sufficiently small
uncertainty.  Here, we discuss the determination of $h/m(X)$ for the neutron n,
the $^{133}$Cs atom, and the $^{87}$Rb atom. The results, including the
inferred values of $\alpha$, are summarized in Table~\ref{tab:dhom}.

\begin{table*}

\caption{Summary of data related to the quotients $h/m_{\rm n}d_{220}({\rm
{\scriptstyle{W04}}})$, $h/m$(Cs), and $h/m$(Rb), together with inferred values
of $\alpha$.}
\label{tab:dhom}
\def\sp{\hbox to 10 pt{}}
\begin{tabular}{l@{\sp}l@{\sp}l@{\sp}l@{\sp}l}
\toprule
\vbox to 10 pt {}
Quantity     &  \hbox to 15pt {} Value
& \hbox to -20pt {} Relative standard
& Identification & Sect. and Eq. \\
& &  \hbox to -5pt {}uncertainty $u_{\rm r}$ &  \\
\colrule
$h/m_{\rm n}d_{220}({\rm {\scriptstyle{W04}}})$ \vbox to 12pt {}&
$ 2060.267\,004(84)$ m s$^{-1}$
& $ 4.1\times 10^{-8}$ & PTB-99 & \ref{sssec:pcnmr} (\ref{eq:homndw04}) \\

\ \ $\alpha^{-1}$ &\ \ $ 137.036\,0077(28)$&$ 2.1\times 10^{-8}$& 
    & \ref{sssec:pcnmr} (\ref{eq:alphinvhom}) \\

$h/m$(Cs)    \vbox to 15pt {}&
$ 3.002\,369\,432(46)\times 10^{-9}$ m$^2$ s$^{-1}$
& $ 1.5\times 10^{-8}$ & StanfU-02 & \ref{sssec:pccsmr} (\ref{eq:homcs02}) \\

\ \ $\alpha^{-1}$ &\ \ $ 137.036\,0000(11)$&$ 7.7\times 10^{-9}$& 
    & \ref{sssec:pccsmr} (\ref{eq:alphinvcs02}) \\

$h/m$(Rb)    \vbox to 15pt {}&
$ 4.591\,359\,287(61)\times 10^{-9}$ m$^2$ s$^{-1}$
& $ 1.3\times 10^{-8}$ & LKB-06 & \ref{sssec:pccsmr} (\ref{eq:homrb06}) \\

\ \ $\alpha^{-1}$ &\ \ $ 137.035\,998\,83(91)$&$ 6.7\times 10^{-9}$& 
    & \ref{sssec:pccsmr} (\ref{eq:alphinvrb06}) \\

\botrule
\end{tabular}
\end{table*}

\subsubsection{Quotient $h/m_{\rm n}$}
\label{sssec:pcnmr}

The PTB determination of $h/m_{\rm n}$ was carried out at the ILL high-flux
reactor.  The de Broglie relation $p=m_{\rm n}v=h/\lambda$ was used to
determine $h/m_{\rm n} = \lambda v$ for the neutron by measuring both its de
Broglie wavelength $\lambda$ and corresponding velocity $v$. More specifically,
the de Broglie wavelength, $\lambda\approx  0.25$ nm, of slow neutrons was
determined using back reflection from a silicon crystal, and the velocity, $v
\approx 1600$ m/s, of the neutrons was determined by a special time-of-flight
method. The final result of the experiment is \cite{1999075}
\begin{eqnarray}
{h\over m_{\rm n}d_{220}({\rm {\scriptstyle W04}})} &=&  2060.267\,004(84) \ {\rm m \ s^{-1}}
\nonumber \\  && \qquad \qquad [ 4.1\times 10^{-8}] \ ,
\label{eq:homndw04}
\end{eqnarray}
where as before, $d_{220}({\rm {\scriptstyle W04}})$ is the \{220\} lattice
spacing of the crystal WASO~04 at $t_{90} = 22.5~^\circ$C in vacuum.  This
result is correlated with the PTB fractional lattice-spacing differences given
in Eqs.~(\ref{eq:drw42w04}) to (\ref{eq:drnr3w04})---the correlation
coefficients are about 0.2.

The equation for the PTB result, which follows from Eq.~(\ref{eq:alhmx}), is 
\begin{eqnarray}
{h\over m_{\rm n}d_{220}({\rm {\scriptstyle W04}})} =
{ A_{\rm r}({\rm e})\over A_{\rm r}({\rm n})} \,
{c\alpha^2 \over 2 R_\infty 
d_{220}({\rm {\scriptstyle W04}})} \ .
\label{eq:obseqhomnw04}
\end{eqnarray}

The value of $\alpha$ that can be inferred from this relation and the PTB value
of $h/m_{\rm n}d_{220}({\rm \scriptstyle W04})$, the 2006 recommended values of
$R_\infty$, $A_{\rm r}$(e), and $A_{\rm r}$(n), the NIST and PTB fractional
lattice-spacing-differences in Table~\ref{tab:d220rel}, and the four XROI
values of $d_{220}({{\scriptstyle X}})$ in Table~\ref{tab:d220abs} for crystals
WASO~4.2a, NRLM3, and MO$^{*}$, is 
\begin{eqnarray}
\alpha^{-1} &=&  137.036\,0077(28) \quad [ 2.1\times 10^{-8}] \ .
\label{eq:alphinvhom}
\end{eqnarray}
This value is included in Table~\ref{tab:dhom} as the first entry; it disagrees
with the $\alpha$ values from the two other $h/m$ results.

It is also of interest to calculate the value of $d_{220}$ implied by the PTB
result for $h/m_{\rm n}d_{220}({\rm \scriptstyle W04})$. Based on
Eq.~(\ref{eq:obseqhomnw04}), the 2006 recommended values of $R_\infty$, $A_{\rm
r}$(e), $A_{\rm r}$(p), $A_{\rm r}$(d), $\alpha$, the NIST and PTB fractional
lattice-spacing-differences in Table~\ref{tab:d220rel}, and the value of
${\lambda_{\rm meas}}/{d_{220}({\rm {\scriptstyle ILL}})}$ given in
Eq.~({\ref{eq:ladill99}), we find
\begin{eqnarray}
d_{220} &=&  192\,015.5982(79)~{\rm fm} \quad [ 4.1\times 10^{-8}] \ .
\label{eq:d220homn}
\end{eqnarray}
This result is included in Table~\ref{tab:d220abs} as the last entry; it agrees
with the NMIJ value, but disagrees with the PTB and INRIM values.

\subsubsection{Quotient $h/m({\rm ^{133}Cs})$}
\label{sssec:pccsmr}

The Stanford University atom interferometry experiment to measure the atomic
recoil frequency shift of photons absorbed and emitted by ${\rm ^{133}Cs}$
atoms, $\Delta\nu_{\rm Cs}$, in order to determine the quotient $h/m({\rm
^{133}Cs})$ is described in CODATA-02.  As discussed there, the expression
applicable to the Stanford experiment is 
\begin{eqnarray}
{h\over m(^{133}{\rm Cs})} = {c^{2}\, \Delta\nu_{\rm Cs}\over 2\, \nu^{2}_{\rm eff}} \ ,
\label{eq:hmdnu}
\end{eqnarray}
where the frequency $\nu_{\rm eff}$ corresponds to the sum of the energy
difference between the ground-state hyperfine level with $F=3$ and the
6P$_{1/2}$ state $F=3$ hyperfine level and the energy difference between the
ground-state hyperfine level with $F=4$ and the same 6P$_{1/2}$ hyperfine
level.  The result for $\Delta\nu_{\rm Cs}/2$ reported in 2002 by the Stanford
researchers is \cite{2002223} 
\begin{eqnarray} 
{\Delta\nu_{\rm Cs}\over 2} =
 15\,006.276\,88(23)~{\rm Hz} \qquad [ 1.5\times 10^{-8}] \ .  \label{eq:dnu02}
\end{eqnarray}

The Stanford effort included an extensive study of corrections due to possible
systematic effects. The largest component of uncertainty by far contributing to
the uncertainty of the final result for $\Delta\nu_{\rm Cs}$, $u_{\rm
r}=14\times10^{-9}$ (Type B), arises from the possible deviation from 1 of the
index of refraction of the dilute background gas of cold cesium atoms that move
with the signal atoms. This component, estimated experimentally, places a lower
limit on the relative uncertainty of the inferred value of $\alpha$ from
Eq.~(\ref{eq:alhmx}) of $u_{\rm r}=7\times10^{-9}$. Without it, $u_{\rm r}$ of
$\alpha$ would be about 3 to 4 parts in $10^{9}$.

In the 2002 adjustment, the value $\nu_{\rm eff}=670\,231\,933\,044(81)~\rm
kHz$ \,[$1.2\times10^{-10}$], based on the measured frequencies of $^{133}{\rm
Cs}$ ${\rm D}_1$-line transitions reported by \textcite{1999036}, was used to
obtain the ratio $h/ m(^{133}{\rm Cs})$ from the Stanford value of
$\Delta\nu_{\rm Cs}/2$. Recently, using a femtosecond laser frequency comb and
a narrow-linewidth diode laser, and eliminating Doppler shift by orienting the
laser beam perpendicular to the $^{133}{\rm Cs}$ atomic beam to within
$5~{\rmmu}{\rm rad}$, \textcite{2006017} remeasured the frequencies of the
required transitions and obtained a value of $\nu_{\rm eff}$ that agrees with
the value used in 2002 but which has a $u_{\rm r}$ 15 times smaller: 
\begin{eqnarray}
\nu_{\rm eff} &=&  670\,231\,932\,889.9(4.8)~{\rm kHz} 
\quad [ 7.2\times 10^{-12}] \ . \qquad
\label{eq:nucseff06}
\end{eqnarray}

Evaluation of Eq.~(\ref{eq:hmdnu}) with this result for $\nu_{\rm eff}$ and the
value of $\Delta\nu_{\rm Cs}/2$ in Eq.~(\ref{eq:dnu02}) yields 
\begin{eqnarray}
{h\over m(^{133}{\rm Cs})} &=&  3.002\,369\,432(46)\times 10^{-9}~{\rm m}^2~{\rm s}^{-1}
\nonumber\\
\label{eq:homcs02}
&&\qquad\qquad[ 1.5\times 10^{-8}] \ ,
\end{eqnarray}
which we take as an input datum in the 2006 adjustment. The observational
equation for this datum is, from Eq.~(\ref{eq:alhmx}),
\begin{eqnarray}
{h\over m(^{133}{\rm Cs})} = {A_{\rm r}({\rm e})\over A_{\rm r}({\rm ^{133}Cs})}
{c\,\alpha^2 \over 2 R_\infty} \ .
\label{eq:homcsoe}
\end{eqnarray}
The value of $\alpha$ that may be inferred from this expression, the Stanford
result for $h/m(^{133}{\rm Cs})$ in Eq.~(\ref{eq:homcs02}), the 2006
recommended values of $R_\infty$ and $A_{\rm r}$(e), and the ASME2003 value of
$A_{\rm r}(^{133}{\rm Cs})$ in Table~\ref{tab:rmass03}, the uncertainties of
which are inconsequential in this application, is
\begin{eqnarray}
\alpha^{-1} =  137.036\,0000(11) \qquad [ 7.7\times 10^{-9}] \ ,
\label{eq:alphinvcs02}
\end{eqnarray}
where the dominant component of uncertainty arises from the measured value of
the recoil frequency shift, in particular, the component of uncertainty due to
a possible index of refraction effect.

In this regard, we note that \textcite{2005008} have experimentally
demonstrated the reality of one aspect of such an effect with a two-pulse light
grating interferometer and have shown that it can have a significant impact on
precision measurements with atom interferometers.  However, theoretical
calculations based on simulations of the Stanford interferometer by
\textcite{pc06scw}, although incomplete, suggest that the experimentally based
uncertainty component $u_{\rm r}=14 \times10^{-9}$ assigned by
\textcite{2002223} to account for this effect is reasonable.  We also note that
\textcite{2005133} have developed an improved theory of momentum transfer when
localized atoms and localized optical fields interact.  The details of such
interactions are relevant to precision atom interferometry.  When
\textcite{2005133} applied the theory to the Stanford experiment to evaluate
possible systematic errors arising from wave-front curvature and distortion, as
well as the Gouy phase shift of gaussian beams, they found that such errors do
not limit the uncertainty of the value of $\alpha$ that can be obtained from
the experiment at the level of a few parts in $10^{9}$, but will play an
important role in future precision atom-interferometer photon-recoil
experiments to measure $\alpha$ with $u_{\rm r}\approx 5 \times{10^{-10}}$,
such as is currently underway at Stanford \cite{2006153}.

\subsubsection{Quotient $h/m({\rm ^{87}Rb})$}
\label{sssec:pcrbmr}

In the LKB experiment \cite{2006001, 2006276}, the quotient $h/m({\rm
^{87}Rb})$, and hence $\alpha$, is determined by accurately measuring the
rubidium recoil velocity $v_{\rm r}=\hbar k/m({\rm ^{87}Rb})$ when a rubidium
atom absorbs or emits a photon of wave vector $k = 2{\rmpi}/\lambda$, where
$\lambda$ is the wavelength of the photon and $\nu = c/\lambda$ is its
frequency. The measurements are based on Bloch oscillations in a vertical
accelerated optical lattice.

The basic principle of the experiment is to precisely measure the variation of
the atomic velocity induced by an accelerated standing wave using velocity
selective Raman transitions between two ground-state hyperfine levels. A Raman
$\rmpi$ pulse of two counter-propagating laser beams selects an initial narrow
atomic velocity class. After the acceleration process, the final atomic
velocity distribution is probed using a second Raman $\rmpi$ pulse of two
counter-propagating laser beams.

The coherent acceleration of the rubidium atoms arises from a succession of
stimulated two photon transitions also using two counter-propagating laser
beams. Each transition modifies the atomic velocity by $2v_{\rm r}$ leaving the
internal state unchanged. The Doppler shift is compensated by linearly sweeping
the frequency difference of the two lasers. This acceleration can conveniently
be interpreted in terms of Bloch oscillations in the fundamental energy band of
an optical lattice created by the standing wave, because the interference of
the two laser beams leads to a periodic light shift of the atomic energy levels
and hence to the atoms experiencing a periodic potential
\cite{1996211,1997190}.

An atom's momentum evolves by steps of $2\hbar{k}$, each one corresponding to a
Bloch oscillation. After $N$ oscillations, the optical lattice is adiabatically
released and the final velocity distribution, which is the initial distribution
shifted by $2Nv_{\rm r}$, is measured. Due to the high efficiency of Bloch
oscillations, for an acceleration of 2000~m~$\rm s^{-2}$, 900 recoil momenta
can be transferred to a rubidium atom in 3~ms with an efficiency of 99.97~\%
per recoil. 

The atoms are alternately accelerated upwards and downwards by reversing the
direction of the Bloch acceleration laser beams, keeping the same delay between
the selection and the measurement Raman $\rmpi$ pulses. The resulting
differential measurement is independent of gravity. In addition, the
contribution of some systematic effects changes sign when the direction of the
selection and measuring Raman beams is exchanged. Hence, for each up and down
trajectory, the selection and measuring Raman beams are reversed and two
velocity spectra are taken. The mean value of these two measurements is free
from systematic errors to first order. Thus each determination of $h/m({\rm
^{87}Rb})$ is obtained from four velocity spectra, each requiring 5 minutes of
integration time, two from reversing the Raman beams when the acceleration is
in the up direction and two when in the down direction. The Raman and Bloch
lasers are stabilized by means of an ultrastable Fabry-P\'erot cavity and the
frequency of the cavity is checked several times during the 20 minute
measurement against a well-known two-photon transition in $^{85}\rm Rb$.

Taking into account a $(-9.2\pm 4)\times10^{-10}$ correction to $h/m({\rm
^{87}Rb})$ not included in the value reported by \textcite{2006001} due to a
nonzero force gradient arising from a difference in the radius of curvature of
the up and down accelerating beams, the result derived from 72 measurements of
$h/m({\rm ^{87}Rb})$ acquired over 4 days, which we take as an input datum in
the 2006 adjustment, is \cite{2006276}
\begin{eqnarray}
{h\over m(^{87}{\rm Rb})} &=&  4.591\,359\,287(61)\times 10^{-9}~{\rm m}^2~{\rm s}^{-1}
\nonumber\\
\label{eq:homrb06}
&&\qquad\qquad[ 1.3\times 10^{-8}] \ ,
\end{eqnarray}
where the quoted $u_{\rm r}$ contains a statistical component from the 72
measurements of $8.8\times10^{-9}$.

\textcite{2006276} examined many possible sources of systematic error, both
theoretically and experimentally, in this rather complex, sophisticated
experiment in order to ensure that their result was correct. These include
light shifts, index of refraction effects, and the effect of a gravity
gradient, for which the corrections and their uncertainties are in fact
comparatively small. More significant are the fractional corrections of
$(16.8\pm 8)\times10^{-9}$ for wave front curvature and Guoy phase, $(-13.2\pm
4)\times10^{-9}$ for second order Zeeman effect, and $4(4)\times10^{-9}$ for
the alignment of the Raman and Bloch beams. The total of all corrections is
given as $10.98(10.0)\times10^{-9}$. 

From Eq.~(\ref{eq:alhmx}), the observational equation for the LKB value of
$h/m({\rm ^{87}Rb})$ in Eq~(\ref{eq:homrb06}) is 
\begin{eqnarray}
{h\over m(^{87}{\rm Rb})} = {A_{\rm r}({\rm e})\over A_{\rm r}({\rm ^{87}Rb})}
{c\,\alpha^2 \over 2 R_\infty} \ .
\label{eq:homrboe}
\end{eqnarray}
Evaluation of this expression with the LKB result and the 2006 recommended
values of $R_\infty$ and $A_{\rm r}$(e), and the value of $A_{\rm r}(^{87}$Rb)
resulting from the final least-squares adjustment on which the 2006 recommended
values are based, all of whose uncertainties are negligible in this context,
yields
\begin{eqnarray}
\alpha^{-1} =  137.035\,998\,83(91) \qquad [ 6.7\times 10^{-9}] \ ,
\label{eq:alphinvrb06}
\end{eqnarray}
which is included in Table \ref{tab:dhom}.  The uncertainty of this value of
$\alpha^{-1}$ is smaller than the uncertainty of any other value except those
in Table~\ref{tab:gfree} deduced from the measurement of $a_{\rm e}$, exceeding
the smallest uncertainty of the two values of $\alpha^{-1}[a_{\rm e}]$ in that
table by a factor of ten.

\section{Thermal physical quantities}
\label{sec:tpq}

The following sections discuss the molar gas constant, Boltzmann constant, and
Stefan-Boltzmann constant---constants associated with phenomena in the fields
of thermodynamics and/or statistical mechanics.

\subsection{Molar gas constant $\bm{R}$}
\label{ssec:mgc}

The square of the speed of sound $c_{\rm a}^2 (p,T)$ of a real gas at pressure
$p$ and thermodynamic temperature $T$ can be written as \cite{1973020}
\begin{eqnarray}
c_{\rm a}^2 (p,T) &=& A_0(T) + A_1(T)p 
\nonumber\\
&&+ A_2(T)p^2 + A_3(T)p^3 + \cdots \ ,
\label{eq:ca2tp}
\end{eqnarray}
where $A_1(T)$ is the first acoustic virial coefficient, $A_2(T)$ is the
second, {\it etc.}  In the limit $p\rightarrow 0$, Eq.~(\ref{eq:ca2tp}) yields
\begin{eqnarray}
c_{\rm a}^2 (0,T) = A_0(T) = \frac{\gamma_0 R\,T}{A_{\rm r}(X) M_{\rm u}} \ ,
\label{eq:ca2t}
\end{eqnarray}
where the expression on the right-hand side is the square of the speed of sound
for an unbounded ideal gas, and where $\gamma_0 = c_p/c_V$ is the ratio of the
specific heat capacity of the gas at constant pressure to that at constant
volume, $A_{\rm r}(X)$ is the relative atomic mass of the atoms or molecules of
the gas, and $M_{\rm u} = 10^{-3}~{\rm kg}~{\rm mol}^{-1}$.  For a monatomic
ideal gas, $\gamma_0 = 5/3$.

The 2006 recommended value of $R$, like the 2002 and 1998 values, is based on
measurements of the speed of sound in argon carried out in two independent
experiments, one done in the 1970s at NPL and the other done in the 1980s at
NIST.  Values of $c_{\rm a}^2(p,T_{\rm TPW})$, where $T_{\rm TPW} = 273.16~{\rm
K}$ is the triple point of water, were obtained at various pressures and
extrapolated to $p=0$ in order to determine $A_0(T_{\rm TPW}) = c_{\rm
a}^2(0,T_{\rm TPW})$, and hence $R$, from the relation
\begin{eqnarray}
R = \frac{c_{\rm a}^2(0,T_{\rm TPW})A_{\rm r}({\rm Ar})M_{\rm u}}{\gamma_0 T_{\rm TPW}}\ ,
\label{eq:rca}
\end{eqnarray}
which follows from Eq.~(\ref{eq:ca2t}).

Because the work of both NIST and NPL is reviewed in CODATA-98 and CODATA-02
and nothing has occurred in the last 4 years that would change the values of
$R$ implied by their reported values of $c_{\rm a}^2(0,T_{\rm TPW})$, we give
only a brief summary here.  Changes in these values due to the new values of
$A_{\rm r}(^{A}{\rm Ar})$ resulting from the 2003 atomic mass evaluation as
given in Table~\ref{tab:rmass03}, or the new IUPAC compilation of atomic
weights of the elements given by \textcite{2006271}, are negligible. 

Since $R$ cannot be expressed as a function of any other of the 2006 adjusted
constants, $R$ itself is taken as an adjusted constant for the NIST and NPL
measurements.

\subsubsection{NIST: speed of sound in argon}
\label{sssec:mgcnist}

In the NIST experiment of \textcite{1988027}, a spherical acoustic resonator at
a temperature $T = T_{\rm TPW}$ filled with argon was used to determine $c_{\rm
a}^2(p,T_{\rm TPW})$.  The final NIST result for the molar gas constant is
\begin{eqnarray}
R =  8.314\,471(15)~{\rm J~mol^{-1}~K^{-1}} \quad [ 1.8\times 10^{-6}] \ .
\label{eq:rnist88}
\end{eqnarray}

The mercury employed to determine the volume of the spherical resonator was
traceable to the mercury whose density was measured by \textcite{1961008} [see
also \textcite{1957014}].  The mercury employed in the NMI Hg electrometer
determination of $K_{\rm J}$ (see \ref{sssec:nmljc}) was also traceable to the
same mercury. Consequently, the NIST value of $R$ and the NMI value of $K_{\rm
J}$ are correlated with the non-negligible correlation coefficient 0.068.

\subsubsection{NPL: speed of sound in argon}
\label{sssec:mgcnpl}

In contrast to the dimensionally fixed resonator used in the NIST experiment,
the NPL experiment employed a variable path length fixed-frequency cylindrical
acoustic interferometer to measure $c_{\rm a}^2(p,T_{\rm TPW})$.  The final NPL
result for the molar gas constant is \cite{1979024}
\begin{eqnarray}
R =  8.314\,504(70)~{\rm J~mol}^{-1}~{\rm K}^{-1} \quad [ 8.4\times 10^{-6}] \ .
\label{eq:rnpl79}
\end{eqnarray}

Although both the NIST and NPL values of $R$ are based on the same values of
$A_{\rm r}(^{40}{\rm Ar})$, $A_{\rm r}(^{38}{\rm Ar})$, and $A_{\rm
r}(^{36}{\rm Ar})$, the uncertainties of these relative atomic masses are
sufficiently small that the covariance of the two values of $R$ is negligible.

\subsubsection{Other values}
\label{sssec:kov}

The most important of the historical values of $R$ have been reviewed by
\textcite{1984051} [see also \cite{1976022} and CODATA-98].  However, because
of the large uncertainties of these early values, they were not considered for
use in either the 1986, 1998, or 2002 CODATA adjustments, and we do not
consider them for the 2006 adjustment as well.

Also because of its non-competitive uncertainty ($u_{\rm r} = 36 \times
10^{-6}$), we exclude from consideration in the 2006 adjustment, as in the 2002
adjustment, the value of $R$ obtained from measurements of the speed of sound
in argon reported by \textcite{2002221} at the Xi\'an Jiaotong University,
Xi\'an, China (People's Republic of).

\subsection{Boltzmann constant $\bm{k}$}
\label{ssec:bcsbck}

The Boltzmann constant is related to the molar gas constant $R$ and other
adjusted constants by
\begin{eqnarray}
k = \frac{2R_\infty h}{c A_{\rm r}({\rm e}) M_{\rm u} \alpha^2} R
 \,= {R\over N_{\rm A}} \ .
\label{eq:koc}
\end{eqnarray}
No competitive directly measured value of $k$ was available for the 1998 or
2002 adjustments, and the situation remains unchanged for the present
adjustment. Thus, the 2006 recommended value with $u_{\rm r} =
1.7\times10^{-6}$ is obtained from this relation, as were the 1998 and 2002
recommended values.  However, a number of experiments are currently underway
that might lead to competitive values of $k$ (or $R$) in the future; see
\textcite{2006251} for a recent review.

Indeed, one such experiment underway at the PTB based on dielectric constant
gas thermometry (DCGT) was discussed in both CODATA-98 and CODATA-02, but no
experimental result for $A_{\epsilon}/R$, where $A_{\epsilon}$ is the molar
polarizability of  the $^4{\rm He}$ atom, other than that considered in these
two reports, has been published by the PTB group [see also \textcite{2006251}
and \textcite{1996200}]. However, the relative uncertainty of the theoretical
value of the static electric dipole polarizability of the ground state of the
$^4{\rm He}$ atom, which is required to calculate $k$ from $A_{\epsilon}/R$,
has been lowered by more than a factor of ten to below $2 \times 10^{-7}$
\cite{2004107}.  Nevertheless, the change in its value is negligible at the
level of uncertainty of the PTB result for $A_{\epsilon}/R$; hence, the value
$k=1.380\,65(4) \times 10^{-23}\,\rm J\,\,\rm K^{-1}$ [$30 \times 10^{-6}$]
from the PTB experiment given in CODATA-02 is unchanged.

In addition, preliminary results from two other ongoing experiments, the first
being carried out at NIST by \textcite{2007117} and the second at the
University of Paris by \textcite{2007114}, have recently been published.

\textcite{2007117} report $R=8.314\,487(76)$ J mol K$^{-1}$ [$9.1 \times
10^{-6}$], obtained from measurements of the index of refraction $n(p,\,T)$ of
$^{4}\rm He$ gas as a function of $p$ and $T$ by measuring the difference in
the resonant frequencies of a quasispherical microwave resonator when filled
with $^4\rm He$ at a given pressure and when evacuated (that is, at $p=0$).
This experiment has some similarities to the PTB DCGT experiment in that it
determines the quantity $A_{\epsilon}/R$ and hence $k$.  However, in DCGT one
measures the difference in capacitance of a capacitor when filled with $^4\rm
He$ at a given pressure and at $p=0$, and hence one determines the dielectric
constant of the $^4{\rm He}$ gas rather than its index of refraction.  Because
$^4{\rm He}$ is slightly diamagnetic, this means that to obtain
$A_{\epsilon}/R$ in the NIST experiment, a value for $A_{\mu}/R$ is required,
where $A_{\mu}=4\rmpi\chi_{0}/3$ and $\chi_{0}$ is the diamagnetic
susceptibility of a $^4{\rm He}$ atom.

\textcite{2007114} report $k=1.380\,65(26) \times 10^{-23}$ J K$^{-1}$ [$190
\times 10^{-6}$], obtained from measurements as a function of pressure of the
Doppler profile at $T=273.15\,\rm K$ (the ice point) of a well-isolated
rovibrational line in the $\nu_{2}$ band of the ammonium molecule, $^{14}{\rm
N}{\rm H_3}$, and extrapolation to $p=0$. The experiment actually measures
$R=kN_{\rm A}$, because the mass of the ammonium molecule in kilograms is
required but can only be obtained with the requisite accuracy from the molar
masses of $^{14}\rm N$ and $^{1}\rm H$, thereby introducing $N_{\rm A}$.

It is encouraging that the preliminary values of $k$ and $R$ resulting from
these three experiments are consistent with the 2006 recommended values.

\subsection{Stefan-Boltzmann constant $\sigma$}
\label{sssec:sfc}

The Stefan-Boltzmann constant is related to $c$, $h$, and the Boltzmann
constant $k$ by
\begin{eqnarray}
\sigma = \frac{2\rmpi^5 k^4}{15h^3 c^2} \ ,
\label{eq:skhc}
\end{eqnarray}
which, with the aid of Eq.~(\ref{eq:koc}), can be expressed in terms of the
molar gas constant and other adjusted constants as
\begin{eqnarray}
\sigma = \frac{32\rmpi^5 h}{15c^6} \left(\frac{R_\infty R}{A_{\rm r}({\rm e}) 
M_{\rm u} \alpha^2}\right)^4 \ .
\label{eq:shcr4}
\end{eqnarray}
No competitive directly measured value of $\sigma$ was available for the 1998
or 2002 adjustments, and the situation remains unchanged for the 2006
adjustment. Thus, the 2006 recommended value with $u_{\rm r} =
7.0\times10^{-6}$ is obtained from this relation, as were the 1998 and 2002
recommended values.  For a concise summary of experiments that might provide a
competiive value of $\sigma$, see the review by \textcite{2006251}.

\section{Newtonian constant of gravitation  $\bm{G}$}
\label{sec:ncg}

\begin{table*}
\caption{Summary of the results of measurements of the Newtonian constant of
gravitation relevant to the 2006 adjustment together with the 2002 and 2006
CODATA recommended values. }
\label{tab:bg}
\begin{tabular}{llllll}
\toprule
\vbox to 10 pt {}
Item &  Source & Identification$\footnotemark[1] $ & Method & \hbox to 10 pt {} $10^{11} \, G$                        & Rel. stand. \\
     &         &                &        & $\overline{{\rm m}^3 \  {\rm kg}^{-1} \ {\rm s}^{-2}}$  & uncert $u_{\rm r}$ \\
\noalign{\vbox to 2 pt {}}
\colrule
\noalign{\vbox to 5 pt {}}

     & 2002 CODATA Adjustment        & CODATA-02 &                           & $ 6.6742(10)   $ & $ 1.5\times 10^{-4}$  \\
         &                         &           &                           &                 &                     \\

\vsp  $a.$ & \cite{1996199}  & TR\&D-96  & Fiber torsion balance,    & $ 6.672\,9(5)$     & $ 7.5\times 10^{-5}$  \\
         &                               &           & dynamic mode              &                 &                     \\

\vsp  $b.$ & \cite{1997025}      & LANL-97   & Fiber torsion balance,    & $ 6.674\,0(7)$     & $ 1.0\times 10^{-4}$  \\
         &                               &           & dynamic mode              &                 &                     \\

\vsp  $c.$ & \cite{2000088,pc02gm} & UWash-00    & Fiber torsion
balance,            & $ 6.674\,255(92)$   & $ 1.4\times 10^{-5}$  \\
         &                               &           & dynamic compensation      &                 &                     \\

\vsp  $d.$ & \cite{2001089}      & BIPM-01   & Strip torsion balance,    & $ 6.675\,59(27)$     & $ 4.0\times 10^{-5}$  \\
         &                               &    & compensation mode, static deflection          &        &                     \\

\vsp  $e.$ & \cite{02kleinevoss,pc02kmph} & UWup-02   & Suspended body,           & $ 6.674\,22(98)$    & $ 1.5\times 10^{-4}$  \\
         &                               &           & displacement              &                 &                     \\

\vsp  $f.$ & \cite{2003219}      & MSL-03    & Strip torsion balance,    &  $ 6.673\,87(27)$     &  $ 4.0\times 10^{-5}$ \\
         &                               &           & compensation mode         &                 &                     \\

\vsp  $g.$ & \cite{2005292} & HUST-05  & Fiber torsion balance, & $ 6.672\,3(9)$& $ 1.3\times 10^{-4}$  \\
         &                               &           & dynamic mode              &                 &                     \\

\vsp  $h.$ & \cite{2006238}      & UZur-06   & Stationary body,          & $ 6.674\,25(12)$    & $ 1.9\times 10^{-5}$  \\
         &                               &           & weight change             &                 &                     \\

\vsp     & 2006 CODATA Adjustment        & CODATA-06 &                          & $ 6.674\,28(67)$     & $ 1.0\times 10^{-4}$  \\
\botrule
\end{tabular}
\begin{minipage}{7 in}
\footnotetext[1]{
TR\&D: Tribotech Research and Development Company, Moscow, Russian Federation;
LANL: Los Alamos National Laboratory, Los Alamos, New Mexico, USA;
UWash: University of Washington, Seattle, Washington, USA;
BIPM: International Bureau of Weights and Measures, S\`{e}vres, France;
UWup: University of Wuppertal, Wuppertal, Germany;
MSL: Measurement Standards Laboratory, Lower Hutt, New Zealand;
HUST: Huazhong University of Science and Technology, Wuhan, PRC;
UZur: University of Zurich, Zurich, Switzerland.
}
\end{minipage}
\end{table*}

\begin{figure}
\rotatebox{-90}{\resizebox{!}{4.2in}{
\includegraphics[clip,trim=10 40 40 10]{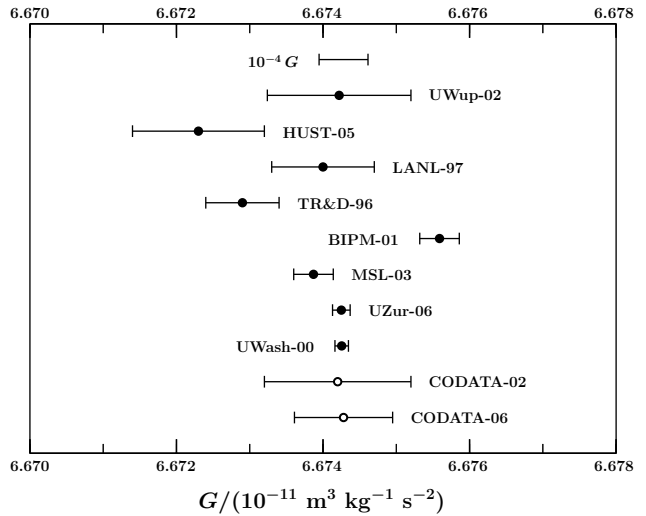}
}}
\caption{Values of the Newtonian constant of gravitation $G$.}
\label{fig:bigg}
\end{figure}

Because there is no known quantitative theoretical relationship between the
Newtonian constant of gravitation $G$ and other fundamental constants, and
because the currently available experimental values of $G$ are independent of
all of the other data relevant to the 2006 adjustment, these experimental
values contribute only to the determination of the 2006 recommended value of
$G$ and can be considered independently from the other data.

The historic difficulty of determining $G$, as demonstrated by the
inconsistencies among different measurements, is described in detail in
CODATA-86, CODATA-98, and CODATA-02.  Although no new competitive independent
result for $G$ has become available in the last 4 years, adjustments to two
existing results considered in 2002 have been made by researchers involved in
the original work.  One of the two results that has changed is from the
Huazhong University of Science and Technology (HUST) and is now identified as
HUST-05; the other is from the University of Zurich (UZur) and is now
identified as UZur-06. These revised results are discussed in some detail
below.

Table~\ref{tab:bg} summarizes the various values of $G$ considered here, which
are the same as in 2002 with the exception of these two revised results, and
Fig.~\ref{fig:bigg} compares them graphically. For reference purposes, both the
table and figure include the 2002 and 2006 CODATA recommended values. The
result now identified as TR\&D-96 was previously identified as TR\&D-98. The
change is because a 1996 reference, \cite{1996199}, was found that reports the
same result as does the 1998 reference \cite{1998091}.

\def\bgs{\,G_0}
For simplicity, in the following text, we write $G$ as a numerical factor
multiplying $\bgs$, where
\begin{eqnarray}
\bgs = 10^{-11}~{\rm  m^{3}~kg^{-1}~s^{-2}} \ .
\end{eqnarray}

\subsection{Updated values}
\label{ssec:uv}

\subsubsection{Huazhong University of Science and Technology}
\label{sssec:hust05}

The HUST group, which determines $G$ by the time-of-swing method using a
high-$Q$ torsion pendulum with two horizontal, 6.25~kg stainless steel
cylindrical source masses labeled A and B positioned on either side of the test
mass, has reported a fractional correction of $+360\times10^{-6}$ to their
original result given by \textcite{1999044}. It arises in part from recently
discovered density inhomogeneities in the source masses, the result of which is
a displacement of the center of mass of each source mass from its geometrical
center \cite{2005292}. Using a ``weighbridge'' with a commercial electronic
balance---a method developed by \textcite{1995135} to locate the center of mass
of a test object with micrometers precision---\textcite{2005292} found that the
axial eccentricities $e_{\rm A}$ and $e_{\rm B}$ of the two source masses were
$(10.3~\pm~2.6)~\rmmu\rm m$ and $(6.3~\pm~3.7)~\rmmu\rm m$, with the result
that the equivalent displacements between the test mass and the source masses
are larger than the values used by \cite{1999044}. Assuming a linear axial
density distribution, the calculated fractional correction to the previous
result is $+210\times10^{-6}$ with an additional component of relative standard
uncertainty of $78\times10^{-6}$ due to the uncertainties of the
eccentricities. 

The remaining $150\times10^{-6}$ portion of the $360\times10^{-6}$ fractional
correction is also discussed by \textcite{2005292} and arises as follows. In
the HUST experiment, $G$ is determined by comparing the period of the torsion
pendulum with and without the source masses.  When the source masses are
removed, they are replaced by air. Since the masses of the source masses used
by \textcite{1999044} are the vacuum masses, a correction for the air, first
suggested by R. S. Davis and T. J. Quinn of the BIPM, is required.  This
correction was privately communicated to the Task Group by the HUST researchers
in 2003 and included in the HUST value of $G$ used in the 2002 adjustment.

The HUST revised value of $G$, including the additional component of
uncertainty due to the measurement of the eccentricities $e_{\rm A}$ and $e_{\rm
B}$, is item $g$ in Table~\ref{tab:bg}.

\subsubsection{University of Zurich}  
\label{sssec:uzur06}

The University of Zurich result for $G$ discussed in CODATA-02 and used in the
2002 adjustment, $G = 6.674\,07(22)\bgs~~[3.3\times10^{-5}]$, was reported by
\textcite{2002123}. It was based on the weighted mean of three highly
consistent values obtained from three series of measurements carried out at the
Paul Scherrer Institute (PSI), Villigen, Switzerland, in 2001 and 2002 and
denoted Cu, Ta~I, and Ta~II.  The designation Cu means that the test masses
were gold plated copper, and the designation Ta means that they were tantalum.
Following the publication of \textcite{2002123}, an extensive reanalysis of the
original data was carried out by these authors together with other University
of Zurich researchers, the result being the value of $G$ in Table~\ref{tab:bg},
item $h$, as given in the detailed final report on the experiment
\cite{2006238}.

In the University of Zurich approach to determining $G$, a modified commercial
single-pan balance is used to measure the change in the difference in weight of
two cylindrical test masses when the relative position of two source masses is
changed. The quantity measured is the 800~$\rmmu$g difference signal obtained
at many different working points in the balance calibration range using two
sets of 16 individual wire weights, allowing an $in$ $situ$ measurement of the
balance non-linearity over the entire 0.2~g balance calibration interval. A
more rigorous analysis using a fitting method with Legendre polynomials has now
allowed the relative standard uncertainty contribution to $G$ from balance
nonlinearity to be reduced from $18\times10^{-6}$ to $6.1\times10^{-6}$ based
on the Cu test-mass data. Various problems with the mass handler for the wire
weights that did not allow the application of the Legendre polynomial fitting
procedure occurred during the Ta test-mass measurements, resulting in large
systematic errors. Therefore, the researchers decided to include only the Cu
data in their final analysis \cite{2006238}.
\vspace{0pt}

Each source mass consisted of a cylindrical tank filled with $7.5\times10^3$~kg
of mercury. Since the mercury represented approximately 94~\% of the total
mass, special care was taken in determining its mass and density. These
measurements were further used to obtain more accurate values for the key tank
dimensions and Hg mass. This was done by minimizing a $\chi^{2}$ function that
depended on the tank dimensions and the Hg mass and density, and using the
dependence of the density on these dimensions and the Hg mass as a constraint.
Calculation of the mass integration constant with these improved values reduced
the $u_{\rm r}$ of this critical quantity from $20.6\times10^{-6}$ to
$6.7\times10^{-6}$.

Although the analysis of \textcite{2002123} assumed a linear temporal drift of
the balance zero point, a careful examination by \textcite{2006238} found that
the drift was significantly nonlinear and was influenced by the previous load
history of the balance. A series of Legendre polynomials and a sawtooth
function, respectively, were therefore used to describe the slow and rapid
variations of the observed balance zero-point with time. 

The 2002 value of $G$ obtained from the Cu data was 6.674\,03$\bgs$, consistent
with the Ta~I, and Ta~II values of 6.674\,09$\bgs$ and 6.674\,10$\bgs$
\cite{2002123}, whereas the value from the present Cu data analysis is
$6.674\,25(12)\bgs$, with the $3.3\times 10^{-5}$ fractional increase being due
primarily to the application of the nonlinear zero point drift correction. A
minor contributor to the difference is the inclusion of the very first Cu data
set that was omitted in the 2002 analysis due to a large start-up zero-point
drift that is now correctable with the new Legendre polynomial-sawtooth
function analysis technique, and the exclusion of a data set that had a
temperature stabilization system failure that went undetected by the old data
analysis method \cite{pc07ss}.

\subsection{Determination of 2006 recommended value of $G$}
\label{sssec:detofbg}

The overall agreement of the eight values of $G$ in Table~\ref{tab:bg} (items
$a$ to $h$) has improved somewhat since the 2002 adjustment, but the situation
is still far from satisfactory. Their weighted mean is $G=6.674\,275(68)\bgs$
with $\chi^2=38.6$ for degrees of freedom $\nu=N-M=8-1=7$ , Birge ratio $R_{\rm
B} = \sqrt{\chi^2/\nu} = 2.35$, and normalized residuals $r_{i}$ of $-2.75$,
$-0.39$, $-0.22$, $4.87$, $-0.56$, $-1.50$, $-2.19$, and $-0.19$, respectively
(see Appendix E of CODATA-98). The BIPM-01 value with $r_{i} = 4.87$ is clearly
the most problematic. For comparison, the 2002 weighted mean was $G =
6.674\,232(75)\bgs$ with $\chi^2=57.7$ for $\nu=7$ and $R_{\rm B} = 2.87$.

If the BIPM value is deleted, the weighted mean is reduced by 1.3 standard
uncertainties to $G=6.674\,187(70)\bgs$, and $\chi^2=13.3$, $\nu=6$, and
$R_{\rm B} = 1.49$. In this case, the two remaining data with significant
normalized residuals are the the TR\&D-96 and the HUST-05 results with $r_i =
-2.57$ and $-2.10$, respectively.  If these two data, which agree with each
other, are deleted, the weighted mean is $G=6.674\,225(71)\bgs$ with
$\chi^2=2.0$, $\nu=4$, $R_{\rm B} = 0.70$, and with all normalized residuals
less than one except $r_i=-1.31$ for datum MSL-03.  Finally, if the UWash-00
and UZur-06 data, which have the smallest assigned uncertainties of the initial
eight values and which are in excellent agreement with each other, are deleted
from the initial group of eight data, the weighted mean of the remaining six
data is $G=6.674\,384(167)\bgs$ with $\chi^2=38.1$, $\nu=6$, and $R_{\rm
B}=2.76$.  The normalized residuals for these six data, TR\&D-96, LANL-97,
BIPM-01, UWup-02, MSL-03, and HUST-05, are $-2.97$, $-0.55$, 4.46, $-0.17$,
$-1.91$ and $-2.32$, respectively.

Finally, if the uncertainties of each of the eight values of $G$ are multiplied
by the Birge ratio associated with their weighted mean, $R_{\rm B} = 2.35$, so
that $\chi^2$ of their weighted mean becomes equal to its expected value of
$\nu=7$ and $R_{\rm B} = 1$, the normalized residual of the datum BIPM-01 would
still be larger than two. 

Based on the results of the above calculations, the historical difficulty of
determining $G$, the fact that all eight values of $G$ in Table~\ref{tab:bg}
are credible, and that the two results with the smallest uncertainties,
UWash-00 and UZur-06, are highly consistent with one another, the Task Group
decided to take as the 2006 CODATA recommended value of $G$ the weighted mean
of all of the data, but with an uncertainty of 0.000\,67$\bgs$, corresponding
to $u_{\rm r}=1.0\times10^{-4}$:
\begin{eqnarray}
G =  6.674\,28(67) \times 10^{-11}~{\rm  m^{3}~kg^{-1}~s^{-2}} \ .
\quad [ 1.0\times 10^{-4}] \nonumber\\
\end{eqnarray}
This value exceeds the 2002 recommended value by the fractional amount $1.2
\times 10^{-5}$, which is less than one tenth of the uncertainty $u_{\rm r}=1.5
\times 10^{-4}$ of the 2002 value. Further, the uncertainty of the 2006 value,
$u_{\rm r}=1.0 \times 10^{-4}$, is two thirds that of the 2002 value. 

In assigning this uncertainty to the 2006 recommended value of $G$, the Task
Group recognized that if the uncertainty was smaller than really justified by
the data, taking into account the history of measurements of $G$, it might
discourage the initiation of new research efforts to determine $G$, if not the
continuation of some of the research efforts already underway.  Such efforts
need to be encouraged in order to provide a more solid and redundant data set
upon which to base future recommended values. On the other hand, if the
uncertainty was too large, for example, if the uncertainty of the 2002
recommended value had been retained for the 2006 value, then the recommended
value would not have reflected the fact that we now have two data that are in
excellent agreement, have $u_{\rm r}$ less than $2 \times 10^{-5}$, and are the
two most accurate values available.

\subsection{Prospective values}
\label{sssec:otherg}

New techniques to measure $G$ using atom interferometry are currently under
development in at least two laboratories---the Universit\`a de Firenze in Italy
and Stanford University in the United States. This comes as no surprise since
atom interferometry is also being developed to measure the local acceleration
due to gravity $g$ (see the last paragraph of Sec.~\ref{sec:squ}).  Recent
proof of principle experiments combine two vertically separated atomic clouds
forming an atom-interferometer-gravity-gradiometer that measures the change in
the gravity gradient when a well characterized source mass is displaced.
Measuring the change in the gravity gradient allows the rejection of many
possible systematic errors.  \textcite{2006244} at the Universit\`a de Firenze
used a Rb fountain and a fast launch juggling sequence of two atomic clouds to
measure $G$ to 1~\%, obtaining the value $6.64(6)\bgs$; they hope to reach a
final uncertainty of 1 part in $10^4$.  \textcite{2007006} at Stanford used two
separate Cs atom interferometer gravimeters to measure $G$ and obtained the
value $6.693(34)\bgs$. The two largest uncertainties from systematic effects
were the determination of the initial atom velocity and the initial atom
position. The Stanford researchers also hope  to achieve a final uncertainty of
1 part in $10^4$. Although neither of these results is significant for the
current analysis of $G$, future results could be of considerable interest.

\section{X-ray and electroweak quantities}
\label{sec:xeq}

\subsection{X-ray units}
\label{ssec:xru}

Historically, units that have been used to express the wavelengths of x-ray
lines are the copper ${\rm K\rmalpha_1}$ x unit, symbol ${\rm
xu(CuK\rmalpha_1)}$, the molybdenum K${\rm \rmalpha}_1$ x unit, symbol ${\rm
xu(MoK\rmalpha_1)}$, and the {\aa}ngstrom star, symbol \AA$^*$.  They are
defined by assigning an exact, conventional value to the wavelength of the
${\rm CuK\rmalpha_1}$, ${\rm MoK\rmalpha_1}$, and ${\rm WK\rmalpha_1}$ x-ray
lines when each is expressed in its corresponding unit:
\begin{eqnarray}
\lambda({\rm CuK\rmalpha_1}) &=& 1\,537.400 \ 
{\rm xu(CuK\rmalpha_1)}  \label{eq:cuxudef} \\ 
\lambda({\rm MoK\rmalpha_1}) &=& 707.831 \ 
{\rm xu(MoK\rmalpha_1)}  \label{eq:moxudef} \\ 
\lambda({\rm WK\rmalpha_1}) &=& 0.209\,010\,0 \ 
{\rm \AA^*} \ . \label{eq:wxudef}
\end{eqnarray}

The experimental work that determines the best values of these three units was
reviewed in CODATA-98, and the relevant data may be summarized as follows:
\begin{eqnarray}
{\lambda({\rm CuK\rmalpha_1})\over d_{220}({\rm {\scriptstyle W4.2a}})} &=&
 0.802\,327\,11(24) \quad [ 3.0\times 10^{-7}]
\label{eq:lcusdw42aptb}
\\
{\lambda({\rm WK\rmalpha_1})\over d_{220}({\rm {\scriptstyle N}})} &=&
 0.108\,852\,175(98) \quad [ 9.0\times 10^{-7}]
\label{eq:lamnsdwnist}
\\
{\lambda({\rm MoK\rmalpha_1})\over d_{220}({\rm {\scriptstyle N}})} &=& 
 0.369\,406\,04(19) \quad [ 5.3\times 10^{-7}] 
\label{eq:lmodn73}
\\
{\lambda({\rm CuK\rmalpha_1})\over d_{220}({\rm {\scriptstyle N}})} &=& 
 0.802\,328\,04(77) \quad [ 9.6\times 10^{-7}] 
\label{eq:lcudn73} \ ,~~~~~
\end{eqnarray}
where ${d_{220}({\rm {\scriptstyle W4.2a}})}$ and ${d_{220}({\rm {\scriptstyle
N}})}$ denote the \{220\} lattice spacings, at the standard reference
conditions $p=0$ and $t_{90}=22.5~^\circ{\rm C}$, of particular silicon
crystals used in the measurements.  The result in Eq.~(\ref{eq:lcusdw42aptb})
is from a collaboration between researchers from Friedrich-Schiller University
(FSU), Jena, Germany and the PTB \cite{1991096}.  The lattice spacing
${d_{220}({\rm {\scriptstyle N}})}$ is related to crystals of known lattice
spacing through Eq.~(\ref{eq:drw17n}).

In order to obtain best values in the least-squares sense for ${\rm
xu(CuK\rmalpha_1)}$, ${\rm xu(MoK\rmalpha_1)}$, and ${\rm \AA^*}$, we take
these units to be adjusted constants.  Thus, the observational equations for
the data of Eqs.~(\ref{eq:lcusdw42aptb}) to (\ref{eq:lcudn73}) are 
\begin{eqnarray}
\frac{\lambda({\rm CuK\rmalpha_1})}{d_{220}({\rm {\scriptstyle{W4.2a}})}} &=& 
\frac{\rm 1\,537.400 ~xu(CuK\rmalpha_1)}{d_{220}({\rm {\scriptstyle{W4.2a}})}} \\
\frac{\lambda({\rm WK\rmalpha_1})}{d_{220}({\rm {\scriptstyle N}})} &=& 
\frac{\rm 0.209\,010\,0 ~\AA^*}{d_{220}({\rm {\scriptstyle N}})} 
\label{eq:lwdn79obeq} \\
\frac{\lambda({\rm MoK\rmalpha_1})}{d_{220}({\rm {\scriptstyle N}})} &=& 
\frac{\rm 707.831 ~xu(MoK\rmalpha_1)}{d_{220}({\rm {\scriptstyle N}})} 
\label{eq:lmodn73obeq} \\
\frac{\lambda({\rm CuK\rmalpha_1})}{d_{220}({\rm {\scriptstyle N}})} &=& 
\frac{\rm 1\,537.400 ~xu(CuK\rmalpha_1)}{d_{220}({\rm {\scriptstyle N}})} 
\label{eq:lcudn73obeq} \, ,
\label{eq:lcud42a79obeq}
\end{eqnarray}
where $d_{220}({\rm {\scriptstyle N}})$ is taken to be an adjusted constant and
$d_{220}({\rm {\scriptstyle W17}})$ and $d_{220}({\rm {\scriptstyle{W4.2a}}})$
are adjusted constants as well.

\subsection{Particle Data Group input}
\label{ssec:lisq}
\shortcites{2002170}

There are a few cases in the 2006 adjustment where an inexact constant that is
used in the analysis of input data is not treated as an adjusted quantity,
because the adjustment has a negligible effect on its value.  Three such
constants, used in the calculation of the theoretical expressions for the
electron and muon magnetic moment anomalies $a_{\rm e}$ and $a_{\rmssmu}$, are
the mass of the tau lepton $m_{\rmsstau}$, the Fermi coupling constant $G_{\rm
F}$, and sine squared of the weak mixing angle sin$^{2}{\theta}_{\rm W}$, and
are obtained from the most recent report of the Particle Data Group
\cite{2006110}:
\begin {eqnarray}
m_{\rmsstau}c^{2} &=&  1776.99(29) \ {\rm MeV}
\qquad [ 1.6\times 10^{-4}]
\label{eq:mtaumev}
 \\
{G_{\rm F}\over(\hbar c)^{3}} &=&  1.166\,37(1)\times 10^{-5} \ {\rm GeV}^{-2}
\quad [ 8.6\times 10^{-6}]  \nonumber\\
\label{eq:gf} 
 \\
{\rm sin}^{2}{\theta}_{\rm W} &=&  0.222\,55(56)
\qquad [ 2.5\times 10^{-3}] \ .
\label{eq:sin2thw}
\end{eqnarray}
To facilitate the calculations, the uncertainty of $m_{\rmsstau}c^{2}$ is
symmetrized and taken to be 0.29~MeV rather than $+0.29~{\rm MeV}$, $-0.26~{\rm
MeV}$.  We use the definition sin$^{2}{\theta}_{\rm W} = 1 - (m_{\rm W}/m_{\rm
Z})^{2}$, where $m_{\rm W}$ and $m_{\rm Z}$ are, respectively, the masses of
the ${\rm W}^{\pm}$ and ${\rm Z}^{0}$ bosons, because it is employed in the
calculation of the electroweak contributions to $a_{\rm e}$ and $a_{\rmssmu}$
\cite{1996033}.  The Particle Data Group's recommended value for the mass ratio
of these bosons is $m_{\rm W}/m_{\rm Z} =  0.881\,73(32)$, which leads to the
value of sin$^{2}{\theta}_{\rm W}$ given above.

\section{Analysis of Data}
\label{sec:ad}

The previously discussed input data are examined in this section for their
mutual compatibility and their potential role in determining the 2006
recommended values of the constants.  Based on this analysis, the data are
selected for the final least-squares adjustment from which the recommended
values are obtained.  Because the data on the Newtonian constant of
gravitation $G$ are independent of the other data and are analyzed in
Sec.~\ref{sec:ncg}, they are not examined further.  The consistency of the
input data is evaluated by directly comparing different measurements of the
same quantity, and by directly comparing the values of a single fundamental
constant inferred from measurements of different quantities.  As noted in the
outline section of this paper, the inferred value is for comparison purposes
only; the datum from which it is obtained, not the inferred value, is the input
datum in the adjustment.  The potential role of a particular input datum is
gauged by carrying out a least-squares adjustment using all of the initially
considered data.  A particular measurement of a quantity is included in the
final adjustment if its uncertainty is not more than about ten times the
uncertainty of the value of that quantity provided by other data in the
adjustment.  The measure we use is the ``self sensitivity coefficient'' of an
input datum $S_{\rm c}$ (see CODATA-98), which must be greater than 0.01 in
order for the datum to be included.

The input data are given in Tables~\ref{tab:rdata}, \ref{tab:pdata}, and
\ref{tab:cdata} and their covariances are given as correlation coefficients in
Tables~\ref{tab:rdcc}, \ref{tab:pdcc}, and \ref{tab:cdcc}.  The $\delta$s given
in Tables~\ref{tab:rdata}, \ref{tab:pdata}, and \ref{tab:cdata} are quantities
added to corresponding theoretical expressions to account for the uncertainties
of those expressions, as previously discussed (see, for example,
Sec.~\ref{par:teu}).  Note that the value of the Rydberg constant $R_\infty$
depends only weakly on changes, at the level of the uncertainties, of the data
in Tables~\ref{tab:pdata} and \ref{tab:cdata}.

\setcounter{topnumber}{3}
\def\m{\phantom{-}}
\def\fixh{\vbox to 9pt {}}
\def\vsp{\vbox to 10pt{}}
\def\sp{\hbox to 10 pt {}}
\begin{table*}[t]
\caption{Summary of principal input data for the determination of the 2006
recommended value of the Rydberg constant $R_\infty$.  [The notation for the
additive corrections $\delta_{X}(n{\rm L}_j)$ in this table has the same
meaning as the notation $\delta^{X}_{n{\rm L}j}$ in Sec.~\ref{par:teu}.] }
\label{tab:rdata}
\begin{tabular}{l@{\sp}l@{\sp}l@{\sp}l@{\sp}l@{\sp}l}
\toprule
\vbox to 10 pt {}
Item &  Input datum    &  \hbox to 15pt {} Value
& \hbox to -20pt {} Relative standard
& Identification & Sec. \\
number& & & \hbox to -5pt {}uncertainty\footnotemark[1]
 $u_{\rm r}$ &  \\
\colrule
 $A1   $ \vsp & $\delta_{\rm H}({\rm 1S_{1/2}})$ & $ 0.0(3.7)$  kHz & $[ 1.1\times 10^{-12}]$ & theory & \ref{par:teu}  \\
 $A2   $ & $\delta_{\rm H}({\rm 2S_{1/2}})$ & $ 0.00(46)$  kHz & $[ 5.6\times 10^{-13}]$ & theory & \ref{par:teu}  \\
 $A3   $ & $\delta_{\rm H}({\rm 3S_{1/2}})$ & $ 0.00(14)$  kHz & $[ 3.7\times 10^{-13}]$ & theory & \ref{par:teu}  \\
 $A4   $ & $\delta_{\rm H}({\rm 4S_{1/2}})$ & $ 0.000(58)$  kHz & $[ 2.8\times 10^{-13}]$ & theory & \ref{par:teu}  \\
 $A5   $ & $\delta_{\rm H}({\rm 6S_{1/2}})$ & $ 0.000(20)$  kHz & $[ 2.1\times 10^{-13}]$ & theory & \ref{par:teu}  \\
 $A6   $ & $\delta_{\rm H}({\rm 8S_{1/2}})$ & $ 0.0000(82)$  kHz & $[ 1.6\times 10^{-13}]$ & theory & \ref{par:teu}  \\
 $A7   $ & $\delta_{\rm H}({\rm 2P_{1/2}})$ & $ 0.000(69)$  kHz & $[ 8.4\times 10^{-14}]$ & theory & \ref{par:teu}  \\
 $A8   $ & $\delta_{\rm H}({\rm 4P_{1/2}})$ & $ 0.0000(87)$  kHz & $[ 4.2\times 10^{-14}]$ & theory & \ref{par:teu}  \\
 $A9   $ & $\delta_{\rm H}({\rm 2P_{3/2}})$ & $ 0.000(69)$  kHz & $[ 8.4\times 10^{-14}]$ & theory & \ref{par:teu}  \\
 $A10  $ & $\delta_{\rm H}({\rm 4P_{3/2}})$ & $ 0.0000(87)$  kHz & $[ 4.2\times 10^{-14}]$ & theory & \ref{par:teu}  \\
 $A11  $ & $\delta_{\rm H}({\rm 8D_{3/2}})$ & $ 0.000\,00(48)$  kHz & $[ 9.3\times 10^{-15}]$ & theory & \ref{par:teu}  \\
 $A12  $ & $\delta_{\rm H}({\rm 12D_{3/2}})$ & $ 0.000\,00(15)$  kHz & $[ 6.6\times 10^{-15}]$ & theory & \ref{par:teu}  \\
 $A13  $ & $\delta_{\rm H}({\rm 4D_{5/2}})$ & $ 0.0000(38)$  kHz & $[ 1.9\times 10^{-14}]$ & theory & \ref{par:teu}  \\
 $A14  $ & $\delta_{\rm H}({\rm 6D_{5/2}})$ & $ 0.0000(11)$  kHz & $[ 1.2\times 10^{-14}]$ & theory & \ref{par:teu}  \\
 $A15  $ & $\delta_{\rm H}({\rm 8D_{5/2}})$ & $ 0.000\,00(48)$  kHz & $[ 9.3\times 10^{-15}]$ & theory & \ref{par:teu}  \\
 $A16  $ & $\delta_{\rm H}({\rm 12D_{5/2}})$ & $ 0.000\,00(16)$  kHz & $[ 7.0\times 10^{-15}]$ & theory & \ref{par:teu}  \\
 $A17  $ \vsp & $\delta_{\rm D}({\rm 1S_{1/2}})$ & $ 0.0(3.6)$  kHz & $[ 1.1\times 10^{-12}]$ & theory & \ref{par:teu}  \\
 $A18  $ & $\delta_{\rm D}({\rm 2S_{1/2}})$ & $ 0.00(45)$  kHz & $[ 5.4\times 10^{-13}]$ & theory & \ref{par:teu}  \\
 $A19  $ & $\delta_{\rm D}({\rm 4S_{1/2}})$ & $ 0.000(56)$  kHz & $[ 2.7\times 10^{-13}]$ & theory & \ref{par:teu}  \\
 $A20  $ & $\delta_{\rm D}({\rm 8S_{1/2}})$ & $ 0.0000(80)$  kHz & $[ 1.6\times 10^{-13}]$ & theory & \ref{par:teu}  \\
 $A21  $ & $\delta_{\rm D}({\rm 8D_{3/2}})$ & $ 0.000\,00(48)$  kHz & $[ 9.3\times 10^{-15}]$ & theory & \ref{par:teu}  \\
 $A22  $ & $\delta_{\rm D}({\rm 12D_{3/2}})$ & $ 0.000\,00(15)$  kHz & $[ 6.6\times 10^{-15}]$ & theory & \ref{par:teu}  \\
 $A23  $ & $\delta_{\rm D}({\rm 4D_{5/2}})$ & $ 0.0000(38)$  kHz & $[ 1.9\times 10^{-14}]$ & theory & \ref{par:teu}  \\
 $A24  $ & $\delta_{\rm D}({\rm 8D_{5/2}})$ & $ 0.000\,00(48)$  kHz & $[ 9.3\times 10^{-15}]$ & theory & \ref{par:teu}  \\
 $A25  $ & $\delta_{\rm D}({\rm 12D_{5/2}})$ & $ 0.000\,00(16)$  kHz & $[ 7.0\times 10^{-15}]$ & theory & \ref{par:teu}  \\
 $A26  $ \vsp & $\nu_{\rm H}({\rm 1S_{1/2}}-{\rm 2S_{1/2}})$ & $ 2\,466\,061\,413\,187.074(34)$  kHz & $ 1.4\times 10^{-14}$  & MPQ-04 & \ref{sssec:rydex}  \\
 $A27  $ & $\nu_{\rm H}({\rm 2S_{1/2}}-{\rm 8S_{1/2}})$ & $ 770\,649\,350\,012.0(8.6)$  kHz & $ 1.1\times 10^{-11}$  & LK/SY-97 & \ref{sssec:rydex}  \\
 $A28  $ & $\nu_{\rm H}({\rm 2S_{1/2}}-{\rm 8D_{3/2}})$ & $ 770\,649\,504\,450.0(8.3)$  kHz & $ 1.1\times 10^{-11}$  & LK/SY-97 & \ref{sssec:rydex}  \\
 $A29  $ & $\nu_{\rm H}({\rm 2S_{1/2}}-{\rm 8D_{5/2}})$ & $ 770\,649\,561\,584.2(6.4)$  kHz & $ 8.3\times 10^{-12}$  & LK/SY-97 & \ref{sssec:rydex}  \\
 $A30  $ & $\nu_{\rm H}({\rm 2S_{1/2}}-{\rm 12D_{3/2}})$ & $ 799\,191\,710\,472.7(9.4)$  kHz & $ 1.2\times 10^{-11}$  & LK/SY-98 & \ref{sssec:rydex}  \\
 $A31  $ & $\nu_{\rm H}({\rm 2S_{1/2}}-{\rm 12D_{5/2}})$ & $ 799\,191\,727\,403.7(7.0)$  kHz & $ 8.7\times 10^{-12}$  & LK/SY-98 & \ref{sssec:rydex}  \\
 $A32  $ & $\nu_{\rm H}({\rm 2S_{1/2}}-{\rm 4S_{1/2}}) - {1\over4}\nu_{\rm H}({\rm 1S_{1/2}}-{\rm 2S_{1/2}})$ & $ 4\,797\,338(10)$  kHz & $ 2.1\times 10^{-6}$ & MPQ-95 & \ref{sssec:rydex}  \\
 $A33  $ & $\nu_{\rm H}({\rm 2S_{1/2}}-{\rm 4D_{5/2}}) - {1\over4}\nu_{\rm H}({\rm 1S_{1/2}}-{\rm 2S_{1/2}})$ & $ 6\,490\,144(24)$  kHz & $ 3.7\times 10^{-6}$ & MPQ-95 & \ref{sssec:rydex}  \\
 $A34  $ & $\nu_{\rm H}({\rm 2S_{1/2}}-{\rm 6S_{1/2}}) - {1\over4}\nu_{\rm H}({\rm 1S_{1/2}}-{\rm 3S_{1/2}})$ & $ 4\,197\,604(21)$  kHz & $ 4.9\times 10^{-6}$  & LKB-96 & \ref{sssec:rydex}  \\
 $A35  $ & $\nu_{\rm H}({\rm 2S_{1/2}}-{\rm 6D_{5/2}}) - {1\over4}\nu_{\rm H}({\rm 1S_{1/2}}-{\rm 3S_{1/2}})$ & $ 4\,699\,099(10)$  kHz & $ 2.2\times 10^{-6}$  & LKB-96 & \ref{sssec:rydex}  \\
 $A36  $ & $\nu_{\rm H}({\rm 2S_{1/2}}-{\rm 4P_{1/2}}) - {1\over4}\nu_{\rm H}({\rm 1S_{1/2}}-{\rm 2S_{1/2}})$ & $ 4\,664\,269(15)$  kHz & $ 3.2\times 10^{-6}$  & YaleU-95 & \ref{sssec:rydex}  \\
 $A37  $ & $\nu_{\rm H}({\rm 2S_{1/2}}-{\rm 4P_{3/2}}) - {1\over4}\nu_{\rm H}({\rm 1S_{1/2}}-{\rm 2S_{1/2}})$ & $ 6\,035\,373(10)$  kHz & $ 1.7\times 10^{-6}$  & YaleU-95 & \ref{sssec:rydex}  \\
 $A38  $ & $\nu_{\rm H}({\rm 2S_{1/2}}-{\rm 2P_{3/2}})$ & $ 9\,911\,200(12)$  kHz & $ 1.2\times 10^{-6}$  & HarvU-94 & \ref{sssec:rydex}  \\
 $A39.1$ & $\nu_{\rm H}({\rm 2P_{1/2}}-{\rm 2S_{1/2}})$ & $ 1\,057\,845.0(9.0)$  kHz & $ 8.5\times 10^{-6}$  & HarvU-86 & \ref{sssec:rydex}  \\
 $A39.2$ & $\nu_{\rm H}({\rm 2P_{1/2}}-{\rm 2S_{1/2}})$ & $ 1\,057\,862(20)$  kHz & $ 1.9\times 10^{-5}$& USus-79 & \ref{sssec:rydex}  \\
 $A40  $ \vsp & $\nu_{\rm D}({\rm 2S_{1/2}}-{\rm 8S_{1/2}})$ & $ 770\,859\,041\,245.7(6.9)$  kHz & $ 8.9\times 10^{-12}$  & LK/SY-97 & \ref{sssec:rydex}  \\
 $A41  $ & $\nu_{\rm D}({\rm 2S_{1/2}}-{\rm 8D_{3/2}})$ & $ 770\,859\,195\,701.8(6.3)$  kHz & $ 8.2\times 10^{-12}$  & LK/SY-97 & \ref{sssec:rydex}  \\
 $A42  $ & $\nu_{\rm D}({\rm 2S_{1/2}}-{\rm 8D_{5/2}})$ & $ 770\,859\,252\,849.5(5.9)$  kHz & $ 7.7\times 10^{-12}$  & LK/SY-97 & \ref{sssec:rydex}  \\
 $A43  $ & $\nu_{\rm D}({\rm 2S_{1/2}}-{\rm 12D_{3/2}})$ & $ 799\,409\,168\,038.0(8.6)$  kHz & $ 1.1\times 10^{-11}$  & LK/SY-98 & \ref{sssec:rydex}  \\
 $A44  $ & $\nu_{\rm D}({\rm 2S_{1/2}}-{\rm 12D_{5/2}})$ & $ 799\,409\,184\,966.8(6.8)$  kHz & $ 8.5\times 10^{-12}$  & LK/SY-98 & \ref{sssec:rydex}  \\
 $A45  $ & $\nu_{\rm D}({\rm 2S_{1/2}}-{\rm 4S_{1/2}}) - {1\over4}\nu_{\rm D}({\rm 1S_{1/2}}-{\rm 2S_{1/2}})$ & $ 4\,801\,693(20)$  kHz & $ 4.2\times 10^{-6}$ & MPQ-95 & \ref{sssec:rydex}  \\
 $A46  $ & $\nu_{\rm D}({\rm 2S_{1/2}}-{\rm 4D_{5/2}}) - {1\over4}\nu_{\rm D}({\rm 1S_{1/2}}-{\rm 2S_{1/2}})$ & $ 6\,494\,841(41)$  kHz & $ 6.3\times 10^{-6}$ & MPQ-95 & \ref{sssec:rydex}  \\
 $A47  $ \vsp & $\nu_{\rm D}({\rm 1S_{1/2}} -{\rm 2S_{1/2}}) - \nu_{\rm H}({\rm 1S_{1/2}} - {\rm 2S_{1/2}})$ & $ 670\,994\,334.64(15)$  kHz & $ 2.2\times 10^{-10}$ & MPQ-98 & \ref{sssec:rydex}  \\
 $A48  $ \vsp & $R_{\rm p}$ & $ 0.895(18)$ fm & $ 2.0\times 10^{-2}$ & Rp-03 & \ref{sssec:nucrad} \\
 $A49  $ \vsp & $R_{\rm d}$ & $ 2.130(10)$ fm & $ 4.7\times 10^{-3}$ & Rd-98 & \ref{sssec:nucrad} \\
\botrule
\end{tabular}
\begin{minipage}{7 in}
\footnotetext[1]{
The values in brackets are relative to the frequency equivalent
of the binding energy of the indicated level.
}\end{minipage}
\end{table*}

\def\m{\phantom{-}}
\def\fixh{\vbox to 9pt {}}
\def\vsp{\vbox to 10pt{}}
\begin{table*}[t]
\caption{Correlation coefficients
$r(x_i,x_j) \ge 0.0001$
of the input data related to $R_\infty$ in Table~\ref{tab:rdata}.
For simplicity, the two items of data to which a particular correlation coefficient
corresponds are identified by their item numbers in 
Table~\ref{tab:rdata}.}
\label{tab:rdcc}
\def\sp{\hbox to 50 pt{}}
\begin{tabular}{ r @{~} l @{\sp} r @{~} l @{\sp} r @{~} l @{\sp} r @{~} l }
\toprule
\vbox to 10 pt {}
$r$($A1$, $A2$)& = $ 0.9958 $ & $r$($A6$, $A19$)& = $ 0.8599 $ & $r$($A27$, $A28$)& = $ 0.3478 $ & $r$($A30$, $A44$)& = $ 0.1136 $ \\ 
$r$($A1$, $A3$)& = $ 0.9955 $ & $r$($A6$, $A20$)& = $ 0.9913 $ & $r$($A27$, $A29$)& = $ 0.4532 $ & $r$($A31$, $A34$)& = $ 0.0278 $ \\ 
$r$($A1$, $A4$)& = $ 0.9943 $ & $r$($A7$, $A8$)& = $ 0.0043 $ & $r$($A27$, $A30$)& = $ 0.0899 $ & $r$($A31$, $A35$)& = $ 0.0553 $ \\ 
$r$($A1$, $A5$)& = $ 0.8720 $ & $r$($A9$, $A10$)& = $ 0.0043 $ & $r$($A27$, $A31$)& = $ 0.1206 $ & $r$($A31$, $A40$)& = $ 0.1512 $ \\ 
$r$($A1$, $A6$)& = $ 0.8711 $ & $r$($A11$, $A12$)& = $ 0.0005 $ & $r$($A27$, $A34$)& = $ 0.0225 $ & $r$($A31$, $A41$)& = $ 0.1647 $ \\ 
$r$($A1$, $A17$)& = $ 0.9887 $ & $r$($A11$, $A21$)& = $ 0.9999 $ & $r$($A27$, $A35$)& = $ 0.0448 $ & $r$($A31$, $A42$)& = $ 0.1750 $ \\ 
$r$($A1$, $A18$)& = $ 0.9846 $ & $r$($A11$, $A22$)& = $ 0.0003 $ & $r$($A27$, $A40$)& = $ 0.1225 $ & $r$($A31$, $A43$)& = $ 0.1209 $ \\ 
$r$($A1$, $A19$)& = $ 0.9830 $ & $r$($A12$, $A21$)& = $ 0.0003 $ & $r$($A27$, $A41$)& = $ 0.1335 $ & $r$($A31$, $A44$)& = $ 0.1524 $ \\ 
$r$($A1$, $A20$)& = $ 0.8544 $ & $r$($A12$, $A22$)& = $ 0.9999 $ & $r$($A27$, $A42$)& = $ 0.1419 $ & $r$($A32$, $A33$)& = $ 0.1049 $ \\ 
$r$($A2$, $A3$)& = $ 0.9954 $ & $r$($A13$, $A14$)& = $ 0.0005 $ & $r$($A27$, $A43$)& = $ 0.0980 $ & $r$($A32$, $A45$)& = $ 0.2095 $ \\ 
$r$($A2$, $A4$)& = $ 0.9942 $ & $r$($A13$, $A15$)& = $ 0.0005 $ & $r$($A27$, $A44$)& = $ 0.1235 $ & $r$($A32$, $A46$)& = $ 0.0404 $ \\ 
$r$($A2$, $A5$)& = $ 0.8719 $ & $r$($A13$, $A16$)& = $ 0.0004 $ & $r$($A28$, $A29$)& = $ 0.4696 $ & $r$($A33$, $A45$)& = $ 0.0271 $ \\ 
$r$($A2$, $A6$)& = $ 0.8710 $ & $r$($A13$, $A23$)& = $ 0.9999 $ & $r$($A28$, $A30$)& = $ 0.0934 $ & $r$($A33$, $A46$)& = $ 0.0467 $ \\ 
$r$($A2$, $A17$)& = $ 0.9846 $ & $r$($A13$, $A24$)& = $ 0.0002 $ & $r$($A28$, $A31$)& = $ 0.1253 $ & $r$($A34$, $A35$)& = $ 0.1412 $ \\ 
$r$($A2$, $A18$)& = $ 0.9887 $ & $r$($A13$, $A25$)& = $ 0.0002 $ & $r$($A28$, $A34$)& = $ 0.0234 $ & $r$($A34$, $A40$)& = $ 0.0282 $ \\ 
$r$($A2$, $A19$)& = $ 0.9829 $ & $r$($A14$, $A15$)& = $ 0.0005 $ & $r$($A28$, $A35$)& = $ 0.0466 $ & $r$($A34$, $A41$)& = $ 0.0307 $ \\ 
$r$($A2$, $A20$)& = $ 0.8543 $ & $r$($A14$, $A16$)& = $ 0.0005 $ & $r$($A28$, $A40$)& = $ 0.1273 $ & $r$($A34$, $A42$)& = $ 0.0327 $ \\ 
$r$($A3$, $A4$)& = $ 0.9939 $ & $r$($A14$, $A23$)& = $ 0.0002 $ & $r$($A28$, $A41$)& = $ 0.1387 $ & $r$($A34$, $A43$)& = $ 0.0226 $ \\ 
$r$($A3$, $A5$)& = $ 0.8717 $ & $r$($A14$, $A24$)& = $ 0.0003 $ & $r$($A28$, $A42$)& = $ 0.1475 $ & $r$($A34$, $A44$)& = $ 0.0284 $ \\ 
$r$($A3$, $A6$)& = $ 0.8708 $ & $r$($A14$, $A25$)& = $ 0.0002 $ & $r$($A28$, $A43$)& = $ 0.1019 $ & $r$($A35$, $A40$)& = $ 0.0561 $ \\ 
$r$($A3$, $A17$)& = $ 0.9843 $ & $r$($A15$, $A16$)& = $ 0.0005 $ & $r$($A28$, $A44$)& = $ 0.1284 $ & $r$($A35$, $A41$)& = $ 0.0612 $ \\ 
$r$($A3$, $A18$)& = $ 0.9842 $ & $r$($A15$, $A23$)& = $ 0.0002 $ & $r$($A29$, $A30$)& = $ 0.1209 $ & $r$($A35$, $A42$)& = $ 0.0650 $ \\ 
$r$($A3$, $A19$)& = $ 0.9827 $ & $r$($A15$, $A24$)& = $ 0.9999 $ & $r$($A29$, $A31$)& = $ 0.1622 $ & $r$($A35$, $A43$)& = $ 0.0449 $ \\ 
$r$($A3$, $A20$)& = $ 0.8541 $ & $r$($A15$, $A25$)& = $ 0.0002 $ & $r$($A29$, $A34$)& = $ 0.0303 $ & $r$($A35$, $A44$)& = $ 0.0566 $ \\ 
$r$($A4$, $A5$)& = $ 0.8706 $ & $r$($A16$, $A23$)& = $ 0.0002 $ & $r$($A29$, $A35$)& = $ 0.0602 $ & $r$($A36$, $A37$)& = $ 0.0834 $ \\ 
$r$($A4$, $A6$)& = $ 0.8698 $ & $r$($A16$, $A24$)& = $ 0.0002 $ & $r$($A29$, $A40$)& = $ 0.1648 $ & $r$($A40$, $A41$)& = $ 0.5699 $ \\ 
$r$($A4$, $A17$)& = $ 0.9831 $ & $r$($A16$, $A25$)& = $ 0.9999 $ & $r$($A29$, $A41$)& = $ 0.1795 $ & $r$($A40$, $A42$)& = $ 0.6117 $ \\ 
$r$($A4$, $A18$)& = $ 0.9830 $ & $r$($A17$, $A18$)& = $ 0.9958 $ & $r$($A29$, $A42$)& = $ 0.1908 $ & $r$($A40$, $A43$)& = $ 0.1229 $ \\ 
$r$($A4$, $A19$)& = $ 0.9888 $ & $r$($A17$, $A19$)& = $ 0.9942 $ & $r$($A29$, $A43$)& = $ 0.1319 $ & $r$($A40$, $A44$)& = $ 0.1548 $ \\ 
$r$($A4$, $A20$)& = $ 0.8530 $ & $r$($A17$, $A20$)& = $ 0.8641 $ & $r$($A29$, $A44$)& = $ 0.1662 $ & $r$($A41$, $A42$)& = $ 0.6667 $ \\ 
$r$($A5$, $A6$)& = $ 0.7628 $ & $r$($A18$, $A19$)& = $ 0.9941 $ & $r$($A30$, $A31$)& = $ 0.4750 $ & $r$($A41$, $A43$)& = $ 0.1339 $ \\ 
$r$($A5$, $A17$)& = $ 0.8622 $ & $r$($A18$, $A20$)& = $ 0.8640 $ & $r$($A30$, $A34$)& = $ 0.0207 $ & $r$($A41$, $A44$)& = $ 0.1687 $ \\ 
$r$($A5$, $A18$)& = $ 0.8621 $ & $r$($A19$, $A20$)& = $ 0.8627 $ & $r$($A30$, $A35$)& = $ 0.0412 $ & $r$($A42$, $A43$)& = $ 0.1423 $ \\ 
$r$($A5$, $A19$)& = $ 0.8607 $ & $r$($A21$, $A22$)& = $ 0.0001 $ & $r$($A30$, $A40$)& = $ 0.1127 $ & $r$($A42$, $A44$)& = $ 0.1793 $ \\ 
$r$($A5$, $A20$)& = $ 0.7481 $ & $r$($A23$, $A24$)& = $ 0.0001 $ & $r$($A30$, $A41$)& = $ 0.1228 $ & $r$($A43$, $A44$)& = $ 0.5224 $ \\ 
$r$($A6$, $A17$)& = $ 0.8613 $ & $r$($A23$, $A25$)& = $ 0.0001 $ & $r$($A30$, $A42$)& = $ 0.1305 $ & $r$($A45$, $A46$)& = $ 0.0110 $ \\ 
$r$($A6$, $A18$)& = $ 0.8612 $ & $r$($A24$, $A25$)& = $ 0.0001 $ & $r$($A30$, $A43$)& = $ 0.0901 $ & &  \\ 
\botrule
\end{tabular}
\end{table*}

\clearpage

 \def\Narahjc{$B1   $}
 \def\Narbhjc{$B2.1 $}
 \def\Narbhjf{$B2.2 $}
 \def\Narchjf{$B3   $}
 \def\Narchejf{$B4   $}
 \def\Nardhejf{$B5   $}
 \def\Narafojf{$B6   $}
 \def\Narhgrbjc{$B7   $}
 \def\Naracccsjc{$B8   $}
 \def\Nareie{$B9   $}
 \def\Ncaeth{$B10  $}
 \def\Naeuwashhg{$B11.1$}
 \def\Naeharvjf{$B11.2$}
 \def\Ncamth{$B12  $}
 \def\Nrbarjf{$B13  $}
 \def\Ncgcth{$B14  $}
 \def\Ncgoth{$B15  $}
 \def\Nrfsfccjb{$B16  $}
 \def\Nrfsfcojb{$B17  $}
 \def\Nmehmphmitgb{$B18  $}
 \def\Nmddmedmitgh{$B19  $}
 \def\Nmuphdsmudhdjc{$B20  $}
 \def\Nsdpjc{$B21  $}
 \def\Nmuthtsmuphtjc{$B22  $}
 \def\Nstpjc{$B23  $}
 \def\Nmuehmuppmitgg{$B24  $}
 \def\Nmuhecmuppnplic{$B25  $}
 \def\Nmunmupphargi{$B26  $}
 \def\Ncmuhfsth{$B27  $}
 \def\NnupLhbkHz{$B28.1$}
 \def\NnupLii{$B28.2$}
 \def\NnumLhbkHz{$B29  $}
 \def\NnumLii{$B30  $}
 \def\Ngppnisthi{$B31.1$}
 \def\Ngpplnimie{$B31.2$}
 \def\Nghkvig{$B32  $}
 \def\Ngpphnimie{$B33.1$}
 \def\Ngppnplgi{$B33.2$}
 \def\Nrknistig{$B34.1$}
 \def\Nrknmlig{$B34.2$}
 \def\Nrknplhh{$B34.3$}
 \def\Nrknimie{$B34.4$}
 \def\Nrklcieja{$B34.5$}
 \def\NkjnmlhiGHz{$B35.1$}
 \def\NkjptbiaGHz{$B35.2$}
 \def\Nkjbrknplij{$B36.1$}
 \def\Nkjbrknistih{$B36.2$}
 \def\Nkjbrknistjg{$B36.3$}
 \def\Nfijnisthj{$B37  $}
 \def\Ndwdbptbha{$B38.1$}
 \def\Ndwdbainrimptbjg{$B38.2$}
 \def\Ndnrcnmijjd{$B39  $}
 \def\Ndmodinrimjg{$B40  $}
 \def\Ndrwagn{$B41  $}
 \def\Ndrillwag{$B42  $}
 \def\Ndrillmod{$B43  $}
 \def\Ndrillnrc{$B44  $}
 \def\Ndrnrcwjdhk{$B45  $}
 \def\Ndrnrdwjdhk{$B46  $}
 \def\Ndrwagwjdhk{$B47  $}
 \def\Ndrwdbwjd{$B48  $}
 \def\Ndrwagwjd{$B49  $}
 \def\Ndrmodwjd{$B50  $}
 \def\Ndrnrcwjd{$B51  $}
 \def\Nddwjd{$B52  $}
 \def\Nvmsilwgacje{$B53  $}
 \def\Nladillii{$B54  $}
 \def\Nhomndwjd{$B55  $}
 \def\Nhomcsjb{$B56  $}
 \def\Nhomrbjf{$B57  $}
 \def\Nrnisthh{$B58.1$}
 \def\Nrnplgi{$B58.2$}
 \def\Nlcusdwdbaptb{$B59  $}
 \def\Nlamnsdwnist{$B60  $}
 \def\Nlamnsdmonist{$B61  $}
 \def\Nlamnsdcunist{$B62  $}
\def\m{\phantom{-}}
\def\fixh{\vbox to 9pt {}}
\begin{longtable*}{ll@{\qquad}llll}
\caption{Summary of principal input data for the determination of the 2006
recommended values of the fundamental constants ($R_\infty$ and $G$ excepted).
\label{tab:pdata}} \\
\toprule
\vbox to 10 pt {}
Item & Input datum  &  \hbox to 15pt {} Value
& \hbox to -20pt {} Relative standard
& Identification & Sec. and Eq. \\
number& & & \hbox to -5pt {}uncertainty$^a$
$u_{\rm r}$ &  \\
\colrule
\endfirsthead

\caption{{\it (Continued).}
Summary of principal input data for the
determination of the 2006
recommended values of the fundamental constants ($R_\infty$ and $G$ excepted).} \\
\colrule
\vbox to 10 pt {}
Item & Input datum  &  \hbox to 15pt {} Value
& \hbox to -20pt {} Relative standard
& Identification & Sec. and Eq. \\
number& & & \hbox to -5pt {}uncertainty$^a$
$u_{\rm r}$ &  \\
\colrule
\endhead
\colrule
\endfoot
\endlastfoot
 $B1   $ &\fixh $A_{\rm r}(^{1}{\rm H})$ \vbox to 10pt{}    & $\phantom{-}  1.007\,825\,032\,07(10)$ &  $ 1.0\times 10^{-10}$ & AMDC-03 & \ref{ssec:rama} \\
 $B2.1 $ &\fixh $A_{\rm r}(^{2}{\rm H})$     & $\phantom{-}  2.014\,101\,777\,85(36)$ & $ 1.8\times 10^{-10}$ & AMDC-03 & \ref{ssec:rama} \\
 $B2.2 $ &\fixh $A_{\rm r}(^{2}{\rm H})$     & $\phantom{-}  2.014\,101\,778\,040(80)$ & $ 4.0\times 10^{-11}$ & UWash-06 & \ref{ssec:rama} \\
 $B3   $ &\fixh $A_{\rm r}(^{3}{\rm H})$     & $\phantom{-}  3.016\,049\,2787(25)$ & $ 4.0\times 10^{-11}$ & MSL-06 & \ref{ssec:rama} \\
 $B4   $ &\fixh $A_{\rm r}(^{3}{\rm He})$    & $\phantom{-}  3.016\,029\,3217(26)$ & $ 8.6\times 10^{-10}$ & MSL-06 & \ref{ssec:rama} \\
 $B5   $ &\fixh $A_{\rm r}(^{4}{\rm He})$    & $\phantom{-}  4.002\,603\,254\,131(62)$ & $ 1.5\times 10^{-11}$ & UWash-06 & \ref{ssec:rama} \\
 $B6   $ &\fixh $A_{\rm r}(^{16}{\rm O})$    & $\phantom{-}  15.994\,914\,619\,57(18)$ & $ 1.1\times 10^{-11}$ & UWash-06 & \ref{ssec:rama} \\
 $B7   $ &\fixh $A_{\rm r}(^{87}{\rm Rb})$    & $\phantom{-}  86.909\,180\,526(12)$ & $ 1.4\times 10^{-10}$ & AMDC-03 & \ref{ssec:rama} \\
 $B8^b$ &\fixh $A_{\rm r}(^{133}{\rm Cs})$    & $\phantom{-}  132.905\,451\,932(24)$ & $ 1.8\times 10^{-10}$ & AMDC-03 & \ref{ssec:rama} \\
 $B9   $ &\fixh $A_{\rm r}({\rm e})$         & $\phantom{-}  0.000\,548\,579\,9111(12)$  & $ 2.1\times 10^{-9}$ & UWash-95 & \ref{ssec:ptmare} (\ref{eq:areexp}) \\
 $B10  $ &\fixh $\delta_{\rm e}$ & $\phantom{-}  0.00(27)\times 10^{-12}$ & [$ 2.4\times 10^{-10}$] & theory & \ref{sssec:ath} (\ref{eq:caeth}) \\
 $B11.1$ &\fixh $a_{\rm e}$ & $\phantom{-}  1.159\,652\,1883(42)\times 10^{-3}$ & $ 3.7\times 10^{-9}$ & UWash-87 & \ref{par:aeuw} (\ref{eq:aeuwash}) \\
 $B11.2$ &\fixh $a_{\rm e}$ & $\phantom{-}  1.159\,652\,180\,85(76)\times 10^{-3}$ & $ 6.6\times 10^{-10}$ & HarvU-06 & \ref{par:harvard} (\ref{eq:aeharv06}) \\
 $B12  $ &\fixh $\delta_{\rmssmu}$ & $\phantom{-}  0.0(2.1)\times 10^{-9}$ & [$ 1.8\times 10^{-6}$] & theory & \ref{sssec:amuth} (\ref{eq:camth}) \\
 $B13  $ &\fixh $\overline{R}$ & $\phantom{-} 0.003\,707\,2064(20)$ & $ 5.4\times 10^{-7}$ & BNL-06 & \ref{sssec:amb} (\ref{eq:rbar06}) \\
 $B14  $ &\fixh $\delta_{\rm C}$ & $\phantom{-}  0.00(27)\times 10^{-10}$ & [$ 1.4\times 10^{-11}$] & theory & \ref{sssec:thbegf} (\ref{eq:cgcth}) \\
 $B15  $ &\fixh $\delta_{\rm O}$ & $\phantom{-}  0.0(1.1)\times 10^{-10}$ & [$ 5.3\times 10^{-11}$] & theory & \ref{sssec:thbegf} (\ref{eq:cgoth}) \\
 $B16  $ &\fixh $f_{\rm s}({\rm ^{12}C^{5+}})/f_{\rm c}({\rm ^{12}C^{5+}})$ & $\phantom{-} 4376.210\,4989(23)$ & $ 5.2\times 10^{-10}$ & GSI-02 & \ref{par:gec} (\ref{eq:rfsfcc02}) \\
 $B17  $ &\fixh $f_{\rm s}({\rm ^{16}O^{7+}})/f_{\rm c}({\rm ^{16}O^{7+}})$ & $\phantom{-} 4164.376\,1837(32)$ & $ 7.6\times 10^{-10}$ & GSI-02 & \ref{par:geo} (\ref{eq:rfsfco02}) \\
 $B18  $ &\fixh $\mu_{\rm e^-}({\rm H})/\mu_{\rm p}({\rm H})$ & $ -658.210\,7058(66)$ & $ 1.0\times 10^{-8}$ & MIT-72 & \ref{par:epmmr} (\ref{eq:wkmw}) \\
 $B19  $ &\fixh $\mu_{\rm d}({\rm D})/\mu_{\rm e^-}({\rm D})$ & $ -4.664\,345\,392(50)\times 10^{-4}$ & $ 1.1\times 10^{-8}$ & MIT-84 & \ref{par:demmr} (\ref{eq:mddmedmit78}) \\
 $B20  $ &\fixh $\mu_{\rm p}({\rm HD})/\mu_{\rm d}({\rm HD})$ & $\phantom{-}  3.257\,199\,531(29)$ & $ 8.9\times 10^{-9}$ & StPtrsb-03 & \ref{par:dpmmr} (\ref{eq:muphd}) \\
 $B21  $ &\fixh $\sigma_{\rm dp}$ & $\phantom{-}  15(2)\times 10^{-9}$ & & StPtrsb-03 & \ref{par:dpmmr} (\ref{eq:sdp03}) \\
 $B22  $ &\fixh $\mu_{\rm t}({\rm HT})/\mu_{\rm p}({\rm HT})$ & $\phantom{-}  1.066\,639\,887(10)$ & $ 9.4\times 10^{-9}$ & StPtrsb-03 & \ref{par:dpmmr} (\ref{eq:mutht}) \\
 $B23  $ &\fixh $\sigma_{\rm tp}$ & $\phantom{-}  20(3)\times 10^{-9}$ & & StPtrsb-03 & \ref{par:dpmmr} (\ref{eq:stp03}) \\
 $B24  $ &\fixh $\mu_{\rm e^-}({\rm H})/\mu_{\rm p}^\prime$ & $ -658.215\,9430(72)$ & $ 1.1\times 10^{-8}$ & MIT-77 & \ref{par:espmm} (\ref{eq:muehmuppmit77}) \\
 $B25  $ &\fixh $\mu_{\rm h}^\prime/\mu_{\rm p}^\prime$ & $ -0.761\,786\,1313(33)$ & $ 4.3\times 10^{-9}$ & NPL-93 & \ref{par:hspmmr} (\ref{eq:muhe3muppnpl93}) \\
 $B26  $ &\fixh $\mu_{\rm n}/\mu_{\rm p}^\prime$ & $ -0.684\,996\,94(16)$ & $ 2.4\times 10^{-7}$ & ILL-79 & \ref{par:nsprmr} (\ref{eq:munmupphar79}) \\
 $B27  $ &\fixh $\delta_{\rm Mu}$ & $\phantom{-}  0(101)$ Hz & [$ 2.3\times 10^{-8}$] & theory & \ref{sssec:muhfs} (\ref{eq:cmu}) \\
 $B28.1$ &\fixh $\Delta\nu_{\rm Mu}$ & $\phantom{-} 4\,463\,302.88(16)$ kHz & $ 3.6\times 10^{-8}$ & LAMPF-82 & \ref{par:lampf1982} (\ref{eq:nupL82}) \\
 $B28.2$ &\fixh $\Delta\nu_{\rm Mu}$ & $\phantom{-} 4\,463\,302\,765(53)$ Hz & $ 1.2\times 10^{-8}$ & LAMPF-99 & \ref{par:lampf1999} (\ref{eq:nupL99}) \\
 $B29  $ &\fixh $\nu(58~{\rm MHz})$ & $\phantom{-} 627\,994.77(14)$ kHz & $ 2.2\times 10^{-7}$ & LAMPF-82 & \ref{par:lampf1982} (\ref{eq:numL82}) \\
 $B30  $ &\fixh $\nu(72~{\rm MHz})$ & $\phantom{-} 668\,223\,166(57)$ Hz & $ 8.6\times 10^{-8}$ & LAMPF-99 & \ref{par:lampf1999} (\ref{eq:numL99}) \\
 $B31.1^b$ &\fixh ${\it\Gamma}_{\rm p-90}^{\,\prime}({\rm lo})$ & $\phantom{-}  2.675\,154\,05(30)\times 10^{8}$ s$^{-1}$ T$^{-1}$ & $ 1.1\times 10^{-7}$ & NIST-89 & \ref{par:nistlf} (\ref{eq:gppnist89}) \\
 $B31.2^b$ &\fixh ${\it\Gamma}_{\rm p-90}^{\,\prime}({\rm lo})$ & $\phantom{-}  2.675\,1530(18)\times 10^{8}$ s$^{-1}$ T$^{-1}$ & $ 6.6\times 10^{-7}$ & NIM-95 & \ref{par:nimlo} (\ref{eq:gpplnim95}) \\
 $B32^b  $ &\fixh ${\it\Gamma}_{\rm h-90}^{\,\prime}({\rm lo})$ & $\phantom{-}  2.037\,895\,37(37)\times 10^{8}$ s$^{-1}$ T$^{-1}$ & $ 1.8\times 10^{-7}$ & KR/VN-98 & \ref{par:kvlf} (\ref{eq:ghkv97}) \\
 $B33.1^b$ &\fixh ${\it\Gamma}_{\rm p-90}^{\,\prime}({\rm hi})$ & $\phantom{-}  2.675\,1525(43)\times 10^{8}$ s$^{-1}$ T$^{-1}$ & $ 1.6\times 10^{-6}$ & NIM-95 & \ref{par:nimhi} (\ref{eq:gpphnim95}) \\
 $B33.2^b$ &\fixh ${\it\Gamma}_{\rm p-90}^{\,\prime}({\rm hi})$ & $\phantom{-}  2.675\,1518(27)\times 10^{8}$ s$^{-1}$ T$^{-1}$ & $ 1.0\times 10^{-6}$ & NPL-79 & \ref{par:nplhf} (\ref{eq:gppnpl79}) \\
 $B34.1^b$ &\fixh $R_{\rm K}$ & $\phantom{-}  25\,812.808\,31(62) \ {\rm \Omega}$ & $ 2.4\times 10^{-8}$ & NIST-97 & \ref{sssec:nistvkc} (\ref{eq:rknist97}) \\
 $B34.2^b$ &\fixh $R_{\rm K}$ & $\phantom{-}  25\,812.8071(11) \ {\rm \Omega}$ & $ 4.4\times 10^{-8}$ & NMI-97 & \ref{sssec:nmlvkc} (\ref{eq:rknml97}) \\
 $B34.3^b$ &\fixh $R_{\rm K}$ & $\phantom{-}  25\,812.8092(14) \ {\rm \Omega}$ & $ 5.4\times 10^{-8}$ & NPL-88 & \ref{sssec:nplvkc} (\ref{eq:rknpl88}) \\
 $B34.4^b$ &\fixh $R_{\rm K}$ & $\phantom{-}  25\,812.8084(34) \ {\rm \Omega}$ & $ 1.3\times 10^{-7}$ & NIM-95 & \ref{sssec:nimvkc} (\ref{eq:rknim95}) \\
 $B34.5^b$ &\fixh $R_{\rm K}$ & $\phantom{-}  25\,812.8081(14) \ {\rm \Omega}$ & $ 5.3\times 10^{-8}$ & LNE-01 & \ref{sssec:lcievkc} (\ref{eq:rklcie01}) \\
 $B35.1^b$ &\fixh $K_{\rm J}$  & $\phantom{-}  483\,597.91(13) \ {\rm GHz \ V^{-1}}$ & $ 2.7\times 10^{-7}$ & NMI-89 & \ref{sssec:nmljc} (\ref{eq:kjnml89}) \\
 $B35.2^b$ &\fixh $K_{\rm J}$ & $\phantom{-}  483\,597.96(15) \ {\rm GHz \ V^{-1}}$ & $ 3.1\times 10^{-7}$ & PTB-91 & \ref{sssec:ptbjc} (\ref{eq:kjptb91}) \\
 $B36.1^c$ &\fixh $K_{\rm J}^2R_{\rm K}$ & $\phantom{-}  6.036\,7625(12)\times 10^{33}\ {\rm J^{-1} \ s^{-1}}$ & $ 2.0\times 10^{-7}$ & NPL-90 & \ref{sssec:kj2rknpl} (\ref{eq:kj2rknpl90}) \\
 $B36.2^c$ &\fixh $K_{\rm J}^2R_{\rm K}$ & $\phantom{-}  6.036\,761\,85(53)\times 10^{33}\ {\rm J^{-1} \ s^{-1}}$ & $ 8.7\times 10^{-8}$ & NIST-98 & \ref{par:kj2rk98} (\ref{eq:kj2rknist98}) \\
 $B36.3^c$ &\fixh $K_{\rm J}^2R_{\rm K}$ & $\phantom{-}  6.036\,761\,85(22)\times 10^{33}\ {\rm J^{-1} \ s^{-1}}$ & $ 3.6\times 10^{-8}$ & NIST-07 & \ref{par:kj2rk07} (\ref{eq:kj2rknist07}) \\
 $B37^b  $ &\fixh ${\cal F}_{90}$ & $\phantom{-}  96\,485.39(13)\ {\rm C \ mol^{-1}}$ & $ 1.3\times 10^{-6}$ & NIST-80 & \ref{sssec:fcnist} (\ref{eq:f90nist80}) \\
 $B38.1^c$ &\fixh $d_{220}({\rm {\scriptstyle{W4.2a}}})$ & $\phantom{-}  192\,015.563(12)$ fm & $ 6.2\times 10^{-8}$ & PTB-81 & \ref{par:ptblat} (\ref{eq:dw42ptb81}) \\
 $B38.2^c$ &\fixh $d_{220}({\rm {\scriptstyle{W4.2a}}})$ & $\phantom{-}  192\,015.5715(33)$ fm & $ 1.7\times 10^{-8}$ & INRIM-07 & \ref{par:inrimlat} (\ref{eq:dw42ainrimptb07}) \\
 $B39^c  $ &\fixh $d_{220}({\rm {\scriptstyle{NR3}}})$ & $\phantom{-}  192\,015.5919(76)$ fm & $ 4.0\times 10^{-8}$ & NMIJ-04 & \ref{par:nmijlat} (\ref{eq:dnr3nmij04}) \\
 $B40^c  $ &\fixh $d_{220}({\rm {\scriptstyle{MO^*}}})$ & $\phantom{-}  192\,015.5498(51)$ fm & $ 2.6\times 10^{-8}$ & INRIM-07 & \ref{par:inrimlat} (\ref{eq:dmo4inrim07}) \\
 $B41  $ &\fixh $1-d_{220}({\rm {\scriptstyle{N}}})/d_{220}({\rm {\scriptstyle W17}})$  & $ \phantom{-}  7(22)\times 10^{-9}$ & & NIST-97 & \ref{par:nistdiffs} (\ref{eq:drw17n}) \\
 $B42  $ &\fixh $1-d_{220}({\rm {\scriptstyle{W17}}})/d_{220}({\rm {\scriptstyle ILL}})$  & $  -8(22)\times 10^{-9}$ & & NIST-99 & \ref{par:nistdiffs} (\ref{eq:drillw17}) \\
 $B43  $ &\fixh $1-d_{220}({\rm {\scriptstyle{MO^*}}})/d_{220}({\rm {\scriptstyle ILL}})$  & $\phantom{-}  86(27)\times 10^{-9}$ & & NIST-99 & \ref{par:nistdiffs} (\ref{eq:drillmo4}) \\
 $B44  $ &\fixh $1-d_{220}({\rm {\scriptstyle{NR3}}})/d_{220}({\rm {\scriptstyle ILL}})$  & $\phantom{-}  34(22)\times 10^{-9}$ & & NIST-99 & \ref{par:nistdiffs} (\ref{eq:drillnr3}) \\
 $B45  $ &\fixh $d_{220}({\rm {\scriptstyle{NR3}}})/d_{220}({\rm {\scriptstyle W04}})-1$  & $  -11(21)\times 10^{-9}$ & & NIST-06 & \ref{par:nistdiffs} (\ref{eq:drnr3w04hk}) \\
 $B46  $ &\fixh $d_{220}({\rm {\scriptstyle{NR4}}})/d_{220}({\rm {\scriptstyle W04}})-1$  & $ \phantom{-}  25(21)\times 10^{-9}$ & & NIST-06 & \ref{par:nistdiffs} (\ref{eq:drnr4w04hk}) \\
 $B47  $ &\fixh $d_{220}({\rm {\scriptstyle{W17}}})/d_{220}({\rm {\scriptstyle W04}})-1$  & $\phantom{-}  11(21)\times 10^{-9}$ & & NIST-06 & \ref{par:nistdiffs} (\ref{eq:drw17w04hk}) \\
 $B48  $ &\fixh $d_{220}({\rm {\scriptstyle{W4.2a}}})/d_{220}({\rm {\scriptstyle W04}})-1$  & $  -1(21)\times 10^{-9}$ & & PTB-98 & \ref{par:ptbdiffs} (\ref{eq:drw42w04}) \\
 $B49  $ &\fixh $d_{220}({\rm {\scriptstyle{W17}}})/d_{220}({\rm {\scriptstyle W04}})-1$  & $\phantom{-}  22(22)\times 10^{-9}$ & & PTB-98 & \ref{par:ptbdiffs} (\ref{eq:drw17w04}) \\
 $B50  $ &\fixh $d_{220}({\rm {\scriptstyle{MO^*}}})/d_{220}({\rm {\scriptstyle W04}})-1$  & $  -103(28)\times 10^{-9}$ & & PTB-98 & \ref{par:ptbdiffs} (\ref{eq:drmo4w04}) \\
 $B51  $ &\fixh $d_{220}({\rm {\scriptstyle{NR3}}})/d_{220}({\rm {\scriptstyle W04}})-1$  & $  -23(21)\times 10^{-9}$ & & PTB-98 & \ref{par:ptbdiffs} (\ref{eq:drnr3w04}) \\
 $B52  $ &\fixh $d_{220}/d_{220}({\rm {\scriptstyle W04}})-1$  & $\phantom{-}  10(11)\times 10^{-9}$ & & PTB-03 & \ref{par:ptbdiffs} (\ref{eq:ddw04}) \\
 $B53^c  $ &\fixh $V_{\rm m}({\rm Si})$ & $\phantom{-}  12.058\,8254(34)\times 10^{-6}$ m$^3$ mol$^{-1}$ & $ 2.8\times 10^{-7}$ & N/P/I-05 & \ref{ssec:mvsi} (\ref{eq:vmsilwgac05}) \\
 $B54  $ &\fixh $\lambda_{\rm meas}/d_{220}({\rm {\scriptstyle ILL}})$ & $\phantom{-}  0.002\,904\,302\,46(50)$ m s$^{-1}$ & $ 1.7\times 10^{-7}$ & NIST-99 & \ref{ssec:arn} (\ref{eq:ladill99}) \\
 $B55^c  $ &\fixh $h/m_{\rm n}d_{220}({\rm {\scriptstyle W04}})$ & $\phantom{-}  2060.267\,004(84)$ m s$^{-1}$ & $ 4.1\times 10^{-8}$ & PTB-99 & \ref{sssec:pcnmr} (\ref{eq:homndw04}) \\
 $B56^b  $ &\fixh $h/m({\rm ^{133}Cs})$ & $\phantom{-}  3.002\,369\,432(46)\times 10^{-9}$ m$^2$ s$^{-1}$ & $ 1.5\times 10^{-8}$ & StanfU-02 & \ref{sssec:pccsmr} (\ref{eq:homcs02}) \\
 $B57  $ &\fixh $h/m({\rm ^{87}Rb})$ & $\phantom{-}  4.591\,359\,287(61)\times 10^{-9}$ m$^2$ s$^{-1}$ & $ 1.3\times 10^{-8}$ & LKB-06 & \ref{sssec:pcrbmr} (\ref{eq:homrb06}) \\
 $B58.1$ &\fixh $R$ & $\phantom{-}  8.314\,471(15)$ J mol$^{-1}$ K$^{-1}$  & $ 1.8\times 10^{-6}$ & NIST-88 & \ref{sssec:mgcnist} (\ref{eq:rnist88}) \\
 $B58.2$ &\fixh $R$ & $\phantom{-}  8.314\,504(70)$ J mol$^{-1}$ K$^{-1}$ & $ 8.4\times 10^{-6}$ & NPL-79 & \ref{sssec:mgcnpl} (\ref{eq:rnpl79}) \\
 $B59  $ &\fixh $\lambda({\rm CuK\alpha_1})/d_{220}({\rm {\scriptstyle{W4.2a}}})$ & $\phantom{-}  0.802\,327\,11(24)$ & $ 3.0\times 10^{-7}$ & FSU/PTB-91 & \ref{ssec:xru} (\ref{eq:lcusdw42aptb}) \\
 $B60  $ &\fixh $\lambda({\rm WK\alpha_1})/d_{220}({\rm {\scriptstyle{N}}})$ & $\phantom{-}  0.108\,852\,175(98)$ & $ 9.0\times 10^{-7}$ & NIST-79 & \ref{ssec:xru} (\ref{eq:lamnsdwnist}) \\
 $B61  $ &\fixh $\lambda({\rm MoK\alpha_1})/d_{220}({\rm {\scriptstyle{N}}})$ & $\phantom{-}  0.369\,406\,04(19)$ & $ 5.3\times 10^{-7}$ & NIST-73 & \ref{ssec:xru} (\ref{eq:lmodn73}) \\
 $B62  $ &\fixh $\lambda({\rm CuK\alpha_1})/d_{220}({\rm {\scriptstyle{N}}})$ & $\phantom{-}  0.802\,328\,04(77)$ & $ 9.6\times 10^{-7}$ & NIST-73 & \ref{ssec:xru} (\ref{eq:lcudn73}) \\
\botrule
\multicolumn{6}{l}{\footnotesize $^a$The values in brackets are relative
to the quantities $a_{\rm e}$, $a_{\rmssmu}$, $g_{\rm e^-}(^{12}{\rm C}^{5+})$,
$g_{\rm e^-}(^{16}{\rm O}^{7+})$, or $\Delta\nu_{\rm Mu}$ as appropriate.\fixh}
\\
\multicolumn{6}{l}{\footnotesize $^b$Datum not included in the final
least-squares adjustment that provides the recommended values of the
constants.} \\
\multicolumn{6}{l}{\footnotesize $^c$Datum included in the final
least-squares adjustment with an expanded uncertainty.} \\
\end{longtable*}

\begin{table*}
\caption{Non-negligible correlation coefficients $r(x_i,x_j)$ of the input data
in Table~\ref{tab:pdata}.  For simplicity, the two items of data to which a
particular correlation coefficient corresponds are identified by their item
numbers in Table~\ref{tab:pdata}.}
\label{tab:pdcc}
\def\sp{\hbox to 38 pt{}}
\begin{tabular}{ r @{~} l @{\sp} r @{~} l @{\sp} r @{~} l @{\sp} r @{~} l }
\toprule
\vbox to 10 pt {}
%
$r$($B1$, $B2.1$)& = $ 0.073 $ & $r$($B38.1$, $B38.2$)& = $ 0.191 $ & $r$($B42$, $B46$)& = $ 0.065 $ & $r$($B46$, $B47$)& = $ 0.509 $ \\ 
$r$($B2.2$, $B5$)& = $ 0.127 $ & $r$($B38.2$, $B40$)& = $ 0.057 $ & $r$($B42$, $B47$)& = $ -0.367 $ & $r$($B48$, $B49$)& = $ 0.469 $ \\ 
$r$($B2.2$, $B6$)& = $ 0.089 $ & $r$($B41$, $B42$)& = $ -0.288 $ & $r$($B43$, $B44$)& = $ 0.421 $ & $r$($B48$, $B50$)& = $ 0.372 $ \\ 
$r$($B5$, $B6$)& = $ 0.181 $ & $r$($B41$, $B43$)& = $ 0.096 $ & $r$($B43$, $B45$)& = $ 0.053 $ & $r$($B48$, $B51$)& = $ 0.502 $ \\ 
$r$($B14$, $B15$)& = $ 0.919 $ & $r$($B41$, $B44$)& = $ 0.117 $ & $r$($B43$, $B46$)& = $ 0.053 $ & $r$($B48$, $B55$)& = $ 0.258 $ \\ 
$r$($B16$, $B17$)& = $ 0.082 $ & $r$($B41$, $B45$)& = $ 0.066 $ & $r$($B43$, $B47$)& = $ 0.053 $ & $r$($B49$, $B50$)& = $ 0.347 $ \\ 
$r$($B28.1$, $B29$)& = $ 0.227 $ & $r$($B41$, $B46$)& = $ 0.066 $ & $r$($B44$, $B45$)& = $ -0.367 $ & $r$($B49$, $B51$)& = $ 0.469 $ \\ 
$r$($B28.2$, $B30$)& = $ 0.195 $ & $r$($B41$, $B47$)& = $ 0.504 $ & $r$($B44$, $B46$)& = $ 0.065 $ & $r$($B49$, $B55$)& = $ 0.241 $ \\ 
$r$($B31.2$, $B33.1$)& = $ -0.014 $ & $r$($B42$, $B43$)& = $ 0.421 $ & $r$($B44$, $B47$)& = $ 0.065 $ & $r$($B50$, $B51$)& = $ 0.372 $ \\ 
$r$($B35.1$, $B58.1$)& = $ 0.068 $ & $r$($B42$, $B44$)& = $ 0.516 $ & $r$($B45$, $B46$)& = $ 0.509 $ & $r$($B50$, $B55$)& = $ 0.192 $ \\ 
$r$($B36.2$, $B36.3$)& = $ 0.140 $ & $r$($B42$, $B45$)& = $ 0.065 $ & $r$($B45$, $B47$)& = $ 0.509 $ & $r$($B51$, $B55$)& = $ 0.258 $ \\ 
\botrule
\end{tabular}
\end{table*}

\def\m{\phantom{-}}
\def\fixh{\vbox to 9pt {}}
\def\vsp{\vbox to 10pt{}}
\begin{table*}[t]
\caption{Summary of principal input data for the determination of the relative
atomic mass of the electron from antiprotonic helium transitions.  The numbers
in parentheses $(n,l:n^\prime,l^\prime)$ denote the transition
$(n,l)\rightarrow(n^\prime,l^\prime)$.}
\label{tab:cdata}
\def\sp{\hbox to 33 pt{}}
\begin{tabular}{l@{\sp}l@{\sp}l@{\sp}l@{\sp}l@{\sp}l}
\toprule
Item \T &  Input Datum    &  \hbox to 15pt {} Value
& \hbox to -20pt {} Relative standard
& Identification & Sec. \\
number& & & \hbox to -5pt {}uncertainty\footnotemark[1]
 $u_{\rm r}$ &  \\
\colrule
 $C1   $ \vsp & $\delta_{\bar{\rm p}^4{\rm He}^+}(32,31:31,30)$ & $ 0.00(82)$  MHz & $[ 7.3\times 10^{-10}]$ & JINR-06 & \ref{ssec:aph}  \\
 $C2   $ & $\delta_{\bar{\rm p}^4{\rm He}^+}(35,33:34,32)$ & $ 0.0(1.0)$  MHz & $[ 1.3\times 10^{-9}]$ & JINR-06 & \ref{ssec:aph}  \\
 $C3   $ & $\delta_{\bar{\rm p}^4{\rm He}^+}(36,34:35,33)$ & $ 0.0(1.2)$  MHz & $[ 1.6\times 10^{-9}]$ & JINR-06 & \ref{ssec:aph}  \\
 $C4   $ & $\delta_{\bar{\rm p}^4{\rm He}^+}(39,35:38,34)$ & $ 0.0(1.1)$  MHz & $[ 1.8\times 10^{-9}]$ & JINR-06 & \ref{ssec:aph}  \\
 $C5   $ & $\delta_{\bar{\rm p}^4{\rm He}^+}(40,35:39,34)$ & $ 0.0(1.2)$  MHz & $[ 2.4\times 10^{-9}]$ & JINR-06 & \ref{ssec:aph}  \\
 $C6   $ & $\delta_{\bar{\rm p}^4{\rm He}^+}(32,31:31,30)$ & $ 0.0(1.3)$  MHz & $[ 2.9\times 10^{-9}]$ & JINR-06 & \ref{ssec:aph}  \\
 $C7   $ & $\delta_{\bar{\rm p}^4{\rm He}^+}(37,35:38,34)$ & $ 0.0(1.8)$  MHz & $[ 4.4\times 10^{-9}]$ & JINR-06 & \ref{ssec:aph}  \\
 $C8   $ & $\delta_{\bar{\rm p}^3{\rm He}^+}(32,31:31,30)$ & $ 0.00(91)$  MHz & $[ 8.7\times 10^{-10}]$ & JINR-06 & \ref{ssec:aph}  \\
 $C9   $ & $\delta_{\bar{\rm p}^3{\rm He}^+}(34,32:33,31)$ & $ 0.0(1.1)$  MHz & $[ 1.4\times 10^{-9}]$ & JINR-06 & \ref{ssec:aph}  \\
 $C10  $ & $\delta_{\bar{\rm p}^3{\rm He}^+}(36,33:35,32)$ & $ 0.0(1.2)$  MHz & $[ 1.8\times 10^{-9}]$ & JINR-06 & \ref{ssec:aph}  \\
 $C11  $ & $\delta_{\bar{\rm p}^3{\rm He}^+}(38,34:37,33)$ & $ 0.0(1.1)$  MHz & $[ 2.3\times 10^{-9}]$ & JINR-06 & \ref{ssec:aph}  \\
 $C12  $ & $\delta_{\bar{\rm p}^3{\rm He}^+}(36,34:37,33)$ & $ 0.0(1.8)$  MHz & $[ 4.4\times 10^{-9}]$ & JINR-06 & \ref{ssec:aph}  \\
 $C13  $ \vsp & $\nu_{\bar{\rm p}{\rm ^4He}^+}(32,31:31,30)$ & $ 1\,132\,609\,209(15)$  MHz & $ 1.4\times 10^{-8}$  & CERN-06 & \ref{ssec:aph}  \\
 $C14  $ & $\nu_{\bar{\rm p}{\rm ^4He}^+}(35,33:34,32)$ & $ 804\,633\,059.0(8.2)$  MHz & $ 1.0\times 10^{-8}$  & CERN-06 & \ref{ssec:aph}  \\
 $C15  $ & $\nu_{\bar{\rm p}{\rm ^4He}^+}(36,34:35,33)$ & $ 717\,474\,004(10)$  MHz & $ 1.4\times 10^{-8}$  & CERN-06 & \ref{ssec:aph}  \\
 $C16  $ & $\nu_{\bar{\rm p}{\rm ^4He}^+}(39,35:38,34)$ & $ 636\,878\,139.4(7.7)$  MHz & $ 1.2\times 10^{-8}$  & CERN-06 & \ref{ssec:aph}  \\
 $C17  $ & $\nu_{\bar{\rm p}{\rm ^4He}^+}(40,35:39,34)$ & $ 501\,948\,751.6(4.4)$  MHz & $ 8.8\times 10^{-9}$  & CERN-06 & \ref{ssec:aph}  \\
 $C18  $ & $\nu_{\bar{\rm p}{\rm ^4He}^+}(32,31:31,30)$ & $ 445\,608\,557.6(6.3)$  MHz & $ 1.4\times 10^{-8}$  & CERN-06 & \ref{ssec:aph}  \\
 $C19  $ & $\nu_{\bar{\rm p}{\rm ^4He}^+}(37,35:38,34)$ & $ 412\,885\,132.2(3.9)$  MHz & $ 9.4\times 10^{-9}$  & CERN-06 & \ref{ssec:aph}  \\
 $C20  $ & $\nu_{\bar{\rm p}{\rm ^3He}^+}(32,31:31,30)$ & $ 1\,043\,128\,608(13)$  MHz & $ 1.3\times 10^{-8}$  & CERN-06 & \ref{ssec:aph}  \\
 $C21  $ & $\nu_{\bar{\rm p}{\rm ^3He}^+}(34,32:33,31)$ & $ 822\,809\,190(12)$  MHz & $ 1.5\times 10^{-8}$  & CERN-06 & \ref{ssec:aph}  \\
 $C22  $ & $\nu_{\bar{\rm p}{\rm ^3He}^+}(36,33:34,32)$ & $ 646\,180\,434(12)$  MHz & $ 1.9\times 10^{-8}$  & CERN-06 & \ref{ssec:aph}  \\
 $C23  $ & $\nu_{\bar{\rm p}{\rm ^3He}^+}(38,34:37,33)$ & $ 505\,222\,295.7(8.2)$  MHz & $ 1.6\times 10^{-8}$  & CERN-06 & \ref{ssec:aph}  \\
 $C24  $ & $\nu_{\bar{\rm p}{\rm ^3He}^+}(36,34:37,33)$ & $ 414\,147\,507.8(4.0)$  MHz & $ 9.7\times 10^{-9}$  & CERN-06 & \ref{ssec:aph}  \\
\botrule
\end{tabular}
\begin{minipage}{7 in}
\footnotetext[1]{The values in brackets are relative to the corresponding transition frequency.}
\end{minipage}
\end{table*}

\begin{table*}
\caption{Non-negligible correlation coefficients $r(x_i,x_j)$ of the input data
in Table~\ref{tab:cdata}.  For simplicity, the two items of data to which a
particular correlation coefficient corresponds are identified by their item
numbers in Table~\ref{tab:cdata}.}
\label{tab:cdcc}
\def\sp{\hbox to 57 pt{}}
\begin{tabular}{ r @{~} l @{\sp} r @{~} l @{\sp} r @{~} l @{\sp} r @{~} l }
\toprule
\vbox to 10 pt {}
%
$r$($C1$, $C2$)& = $ 0.929 $ & $r$($C9$, $C10$)& = $ 0.925 $ & $r$($C14$, $C23$)& = $ 0.132 $ & $r$($C17$, $C24$)& = $ 0.287 $ \\ 
$r$($C1$, $C3$)& = $ 0.912 $ & $r$($C9$, $C11$)& = $ 0.949 $ & $r$($C14$, $C24$)& = $ 0.271 $ & $r$($C18$, $C19$)& = $ 0.235 $ \\ 
$r$($C1$, $C4$)& = $ 0.936 $ & $r$($C9$, $C12$)& = $ 0.978 $ & $r$($C15$, $C16$)& = $ 0.223 $ & $r$($C18$, $C20$)& = $ 0.107 $ \\ 
$r$($C1$, $C5$)& = $ 0.883 $ & $r$($C10$, $C11$)& = $ 0.907 $ & $r$($C15$, $C17$)& = $ 0.198 $ & $r$($C18$, $C21$)& = $ 0.118 $ \\ 
$r$($C1$, $C6$)& = $ 0.758 $ & $r$($C10$, $C12$)& = $ 0.934 $ & $r$($C15$, $C18$)& = $ 0.140 $ & $r$($C18$, $C22$)& = $ 0.122 $ \\ 
$r$($C1$, $C7$)& = $ 0.957 $ & $r$($C11$, $C12$)& = $ 0.959 $ & $r$($C15$, $C19$)& = $ 0.223 $ & $r$($C18$, $C23$)& = $ 0.112 $ \\ 
$r$($C2$, $C3$)& = $ 0.900 $ & $r$($C13$, $C14$)& = $ 0.210 $ & $r$($C15$, $C20$)& = $ 0.128 $ & $r$($C18$, $C24$)& = $ 0.229 $ \\ 
$r$($C2$, $C4$)& = $ 0.924 $ & $r$($C13$, $C15$)& = $ 0.167 $ & $r$($C15$, $C21$)& = $ 0.142 $ & $r$($C19$, $C20$)& = $ 0.170 $ \\ 
$r$($C2$, $C5$)& = $ 0.872 $ & $r$($C13$, $C16$)& = $ 0.224 $ & $r$($C15$, $C22$)& = $ 0.141 $ & $r$($C19$, $C21$)& = $ 0.188 $ \\ 
$r$($C2$, $C6$)& = $ 0.748 $ & $r$($C13$, $C17$)& = $ 0.197 $ & $r$($C15$, $C23$)& = $ 0.106 $ & $r$($C19$, $C22$)& = $ 0.191 $ \\ 
$r$($C2$, $C7$)& = $ 0.945 $ & $r$($C13$, $C18$)& = $ 0.138 $ & $r$($C15$, $C24$)& = $ 0.217 $ & $r$($C19$, $C23$)& = $ 0.158 $ \\ 
$r$($C3$, $C4$)& = $ 0.907 $ & $r$($C13$, $C19$)& = $ 0.222 $ & $r$($C16$, $C17$)& = $ 0.268 $ & $r$($C19$, $C24$)& = $ 0.324 $ \\ 
$r$($C3$, $C5$)& = $ 0.856 $ & $r$($C13$, $C20$)& = $ 0.129 $ & $r$($C16$, $C18$)& = $ 0.193 $ & $r$($C20$, $C21$)& = $ 0.109 $ \\ 
$r$($C3$, $C6$)& = $ 0.734 $ & $r$($C13$, $C21$)& = $ 0.142 $ & $r$($C16$, $C19$)& = $ 0.302 $ & $r$($C20$, $C22$)& = $ 0.108 $ \\ 
$r$($C3$, $C7$)& = $ 0.927 $ & $r$($C13$, $C22$)& = $ 0.141 $ & $r$($C16$, $C20$)& = $ 0.172 $ & $r$($C20$, $C23$)& = $ 0.081 $ \\ 
$r$($C4$, $C5$)& = $ 0.878 $ & $r$($C13$, $C23$)& = $ 0.106 $ & $r$($C16$, $C21$)& = $ 0.190 $ & $r$($C20$, $C24$)& = $ 0.166 $ \\ 
$r$($C4$, $C6$)& = $ 0.753 $ & $r$($C13$, $C24$)& = $ 0.216 $ & $r$($C16$, $C22$)& = $ 0.189 $ & $r$($C21$, $C22$)& = $ 0.120 $ \\ 
$r$($C4$, $C7$)& = $ 0.952 $ & $r$($C14$, $C15$)& = $ 0.209 $ & $r$($C16$, $C23$)& = $ 0.144 $ & $r$($C21$, $C23$)& = $ 0.090 $ \\ 
$r$($C5$, $C6$)& = $ 0.711 $ & $r$($C14$, $C16$)& = $ 0.280 $ & $r$($C16$, $C24$)& = $ 0.294 $ & $r$($C21$, $C24$)& = $ 0.184 $ \\ 
$r$($C5$, $C7$)& = $ 0.898 $ & $r$($C14$, $C17$)& = $ 0.247 $ & $r$($C17$, $C18$)& = $ 0.210 $ & $r$($C22$, $C23$)& = $ 0.091 $ \\ 
$r$($C6$, $C7$)& = $ 0.770 $ & $r$($C14$, $C18$)& = $ 0.174 $ & $r$($C17$, $C19$)& = $ 0.295 $ & $r$($C22$, $C24$)& = $ 0.186 $ \\ 
$r$($C8$, $C9$)& = $ 0.978 $ & $r$($C14$, $C19$)& = $ 0.278 $ & $r$($C17$, $C20$)& = $ 0.152 $ & $r$($C23$, $C24$)& = $ 0.154 $ \\ 
$r$($C8$, $C10$)& = $ 0.934 $ & $r$($C14$, $C20$)& = $ 0.161 $ & $r$($C17$, $C21$)& = $ 0.167 $ & &  \\ 
$r$($C8$, $C11$)& = $ 0.959 $ & $r$($C14$, $C21$)& = $ 0.178 $ & $r$($C17$, $C22$)& = $ 0.169 $ & &  \\ 
$r$($C8$, $C12$)& = $ 0.988 $ & $r$($C14$, $C22$)& = $ 0.177 $ & $r$($C17$, $C23$)& = $ 0.141 $ & &  \\ 
\botrule
\end{tabular}
\end{table*}

\subsection{Comparison of data}
\label{ssec:cod}

The classic Lamb shift is the only quantity among the Rydberg constant data
with more than one measured value, but there are ten different quantities with
more than one measured value among the other data.  The item numbers given in
Tables~\ref{tab:rdata} and \ref{tab:pdata} for the members of such groups of
data ($A39$, $B2$, $B11$, $B28$, $B31$, $B33$-$B36$, $B38$, and $B58$) have a
decimal point with an additional digit to label each member.

In fact, all of the data for which there is more than one measurement were
directly compared in either the 1998 or 2002 adjustments except the following
new data: the University of Washington result for $A_{\rm r}$($^2$H), item
$B2.2$, the Harvard University result for $a_{\rm e}$, item $B11.2$, the NIST
watt-balance result for $K_{\rm J}^2\,R_{\rm K}$ item $B36.3$, and the INRIM
result for $d_{220}({\scriptstyle{\rm W4.2a}})$, item $B38.2$.  The two values
of $A_{\rm r}$($^2$H) agree well---they differ by only 0.5$u_{\rm diff}$; the
two values of $a_{\rm e}$ are in acceptable agreement---they differ by
1.7$u_{\rm diff}$; the two values of $d_{220}({\scriptstyle{\rm W4.2a}})$ also
agree well--they differ by 0.7$u_{\rm diff}$; and the three values of $K_{\rm
J}^2\,R_{\rm K}$ are highly consistent---their mean and implied value of $h$
are
\begin{eqnarray}
K_{\rm J}^2\,R_{\rm K} &=& 6.036\,761\,87(21) \times 10^{33} \ \mbox{J}^{-1}~\mbox{s}^{-1}
\\
h &=& 6.626\,068\,89(23) \times 10^{-34} \ \mbox{J~s}
\label{eq:k2rh}
\end{eqnarray}
with $\chi^2= 0.27$ for $\nu = N-M = 2$ degrees of freedom, where $N$ is the
number of measurements and $M$ is the number of unknowns, and with Birge ratio
$R_{\rm B} = \sqrt{\chi^2/\nu} = 0.37$ (see Appendix E of CODATA-98).  The
normalized residuals for the three values are $0.52$, $-0.04$, and $-0.09$, and
their weights in the calculation of the weighted mean are 0.03, 0.10, and 0.87.

Data for quantities with more than one directly measured value used in earlier
adjustments are consistent, with the exception of the VNIIM 1989 result for
${\it \Gamma}_{{\rm h}-90}^\prime$(lo), which is not included in the present
adjustment (see Sec.~\ref{sec:elmeas}).  We also note that none of these data
has a weight of less than 0.02 in the weighted mean of measurements of the same
quantity.

The consistency of measurements of various quantities of different types is
shown mainly by comparing the values of the fine-structure constant $\alpha$ or
the Planck constant $h$ inferred from the measured values of the quantities.
Such inferred values of $\alpha$ and $h$ are given throughout the data review
sections, and the results are summarized and discussed further here.

\def\m{\phantom{-}}
\def\fixh{\vbox to 10pt {}}
\begin{table*}
\caption{Comparison of the input data in Table~\ref{tab:pdata} through inferred
values of the fine-structure constant $\alpha$ in order of increasing standard
uncertainty.}
\label{tab:alphas}
\def\sp{\hbox to 26 pt{}}
\begin{tabular}{l@{\sp}l@{\sp}l@{\sp}l@{\sp}l@{\sp}l}
\toprule
Primary & Item & Identification \fixh
& Sec. and Eq. & \hbox to 25pt{} $\alpha^{-1}$ & \hbox to -5pt {} Relative standard \\
source & number &
& & & \hbox to -5pt {} uncertainty $u_{\rm r}$  \\
\colrule

$a_{\rm e}$ & \fixh \Naeharvjf &
HarvU-06 & \ref{sssec:alphaae} (\ref{eq:alphinvharvu06}) & 
$ 137.035\,999\,711(96)$& $ 7.0\times 10^{-10}$\\

$a_{\rm e}$ & \fixh \Naeuwashhg&
UWash-87 & \ref{sssec:alphaae} (\ref{eq:alphinvuwash87}) & 
$ 137.035\,998\,83(50)$& $ 3.7\times 10^{-9}$\\

$h/m({\rm Rb})$ & \fixh \Nhomrbjf &
LKB-06 & \ref{sssec:pcrbmr} (\ref{eq:alphinvrb06}) & 
$ 137.035\,998\,83(91)$& $ 6.7\times 10^{-9}$\\

$h/m({\rm Cs})$ & \fixh \Nhomcsjb &
StanfU-02 & \ref{sssec:pccsmr} (\ref{eq:alphinvcs02}) & 
$ 137.036\,0000(11)$& $ 7.7\times 10^{-9}$\\

$h/m_{\rm n}d_{220}({\rm {\scriptstyle W04}})$ & \fixh \Nhomndwjd &
PTB-99 & & & \\

\ \ $\overline{d_{220}}$ & \fixh  \Ndwdbptbha-\Ndmodinrimjg &
Mean & \ref{sssec:pcnmr} (\ref{eq:alphinvhom}) & 
$ 137.036\,0077(28)$&$ 2.1\times 10^{-8}$ \\

$R_{\rm K}$ & \fixh \Nrknistig &
NIST-97 & \ref{sssec:nistvkc} (\ref{eq:airknist97}) & 
$ 137.036\,0037(33)$&$ 2.4\times 10^{-8}$ \\

${\it\Gamma}_{\rm p-90}^{\,\prime}({\rm lo})$ & \fixh \Ngppnisthi &
NIST-89 & \ref{par:nistlf} (\ref{eq:alphinistg89}) & 
$ 137.035\,9879(51)$&$ 3.7\times 10^{-8}$ \\

$R_{\rm K}$ & \fixh \Nrknmlig &
NMI-97 & \ref{sssec:nmlvkc} (\ref{eq:airknml97}) & 
$ 137.035\,9973(61)$&$ 4.4\times 10^{-8}$ \\

$R_{\rm K}$ & \fixh \Nrklcieja &
LNE-01 & \ref{sssec:lcievkc} (\ref{eq:airklcie01}) & 
$ 137.036\,0023(73)$&$ 5.3\times 10^{-8}$ \\

$R_{\rm K}$ & \fixh \Nrknplhh &
NPL-88 & \ref{sssec:nplvkc} (\ref{eq:airknpl88}) & 
$ 137.036\,0083(73)$&$ 5.4\times 10^{-8}$ \\

$\Delta\nu_{\rm Mu}$ & \fixh \NnupLhbkHz,\NnupLii &
LAMPF & \ref{par:clampf} (\ref{eq:alphiL}) & 
$ 137.036\,0017(80)$ & $ 5.8\times 10^{-8}$ \\

${\it\Gamma}_{\rm h-90}^{\,\prime}({\rm lo})$ & \fixh \Nghkvig &
KR/VN-98 & \ref{par:kvlf} (\ref{eq:alphinvkv97}) & 
$ 137.035\,9852(82)$&$ 6.0\times 10^{-8}$ \\

$R_{\rm K}$ & \fixh \Nrknimie &
NIM-95 & \ref{sssec:nimvkc} (\ref{eq:airknim95}) & 
$ 137.036\,004(18)$&$ 1.3\times 10^{-7}$ \\

${\it\Gamma}_{\rm p-90}^{\,\prime}({\rm lo})$ & \fixh \Ngpplnimie &
NIM-95 & \ref{par:nimlo} (\ref{eq:alignim95}) & 
$ 137.036\,006(30)$&$ 2.2\times 10^{-7}$ \\

$\nu_{\rm H},\nu_{\rm D}$ & \fixh $A26$-$A47$ &
Various & \ref{par:trfreq} (\ref{eq:alphinvhd}) &
$ 137.036\,002(48)$&$ 3.5\times 10^{-7}$ \\

$\overline{R}$ & \fixh \Nrbarjf &
BNL-02 & \ref{par:tamual} (\ref{eq:alphiam06}) & 
$ 137.035\,67(26)$&$ 1.9\times 10^{-6}$ \\

\botrule
\end{tabular}
\end{table*}

\begin{figure} 
\rotatebox{-90}{\resizebox{!}{4.4in}{
\includegraphics[clip,trim=10 40 40 10]{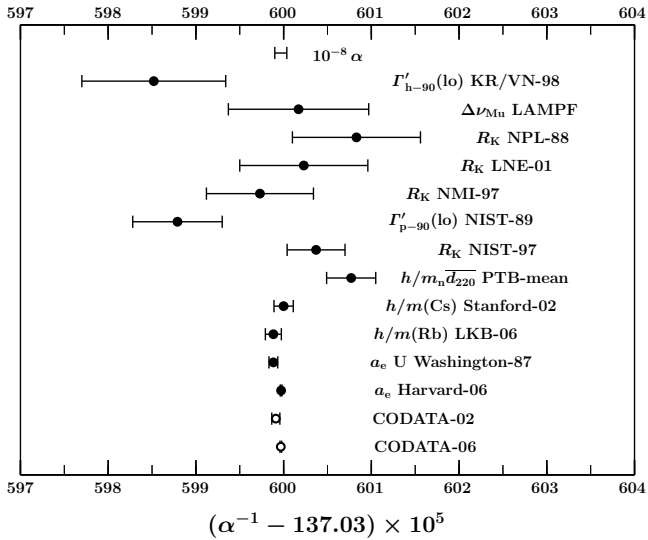} }} 
\caption { Values of the fine-structure constant $\alpha$ with $u_{\rm
r}<10^{-7}$ implied by the input data in Table~\ref{tab:pdata}, in order of
decreasing uncertainty from top to bottom, and the 2002 and 2006 CODATA
recommended values of $\alpha$.  (See Table~\ref{tab:alphas}.)  Here ``mean''
indicates the PTB-99 result for $h/m_{\rm n}d_{220}({\scriptstyle{\rm W04}})$
using the value of $d_{220}({\scriptstyle{\rm W04}})$ implied by the
four XROI lattice-spacing measurements.}
\label{fig:aall}
\end{figure}

\begin{figure}
\rotatebox{-90}{\resizebox{!}{4.2in}{
\includegraphics[clip,trim=10 40 40 10]{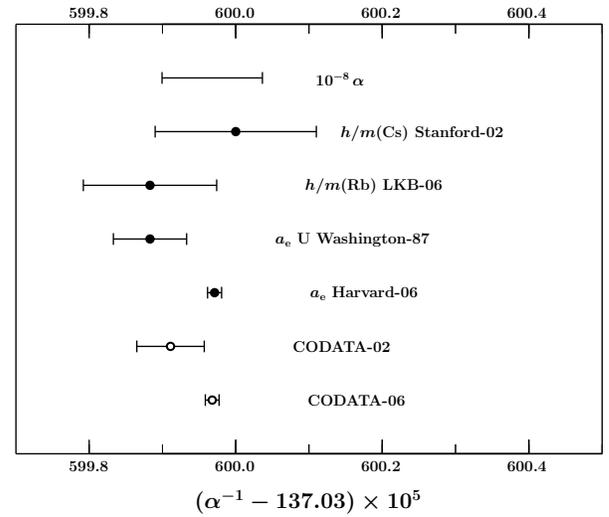}
}}
\caption{ Values of the fine-structure constant $\alpha$ with $u_{\rm r} <
10^{-8}$ implied by the input data in Table~\ref{tab:pdata}, in order of
decreasing uncertainty from top to bottom.  (See Table~\ref{tab:alphas}.)}
\label{fig:aallx}
\end{figure}

\begin{figure}
\rotatebox{-90}{\resizebox{!}{4.2in}{
\includegraphics[clip,trim= 10 40 40 10]{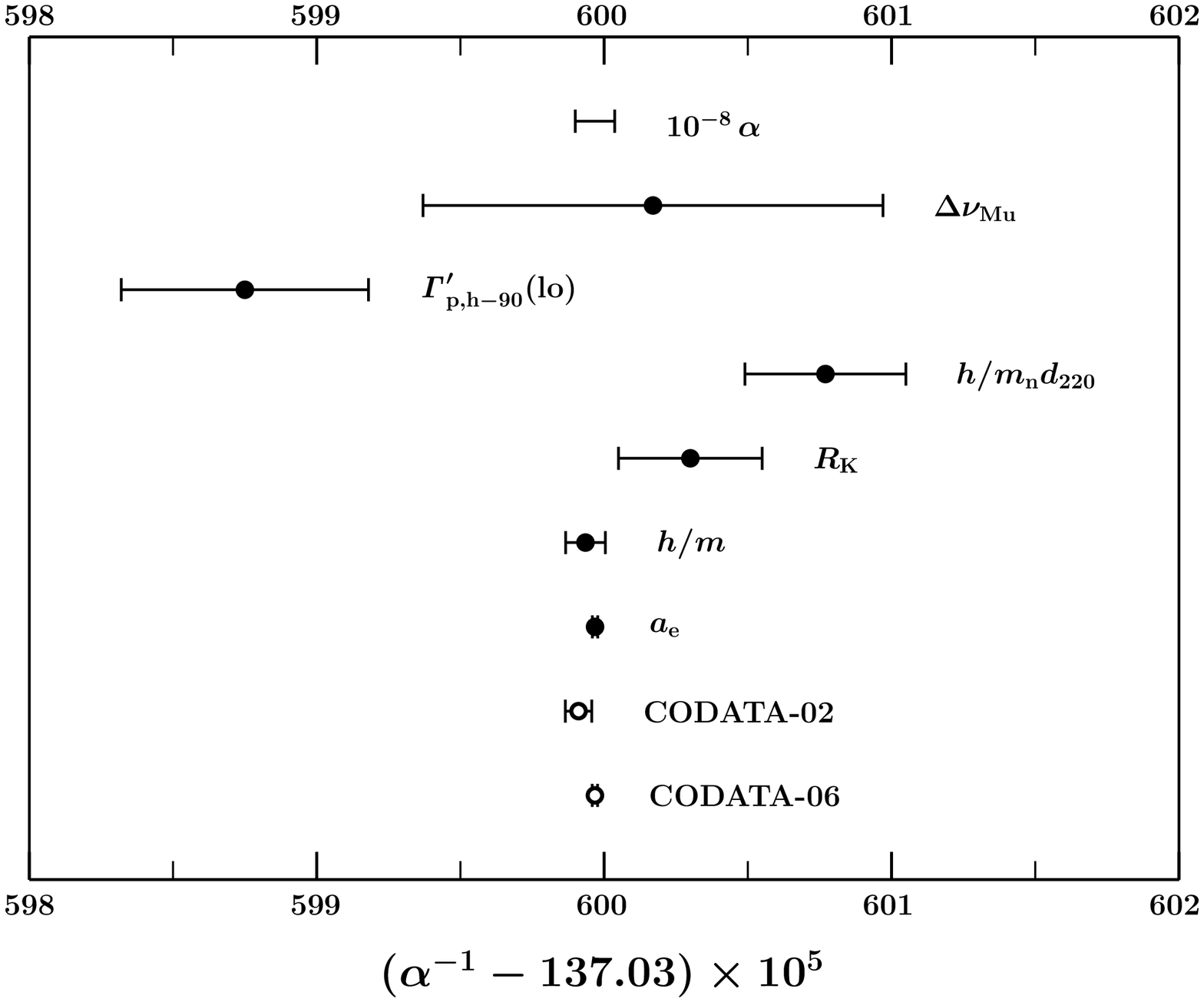}
}}
\caption{ Values of the fine-structure constant $\alpha$ with $u_{\rm r} <
10^{-7}$ implied by the input data in Table~\ref{tab:pdata}, taken as a
weighted mean when more than one measurement of a given type is considered [see
Eqs.~(\ref{eq:alphasfirst}) to (\ref{eq:alphaslast})], in order of decreasing
uncertainty from top to bottom.}
\label{fig:aav}
\end{figure}

The consistency of a significant fraction of the data of Tables~\ref{tab:rdata}
and \ref{tab:pdata} is indicated in Table~\ref{tab:alphas} and
Figs.~\ref{fig:aall}, \ref{fig:aallx}, and \ref{fig:aav}, which give and
graphically compare the values of $\alpha$ inferred from that data.
Figures~\ref{fig:aall} and \ref{fig:aallx} compare the data that yield values
of $\alpha$ with $u_{\rm r}<10^{-7}$ and $u_{\rm r}<10^{-8}$, respectively;
Fig.~\ref{fig:aav} also compares the data that yield values of $\alpha$ with
$u_{\rm r}<10^{-7}$, but does so through combined values of $\alpha$ obtained
from similar experiments.  Most of the values of $\alpha$ are in reasonable
agreement, implying that most of the data from which they are obtained are
reasonably consistent.  There are, however, two important exceptions.

The value of $\alpha$ inferred from the PTB measurement of $h/m_{\rm
n}d_{220}({\scriptstyle{\rm W04}})$, item $B55$, is based on the mean value
$\overline{d_{220}}$ of $d_{220}({\scriptstyle{\rm W04}})$ implied by the four
direct \{220\} XROI lattice spacing measurements, items $B38.1$-$B40$.
It disagrees by about 2.8$u_{\rm diff}$ with the value of $\alpha$ with the
smallest uncertainty, that inferred from the Harvard University measurement of
$a_{\rm e}$.  Also, the value of $\alpha$ inferred from the NIST measurement of
$\iG_{\rm p-90}^\prime$(lo) disagrees with the latter by about 2.3$u_{\rm
diff}$.  But it is also worth noting that the value $\alpha^{-1} =
137.036\,0000(38)~[2.8\times 10^{-8}]$ implied by $h/m_{\rm
n}d_{220}({\scriptstyle{\rm W04}})$ together with item $B39$ alone, the NMIJ
XROI measurement of $d_{220}({\scriptstyle{\rm NR3}})$, agrees well with the
Harvard $a_{\rm e}$ value of $\alpha$.  If instead one uses the three other
direct XROI lattice spacing measurements, items $B38.1$, $B38.2$, and $B40$,
which agree among themselves, one finds $\alpha^{-1} =
137.036\,0092(28)~[2.1\times 10^{-8}]$.  This value disagrees with $\alpha$
from the Harvard $a_{\rm e}$ by 3.3$u_{\rm diff}$.  

The values of $\alpha$ compared in Fig.~\ref{fig:aav} follow from
Table~\ref{tab:alphas} and are, again in order of increasing uncertainty,
\begin{eqnarray} \label{eq:alphasfirst} \alpha^{-1}[a_{\rm e}] &=&
137.035\,999\,683(94) \nonumber\\ && \qquad\qquad [ 6.9\times 10^{-10}]\quad \\
\alpha^{-1}[h/m] &=&  137.035\,999\,35(69) \nonumber\\ && \qquad\qquad ~ [
5.0\times 10^{-9}] \\ \alpha^{-1}[R_{\rm K}] &=&  137.036\,0030(25) \nonumber\\
&& \qquad\qquad ~ [ 1.8\times 10^{-8}] \\ \alpha^{-1}[h/m_{\rm n}d_{220}] &=&
137.036\,0077(28) \nonumber\\ && \qquad\qquad ~ [ 2.1\times 10^{-8}] \\
\alpha^{-1}[{\it \Gamma}^{\,\prime}_{\rm p,h-90}({\rm lo})] &=&
137.035\,9875(43) \nonumber\\ && \qquad\qquad ~ [ 3.1\times 10^{-8}]  \\
\alpha^{-1}[\Delta\nu_{\rm Mu}] &=&   137.036\,0017(80) \nonumber\\ &&
\qquad\qquad ~ [ 5.8\times 10^{-8}] \ .  \label{eq:alphaslast} \end{eqnarray}
Here $\alpha^{-1}[a_{\rm e}]$ is the weighted mean of the two $a_{\rm e}$
values of $\alpha$; $\alpha^{-1}[h/m]$ is the weighted mean of the
$h/m(^{87}{\rm Rb})$ and $h/m(^{133}{\rm Cs})$ values; $\alpha^{-1}[R_{\rm K}]$
is the weighted mean of the five quantum Hall effect-calculable capacitor
values; $\alpha^{-1}[h/m_{\rm n}d_{220}]$ is the value as given in
Table~\ref{tab:alphas} and is based on the measurement of $h/m_{\rm
n}d_{220}({\scriptstyle{\rm W04}})$ and the value of $d_{220}({\scriptstyle{\rm
W04}})$ inferred from the four XROI determinations of the \{220\} lattice
spacing of three different silicon crystals; $\alpha^{-1}[{\it
\Gamma}^{\,\prime}_{\rm p,h-90}({\rm lo})]$ is the weighted mean of the two
values of $\alpha^{-1}[{\it \Gamma}^{\,\prime}_{\rm p-90}({\rm lo})]$ and one
value of $\alpha^{-1}[{\it \Gamma}^{\,\prime}_{\rm h-90}({\rm lo})]$; and
$\alpha^{-1}[\Delta\nu_{\rm Mu}]$ is the value as given in
Table~\ref{tab:alphas} and is based on the 1982 and 1999 measurements at LAMPF
on muonium.

Figures~\ref{fig:aall}, \ref{fig:aallx}, and \ref{fig:aav} show that even if
all of the data of Table~\ref{tab:pdata} were retained, the 2006 recommended
value of $\alpha$ would be determined to a great extent by $a_{\rm e}$, and in
particular, the Harvard University determination of $a_{\rm e}$.

The consistency of a significant fraction of the data of Table~\ref{tab:pdata}
is indicated in Table~\ref{tab:plancks} and Figs.~\ref{fig:hall} and
\ref{fig:hav}, which give and graphically compare the values of $h$ inferred
from those data.  Figure~\ref{fig:hall} compares the data by showing each
inferred value of $h$ in the table, while Fig.~\ref{fig:hav} compares the data
through combined values of $h$ from similar experiments.  The values of $h$ are
in good agreement, implying that the data from which they are obtained are
consistent, with one important exception.  The value of $h$ inferred from
$V_{\rm m}$(Si), item $B53$, disagrees by 2.9$u_{\rm diff}$ with the value of
$h$ from the weighted mean of the three watt-balance values of $K_{\rm
J}^2\,R_{\rm K}$ [uncertainty $u_{\rm r} = 3.4 \times 10^{-8}$---see
Eq.~(\ref{eq:k2rh})].

In this regard, it is worth noting that a value of $d_{220}$ of an ideal
silicon crystal is required to obtain a value of $h$ from $V_{\rm m}$(Si) [see
Eq.~(\ref{eq:vmsith})], and the value used to obtain the inferred value of $h$
given in Eq.~(\ref{eq:hvmsilwgac05}) and Table~\ref{tab:plancks} is based on
all four XROI lattice spacing measurements, items $B38.1$-$B40$, plus the
indirect value from $h/m_{\rm n}d_{220}({\scriptstyle{\rm W04}})$ (see
Table~\ref{tab:d220abs} and Fig.~\ref{fig:silicon}).
However, the NMIJ measurement of $d_{220}({\scriptstyle{\rm NR3}})$, item
$B39$, and the indirect value of $d_{220}$ from $h/m_{\rm
n}d_{220}({\scriptstyle{\rm W04}})$, yield values of $h$ from $V_{\rm m}$(Si)
that are less consistent with the watt-balance mean value than the three other
direct XROI lattice spacing measurements, items $B38.1$, $B38.2$, and $B40$,
which agree among themselves (a disagreement of about 3.8$u_{\rm diff}$
compared to 2.5$u_{\rm diff}$).  In contrast, the NMIJ measurement of
$d_{220}({\scriptstyle{\rm NR3}})$ yields a value of $\alpha$ from $h/m_{\rm
n}d_{220}({\scriptstyle{\rm W04}})$ that is in excellent agreement with the
Harvard University value from $a_{\rm e}$, while the three other lattice
spacing measurements yield a value of $\alpha$ in poor agreement with alpha
from $a_{\rm e}$ (3.3$u_{\rm diff}$).

\def\m{\phantom{-}}
\def\fixh{\vbox to 9pt {}}
\begin{table*}
\caption{Comparison of the input data in Table~\ref{tab:pdata} through inferred
values of the Planck constant $h$ in order of increasing standard uncertainty.}
\label{tab:plancks}
\def\sp{\hbox to 30 pt {}}
\begin{tabular}{l@{\sp}l@{\sp}l@{\sp}l@{\sp}l@{\sp}l}
\toprule
\fixh Primary & Item & Identification
& Sec. and Eq. & \hbox to 20pt{}$h/({\rm J \ s})$ 
& Relative standard \\
source & number &
& & & \hbox to -5pt {} uncertainty $u_{\rm r}$  \\
\colrule

$K_{\rm J}^2R_{\rm K}$ & \fixh \Nkjbrknistjg &
NIST-07 & \ref{par:kj2rk07} (\ref{eq:hwbnist07}) &
$ 6.626\,068\,91(24)\times 10^{-34}$ & $ 3.6\times 10^{-8}$ \\

$K_{\rm J}^2R_{\rm K}$ & \fixh \Nkjbrknistih &
NIST-98 & \ref{par:kj2rk98} (\ref{eq:hwbnist98}) &
$ 6.626\,068\,91(58)\times 10^{-34}$ & $ 8.7\times 10^{-8}$ \\

$K_{\rm J}^2R_{\rm K}$ & \fixh \Nkjbrknplij & 
NPL-90 & \ref{sssec:kj2rknpl} (\ref{eq:hwbnpl90}) &
$ 6.626\,0682(13)\times 10^{-34}$ & $ 2.0\times 10^{-7}$ \\

$V_{\rm m}$(Si) & \fixh \Nvmsilwgacje &
N/P/I-05 & \ref{ssec:mvsi} (\ref{eq:hvmsilwgac05}) &
$ 6.626\,0745(19)\times 10^{-34}$ & $ 2.9\times 10^{-7}$ \\

$K_{\rm J}$ & \fixh \NkjnmlhiGHz &
NMI-89 & \ref{sssec:nmljc} (\ref{eq:hnml89}) &
$ 6.626\,0684(36)\times 10^{-34}$ & $ 5.4\times 10^{-7}$ \\

$K_{\rm J}$ & \fixh \NkjptbiaGHz &
PTB-91 & \ref{sssec:ptbjc} (\ref{eq:hptb91}) &
$ 6.626\,0670(42)\times 10^{-34}$ & $ 6.3\times 10^{-7}$ \\

${\it\Gamma}_{\rm p-90}^{\,\prime}({\rm hi})$ & \fixh \Ngppnplgi &
NPL-79 & \ref{par:nplhf} (\ref{eq:hnplg79}) &
$ 6.626\,0729(67)\times 10^{-34}$&$ 1.0\times 10^{-6}$ \\

${\cal F}_{90}$ & \fixh \Nfijnisthj &
NIST-80 & \ref{sssec:fcnist} (\ref{eq:hfnist80}) &
$ 6.626\,0657(88)\times 10^{-34}$ & $ 1.3\times 10^{-6}$ \\

${\it\Gamma}_{\rm p-90}^{\,\prime}({\rm hi})$ & \fixh \Ngpphnimie &
NIM-95 & \ref{par:nimhi} (\ref{eq:hnimg95}) &
$ 6.626\,071(11)\times 10^{-34}$&$ 1.6\times 10^{-6}$ \\

\botrule
\end{tabular}
\end{table*}

\begin{figure}
\rotatebox{-90}{\resizebox{!}{4.2in}{
\includegraphics[clip,trim= 10 40 40 10]{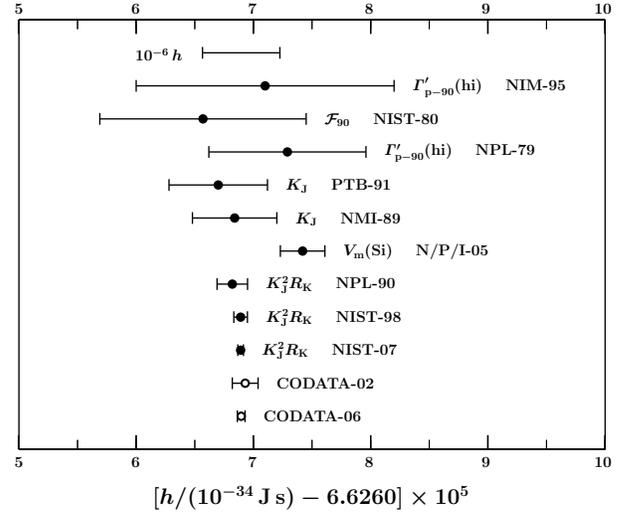}
}}
\caption{ Values of the Planck constant $h$ implied by the input data in
Table~\ref{tab:pdata}, in order of decreasing uncertainty from top to bottom.
(See Table~\ref{tab:plancks}.)}
\label{fig:hall}
\end{figure}

\begin{figure}
\rotatebox{-90}{\resizebox{!}{4.2in}{
\includegraphics[clip,trim= 10 40 40 10]{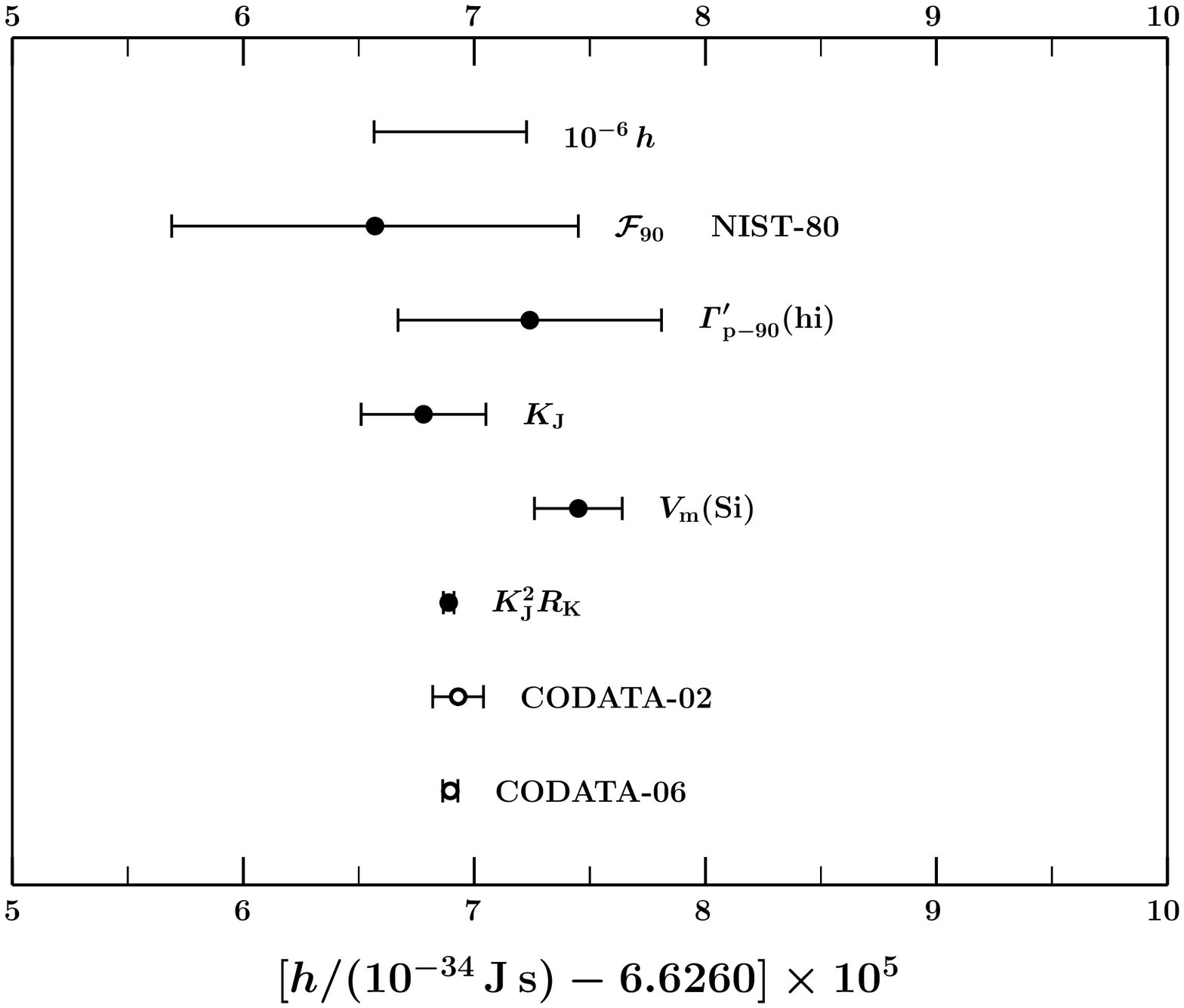}
}}
\caption{ Values of the Planck constant $h$ implied by the input data in
Table~\ref{tab:pdata}, taken as a weighted mean when more than one measurement
of a given type is considered [see Eqs.~(\ref{eq:hsfirst}) to
(\ref{eq:hslast})], in order of decreasing uncertainty from top to bottom.}
\label{fig:hav}
\end{figure}

The values of $h$ compared in Fig.~\ref{fig:hav} follow from
Table~\ref{tab:plancks} and are, again in order of increasing uncertainty,
\begin{eqnarray}
\label{eq:hsfirst}
h[K_{\rm J}^2R_{\rm K}] &=&  6.626\,068\,89(23)\times 10^{-34} 
\nonumber\\ && \qquad\qquad\qquad 
~ [ 3.4\times 10^{-8}]
\\
h[V_{\rm m}{\rm (Si)}] &=&  6.626\,0745(19)\times 10^{-34} 
\nonumber\\ && \qquad\qquad\qquad 
~ [ 2.9\times 10^{-7}]
\\
h[K_{\rm J}] &=&  6.626\,0678(27)\times 10^{-34} 
\nonumber\\ && \qquad\qquad\qquad 
~ [ 4.1\times 10^{-7}]
\\
h[{\it \Gamma}^\prime_{\rm p-90}{\rm (hi)}] &=&  6.626\,0724(57)\times 10^{-34} 
\nonumber\\ && \qquad\qquad\qquad 
~ [ 8.6\times 10^{-7}]
\\
h[{\cal F}_{90}] &=&   6.626\,0657(88)\times 10^{-34} 
\nonumber\\ && \qquad\qquad\qquad 
~ [ 1.3\times 10^{-6}] \ .
\label{eq:hslast}
\end{eqnarray}
Here $h[K_{\rm J}^2R_{\rm K}]$ is the weighted mean of the three values of $h$
from the three watt-balance measurements of $K_{\rm J}^2R_{\rm K}$; $h[V_{\rm
m}{\rm (Si)}]$ is the value as given in Table~\ref{tab:plancks} and is based on
all four XROI $d_{220}$ lattice spacing measurements plus the indirect lattice
spacing value from $h/m_{\rm n}d_{220}({\scriptstyle{\rm W04}})$; $h[K_{\rm
J}]$ is the weighted mean of the two direct Josephson effect measurements of
$K_{\rm J}$; $h[{\it \Gamma}_{\rm p-90}$(hi)] is the weighted mean of the two
values of $h$ from the two measurements of ${\it \Gamma}_{\rm p-90}$(hi); and
$h[{\cal F}_{90}]$ is the value as given in Table~\ref{tab:plancks} and comes
from the silver coulometer measurement of ${\cal F}_{90}$.
Figures~\ref{fig:hall} and \ref{fig:hav} show that even if all of the data of
Table~\ref{tab:pdata} were retained, the 2006 recommended value of $h$ would be
determined to a large extent by $K_{\rm J}^2\,R_{\rm K}$, and in particular,
the NIST 2007 determination of this quantity.

We conclude our data comparisons by listing in Table~\ref{tab:ares} the four
available values of $A_{\rm r}({\rm e})$.  The reasonable agreement of these
values shows that the corresponding input data are consistent.  The most
important of these data are the University of Washington value of $A_{\rm
r}({\rm e})$, $\delta_{\rm C}$, $\delta_{\rm O}$, $f_{\rm s}(^{12}{\rm
C}^{5+})/f_{\rm c}(^{12}C^{5+})$, $f_{\rm s}(^{12}{\rm O}^{7+})/f_{\rm
c}(^{12}{\rm O}^{7+})$, and the antiprotonic helium data, items $B9$,
$B14$-$B17$, and $C1$-$C24$.

In summary, the data comparisons of this section of the paper have identified
the following potential problems: (i) the measurement of $V_{\rm m}$(Si), item
$B53$, is inconsistent with the watt-balance measurements of $K_{\rm
J}^2\,R_{\rm K}$, items $B36.1$-$B36.3$, and somewhat inconsistent with the
mercury-electrometer and voltage-balance measurements of $K_{\rm J}$; (ii) the
three XROI $\{220\}$ lattice spacing values $d_{220}({\scriptstyle{\rm
W4.2a}})$, $d_{220}({\scriptstyle{\rm W4.2a}})$, and $d_{220}({\scriptstyle{\rm
MO^*}})$, items $B38.1$, $B38.2$, and $B40$, are inconsistent with the value of
$d_{220}({\scriptstyle{\rm NR3}})$, item $B39$, and the measurement of
$h/m_{\rm n}d_{220}({\scriptstyle{\rm W04}})$, item $B55$; (iii) the NIST-89
measurement of ${\it \Gamma}^\prime_{\rm p-90}$(lo), item $B33.1$, is
inconsistent with the most accurate data that also determine the value of the
fine-structure constant; (iv) although not a problem in the sense of (i)-(iii),
there are a number of input data with uncertainties so large that they are
unlikely to make a contribution to the determination of the 2006 CODATA
recommended values.

Furthermore, we note that some of the inferred values of $\alpha$ in
Table~\ref{tab:alphas} and most of the inferred values of $h$ in
Table~\ref{tab:plancks} depend on either one or both of the relations $K_{\rm
J} = 2e/h$ and $R_{\rm K} = h/e^2$.  The question of whether relaxing the
assumption that these relations are exact would reduce or possibly even
eliminate some of the observed inconsistencies, considered in Appendix F of
CODATA-02, is addressed in the section following the next section.  This study
indeed confirms the Josephson and quantum Hall effect relations.

\subsection{Multivariate analysis of data}
\label{ssec:mada}

The multivariate analysis of the data is based on the fact that measured
quantities can be expressed as theoretical functions of fundamental constants.
These expressions, or observational equations, are written in terms of a
particular independent subset of the constants whose members are here called
{\it adjusted constants}.  The goal of the analysis is to find the values of
the adjusted constants that predict values for the measured data that best
agree with the data themselves in the least-squares sense (see Appendix E of
CODATA-98).

The symbol $\doteq$ is used to indicate that an observed value of an input
datum of the particular type shown on the left-hand side is ideally given by
the function of the adjusted constants on the right-hand side; however, the two
sides are not necessarily equal, because the equation is one of an
overdetermined set relating the data to the adjusted constants.  The best
estimate of a quantity is given by its observational equation evaluated with
the least-squares estimated values of the adjusted constants on which it
depends.

In essence, we follow the least-squares approach of \textcite{1934003} [see
also \textcite{1912002}], who treated the case where the input data are
correlated.  The 150 input data of Tables~\ref{tab:rdata}, \ref{tab:pdata}, and
\ref{tab:cdata} are of 135 distinct types and are expressed as functions of the
79 adjusted constants listed in Tables~\ref{tab:adjcona}, \ref{tab:adjconb},
and \ref{tab:adjconc}.  The observational equations that relate the input data
to the adjusted constants are given in Tables~\ref{tab:pobseqsa},
\ref{tab:pobseqsb1}, and \ref{tab:pobseqsc}.

Note that the various binding energies $E_{\rm b}(X)/m_{\rm u}c^2$ in
Table~\ref{tab:pobseqsb1}, such as in the equation for item $B$1, are treated
as fixed quantities with negligible uncertainties.  Similarly, the bound-state
$g$-factor ratios in this table, such as in the equation for item $B$18, are
treated in the same way.  Further, the frequency $f_{\rm p}$ is not an adjusted
constant but is included in the equation for items $B$29 and $B$30 to indicate
that they are functions of $f_{\rm p}$.  Finally, the observational equation
for items $B$29 and $B$30, based on Eqs.~(\ref{eq:murat}), (\ref{eq:se}), and
(\ref{eq:mumemump}) of Sec.~\ref{ssec:muhfs}, includes the functions $a_{\rm
e}(\alpha,\delta_{\rm e})$ and $a_{\rmssmu}(\alpha,\delta_{\rmssmu})$ as well
as the theoretical expression for input data of type $B$28, $\Delta\nu_{\rm
Mu}$.  The latter expression is discussed in Sec.~\ref{sssec:muhfs} and is a
function of $R_\infty, \alpha, m_{\rm e}/m_{\rmssmu}, a_{\rmssmu}(\alpha,
\delta_{\rmssmu})$, and $\delta _{\rm Mu}$.

\subsubsection{Summary of adjustments}

A number of adjustments were carried out to gauge the compatibility of the
input data in Tables~\ref{tab:rdata}, \ref{tab:pdata}, and \ref{tab:cdata}
(together with their covariances in Tables~\ref{tab:rdcc}, \ref{tab:pdcc}, and
\ref{tab:cdcc}) and to assess their influence on the values of the adjusted
constants.  The results of 11 of these are given in Tables~\ref{tab:adjustsall}
to \ref{tab:adjustsa} and are discussed in the following paragraphs.  Because
the adjusted value of the Rydberg constant $R_\infty $ is essentially the same
for all six adjustments summarized in Table~\ref{tab:adjustsall} and equal to
that of adjustment~4 of Table~\ref{tab:adjustsa}, the value of $R_\infty$ is
not listed in Table~\ref{tab:adjustsall}. It should also be noted that
adjustment~4 of all three tables is the same adjustment.

\textit{Adjustment 1}.  This initial adjustment includes all of the input data,
four of which have normalized residuals $r_i$ with absolute magnitudes
significantly greater than 2; the values of $r_i$ for these four data resulting
from adjustments~1-6 are given in Table~\ref{tab:adjres}.  Consistent with the
previous discussion, the four most inconsistent items are the molar volume of
silicon $V_{\rm m}$(Si), the quotient $h/m_{\rm n}d_{220}({\scriptstyle{\rm
W04}})$, the XROI measurement of the \{220\} lattice spacing
$d_{220}({\scriptstyle{\rm NR3}})$, and the NIST-89 value of ${\it
\Gamma}^\prime_{\rm p-90}$(lo).  All other input data have values of $r_i$
considerably less than 2, except those for $\nu_{\bar p ^{\,3}{\rm
He}}(32,31:31,30)$ and $\nu_{\bar p ^{\,3}{\rm He}}(36,33:34,32)$, items $C20$
and $C22$, for which $r_{20} = 2.09$ and $r_{22} = 2.06$.  However, the self
sensitivity coefficients $S_{\rm c}$ for these input data are considerably less
than 0.01; hence, because their contribution to the adjustment is small, their
marginally large normalized residuals are of little concern.  In this regard,
we see from Table~\ref{tab:adjres} that three of the four inconsistent data
have values of $S_{\rm c}$ considerably larger than 0.01; the exception is
${\it \Gamma}^\prime_{\rm p-90}$(lo) with $S_{\rm c}= 0.0099$, which is rounded
to 0.010 in the table.

\textit{Adjustment 2}.  Since the four direct lattice spacing measurements,
items $B38.1$-$B40$, are credible, as is the measurement of $h/m_{\rm
n}d_{220}({\scriptstyle{\rm W04}})$, item $B55$, after due consideration the
CODATA Task Group on Fundamental Constants decided that all five of these input
data should be considered for retention, but that each of their {\it a priori}
assigned uncertainties should be weighted by the multiplicative factor 1.5 to
reduce $|r_{\rm i}|$ of $h/m_{\rm n}d_{220}({\scriptstyle{\rm W04}})$ and of
$d_{220}({\scriptstyle{\rm NR3}})$ to a more acceptable level, that is, to
about 2, while maintaining their relative weights.  This has been done in
adjustment 2.  As can be seen from Table~\ref{tab:adjustsall}, this increase of
uncertainties has an inconsequential impact on the value of $\alpha$, and no
impact on the value of $h$.  It does reduce $R_{\rm B}$, as would be expected.

\textit{Adjustment 3}.  Again, since the measurement of $V_{\rm m}$(Si), item
$B53$, as well as the three measurements of $K_{\rm J}^2\,R_{\rm K}$, items
$B36.1$-$B36.3$, and the two measurements of $K_{\rm J}$, items $B35.1$ and
$B35.2$, are credible, the Task Group decided that all six should be considered
for retention, but that each of their {\it a priori} assigned uncertainties
should be weighted by the multiplicative factor 1.5 to reduce $|r_i|$ of
$V_{\rm m}$(Si) to about 2, while maintaining their relative weights.  This has
been done in adjustment 3.  Note that this also reduces $|r_i|$ of $h/m_{\rm
n}d_{220}({\scriptstyle{\rm W04}})$ from $2.03$ in adjustment 2 to $1.89$ in
adjustment 3.  We see from Table~\ref{tab:adjustsall} that this increase in
uncertainty has negligible consequences for the value of $\alpha$, but it does
increase the uncertainty of $h$ by about the same factor, as would be expected.
Also as would be expected, $R_{\rm B}$ is further reduced.

It may be recalled that faced with a similar situation in the 2002 adjustment,
the Task Group decided to use a multiplicative weighting factor of 2.325 in
order to reduce $|r_i|$ of $V_{\rm m}$(Si) to 1.50.  The reduced weighting
factor of 1.5 in the 2006 adjustment recognizes the new value of $K_{\rm
J}^2\,R_{\rm K}$ now available and the excellent agreement with the two earlier
values.

\textit{Adjustment 4}.  In adjustment 3, a number of input data, as measured by
their self-sensitivity coefficients $S_{\rm c}$, do not contribute in a
significant way to the determination of the adjusted constants.  We therefore
omit in adjustment 4 those input data with $S_{\rm c} < 0.01$ in adjustment 3
unless they are a subset of the data of an experiment that provides other input
data with $S_{\rm c} > 0.01$.  The 14 input data deleted in adjustment 4 for
this reason are $B31.1$-$B35.2$, $B37$, and $B56$, which are the five low- and
high-field proton and helion gyromagnetic ratio results; the five calculable
capacitor values of $R_{\rm K}$; both values of $K_{\rm J}$ as obtained using a
Hg electrometer and a voltage balance; the Ag coulometer result for the Faraday
constant; and the recoil/atom interferometry result for the quotient of the
Planck constant and mass of the cesium-133 atom.  The respective values of
$S_{\rm c}$ for these data in adjustment 3 are in the range 0.0000 to 0.0099.
Deleting such marginal data is in keeping with the practice followed in the
1998 and 2002 adjustments; see Sec.~I.D of CODATA-98.

Because $h/m(^{133}{\rm Cs})$, item $B56$, has been deleted as an input datum
due to its low weight, $A_{\rm r}(^{133}{\rm Cs})$, item $B8$, which is not
coupled to any other input datum, has also been omitted as an input datum and
as an adjusted constant from adjustment 4.  This brings the total number of
omitted items to 15.  Table~\ref{tab:adjustsall} shows that deleting these 15
data has virtually no impact on the values of $\alpha$ and $h$.

Adjustment 4 is the adjustment on which the 2006 CODATA recommended values are
based, and as such it is referred to as the ``final adjustment.''

\textit{Adjustments 5 and 6}.  These adjustments are intended to check the
robustness of adjustment 4, the final adjustment, while adjustments 7-11, which
are summarized in Table~\ref{tab:adjustsa}, probe various aspects of the
$R_\infty$ data in Table~\ref{tab:rdata}.

Adjustment 5 only differs from adjustment 3 in that it does not include the
input data that lead to the four most accurate values of $\alpha$: the two
measurements of $a_{\rm e}$, items $B11.1$ and $B11.2$, the measurement of
$h/m({\rm ^{133}Cs})$, item $B56$, and the measurement of $h/m({\rm ^{87}Rb})$,
item $B57$.  The $u_{\rm r}$ of the inferred values of $\alpha$ from these data
are $7.0\times10^{-10}$, $3.7\times10^{-9}$, $7.7\times10^{-9}$, and
$6.7\times10^{-9}$.  We see from Table~\ref{tab:adjustsall} that the value of
$\alpha$ from adjustment 5 is consistent with the 2006 recommended value from
adjustment 4 (the difference is 0.8$u_{\rm diff}$), but its uncertainty is
about 20 times larger.  Moreover, the resulting value of $h$ is the same as the
recommended value.

Adjustment 6 only differs from adjustment 3 in that it does not include the
input data that yield the three most accurate values of $h$, namely, the
watt-balance measurements of $K_{\rm J}^2\,R_{\rm K}$, items $B36.1$-$B36.3$.
The $u_{\rm r}$ of the inferred values of $h$ from these data, as they are used
in adjustment 3 (that is, after their uncertainties are multiplied by the
weighting factor 1.5), are $5.4\times10^{-8}$, $1.3\times10^{-7}$, and
$3.0\times10^{-7}$.  From Table~\ref{tab:adjustsall}, we see that the value of
$h$ from adjustment 6 is consistent with the 2006 recommended value from
adjustment 4 (the difference is 1.4$u_{\rm diff}$), but its uncertainty is well
over 6 times larger.  Furthermore, the resulting value of $\alpha$ is the same
as the recommended value.  Therefore adjustments 5 and 6 suggest that the less
accurate input data are consistent with the more accurate data, thereby
providing a consistency check on the 2006 recommended values of the constants.

\textit{Adjustments 7-11}.  These adjustments differ from adjustment 4, the
final adjustment, in the following ways.  In adjustment 7, the scattering-data
input values for both $R_{\rm p}$ and $R_{\rm d}$, items $A48$ and $A49$, are
omitted; in adjustment 8, only $R_{\rm p}$ is omitted, and in adjustment 9,
only $R_{\rm d}$ is omitted; adjustment 10 includes only the hydrogen data, and
adjustment 11 includes only the deuterium data, but for both, the H-D isotope
shift, item $A47$, is omitted.  Although a somewhat improved value of the
1S$_{1/2}$-2S$_{1/2}$ hydrogen transition frequency and improvements in the
theory of H and D energy levels have become available since the completion of
the 2002 adjustment, the value of $R_\infty$, which is determined almost
entirely by these data, has changed very little.  The values of $R_{\rm p}$ and
$R_{\rm d}$, which are also determined mainly by these data, have changed by
less than one third of their uncertainties.  The experimental and theoretical H
and D data remain highly consistent.

\subsubsection{Test of the Josephson and quantum Hall effect relations}
\label{sssec:epstests}

Investigation of the exactness of the relations $K_{\rm J} = 2e/h$ and $R_{\rm
K} = h/e^2$ is carried out, as in CODATA-02, by writing
\begin{eqnarray}
K_{\rm J} &=&\frac{2e}{h}\left( {1+\varepsilon _{\rm J} } \right)=\left( 
{\frac{8\alpha }{\mu _0 ch}} \right)^{1/2}\left( {1+\varepsilon _{\rm J} } 
\right)
\label{eq:kjeps}
\\
R_{\rm K} &=&\frac{h}{e^2}\left( {1+\varepsilon_{\rm K} } \right)=\frac{\mu 
_0 c}{2\alpha }\left( {1+\varepsilon_{\rm K} } \right) \ ,
\label{eq:rkeps}
\end{eqnarray}
where $\varepsilon_{\rm J}$ and $\varepsilon_{\rm K}$ are unknown correction
factors taken to be additional adjusted constants determined by least-squares
calculations.  Replacing the relations $K_{\rm J} = 2e/h$ and $R_{\rm K} =
h/e^2$ with the generalizations in Eqs.~(\ref{eq:kjeps}) and (\ref{eq:rkeps})
in the analysis leading to the observational equations in
Table~\ref{tab:pobseqsb1} leads to the modified observational equations given
in Table~\ref{tab:pobseqseps}.

The results of seven different adjustments are presented in
Table~\ref{tab:epsilons}.  In addition to the adjusted values of $\alpha$, $h$,
$\varepsilon_{\rm J}$, and $\varepsilon_{\rm K}$, we also give the normalized
residuals $r_i$ of the four input data with the largest values of $|r_i|$:
$V_{\rm m}$(Si), item $B53$, $h/m_{\rm n}d_{220}({\scriptstyle{\rm W04}})$,
item $B55$, $d_{220}({\scriptstyle{\rm NR3}})$, item $B39$, and the NIST-89
value for ${\it \Gamma}^\prime_{\rm p-90}$(lo), item $B31.1$.  The residuals
are included as additional indicators of whether relaxing the assumption
$K_{\rm J} = 2e/h$ and $R_{\rm K} = h/e^2$ reduces the disagreements among the
data.

The adjusted value of $R_\infty$ is not included in Table~\ref{tab:epsilons},
because it remains essentially unchanged from one adjustment to the next and
equal to the 2006 recommended value.  An entry of 0 in the $\varepsilon_{\rm
K}$ column means that it is assumed that $R_{\rm K} = h/e^2$ in the
corresponding adjustment; similarly, an entry of 0 in the $\varepsilon_{\rm J}$
column means that it is assumed that $K_{\rm J} = 2e/h$ in the corresponding
adjustment.  The following comments apply to the adjustments of
Table~\ref{tab:epsilons}.

Adjustment (i) is identical to adjustment 1 of Tables~\ref{tab:adjustsall} and
\ref{tab:adjres} in the previous section and is included here simply for
reference; all of the input data are included and multiplicative weighting
factors have not been applied to any uncertainties.  For this adjustment, $N =
150$, $M = 79$, $\nu = 71$, and $\chi^2 = 92.1$.

The next three adjustments differ from adjustment (i) in that in adjustment
(ii) the relation $K_{\rm J} = 2e/h$ is relaxed, in adjustment (iii) the
relation $R_{\rm K} = h/e^2$ is relaxed, and in adjustment (iv) both of the
relations are relaxed.  For these three adjustments, $N = 150$, $M = 80$, $\nu
= 70$, and $\chi^2 = 91.5$; $N = 150$, $M = 80$, $\nu = 70$, and $\chi^2 =
91.3$; and $N = 150$, $M = 81$, $\nu = 69$, and $\chi^2 = 90.4$, respectively.

It is clear from Table~\ref{tab:epsilons} that there is no evidence for the
inexactness of either of the relations $K_{\rm J} = 2e/h$ or $R_{\rm K} =
h/e^2$.  This conclusion is also true if instead of taking adjustment 1 of
Table~\ref{tab:adjustsall} as our starting point, we had taken adjustment 2 in
which the uncertainties of the five x-ray related data are multiplied by the
factor 1.5.  That is, none of the numbers in Table~\ref{tab:epsilons} would
change significantly, except $R_{\rm B}$ would be reduced from 1.14 to about
1.08.  The reason adjustments (iii)-(vii) summarized in
Table~\ref{tab:epsilons} give values of $\epsilon_{\rm K}$ consistent with zero
within about 2 parts in $10^8$ is mainly because the value of alpha inferred
from the mean of the five measured values of $R_{\rm K}$ under the assumption
$R_{\rm K} = h/e^2$, which has $u_{\rm r} = 1.8 \times 10^{-8}$, agrees with
the value of $\alpha$ with $u_{\rm r} = 7.0 \times 10^{-10}$ inferred from the
Harvard University measured value of $a_{\rm e}$.

Table~\ref{tab:pobseqseps} and the uncertainties of the 2006 input data
indicate that the values of $\epsilon_{\rm J}$ from adjustments (ii) and (iv)
are determined mainly by the input data for the quantities ${\it
\Gamma}^\prime_{\rm p-90}$(lo) and ${\it \Gamma}^\prime_{\rm h-90}$(lo) with
observational equations that depend on $\epsilon_{\rm J}$ but not on $h$; and
by the input data for the quantities ${\it \Gamma}^\prime_{\rm p-90}$(hi),
$K_{\rm J}$, $K_{\rm J}^2\,R_{\rm K}$, and ${\cal F}_{90}$, with observational
equations that depend on both $\epsilon_{\rm J}$ and $h$.  Because the value of
$h$ in these least-squares calculations arises primarily from the measured
value of the molar volume of silicon, $V_{\rm m}$(Si), the values of
$\epsilon_{\rm J}$ in adjustments (ii) and (iv) arise mainly from a combination
of individual values of $\epsilon_{\rm J}$ that either depend on $V_{\rm
m}$(Si) or on ${\it \Gamma}^\prime_{\rm p-90}$(lo) and ${\it
\Gamma}^\prime_{\rm h-90}$(lo).  It is therefore of interest to repeat
adjustment (iv), first with $V_{\rm m}$(Si) deleted but with the ${\it
\Gamma}^\prime_{\rm p-90}$(lo) and ${\it \Gamma}^\prime_{\rm h-90}$(lo) data
included, and then with the latter deleted but with $V_{\rm m}$(Si) included.
These are, in fact, adjustments (v) and (vi) of Table~\ref{tab:epsilons}.

In each of these adjustments, the absolute values of $\epsilon_{\rm J}$ are
comparable and significantly larger than the uncertainties, which are also
comparable, but the values have different signs.  Consequently, when $V_{\rm
m}$(Si) and the ${\it \Gamma}^\prime_{\rm p-90}$(lo) and ${\it
\Gamma}^\prime_{\rm h-90}$(lo) data are included at the same time as in
adjustment (iv), the result for $\epsilon_{\rm J}$ is consistent with zero.

The values of $\epsilon_{\rm J}$ from adjustments (v) and (vi) reflect some of
the inconsistencies among the data: the disagreement of the values of $h$
implied by $V_{\rm m}$(Si) and $K_{\rm J}^2\,R_{\rm K}$ when it is assumed that
the relations $K_{\rm J} = 2e/h$ and $R_{\rm K} = h/e^2$ are exact; and the
disagreement of the values of $\alpha$ implied by the electron magnetic moment
anomaly $a_{\rm e}$ and by ${\it \Gamma}^\prime_{\rm p-90}$(lo) and ${\it
\Gamma}^\prime_{\rm h-90}$(lo) under the same assumption.

In adjustment (vii), the problematic input data for $V_{\rm m}$(Si), ${\it
\Gamma}^\prime_{\rm p-90}$(lo), and ${\it \Gamma}^\prime_{\rm h-90}$(lo) are
simultaneously deleted from the calculation.  Then the value of $\epsilon_{\rm
J}$ arises mainly from the input data for ${\it \Gamma}^\prime_{\rm p-90}$(hi),
$K_{\rm J}$, $K_{\rm J}^2R_{\rm K}$, and ${\cal F}_{90}$.  Like adjustment
(iv), adjustment (vii) shows that $\epsilon_{\rm J}$ is consistent with zero,
although not within 8 parts in $10^8$ but within 7 parts in $10^7$.  However,
adjustment (vii) has the advantage of being based on consistent data.

The comparatively narrow range of values of alpha in Table~\ref{tab:epsilons}
is due to the fact that the input data that mainly determine alpha do not
depend on the Josephson or quantum Hall effects.  This is not the case for the
input data that primarily determine $h$, hence the values of $h$ vary over a
wide range.

\def\m{\phantom{-}}
\def\fixh{\vbox to 10pt {}}
\def\fixhs{\vbox to 4pt {}}
\begin{table*}
\caption{Values of $A_{\rm r}({\rm e})$ implied by the input data in
Table~\ref{tab:pdata} in order of increasing standard uncertainty.}
\label{tab:ares}
\def\sp{\hbox to 17 pt {}}
\begin{tabular}{l@{\sp}l@{\sp}l@{\sp}l@{\sp}l@{\sp}l}
\toprule
Primary & Item & Identification \fixh
& Sec. and Eq. & \hbox to 25pt{} $A_{\rm r}({\rm e})$ & \hbox to -5pt {} Relative standard \\
source & number &
& & & \hbox to -5pt {} uncertainty $u_{\rm r}$  \\
\colrule

$f_{\rm s}({\rm C})/f_{\rm c}({\rm C})$ \hbox to 20pt{} & \fixh \Nrfsfccjb &
GSI-02 & \ref{par:gec} (\ref{eq:arec02}) & 
$ 0.000\,548\,579\,909\,32(29)$ \hbox to 20pt {} & $ 5.2\times 10^{-10}$\\

$f_{\rm s}({\rm O})/f_{\rm c}({\rm O})$ & \fixh \Nrfsfcojb &
GSI-02 & \ref{par:geo} (\ref{eq:areo02}) & 
$ 0.000\,548\,579\,909\,58(42)$& $ 7.6\times 10^{-10}$\\

$\Delta \nu_{\bar{\rm p}\,{\rm He^+}} $ & \fixh $C1-C24$ &
JINR/CERN-06 & \ref{sssec:apheare} (\ref{eq:areaphe}) & 
$ 0.000\,548\,579\,908\,81(91)$& $ 1.7\times 10^{-9}$ \rule[-6pt]{0pt}{6pt}\\

$A_{\rm r}({\rm e})$ & \fixhs \Nareie &
UWash-95 & \ref{ssec:ptmare} (\ref{eq:areexp}) & 
$ 0.000\,548\,579\,9111(12)$& $ 2.1\times 10^{-9}$\\

\botrule
\end{tabular}
\end{table*}

\def\fixh{\vbox to 9pt {}}
\begin{table}
\caption{The 28 adjusted constants (variables) used in the least-squares
multivariate analysis of the Rydberg-constant data given in
Table~\ref{tab:rdata}.  These adjusted constants appear as arguments of the
functions on the right-hand side of the observational equations of
Table~\ref{tab:pobseqsa}.  The notation for hydrogenic energy levels
$E_{X}(n{\rm L}_j)$ and for additive corrections $\delta_{X}(n{\rm L}_j)$ in
this table have the same meaning as the notations $E^{X}_{n{\rm L}j}$ and
$\delta^{X}_{n{\rm L}j}$ in Sec.~\ref{par:teu}.}
\label{tab:adjcona}
\def\sp{\hbox to 18 pt{}}
\begin{tabular}{l@{\sp}l}
\toprule
Adjusted constant & Symbol \T \\
\colrule
 \fixh  Rydberg constant & $R_\infty$ \\
 \fixh  bound-state proton rms charge radius & $R_{\rm p}$ \\
 \fixh  bound-state deuteron rms charge radius~~~~~ & $R_{\rm d}$ \\
 \fixh  additive correction to $E_{\rm H}(1{\rm S}_{1/2})/h$ & $\delta_{\rm H}(1{\rm S}_{1/2})$ \\
 \fixh  additive correction to $E_{\rm H}(2{\rm S}_{1/2})/h$ & $\delta_{\rm H}(2{\rm S}_{1/2})$ \\
 \fixh  additive correction to $E_{\rm H}(3{\rm S}_{1/2})/h$ & $\delta_{\rm H}(3{\rm S}_{1/2})$ \\
 \fixh  additive correction to $E_{\rm H}(4{\rm S}_{1/2})/h$ & $\delta_{\rm H}(4{\rm S}_{1/2})$ \\
 \fixh  additive correction to $E_{\rm H}(6{\rm S}_{1/2})/h$ & $\delta_{\rm H}(6{\rm S}_{1/2})$ \\
 \fixh  additive correction to $E_{\rm H}(8{\rm S}_{1/2})/h$ & $\delta_{\rm H}(8{\rm S}_{1/2})$ \\
 \fixh  additive correction to $E_{\rm H}(2{\rm P}_{1/2})/h$ & $\delta_{\rm H}(2{\rm P}_{1/2})$ \\
 \fixh  additive correction to $E_{\rm H}(4{\rm P}_{1/2})/h$ & $\delta_{\rm H}(4{\rm P}_{1/2})$ \\
 \fixh  additive correction to $E_{\rm H}(2{\rm P}_{3/2})/h$ & $\delta_{\rm H}(2{\rm P}_{3/2})$ \\
 \fixh  additive correction to $E_{\rm H}(4{\rm P}_{3/2})/h$ & $\delta_{\rm H}(4{\rm P}_{3/2})$ \\
 \fixh  additive correction to $E_{\rm H}(8{\rm D}_{3/2})/h$ & $\delta_{\rm H}(8{\rm D}_{3/2})$ \\
 \fixh  additive correction to $E_{\rm H}(12{\rm D}_{3/2})/h$ & $\delta_{\rm H}(12{\rm D}_{3/2})$ \\
 \fixh  additive correction to $E_{\rm H}(4{\rm D}_{5/2})/h$ & $\delta_{\rm H}(4{\rm D}_{5/2})$ \\
 \fixh  additive correction to $E_{\rm H}(6{\rm D}_{5/2})/h$ & $\delta_{\rm H}(6{\rm D}_{5/2})$ \\
 \fixh  additive correction to $E_{\rm H}(8{\rm D}_{5/2})/h$ & $\delta_{\rm H}(8{\rm D}_{5/2})$ \\
 \fixh  additive correction to $E_{\rm H}(12{\rm D}_{5/2})/h$ & $\delta_{\rm H}(12{\rm D}_{5/2})$ \\
 \fixh  additive correction to $E_{\rm D}(1{\rm S}_{1/2})/h$ & $\delta_{\rm D}(1{\rm S}_{1/2})$ \\
 \fixh  additive correction to $E_{\rm D}(2{\rm S}_{1/2})/h$ & $\delta_{\rm D}(2{\rm S}_{1/2})$ \\
 \fixh  additive correction to $E_{\rm D}(4{\rm S}_{1/2})/h$ & $\delta_{\rm D}(4{\rm S}_{1/2})$ \\
 \fixh  additive correction to $E_{\rm D}(8{\rm S}_{1/2})/h$ & $\delta_{\rm D}(8{\rm S}_{1/2})$ \\
 \fixh  additive correction to $E_{\rm D}(8{\rm D}_{3/2})/h$ & $\delta_{\rm D}(8{\rm D}_{3/2})$ \\
 \fixh  additive correction to $E_{\rm D}(12{\rm D}_{3/2})/h$ & $\delta_{\rm D}(12{\rm D}_{3/2})$ \\
 \fixh  additive correction to $E_{\rm D}(4{\rm D}_{5/2})/h$ & $\delta_{\rm D}(4{\rm D}_{5/2})$ \\
 \fixh  additive correction to $E_{\rm D}(8{\rm D}_{5/2})/h$ & $\delta_{\rm D}(8{\rm D}_{5/2})$ \\
 \fixh  additive correction to $E_{\rm D}(12{\rm D}_{5/2})/h$ & $\delta_{\rm D}(12{\rm D}_{5/2})$ \\
\botrule
\end{tabular}
\end{table}

\def\tfrac#1#2{{\vbox to 9pt {}\phantom{_I}\textstyle{#1}\over\vbox to 9pt {}\textstyle{#2}}}
\def\vh{\vbox to 15pt {}}
\def\vhh{\vbox to 10pt {}}
\begin{table*}
\caption{Observational equations that express the input data related to
$R_\infty$ in Table~\ref{tab:rdata} as functions of the adjusted constants in
Table~\ref{tab:adjcona}.  The numbers in the first column correspond to the
numbers in the first column of Table~\ref{tab:rdata}.  The expressions for the
energy levels of hydrogenic atoms are discussed in Sec.~\ref{sssec:rydth}.  As
pointed out in Sec.~\ref{par:teu}, $E_{X}(n{\rm L}_j)/h$ is in fact
proportional to $cR_\infty$ and independent of $h$, hence $h$ is not an
adjusted constant in these equations.  The notation for hydrogenic energy
levels $E_{X}(n{\rm L}_j)$ and for additive corrections $\delta_{X}(n{\rm
L}_j)$ in this table have the same meaning as the notations $E^{X}_{n{\rm L}j}$
and $\delta^{X}_{n{\rm L}j}$ in Sec.~\ref{par:teu}.  See Sec.~\ref{ssec:mada}
for an explanation of the symbol $\doteq$.  }
\label{tab:pobseqsa}
\begin{tabular}{l@{\quad~~}rcll}
\toprule
Type of input & &
\multicolumn{2}{l}{Observational equation} \T & \\
datum      & & &  \\
\colrule

\vh$A1$--$A16 \quad $&$ \delta_{\rm H}(n{\rm L}_j) $&$\doteq$&$ \delta_{\rm H}(n{\rm L}_j) $ \\

\vh$A17$--$A25 \quad $&$ \delta_{\rm D}(n{\rm L}_j) $&$\doteq$&$ \delta_{\rm D}(n{\rm L}_j) $ \\

\vh$A26$--$A31 \quad $
&$ \nu_{\rm H}(n_1{\rm L_1}_{j_1} - n_2{\rm L_2}_{j_2}) $&$\doteq$&$ 
\big[E_{\rm H}\big(n_2{\rm L_2}_{j_2};R_\infty,\alpha,A_{\rm r}({\rm e}),A_{\rm r}({\rm p}),R_{\rm p},\delta_{\rm H}(n_2{\rm L_2}_{j_2})\big) $  \\
\vhh$A38,A39$&&&$-E_{\rm H}\big(n_1{\rm L_1}_{j_1};R_\infty,\alpha,A_{\rm r}({\rm e}),A_{\rm r}({\rm p}),R_{\rm p},\delta_{\rm H}(n_1{\rm L_1}_{j_1})\big)\big]/h $  \\

\vh$A32$--$A37 \quad $
&$ \nu_{\rm H}(n_1{\rm L_1}_{j_1} - n_2{\rm L_2}_{j_2}) 
-\fr{1}{4}\nu_{\rm H}(n_3{\rm L_3}_{j_3} - n_4{\rm L_4}_{j_4})$&$\doteq$&$ 
\Big\{E_{\rm H}\big(n_2{\rm L_2}_{j_2};R_\infty,\alpha,A_{\rm r}({\rm e}),A_{\rm r}({\rm p}),R_{\rm p},\delta_{\rm H}(n_2{\rm L_2}_{j_2})\big) $  \\
\vhh&&&$ \ -E_{\rm H}\big(n_1{\rm L_1}_{j_1};R_\infty,\alpha,A_{\rm r}({\rm e}),A_{\rm r}({\rm p}),R_{\rm p},\delta_{\rm H}(n_1{\rm L_1}_{j_1})\big) $  \\
\vhh
&&&$  \ -\fr{1}{4} \big[E_{\rm H}\big(n_4{\rm L_4}_{j_4};R_\infty,\alpha,A_{\rm r}({\rm e}),A_{\rm r}({\rm p}),R_{\rm p},\delta_{\rm H}(n_4{\rm L_4}_{j_4})\big) $  \\
\vhh&&&$ \quad -E_{\rm H}\big(n_3{\rm L_3}_{j_3};R_\infty,\alpha,A_{\rm r}({\rm e}),A_{\rm r}({\rm p}),R_{\rm p},\delta_{\rm H}(n_3{\rm L_3}_{j_3})\big)\big]\Big\}/h $  \\

\vh$A40$--$A44 \quad $
&$ \nu_{\rm D}(n_1{\rm L_1}_{j_1} - n_2{\rm L_2}_{j_2}) $&$\doteq$&$ 
\big[E_{\rm D}\big(n_2{\rm L_2}_{j_2};R_\infty,\alpha,A_{\rm r}({\rm e}),A_{\rm r}({\rm d}),R_{\rm d},\delta_{\rm D}(n_2{\rm L_2}_{j_2})\big) $  \\
\vhh&&&$-E_{\rm D}\big(n_1{\rm L_1}_{j_1};R_\infty,\alpha,A_{\rm r}({\rm e}),A_{\rm r}({\rm d}),R_{\rm d},\delta_{\rm D}(n_1{\rm L_1}_{j_1})\big)\big]/h $  \\

\vh$A45$--$A46 \quad $
&$ \nu_{\rm D}(n_1{\rm L_1}_{j_1} - n_2{\rm L_2}_{j_2}) 
-\fr{1}{4}\nu_{\rm D}(n_3{\rm L_3}_{j_3} - n_4{\rm L_4}_{j_4})$&$\doteq$&$ 
\Big\{E_{\rm D}\big(n_2{\rm L_2}_{j_2};R_\infty,\alpha,A_{\rm r}({\rm e}),A_{\rm r}({\rm d}),R_{\rm d},\delta_{\rm D}(n_2{\rm L_2}_{j_2})\big) $  \\
\vhh&&&$ \ -E_{\rm D}\big(n_1{\rm L_1}_{j_1};R_\infty,\alpha,A_{\rm r}({\rm e}),A_{\rm r}({\rm d}),R_{\rm d},\delta_{\rm D}(n_1{\rm L_1}_{j_1})\big) $  \\
\vhh
&&&$  \ -\fr{1}{4} \big[E_{\rm D}\big(n_4{\rm L_4}_{j_4};R_\infty,\alpha,A_{\rm r}({\rm e}),A_{\rm r}({\rm d}),R_{\rm d},\delta_{\rm D}(n_4{\rm L_4}_{j_4})\big) $  \\
\vhh&&&$ \quad -E_{\rm D}\big(n_3{\rm L_3}_{j_3};R_\infty,\alpha,A_{\rm r}({\rm e}),A_{\rm r}({\rm d}),R_{\rm d},\delta_{\rm D}(n_3{\rm L_3}_{j_3})\big)\big]\Big\}/h $  \\

\vh$A47 \quad $
&$ \nu_{\rm D}(1{\rm S}_{1/2} - 2{\rm S}_{1/2}) 
-\nu_{\rm H}(1{\rm S}_{1/2} - 2{\rm S}_{1/2})$&$\doteq$&$ 
\Big\{E_{\rm D}\big(2{\rm S}_{1/2};R_\infty,\alpha,A_{\rm r}({\rm e}),A_{\rm r}({\rm d}),R_{\rm d},\delta_{\rm D}(2{\rm S}_{1/2})\big) $  \\
\vhh&&&$ \ -E_{\rm D}\big(1{\rm S}_{1/2};R_\infty,\alpha,A_{\rm r}({\rm e}),A_{\rm r}({\rm d}),R_{\rm d},\delta_{\rm D}(1{\rm S}_{1/2})\big) $  \\
\vhh
&&&$  \ - \big[E_{\rm H}\big(2{\rm S}_{1/2};R_\infty,\alpha,A_{\rm r}({\rm e}),A_{\rm r}({\rm p}),R_{\rm p},\delta_{\rm H}(2{\rm S}_{1/2})\big) $  \\
\vhh&&&$ \quad -E_{\rm H}\big(1{\rm S}_{1/2};R_\infty,\alpha,A_{\rm r}({\rm e}),A_{\rm r}({\rm p}),R_{\rm p},\delta_{\rm H}(1{\rm S}_{1/2})\big)\big]\Big\}/h $  \\

\vh$A48 \quad $&$ R_{\rm p} $&$\doteq$&$ R_{\rm p} $  \\

\vh$A49 \quad $&$ R_{\rm d} $&$\doteq$&$ R_{\rm d} $  \\

\botrule
\end{tabular}
\end{table*}

\def\fixh{\vbox to 9pt {}}
\begin{table}
\caption{The 39 adjusted constants 
(variables) used in the least-squares multivariate analysis of
the input data
in Table~\ref{tab:pdata}.
These adjusted constants appear as arguments
of the functions on the right-hand side of
the observational equations of Table~\ref{tab:pobseqsb1}.
}
\label{tab:adjconb}
\def\sp{\hbox to 24 pt{}}
\begin{tabular}{l@{\sp}l}
\toprule
\vbox to 10 pt {}
Adjusted constant & Symbol \\
\colrule

 \fixh  electron relative atomic mass & $A_{\rm r}({\rm e})$ \\
 \fixh  proton relative atomic mass & $A_{\rm r}({\rm p})$ \\
 \fixh  neutron relative atomic mass & $A_{\rm r}({\rm n})$ \\
 \fixh  deuteron relative atomic mass & $A_{\rm r}({\rm d})$ \\
 \fixh  triton relative atomic mass & $A_{\rm r}({\rm t})$ \\
 \fixh  helion relative atomic mass & $A_{\rm r}({\rm h})$ \\
 \fixh  alpha particle relative atomic mass & $A_{\rm r}(\rmalpha)$ \\
 \fixh  $^{16}$O$^{7+}$ relative atomic mass & $A_{\rm r}(^{16}{\rm O}^{7+})$ \\
 \fixh  $^{87}$Rb relative atomic mass  & $A_{\rm r}(^{87}{\rm Rb})$ \\
 \fixh  $^{133}$Cs relative atomic mass & $A_{\rm r}(^{133}{\rm Cs})$ \\
 \fixh  fine-structure constant & $\alpha$ \\
 \fixh  additive correction to $a_{\rm e}$(th) & $\delta_{\rm e}$ \\
 \fixh  additive correction to $a_{\rmssmu}$(th) & $\delta_{\rmssmu}$ \\
 \fixh  additive correction to $g_{\rm C}$(th) & $\delta_{\rm C}$ \\
 \fixh  additive correction to $g_{\rm O}$(th) & $\delta_{\rm O}$ \\
 \fixh  electron-proton magnetic moment ratio & $\mu_{\rm e^-}/\mu_{\rm p}$ \\
 \fixh  deuteron-electron magnetic moment ratio \qquad& $\mu_{\rm d}/\mu_{\rm e^-}$ \\
 \fixh  triton-proton magnetic moment ratio & $\mu_{\rm t}/\mu_{\rm p}$ \\
 \fixh  shielding difference of d and p in HD & $\sigma_{\rm dp}$ \\
 \fixh  shielding difference of t and p in HT & $\sigma_{\rm tp}$ \\
 \fixh  electron to shielded proton & \\
 \fixh  \ \ magnetic moment ratio & $\mu_{\rm e^-}/\mu^\prime_{\rm p}$ \\
 \fixh  shielded helion to shielded proton  & \\ 
 \fixh  \ \ magnetic moment ratio & $\mu^\prime_{\rm h}/\mu^\prime_{\rm p}$ \\
 \fixh  neutron to shielded proton & \\
 \fixh  \ \ magnetic moment ratio & $\mu_{\rm n}/\mu_{\rm p}^\prime$ \\
 \fixh  electron-muon mass ratio & $m_{\rm e}/m_{\rmssmu}$ \\
 \fixh  additive correction to $\Delta\nu_{\rm Mu}({\rm th})$ & $\delta_{\rm Mu}$ \\
 \fixh  Planck constant & $ h $ \\
 \fixh  molar gas constant & $R$ \\
 \fixh  copper K${\rm \alpha}_1$ x unit & xu(CuK${\rm \alpha}_1)$ \\
 \fixh  molybdenum K${\rm \alpha}_1$ x unit & xu(MoK${\rm \alpha}_1)$ \\
 \fixh  \aa ngstrom star & \AA$^*$  \\
 \fixh  $d_{220}$ of Si crystal ILL & $d_{220}({\rm {\scriptstyle ILL}})$ \\
 \fixh  $d_{220}$ of Si crystal N & $d_{220}({\rm {\scriptstyle N}})$ \\
 \fixh  $d_{220}$ of Si crystal WASO 17 & $d_{220}({\rm {\scriptstyle W17}})$ \\
 \fixh  $d_{220}$ of Si crystal WASO 04 & $d_{220}({\rm {\scriptstyle W04}})$ \\
 \fixh  $d_{220}$ of Si crystal WASO 4.2a & $d_{220}({\rm {\scriptstyle W4.2a}})$ \\
 \fixh  $d_{220}$ of Si crystal MO$^*$ & $d_{220}({\rm {\scriptstyle MO^*}})$ \\
 \fixh  $d_{220}$ of Si crystal NR3 & $d_{220}({\rm {\scriptstyle NR3}})$ \\
 \fixh  $d_{220}$ of Si crystal NR4 & $d_{220}({\rm {\scriptstyle NR4}})$ \\
 \fixh  $d_{220}$ of an ideal Si crystal & $d_{220}$ \\
\botrule
\end{tabular}
\end{table}

\def\tfrac#1#2{{\vbox to 9pt {}\phantom{_I}\textstyle{#1}\over\vbox to 9pt {}\textstyle{#2}}}
\def\vh{\vbox to 15pt {}}
\def\vhh{\vbox to 20pt {}}
\def\vhhh{\vbox to 22pt {}}
\begin{table*}
\caption{ Observational equations that express the input data in
Table~\ref{tab:pdata} as functions of the adjusted constants in
Table~\ref{tab:adjconb}.  The numbers in the first column correspond to the
numbers in the first column of Table~\ref{tab:pdata}.  For simplicity, the
lengthier functions are not explicitly given.  See Sec.~\ref{ssec:mada} for an
explanation of the symbol $\doteq$.}
\label{tab:pobseqsb1}
\def\sp{\hbox to 32 pt {}}
\begin{tabular}{l@{\sp}rcl@{\sp}l}
\toprule
Type of input & &\multicolumn{2}{l}{Observational equation}& \hbox to 80pt{}\\
datum      &\hbox to 10pt{} & & \hbox to 240pt{} & Sec. \\
\colrule
\vh$B1 \quad $&$ A_{\rm r}(^1{\rm H})$ &$\doteq$&$ A_{\rm r}({\rm p}) + A_{\rm r}({\rm e}) - E_{\rm b}(^1{\rm H})/m_{\rm u}c^2$  & \ref{ssec:ramnuc} \\

\vh$B2 \quad $&$ A_{\rm r}(^2{\rm H}) $&$\doteq$&$ A_{\rm r}({\rm d}) + A_{\rm r}({\rm e}) - E_{\rm b}(^2{\rm H})/m_{\rm u}c^2$ & \ref{ssec:ramnuc} \\

\vh$B3 \quad $&$ A_{\rm r}(^3{\rm H}) $&$\doteq$&$ A_{\rm r}({\rm t}) + A_{\rm r}({\rm e}) - E_{\rm b}(^3{\rm H})/m_{\rm u}c^2$ & \ref{ssec:ramnuc} \\

\vh$B4 \quad $&$ A_{\rm r}(^3{\rm He}) $&$\doteq$&$ A_{\rm r}({\rm h}) + 2 A_{\rm r}({\rm e}) - E_{\rm b}(^3{\rm He})/m_{\rm u}c^2$ & \ref{ssec:ramnuc} \\

\vh$B5 \quad $&$ A_{\rm r}(^4{\rm He}) $&$\doteq$&$ A_{\rm r}({\rmalpha}) + 2 A_{\rm r}({\rm e}) - E_{\rm b}(^4{\rm He})/m_{\rm u}c^2$ & \ref{ssec:ramnuc} \\

\vh$B6 \quad $&$ A_{\rm r}(^{16}{\rm O}) $&$\doteq$&$ A_{\rm r}(^{16}{\rm O}^{7+}) + 7 A_{\rm r}({\rm e}) - \left[E_{\rm b}(^{16}{\rm O})-E_{\rm b}(^{16}{\rm O}^{7+})\right]\!/m_{\rm u}c^2$ & \ref{par:geo} \\

\vh$B7 \quad $&$ A_{\rm r}(^{87}{\rm Rb}) $&$\doteq$&$ A_{\rm r}(^{87}{\rm Rb})$ & \\

\vh$B8 \quad $&$ A_{\rm r}(^{133}{\rm Cs}) $&$\doteq$&$ A_{\rm r}(^{133}{\rm Cs})$ & \\

\vh$B9 \quad $&$ A_{\rm r}({\rm e}) $&$\doteq$&$ A_{\rm r}({\rm e})$ & \\

\vh$B10 \quad $&$ \delta_{\rm e}$&$\doteq$&$ \delta_{\rm e}$ \\

\vh$B11 \quad $&$ a_{\rm e} $&$\doteq$&$ a_{\rm e}(\alpha,\delta_e)$ & \ref{sssec:ath} \\

\vh$B12 \quad $&$ \delta_{\rmssmu}$&$\doteq$&$ \delta_{\rmssmu}$ \\

\vhh$B13 \quad $&$ \overline{R}$&$\doteq$&$ -\tfrac{a_{\rmssmu}(\alpha,\delta_{\rmssmu})}
{1+a_{\rm e}(\alpha,\delta_{\rm e})}
\tfrac{m_{\rm e}}{m_{\rmssmu}}
\tfrac{\mu_{\rm e^-}}{\mu_{\rm p}} $
&\ref{sssec:amb}
\\

\vh$B14 \quad $&$ \delta_{\rm C}$&$\doteq$&$ \delta_{\rm C}$ \\

\vh$B15 \quad $&$ \delta_{\rm O}$&$\doteq$&$ \delta_{\rm O}$ \\

\vhh$B16 \quad $&$ \tfrac{f_{\rm s}\left(^{12}{\rm C}^{5+}\right)}{f_{\rm c}\left(^{12}{\rm C}^{5+}\right)} $&$\doteq$&$ -\tfrac{g_{\rm C}(\alpha,\delta_{\rm C})}{10 A_{\rm r}({\rm e})} \left[12-5A_{\rm r}({\rm e}) + \tfrac{E_{\rm b}\left(^{12}{\rm C}\right) -E_{\rm b}\left(^{12}{\rm C}^{5+}\right)}{ m_{\rm u}c^2}\right] $ & \ref{par:gec} \\

\vhh$B17 \quad $&$ \tfrac{f_{\rm s}\left(^{16}{\rm O}^{7+}\right)}{f_{\rm c}\left(^{16}{\rm O}^{7+}\right)} $&$\doteq$&$ -\tfrac{g_{\rm O}(\alpha,\delta_{\rm O})}{14 A_{\rm r}({\rm e})} A_{\rm r}(^{16}{\rm O}^{7+}) $ & \ref{par:geo} \\

\vhh$B18 \quad $&$ \tfrac{\mu_{\rm e^-}({\rm H})}{\mu_{\rm p}({\rm H})} $&$\doteq$&$ 
\tfrac{g_{\rm e^-}({\rm H})}{g_{\rm e^-}}
\left(\tfrac{g_{\rm p}({\rm H})}{g_{\rm p}}\right)^{-1}
\tfrac{\mu_{\rm e^-}}{\mu_{\rm p}} $
&\ref{par:epmmr}
\\

\vhhh$B19 \quad $&$ \tfrac{\mu_{\rm d}({\rm D})}{\mu_{\rm e^-}({\rm D})} $&$\doteq$&$ 
\tfrac{g_{\rm d}({\rm D})}{{g_{\rm d}}}
\left(\tfrac{g_{\rm e^-}({\rm D})}{g_{\rm e^-}}\right)^{-1}
\tfrac{\mu_{\rm d}}{\mu_{\rm e^-}} $
&\ref{par:demmr}
\\

\vhhh$B20 \quad $&$ \tfrac{\mu_{\rm p}({\rm HD})}{\mu_{\rm d}({\rm HD})} $&$\doteq$&$ 
\left[1 + \sigma_{\rm dp} \right]
\tfrac{\mu_{\rm p}}{\mu_{\rm e^-}}
\tfrac{\mu_{\rm e^-}}{\mu_{\rm d}} $
&\ref{par:dpmmr}
\\

\vh$B21 \quad $&$ \sigma_{\rm dp}$&$\doteq$&$ \sigma_{\rm dp}$ \\

\vhhh$B22 \quad $&$ \tfrac{\mu_{\rm t}({\rm HT})}{\mu_{\rm p}({\rm HT})} $&$\doteq$&$ 
\left[1 - \sigma_{\rm tp} \right]
\tfrac{\mu_{\rm t}}{\mu_{\rm p}} $
&\ref{par:dpmmr}
\\

\vh$B23 \quad $&$ \sigma_{\rm tp}$&$\doteq$&$ \sigma_{\rm tp}$ \\

\vhhh$B24 \quad $&$ \tfrac{\mu_{\rm e^-}({\rm H})}{\mu_{\rm p}^\prime} $&$\doteq$&$ 
\tfrac{g_{\rm e^-}({\rm H})}{g_{\rm e^-}}
\tfrac{\mu_{\rm e^-}}{\mu_{\rm p}^\prime} $
&\ref{par:espmm}
\\

\vhhh$B25 \quad $&$ \tfrac{\mu_{\rm h}^\prime}{\mu_{\rm p}^\prime} $&$\doteq$&$ 
\tfrac{\mu_{\rm h}^\prime}{\mu_{\rm p}^\prime} 
$\\

\vh$B26 \quad $&$ \tfrac{\mu_{\rm n}}{\mu_{\rm p}^\prime} $&$\doteq$&$ 
\tfrac{\mu_{\rm n}}{\mu_{\rm p}^\prime} 
$  \\ \\

\botrule
\end{tabular}
\end{table*}

\addtocounter{table}{-1}
\def\tfrac#1#2{{\vbox to 9pt {}\phantom{_I}\textstyle{#1}\over\vbox to 9pt {}\textstyle{#2}}}
\def\vh{\vbox to 15pt {}}
\def\vhh{\vbox to 20pt {}}
\def\vhhh{\vbox to 22pt {}}
\begin{table*}
\caption{{\it (Continued).} 
Observational equations that express the input data in Table~\ref{tab:pdata} as
functions of the adjusted constants in Table~\ref{tab:adjconb}.  The numbers in
the first column correspond to the numbers in the first column of
Table~\ref{tab:pdata}.  For simplicity, the lengthier functions are not
explicitly given.  See Sec.~\ref{ssec:mada} for an explanation of the symbol
$\doteq$.}
\label{tab:pobseqsb2}
\def\sp{\hbox to 30 pt {}}
\begin{tabular}{l@{\sp}rcl@{\sp}l}
\toprule
Type of input & &\multicolumn{2}{l}{Observational equation}& \hbox to 80pt{}\\
datum      &\hbox to 10pt{} & & \hbox to 240pt{} & Sec. \\
\colrule
\vh$B27 \quad $&$ \delta_{\rm Mu}$&$\doteq$&$ \delta_{\rm Mu}$ \\

\vhhh$B28 \quad $&$ \Delta \nu_{\rm Mu} $&$\doteq$&$
\Delta \nu_{\rm Mu}\!\!\left(R_\infty,\alpha,\tfrac{m_{\rm e}}{m_{\rmssmu}},
\delta_{\rmssmu}, \delta_{\rm Mu}\right)$
& \ref{sssec:muhfs} \\

\vhhh$B29,B30 \quad $&$ \nu(f_{\rm p}) $&$\doteq$&$
\nu\!\left(f_{\rm p};R_\infty,\alpha,\tfrac{m_{\rm e}}{m_{\rmssmu}},\tfrac{\mu_{\rm e^-}}{\mu_{\rm p}},
\delta_{\rm e}, \delta_{\rmssmu}, \delta_{\rm Mu}\right) $
&\ref{ssec:muhfs}
\\

\vhh$B31 \quad $&$ {\it\Gamma}_{\rm p-90}^{\,\prime}({\rm lo}) $&$\doteq$&
$ -\tfrac{ K_{\rm J-90}R_{\rm K-90}[1+a_{\rm e}(\alpha,\delta_{\rm e})]\alpha^3 }
{ 2\mu_0 R_\infty  } \left(\tfrac{\mu_{\rm e^-}}{\mu_{\rm p}^\prime}\right)^{-1} $ 
&\ref{sssec:logmr}
\\

\vhhh$B32 \quad $&$ {\it\Gamma}_{\rm h-90}^{\,\prime}({\rm lo}) $&$\doteq$&
$ \tfrac{ K_{\rm J-90}R_{\rm K-90}[1+a_{\rm e}(\alpha,\delta_{\rm e})]\alpha^3 }
{ 2\mu_0 R_\infty  } \left(\tfrac{\mu_{\rm e^-}}{\mu_{\rm p}^\prime}\right)^{-1} 
\tfrac{\mu_{\rm h}^\prime}{\mu_{\rm p}^\prime} $ 
&\ref{sssec:logmr}
\\

\vhhh$B33 \quad $&$ {\it\Gamma}_{\rm p-90}^{\,\prime}({\rm hi}) $&$\doteq$&
$ -\tfrac{ c [1+a_{\rm e}(\alpha,\delta_{\rm e})]\alpha^2 }
{ K_{\rm J-90}R_{\rm K-90} R_\infty h } 
\left(\tfrac{\mu_{\rm e^-}}{\mu_{\rm p}^\prime}\right)^{-1} $ 
&\ref{sssec:higmr}
\\

\vh$B34 \quad $&$ R_{\rm K} $&$\doteq$&$ \tfrac{\mu_0c}{2\alpha} $ 
&\ref{ssec:vkc}
\\

\vhhh$B35 \quad $&$ K_{\rm J} $&$\doteq$&$ 
\left(\tfrac{8\alpha}{\mu_0ch}\right)^{1/2} $ 
&\ref{ssec:jc}
\\

\vhh$B36 \quad $&$ K_{\rm J}^2R_{\rm K} $&$\doteq$&$ \tfrac{4}{h} $ 
&\ref{ssec:kj2rk}
\\

\vhhh$B37 \quad $&$ {\cal F}_{90} $&$\doteq$&
$ \tfrac{ c M_{\rm u} A_{\rm r}({\rm e})  \alpha^2 }
{ K_{\rm J-90}R_{\rm K-90} R_\infty h } 
$& \ref{ssec:f} \\

\vh$B38$-$B40  \quad $&$ d_{220}({{\scriptstyle X}}) $&$\doteq$&$ d_{220}({{\scriptstyle X}}) $ \\

\vhh$B41$-$B52 \quad $&$ \tfrac{d_{220}({{\scriptstyle X}}) }{ d_{220}({{\scriptstyle Y}})} - 1 $&$\doteq$&
$\tfrac{d_{220}({{\scriptstyle X}}) }{ d_{220}({{\scriptstyle Y}})}-1 $ & \\

\vhh$B53 \quad $&$ V_{\rm m}({\rm Si)} $&$\doteq$&$ \tfrac{\sqrt{2}\,cM_{\rm u}A_{\rm r}({\rm e})\alpha^{2}d^{\,3}_{220}
}{ R_{\infty}h} $ & \ref{ssec:mvsi} \\

\vhh$B54 \quad $&$\tfrac{\lambda_{\rm meas} }{ d_{220}({\rm {\scriptstyle ILL}})} $&$\doteq$&
$ \tfrac{\alpha^2 A_{\rm r}({\rm e}) }{ R_\infty d_{220}({\rm {\scriptstyle ILL}})} 
\tfrac{A_{\rm r}({\rm n}) + A_{\rm r}({\rm p}) }{ \left[A_{\rm r}({\rm n}) + A_{\rm r}({\rm p})\right]^2 - A_{\rm r}^2({\rm d})}  $& \ref{ssec:arn} \\

\vhhh$B55 \quad $&$ \tfrac{h}{m_{\rm n}d_{220}({\rm {\scriptstyle W04}})} $&$\doteq$&
$ \tfrac{ A_{\rm r}({\rm e}) }
{ A_{\rm r}({\rm n}) } $
$ \tfrac{ c  \alpha^2 }
{ 2 R_\infty d_{220}({\rm {\scriptstyle W04}}) } $
&\ref{sssec:pcnmr}
\\

\vhhh$B56,B57 \quad $&$ \tfrac{h}{m(X)} $&$\doteq$&
$ \tfrac{ A_{\rm r}({\rm e}) }
{  A_{\rm r}(X)}  $
$ \tfrac{ c\alpha^2 }
{ 2 R_\infty }  $
&\ref{ssec:pcpmq}
\\

\vh$B58 \quad $&$ R $&$\doteq$&$ R $ \\

\vhh$B59,B62 \quad $&$\tfrac{\lambda({\rm CuK\rmalpha_1})}{d_{220}({{\scriptstyle X}})} $&$\doteq$&$ \tfrac{\rm 1\,537.400 ~xu(CuK\rmalpha_1)}{d_{220}({{\scriptstyle X}})}$ &\ref{ssec:xru} \\

\vhhh$B60 \quad $&$\tfrac{\lambda({\rm WK\rmalpha_1})}{d_{220}({\rm {\scriptstyle N}})} $&$\doteq$&$ \tfrac{\rm 0.209\,010\,0 ~\AA^*}{d_{220}({\rm {\scriptstyle N}})} $ & \ref{ssec:xru} \\

\vhhh$B61 \quad $&$\tfrac{\lambda({\rm MoK\rmalpha_1})}{d_{220}({\rm {\scriptstyle N}})} $&$\doteq$&$ \tfrac{\rm 707.831 ~xu(MoK\rmalpha_1)}{d_{220}({\rm {\scriptstyle N}})} $ & \ref{ssec:xru} \\
\\
\botrule
\end{tabular}
\end{table*}

\def\fixh{\vbox to 9pt {}}
\begin{table}
\caption{The 12 adjusted constants (variables) relevant to the antiprotonic
helium data given in Table~\ref{tab:cdata}.  These adjusted constants appear as
arguments of the theoretical expressions on the right-hand side of the
observational equations of Table~\ref{tab:pobseqsc}.}
\label{tab:adjconc}
\def\sp{\hbox to 52 pt{}}
\begin{tabular}{l@{\sp}l}
\toprule
\qquad Transition & Adjusted constant \T \\
\colrule

\fixh $\bar{\rm p}^4$He$^+$: $(32,31) \rightarrow (31,30)$ & $\delta_{\bar{\rm p}^4{\rm He}^+}(32,31\!\!:\!31,30)$ \\
\fixh $\bar{\rm p}^4$He$^+$: $(35,33) \rightarrow (34,32)$ & $\delta_{\bar{\rm p}^4{\rm He}^+}(35,33\!\!:\!34,32)$ \\
\fixh $\bar{\rm p}^4$He$^+$: $(36,34) \rightarrow (35,33)$ & $\delta_{\bar{\rm p}^4{\rm He}^+}(36,34\!\!:\!35,33)$ \\
\fixh $\bar{\rm p}^4$He$^+$: $(37,34) \rightarrow (36,33)$ & $\delta_{\bar{\rm p}^4{\rm He}^+}(37,34\!\!:\!36,33)$ \\
\fixh $\bar{\rm p}^4$He$^+$: $(39,35) \rightarrow (38,34)$ & $\delta_{\bar{\rm p}^4{\rm He}^+}(39,35\!\!:\!38,34)$ \\
\fixh $\bar{\rm p}^4$He$^+$: $(40,35) \rightarrow (39,34)$ & $\delta_{\bar{\rm p}^4{\rm He}^+}(40,35\!\!:\!39,34)$ \\
\fixh $\bar{\rm p}^4$He$^+$: $(37,35) \rightarrow (38,34)$ & $\delta_{\bar{\rm p}^4{\rm He}^+}(37,35\!\!:\!38,34)$ \\
&\\
\fixh $\bar{\rm p}^3$He$^+$: $(32,31) \rightarrow (31,30)$ & $\delta_{\bar{\rm p}^3{\rm He}^+}(32,31\!\!:\!31,30)$ \\
\fixh $\bar{\rm p}^3$He$^+$: $(34,32) \rightarrow (33,31)$ & $\delta_{\bar{\rm p}^3{\rm He}^+}(34,32\!\!:\!33,31)$ \\
\fixh $\bar{\rm p}^3$He$^+$: $(36,33) \rightarrow (35,32)$ & $\delta_{\bar{\rm p}^3{\rm He}^+}(36,33\!\!:\!35,32)$ \\
\fixh $\bar{\rm p}^3$He$^+$: $(38,34) \rightarrow (37,33)$ & $\delta_{\bar{\rm p}^3{\rm He}^+}(38,34\!\!:\!37,33)$ \\
\B\fixh $\bar{\rm p}^3$He$^+$: $(36,34) \rightarrow (37,33)$ & $\delta_{\bar{\rm p}^3{\rm He}^+}(36,34\!\!:\!37,33)$ \\

\botrule
\end{tabular}
\end{table}

\def\tfrac#1#2{{\vbox to 9pt {}\phantom{_I}\textstyle{#1}\over\vbox to 9pt {}\textstyle{#2}}}
\def\vh{\vbox to 15pt {}}
\def\vhh{\vbox to 10pt {}}
\begin{table*}
\caption{Observational equations that express the input data related to
antiprotonic helium in Table~\ref{tab:cdata} as functions of adjusted constants
in Tables~\ref{tab:adjconb} and \ref{tab:adjconc}.  The numbers in the first
column correspond to the numbers in the first column of Table~\ref{tab:cdata}.
Definitions of the symbols and values of the parameters in these equations are
given in Sec.~\ref{ssec:aph}.  See Sec.~\ref{ssec:mada} for an explanation of
the symbol $\doteq$.}
\label{tab:pobseqsc}
\def\sp{\hbox to 72 pt{}}
\begin{tabular}{l@{\sp}rcll}
\toprule
Type of input \T & &
\multicolumn{2}{l}{Observational equation}& \\
datum      & & &  \\
\colrule

\vh$C1$--$C7 \quad $&$ \delta_{\bar{\rm p}{\rm ^4He^+}}(n,l:n^\prime,l^\prime) $&$\doteq$&$ \delta_{\bar{\rm p}{\rm ^4He^+}}(n,l:n^\prime,l^\prime) $ \\
\vh$C8$--$C12 \quad $&$ \delta_{\bar{\rm p}{\rm ^3He^+}}(n,l:n^\prime,l^\prime) $&$\doteq$&$ \delta_{\bar{\rm p}{\rm ^3He^+}}(n,l:n^\prime,l^\prime) $ \\

\vh$C13$--$C19 \quad $&$ \nu_{\bar{\rm p}{\rm ^4He^+}}(n,l:n^\prime,l^\prime) $&$
\doteq$&$ \nu_{\bar{\rm p}{\rm ^4He^+}}^{(0)}(n,l:n^\prime,l^\prime) 
+ a_{\bar{\rm p}{\rm ^4He^+}}(n,l:n^\prime,l^\prime)\left[\left(\tfrac{A_{\rm r}({\rm e})}{A_{\rm r}({\rm p)}}\right)^{\!(0)} \!\!
\left(\tfrac{A_{\rm r}({\rm p})}{A_{\rm r}({\rm e})}\right)-1 \right]$
\\ \vbox to 20 pt {}&&&
$+ b_{\bar{\rm p}{\rm ^4He^+}}(n,l:n^\prime,l^\prime)\left[\left(\tfrac{A_{\rm r}({\rm e})}{A_{\rm r}({\rmalpha)}}\right)^{\!(0)} \!\!
\left(\tfrac{A_{\rm r}({\rmalpha})}{A_{\rm r}({\rm e})}\right)-1 \right]
+\delta_{\bar{\rm p}{\rm ^4He^+}}(n,l:n^\prime,l^\prime) $ \\

\\

\vh$C20$--$C24 \quad $&$ \nu_{\bar{\rm p}{\rm ^3He^+}}(n,l:n^\prime,l^\prime) $&$
\doteq$&$ \nu_{\bar{\rm p}{\rm ^3He^+}}^{(0)}(n,l:n^\prime,l^\prime) 
+ a_{\bar{\rm p}{\rm ^3He^+}}(n,l:n^\prime,l^\prime)\left[\left(\tfrac{A_{\rm r}({\rm e})}{A_{\rm r}({\rm p)}}\right)^{\!(0)} \!\!
\left(\tfrac{A_{\rm r}({\rm p})}{A_{\rm r}({\rm e})}\right)-1 \right]$
\\ \vbox to 20 pt {}&&&
$+ b_{\bar{\rm p}{\rm ^3He^+}}(n,l:n^\prime,l^\prime)\left[\left(\tfrac{A_{\rm r}({\rm e})}{A_{\rm r}({\rm h)}}\right)^{\!(0)} \!\!
\left(\tfrac{A_{\rm r}({\rm h})}{A_{\rm r}({\rm e})}\right)-1 \right]
+\delta_{\bar{\rm p}{\rm ^3He^+}}(n,l:n^\prime,l^\prime) $   \\
\\

\botrule
\end{tabular}
\end{table*}

\def\fixh{\vbox to 12pt {}}
\def\sp{\hbox to 17.3 pt {}}
\begin{table*}
\caption{ Summary of the results of some of the least-squares adjustments used
to analyze all of the input data given in Tables~\ref{tab:rdata},
\ref{tab:rdcc}, \ref{tab:pdata}, and \ref{tab:pdcc}.  The values of $\alpha$
and $h$ are those obtained in the adjustment, $N$ is the number of input data,
$M$ is the number of adjusted constants, $\nu=N-M$ is the degrees of freedom,
and $R_{\rm B}=\sqrt{{\it \chi}^2/\nu}$ is the Birge ratio.  See the text for
an explanation and discussion of each adjustment, but in brief, 1 is all the
data; 2 is 1 with the uncertainties of the key x-ray/silicon data multiplied by
1.5; 3 is 2 with the uncertainties of the key electrical data also multiplied
by 1.5; 4 is the final adjustment from which the 2006 recommended values are
obtained and is 3 with the input data with low weights deleted; 5 is 3 with the
four data that provide the most accurate values of $\alpha$ deleted; and 6 is 3
with the three data that provide the most accurate values of $h$ deleted.  }
\label{tab:adjustsall}
\begin{tabular}{c c @{\sp} c @{\sp} c @{\sp} c @{\sp} c @{\sp} c @{\sp} c @{\sp} c @{\sp} c @{\sp} c} 
\toprule
\fixh Adj. &  $N$& $M$& $\nu$ & ${\it \chi}^2$& $R_{\rm B}$& $\alpha^{-1}$ & $u_{\rm r}(\alpha^{-1})$ & $h$/(J s) & $u_{\rm r}(h)$ \\
\colrule
\fixh \ 1 & 150& 79& 71 & 92.1& 1.14  & $137.035\,999\,687(93)$           & $6.8\times10^{-10}$ & $6.626\,068\,96(22)   \times 10^{-34}$               & $3.4\times10^{-8}$  \\

\fixh \ 2 & 150& 79& 71 & 82.0& 1.07  & $137.035\,999\,682(93)$           & $6.8\times10^{-10}$ & $6.626\,068\,96(22)   \times 10^{-34}$               & $3.4\times10^{-8}$  \\

\fixh \ 3 & 150& 79& 71 & 77.5& 1.04  & $137.035\,999\,681(93)$           & $6.8\times10^{-10}$ & $6.626\,068\,96(33)   \times 10^{-34}$               & $5.0\times10^{-8}$ \\

\fixh \ 4 & 135& 78& 57 & 65.0& 1.07  & $137.035\,999\,679(94)$           & $6.8\times10^{-10}$ & $6.626\,068\,96(33)   \times 10^{-34}$               & $5.0\times10^{-8}$ \\

\fixh \ 5 & 144& 77& 67 & 72.9& 1.04  & $137.036\,0012(19)\phantom{\,44}$ & $1.4\times10^{-8~}$ & $6.626\,068\,96(33)   \times 10^{-34}$               & $5.0\times10^{-8}$  \\

\fixh \ 6 & 147& 79& 68 & 75.4& 1.05  & $137.035\,999\,680(93)$           & $6.8\times10^{-10}$ & $6.626\,0719(21)      \times 10^{-34} \phantom{\,2}$ & $3.2\times10^{-7}$  \\
\botrule
\end{tabular}
\end{table*}

\def\sp{\hbox to 5 pt {}}
\def\sm{\phantom{-}}
\begin{table*}
\caption{ Normalized residuals $r_i $ and self-sensitivity coefficients $S_{\rm
c} $ that result from the six least-squares adjustments summarized in
Table~\ref{tab:adjustsall}  for the four input data whose absolute values of
$r_i $ in Adj.~1 exceed 1.50. $S_{\rm c} $ is a measure of how the
least-squares estimated value of a given type of input datum depends on a
particular measured or calculated value of that type of datum; see Appendix E
of CODATA-98.  See the text for an explanation and discussion of each
adjustment; brief explanations are given at the end of the caption to the
previous table.}
\label{tab:adjres}
\begin{tabular}{|lcc|c|c|c|c|c|c|}
\toprule
 Item   & Input     & Identification &     Adj.~1     &     Adj.~2     &     Adj.~3     & Adj.~4     &     Adj.~5     &     Adj.~6   \\
number  & quantity  &                & $~r_i~~~~S_{\rm c}$ & $~r_i~~~~S_{\rm c}$ & $~r_i~~~~S_{\rm c}$ & $~r_i~~~~S_{\rm c}$ & $~r_i~~~~S_{\rm c}$ & $~r_i~~~~S_{\rm c}$ \\
\colrule
 $B53$                        &\fixh $V_{\rm m}({\rm Si})$                         & N/P/I-05 & $  -2.82~~0.065~$ & $  -2.68~~0.085~$ & $  -1.86~~0.046~$ & $  -1.86~~0.047~$ & $  -1.79~~0.053~$ & $  -0.86~~0.556~$ \\
 $B55$                        &\fixh $h/m_{\rm n}d_{220}({\scriptstyle{\rm W04}})$   & PTB-99   & $  -2.71~~0.155~$ & $  -2.03~~0.118~$ & $  -1.89~~0.121~$ & $  -1.89~~0.121~$ & $  -1.57~~0.288~$ & $  -1.82~~0.123~$ \\
 $B39$                        &\fixh $d_{220}({\scriptstyle{\rm NR3}})$              & NMIJ-04  & $\sm2.37~~0.199~$ & $\sm1.86~~0.145~$ & $\sm1.74~~0.148~$ & $\sm1.74~~0.148~$ & $\sm1.78~~0.151~$ & $  -1.00~~0.353~$ \\
 $B31.1$\rule[-6pt]{0pt}{6pt} &\fixh ${\it\Gamma}_{\rm p-90}^{\,\prime}({\rm lo})$ & NIST-89  & $\sm2.31~~0.010~$ & $\sm2.30~~0.010~$ & $\sm2.30~~0.010~$ &    deleted        & $\sm2.60~~0.143~$ & $\sm2.30~~0.010~$ \\
\botrule
\end{tabular}
\end{table*}

\def\fixh{\vbox to 12pt {}}
\def\sv#1{\hbox to #1 pt {}}
\def\sp{\hbox to 22pt {}}
\def\sd{\phantom{3}}
\begin{table*}
\caption{Summary of the results of some of the least-squares adjustments used
to analyze the input data related to $R_\infty$.  The values of $R_\infty$,
$R_{\rm p}$, and $R_{\rm d}$ are those obtained in the indicated adjustment,
$N$ is the number of input data, $M$ is the number of adjusted constants,
$\nu=N-M$ is the degrees of freedom, and $R_{\rm B}=\sqrt{{\it \chi}^2/\nu}$ is
the Birge ratio.  See the text for an explanation and discussion of each
adjustment, but in brief, 4 is the final adjustment; 7 is 4 with the input data
for $R_{\rm p}$ and $R_{\rm d}$ deleted; 8 is 4 with just the $R_{\rm p}$ datum
deleted; 9 is 4 with just the $R_{\rm d}$ datum deleted; 10 is 4 but with only
the hydrogen data included; and 11 is 4 but with only the deuterium data
included. \\ }
\label{tab:adjustsa}
\begin{tabular}{c@{\sp}c@{\sp}l@{\sp}l@{\sp}l@{\sp} c@{\sp}l@{\sp}l@{\sp}l@{\sp}l}
\toprule
\fixh \sv{2} Adj. & $N$& $M$& $\nu$ & ${\it \chi}^2$& $R_{\rm B}$ & \sv{20} $R_\infty/{\rm m}^{-1}$ & \sv{2}$u_{\rm r}(R_\infty)$ & \sv{0} $R_{\rm p}$/fm & \sv{2} $R_{\rm d}$/fm \\
\colrule
\fixh  $\sd 4$ & $135$ & $78$ & 57 & 65.0& 1.07&  $10\,973\,731.568\,527(73)$ & $6.6\times 10^{-12}$ & $0.8768(69)$ & $2.1402(28)$ \\
\fixh  $\sd 7$ & $133$ & $78$ & 55 & 63.0& 1.07&  $10\,973\,731.568\,518(82)$ & $7.5\times 10^{-12}$ & $0.8760(78)$ & $2.1398(32)$ \\
\fixh  $\sd 8$ & $134$ & $78$ & 56 & 63.8& 1.07&  $10\,973\,731.568\,495(78)$ & $7.1\times 10^{-12}$ & $0.8737(75)$ & $2.1389(30)$ \\
\fixh  $\sd 9$ & $134$ & $78$ & 56 & 63.9& 1.07&  $10\,973\,731.568\,549(76)$ & $6.9\times 10^{-12}$ & $0.8790(71)$ & $2.1411(29)$ \\
\fixh  $10$ &    $117$ & $68$ & 49 & 60.8& 1.11&  $10\,973\,731.568\,562(85)$ & $7.8\times 10^{-12}$ & $0.8802(80)$ &              \\
\fixh  $11$ &    $102$ & $61$ & 41 & 54.7& 1.16&  $10\,973\,731.568\,39(13)$  & $1.1\times 10^{-11}$ &              & $2.1286(93)$ \\
\botrule
\end{tabular}
\end{table*}

\def\tfrac#1#2{{\vbox to 9pt {}\phantom{_I}\textstyle{#1}\over\vbox to 9pt {}\textstyle{#2}}}
\def\vh{\vbox to 15pt {}}
\def\vhh{\vbox to 20pt {}}
\def\vhhh{\vbox to 22pt {}}
\begin{table*}
\caption{ Generalized observational equations that express input data
$B31$-$B37$ in Table~\ref{tab:pdata} as functions of the adjusted constants in
Tables~\ref{tab:adjconb} and \ref{tab:adjcona} with the additional adjusted
constants $\varepsilon_{\rm J}$ and $\varepsilon_{\rm K}$ as given in
Eqs.~(\ref{eq:kjeps}) and (\ref{eq:rkeps}).  The numbers in the first column
correspond to the numbers in the first column of Table~\ref{tab:pdata}.  For
simplicity, the lengthier functions are not explicitly given.  See
Sec.~\ref{ssec:mada} for an explanation of the symbol $\doteq$.}
\label{tab:pobseqseps}
\def\sp{\hbox to 20 pt {}}
\begin{tabular}{l@{\sp}rcl}
\toprule
Type of input \T & &\multicolumn{2}{l}{Generalized observational equation}\\
datum      &\hbox to 10pt{} & & \hbox to 80pt{} \\
\colrule

\vhh$B31^* \quad $&$ {\it\Gamma}_{\rm p-90}^{\,\prime}({\rm lo}) $&$\doteq$&
$ -\tfrac{ K_{\rm J-90}R_{\rm K-90}[1+a_{\rm e}(\alpha,\delta_{\rm e})]\alpha^3 }
{ 2\mu_0 R_\infty (1+\varepsilon_{\rm J})(1+\varepsilon_{\rm K})
 } \left(\tfrac{\mu_{\rm e^-}}{\mu_{\rm p}^\prime}\right)^{-1} $ 
\\

\vhhh$B32^* \quad $&$ {\it\Gamma}_{\rm h-90}^{\,\prime}({\rm lo}) $&$\doteq$&
$ \tfrac{ K_{\rm J-90}R_{\rm K-90}[1+a_{\rm e}(\alpha,\delta_{\rm e})]\alpha^3 }
{ 2\mu_0 R_\infty  (1+\varepsilon_{\rm J})(1+\varepsilon_{\rm K})
} \left(\tfrac{\mu_{\rm e^-}}{\mu_{\rm p}^\prime}\right)^{-1} 
\tfrac{\mu_{\rm h}^\prime}{\mu_{\rm p}^\prime} $ 
\\

\vhhh$B33^* \quad $&$ {\it\Gamma}_{\rm p-90}^{\,\prime}({\rm hi}) $&$\doteq$&
$ -\tfrac{ c [1+a_{\rm e}(\alpha,\delta_{\rm e})]\alpha^2 }
{ K_{\rm J-90}R_{\rm K-90} R_\infty h } 
(1+\varepsilon_{\rm J})(1+\varepsilon_{\rm K})
\left(\tfrac{\mu_{\rm e^-}}{\mu_{\rm p}^\prime}\right)^{-1} $ 
\\

\vh$B34^* \quad $&$ R_{\rm K} $&$\doteq$&$ \tfrac{\mu_0c}{2\alpha}
(1+\varepsilon_{\rm K})$ 
\\

\vhhh$B35^* \quad $&$ K_{\rm J} $&$\doteq$&$ 
\left(\tfrac{8\alpha}{\mu_0ch}\right)^{1/2} (1+\varepsilon_{\rm J})$ 
\\

\vhh$B36^* \quad $&$ K_{\rm J}^2R_{\rm K} $&$\doteq$&$ \tfrac{4}{h} 
(1+\varepsilon_{\rm J})^2(1+\varepsilon_{\rm K})$ 
\\

\vhhh$B37^* \quad $&$ {\cal F}_{90} $&$\doteq$&
$ \tfrac{ c M_{\rm u} A_{\rm r}({\rm e})  \alpha^2 }
{ K_{\rm J-90}R_{\rm K-90} R_\infty h } (1+\varepsilon_{\rm J})(1+\varepsilon_{\rm K})$
\\

\vhh$B62^* \quad $&$ \varepsilon_{\rm J} $&$\doteq$&$
\varepsilon_{\rm J}$ 
\\

\vhh$B63^* \quad $&$ \varepsilon_{\rm K} $&$\doteq$&$
\varepsilon_{\rm K}$ 
\\

\botrule
\end{tabular}
\end{table*}

\def\fixh{\vbox to 12pt {}}
\def\sp#1{\hbox to #1 pt {}}
\def\t{@{\sp{20}}}
\begin{table*}
\caption{ Summary of the results of several least-squares adjustments carried
out to investigate the effect of assuming the relations for $K_{\rm J}$ and
$R_{\rm K}$ given in Eqs.~(\ref{eq:kjeps}) and (\ref{eq:rkeps}).  The values of
$\alpha$, $h$, $\varepsilon_{\rm K}$, and $\varepsilon_{\rm J}$ are those
obtained in the indicated adjustments.  The quantity $R_{\rm B}=\sqrt{{\it
\chi}^2/\nu}$ is the Birge ratio and $r_i$ is the normalized residual of the
indicated input datum (see Table~\ref{tab:pdata}).  These four data have the
largest $|r_i|$ of all the input data and are the only data in Adj.~(i) with
$|r_i| > 1.50$.  See the text for an explanation and discussion of each
adjustment, but in brief, (i) assumes $K_{\rm J}=2e/h$ and $R_{\rm K}=h/e^2$
and uses all the data; (ii) is (i) with the relation $K_{\rm J}=2e/h$ relaxed;
(iii) is (i) with the relation $R_{\rm K}=h/e^2$ relaxed; (iv) is (i) with both
relations relaxed; (v) is (iv) with the $V_{\rm m}(\rm Si)$ datum deleted; (vi)
is (iv) with the ${\it \Gamma}_{{\rm p}-90}^{\,\prime}({\rm lo})$ and ${\it
\Gamma}_{{\rm h}-90}^{\,\prime}({\rm lo})$ data deleted; and (vii) is (iv) with
the $V_{\rm m}(\rm Si)$, ${\it \Gamma}_{{\rm p}-90}^{\,\prime}({\rm lo})$, and
${\it \Gamma}_{{\rm h}-90}^{\,\prime}({\rm lo})$ data deleted.  }
\label{tab:epsilons}
\begin{tabular}{c @{\sp{8}} c @{\sp{10}} c @{\sp{10}} l @{\sp{10}} c @{\sp{10}} c @{\sp{10}} c @{\sp{4}} c @{\sp{4}} c @{\sp{4}} c @{\sp{4}} c} \\
\toprule
\fixh Adj. &  $R_{\rm B}$& $\alpha^{-1}$ & \sp{30}$h$/(J s) & $\varepsilon_{\rm K}$ & $\varepsilon_{\rm J}$ &
 $r_{B53}$ & $r_{B55}$ & $r_{B39}$ & $r_{B31.1}$ \\
\colrule
\fixh \ (i)   & 1.14  & $137.035\,999\,687(93)$ & $6.626\,068\,96(22)\times 10^{-34}$ & $0$                   & $0$                       & $-2.82$ & $-2.71$ & $2.37$ & $2.31$  \\
\fixh \ (ii)  & 1.14  & $137.035\,999\,688(93)$ & $6.626\,0682(10)\times 10^{-34}$    & $0$                   & $-61(79)\times 10^{-9}$   & $-3.22$ & $-2.75$ & $2.39$ & $1.77$  \\
\fixh \ (iii) & 1.14  & $137.035\,999\,683(93)$ & $6.626\,069\,06(25)\times 10^{-34}$ & $16(18)\times10^{-9}$ & $0$                       & $-2.77$ & $-2.71$ & $2.36$ & $2.45$  \\
\fixh \ (iv)  & 1.14  & $137.035\,999\,685(93)$ & $6.626\,0681(11)\times 10^{-34}$    & $20(18)\times10^{-9}$ & $-77(80)\times 10^{-9}$   & $-3.27$ & $-2.75$ & $2.39$ & $1.79$  \\
\fixh \ (v)   & 1.05  & $137.035\,999\,686(93)$ & $6.626\,0653(13)\times 10^{-34}$    & $23(18)\times10^{-9}$ & $-281(95)\times 10^{-9}$  & deleted & $-2.45$ & $2.19$ & $0.01$  \\
\fixh \ (vi)  & 1.05  & $137.035\,999\,686(93)$ & $6.626\,0744(19)\times 10^{-34}$    & $24(18)\times10^{-9}$ & $~407(143)\times 10^{-9}$ & $-0.05$ & $-2.45$ & $2.19$ & deleted \\
\fixh \ (vii) & 1.06  & $137.035\,999\,686(93)$ & $6.626\,0722(95)\times 10^{-34}$    & $24(18)\times10^{-9}$ & $~238(720)\times 10^{-9}$  & deleted & $-2.45$ & $2.19$ & deleted \\
\botrule
\end{tabular}
\end{table*}

\clearpage
\section{The 2006 CODATA recommended values}
\label{sec:2006crv}

\subsection{Calculational details} \label{ssec:cd} As indicated in
Sec.~\ref{ssec:mada}, the 2006 recommended values of the constants are based on
adjustment 4 of Tables~\ref{tab:adjustsall} to \ref{tab:adjustsa}.  This
adjustment is obtained by (i) deleting 15 items from the originally considered
150 items of input data of Tables~\ref{tab:rdata}, \ref{tab:pdata}, and
\ref{tab:cdata}, namely, items $B8$, $B31.1$-$B35.2$, $B37$, and $B56$, because
of their low weight (self sensitivity coefficient $S_{\rm c} < 0.01$); and (ii)
weighting the uncertainties of the nine input data $B36.1$-$B36.3$,
$B38.1$-$B40$, $B53$, and $B55$ by the multiplicative factor 1.5 in order to
reduce the absolute values of their normalized residuals $|r_i|$ to less than
2.  The correlation coefficients of the data, as given in Tables~\ref{tab:rdcc},
\ref{tab:pdcc}, and \ref{tab:cdcc}, are also taken into account.  The 135 final
input data are expressed in terms of the 78 adjusted constants of
Tables~\ref{tab:adjcona}, \ref{tab:adjconb}, and \ref{tab:adjconc},
corresponding to $N-M = \nu = 57$ degrees of freedom.  Because $h/m(^{133}{\rm
Cs})$, item $B56$, has been deleted as an input datum due to its low weight,
$A_{\rm r}(^{133}{\rm Cs})$, item $B8$, has also been deleted as an input datum
and as an adjusted constant.

For the final adjustment, $\chi^2 = 65.0$, $\sqrt{\chi^2/\nu} = R_{\rm B} =
1.04$, and $Q(65.0|57) = 0.22$, where $Q(\chi^2|\nu)$ is the probability that
the observed value of $\chi^2$ for degrees of freedom $\nu$ would have exceeded
that observed value (see Appendix E of CODATA-98).  Each input datum in the
final adjustment has $S_{\rm c} > 0.01$, or is a subset of the data of an
experiment that provides an input datum or input data with $S_{\rm c} > 0.01$.
Not counting such input data with $S_{\rm c} < 0.01$, the six input data with
the largest $|r_i|$ are $B55$, $B53$, $B39$, $C18$, $B11.1$, and $B9$;  their
values of $r_i$ are $-1.89$, $-1.86$, $1.74$, $-1.73$, $1.69$, and $1.45$,
respectively. The next largest $r_i$ are $1.22$ and $1.11$.

The output of the final adjustment is the set of best estimated values, in the
least-squares sense, of the 78 adjusted constants and their variances and
covariances.  Together with 
(i) those constants that have exact values such as $\mu_0$ and $c$; 
(ii) the value of $G$ obtained in Sec.~\ref{sec:ncg}; and 
(iii) the values of $m_\tau$, $G_{\rm F}$, and $\sin^2 \theta_{\rm W}$ given in
Sec.~\ref{ssec:lisq}, all of the 2006 recommended values, including their
uncertainties, are obtained from the 78 adjusted constants.  How this is done
can be found in Sec.~V.B of CODATA-98.

\subsection{Tables of values}
\label{ssec:tov}

The 2006 CODATA recommended values of the basic constants and conversion
factors of physics and chemistry and related quantities are given in
Tables~\ref{tab:abbr} to \ref{tab:enconv2}.  These tables are very similar in
form to their 2002 counterparts; the principal difference is that a number of
new recommended values have been included in the 2006 list, in particular, in
Table~\ref{tab:constants}.  These are $m_{\rm P}c^2$ in GeV, where $m_{\rm P}$
is the Planck mass; the $g$-factor of the deuteron $g_{\rm d}$; $b^\prime =
\nu_{\rm max}/T$, the Wien displacement law-constant for frequency; and for the
first time, 14 recommended values of a number of constants that characterize
the triton, including its mass $m_{\rm t}$, magnetic moment $\mu_{\rm t}$,
$g$-factor $g_{\rm t}$, and the magnetic moment ratios $\mu_{\rm t}/\mu_{\rm
e}$ and $\mu_{\rm t}/\mu_{\rm p}$.  The addition of the triton-related
constants is a direct consequence of the improved measurement of $A_{\rm
r}(^3{\rm H})$ (item $B3$ in Table~\ref{tab:pdata}) and the new NMR
measurements on, and re-examined shielding correction differences for, the HT
molecule (items $B22$ and $B23$ in Table~\ref{tab:pdata}).

Table~\ref{tab:abbr} is a highly abbreviated list containing the values of the
constants and conversion factors most commonly used.  Table~\ref{tab:constants}
is a much more extensive list of values categorized as follows: UNIVERSAL;
ELECTROMAGNETIC; ATOMIC AND NUCLEAR; and PHYSICOCHEMICAL.  The ATOMIC AND
NUCLEAR category is subdivided into 11 subcategories: General; Electroweak;
Electron, ${\rm e}^{-}$; Muon, ${\rmmu}^{-}$; Tau, ${\rmtau}^{-}$; Proton,
${\rm p}$; Neutron, ${\rm n}$; Deuteron, ${\rm d}$; Triton, ${\rm t}$; Helion,
${\rm h}$; and Alpha particle, ${\rmalpha}$.  Table~\ref{tab:varmatrix} gives
the variances, covariances, and correlation coefficients of a selected group of
constants.  (Application of the covariance matrix is discussed in Appendix~E of
CODATA-98.)  Table~\ref{tab:adopted} gives the internationally adopted values
of various quantities; Table~\ref{tab:xrayvalues} lists the values of a number
of x-ray related quantities; Table~\ref{tab:units} lists the values of various
non-SI units; and Tables~\ref{tab:enconv1} and \ref{tab:enconv2} give the
values of various energy equivalents.

All of the values given in Tables~\ref{tab:abbr} to \ref{tab:enconv2} are
available on the Web pages of the Fundamental Constants Data Center of the NIST
Physics Laboratory at physics.nist.gov/constants.  This electronic version of
the 2006 CODATA recommended values of the constants also includes a much more
extensive correlation coefficient matrix.  Indeed, the correlation coefficient
of any two constants listed in the tables is accessible on the Web site, as
well as the automatic conversion of the value of an energy-related quantity
expressed in one unit to the corresponding value expressed in another unit (in
essence, an automated version of Tables~\ref{tab:enconv1} and
\ref{tab:enconv2}).

As discussed in Sec.~\ref{sec:mmagf}, well after the 31 December 2006 closing
date of the 2006 adjustment and the 29 March 2007 distribution date of the 2006
recommended values on the Web, \textcite{2007167} reported their discovery of
an error in the coefficient $A_1^{(8)}$ in the theoretical expression for the
electron magnetic moment anomaly $a_{\rm e}$.  Use of the new coefficient would
lead to an increase in the 2006 recommended value of $\alpha$ by 6.8 times its
uncertainty, and an increase of its uncertainty by a factor of 1.02.  The
recommended values and uncertainties of constants that depend solely on
$\alpha$, or on $\alpha$ in combination with other constants with $u_{\rm r}$
no larger than a few parts in $10^{10}$, would change in the same way. However,
the changes in the recommended values of the vast majority of the constants
listed in the tables would lie in the range 0 to $0.5$ times their 2006
uncertainties, and their uncertainties would remain essentially unchanged.

\section{Summary and Conclusion}
\label{sec:c}

We conclude this report by (i) comparing the 2006 and 2002 CODATA recommended
values of the constants and identifying those new results that have contributed
most to the changes from the 2002 values; (ii) presenting some of the
conclusions that can be drawn from the 2006 recommended values and analysis of
the 2006 input data; and (iii) looking to the future and identifying
experimental and theoretical work that can advance our knowledge of the values
of the constants. 

\subsection{Comparison of 2006 and 2002 CODATA recommended values}
\label{ssec:crv}

The 2006 and 2002 recommended values of a representative group of constants are
compared in Table~\ref{tab:06vs02}.  Regularities in the numbers in columns 2-4
arise because many constants are obtained from expressions proportional to
$\alpha$, $h$, or $R$ raised to various powers.  Thus, the first six quantities
in the table are calculated from expressions proportional to $\alpha^a$, where
$|a| = 1$, 2, 3, or 6.  The next 15 quantities, $h$ through $\mu_{\rm p}$, are
calculated from expressions containing the factor $h^a$, where $|a| = 1$ or
$\frac{1}{2}$.  And the five quantities $R$ through $\sigma$ are proportional
to $R^a$, where $|a| = 1$ or 4. 

\def\pad{{\hbox to 10 pt {}}}
\def\pa{{\hbox to 5 pt {}}}
\def\s#1{\hbox to #1 pt {}}
\begin{table}
\caption{Comparison of the 2006 and 2002 CODATA adjustments of the values of
the constants by the comparison of the corresponding recommended values of a
representative group of constants.  Here $D_{\rm r}$ is the 2006 value minus
the 2002 value divided by the standard uncertainty $u$ of the 2002 value ({\it
i.e.,} $D_{\rm r}$ is the change in the value of the constant from 2002 to 2006
relative to its 2002 standard uncertainty).}
\label{tab:06vs02}
\def\sp{\hbox to 22 pt {}}
\begin{tabular}{c@{\sp}l@{\sp}c@{\sp}r}
\toprule
Quantity \T & 2006 rel. std.  & Ratio 2002 $u_{\rm r}$    & $D_{\rm r}$ \pad \\
         & uncert. $u_{\rm r}$ & to 2006 $u_{\rm r}$ \s{4} &    \\
\colrule
\vbox to 10pt{}$\alpha$  &  \pa $ 6.8\times 10^{-10}$ & \pa $ 4.9$ & $ -1.3$\pad  \\
$R_{\rm K}$  &  \pa $ 6.8\times 10^{-10}$ & \pa $ 4.9$ & $ 1.3$\pad  \\
$a_{\rm 0}$  &  \pa $ 6.8\times 10^{-10}$ & \pa $ 4.9$ & $ -1.3$\pad   \\
$\lambda_{\rm C}$  &  \pa $ 1.4\times 10^{-9}$ & \pa $ 4.9$ & $ -1.3$\pad  \\
$r_{\rm e}$  &  \pa $ 2.1\times 10^{-9}$ & \pa $ 4.9$ & $ -1.3$\pad \\
$\sigma_{\rm e}$  &  \pa $ 4.1\times 10^{-9}$ & \pa $ 4.9$ & $ -1.3$\pad   \\
$ h $  &  \pa $ 5.0\times 10^{-8}$ & \pa $ 3.4$ & $ -0.3$\pad  \\
$m_{\rm e}$  &  \pa $ 5.0\times 10^{-8}$ & \pa $ 3.4$ & $ -0.3$\pad \\
$m_{\rm h}$  &  \pa $ 5.0\times 10^{-8}$ & \pa $ 3.4$ & $ -0.3$\pad \\
$m_{\rmssalpha}$  &  \pa $ 5.0\times 10^{-8}$ & \pa $ 3.4$ & $ -0.3$\pad  \\
$N_{\rm A}$  &  \pa $ 5.0\times 10^{-8}$ & \pa $ 3.4$ & $ 0.3$\pad  \\
$E_{\rm h}$  &  \pa $ 5.0\times 10^{-8}$ & \pa $ 3.4$ & $ -0.3$\pad  \\
$c_1$  &  \pa $ 5.0\times 10^{-8}$ & \pa $ 3.4$ & $ -0.3$\pad \\
$e$  &  \pa $ 2.5\times 10^{-8}$ & \pa $ 3.4$ & $ -0.3$\pad \\
$K_{\rm J}$  &  \pa $ 2.5\times 10^{-8}$ & \pa $ 3.4$ & $ 0.3$\pad \\
$F$  &  \pa $ 2.5\times 10^{-8}$ & \pa $ 3.4$ & $ 0.2$\pad \\
$\gamma^{\,\prime}_{\rm p}$  &  \pa $ 2.7\times 10^{-8}$ & \pa $ 3.2$ & $ 0.2$\pad \\
$\mu_{\rm B}$  &  \pa $ 2.5\times 10^{-8}$ & \pa $ 3.4$ & $ -0.4$\pad \\
$\mu_{\rm N}$  &  \pa $ 2.5\times 10^{-8}$ & \pa $ 3.4$ & $ -0.4$\pad \\
$\mu_{\rm e}$  &  \pa $ 2.5\times 10^{-8}$ & \pa $ 3.4$ & $ 0.4$\pad \\
$\mu_{\rm p}$  &  \pa $ 2.6\times 10^{-8}$ & \pa $ 3.3$ & $ -0.4$\pad \\
$R$  &  \pa $ 1.7\times 10^{-6}$ & \pa $ 1.0$ & $ 0.0$\pad  \\
$k$  &  \pa $ 1.7\times 10^{-6}$ & \pa $ 1.0$ & $ 0.0$\pad  \\
$V_{\rm m}$  &  \pa $ 1.7\times 10^{-6}$ & \pa $ 1.0$ & $ 0.0$\pad \\
$c_2$  &  \pa $ 1.7\times 10^{-6}$ & \pa $ 1.0$ & $ 0.0$\pad \\
$\sigma$  &  \pa $ 7.0\times 10^{-6}$ & \pa $ 1.0$ & $ 0.0$\pad \\
$ G $  &  \pa $ 1.0\times 10^{-4}$ & \pa $ 1.5$ & $ 0.1$\pad \\
$R_\infty$  &  \pa $ 6.6\times 10^{-12}$ & \pa $ 1.0$ & $ 0.0$\pad \\
$m_{\rm e}/m_{\rm p}$  &  \pa $ 4.3\times 10^{-10}$ & \pa $ 1.1$ & $ 0.2$\pad \\
$m_{\rm e}/m_{\rmssmu}$  &  \pa $ 2.5\times 10^{-8}$ & \pa $ 1.0$ & $ 0.3$\pad \\
$A_{\rm r}({\rm e})$  &  \pa $ 4.2\times 10^{-10}$ & \pa $ 1.0$ & $ -0.1$\pad \\
$A_{\rm r}({\rm p})$  &  \pa $ 1.0\times 10^{-10}$ & \pa $ 1.3$ & $ -0.9$\pad \\
$A_{\rm r}({\rm n})$  &  \pa $ 4.3\times 10^{-10}$ & \pa $ 1.3$ & $ 0.7$\pad \\
$A_{\rm r}({\rm d})$  &  \pa $ 3.9\times 10^{-11}$ & \pa $ 4.5$ & $ 0.1$\pad \\
$A_{\rm r}({\rm h})$  &  \pa $ 8.6\times 10^{-10}$ & \pa $ 2.3$ & $ 0.7$\pad \\
$A_{\rm r}({\rmalpha})$  &  \pa $ 1.5\times 10^{-11}$ & \pa $ 0.9$ & $ -0.4$\pad \\
$d_{\rm 220}$  &  \pa $ 2.6\times 10^{-8}$ & \pa $ 1.4$ & $ -2.9$\pad \\
$g_{\rm e}$  &  \pa $ 7.4\times 10^{-13}$ & \pa $ 5.0$ & $ 1.3$\pad \\
$g_{\rmssmu}$  &  \pa $ 6.0\times 10^{-10}$ & \pa $ 1.0$ & $ -1.4$\pad \\
$\mu_{\rm p}/\mu_{\rm B}$  &  \pa $ 8.1\times 10^{-9}$ & \pa $ 1.2$ & $ 0.2$\pad \\
$\mu_{\rm p}/\mu_{\rm N}$  &  \pa $ 8.2\times 10^{-9}$ & \pa $ 1.2$ & $ 0.2$\pad \\
$\mu_{\rm n}/\mu_{\rm N}$  &  \pa $ 2.4\times 10^{-7}$ & \pa $ 1.0$ & $ 0.0$\pad \\
$\mu_{\rm d}/\mu_{\rm N}$  &  \pa $ 8.4\times 10^{-9}$ & \pa $ 1.3$ & $ -0.2$\pad \\
$\mu_{\rm e}/\mu_{\rm p}$  &  \pa $ 8.1\times 10^{-9}$ & \pa $ 1.2$ & $ 0.2$\pad \\
$\mu_{\rm n}/\mu_{\rm p}$  &  \pa $ 2.4\times 10^{-7}$ & \pa $ 1.0$ & $ 0.0$\pad \\
$\mu_{\rm d}/\mu_{\rm p}$  &  \pa $ 7.7\times 10^{-9}$ & \pa $ 1.9$ & $ -0.3$\pad \\
\botrule

\end{tabular}
\end{table}

Further comments on the entries in Table~\ref{tab:06vs02} are as follows.

(i) The uncertainty of the 2002 recommended value of $\alpha$ has been reduced
by nearly a factor of five by the measurement of $a_{\rm e}$ at Harvard
University and the improved theoretical expression for $a_{\rm e}$(th).  The
difference between the Harvard result and the earlier University of Washington
result, which played a major role in the determination of $\alpha$ in the 2002
adjustment, accounts for most of the change in the recommended value of
$\alpha$ from 2002 to 2006. 

(ii) The uncertainty of the 2002 recommended value of $h$ has been reduced by
over a factor of three due to the new NIST watt-balance result for $K_{\rm
J}^2R_{\rm K}$ and because the factor used to increase the uncertainties of the
data related to $h$ (applied to reduce the inconsistencies among the data), was
reduced from 2.325 in the 2002 adjustment to 1.5 in the 2006 adjustment.  That
the change in value from 2002 to 2006 is small is due to the excellent
agreement between the new value of $K_{\rm J}^2R_{\rm K}$ and the earlier NIST
and NPL values, which played a major role in the determination of $h$ in the
2002 adjustment. 

(iii) The updating of two measurements that contributed to the determination of
the 2002 recommended value of $G$ reduced the spread in the values and
reinforced the most accurate result, that from the University of Washington.
On this basis, the Task Group reduced the assigned uncertainty from $u_{\rm r}
= 1.5 \times 10^{-5}$ in 2002 to $u_{\rm r} = 1.0 \times 10^{-5}$ in 2006.
This uncertainty reflects the historical difficulty of measuring $G$.  Although
the recommended value is the weighted mean of the eight available values, the
assigned uncertainty is still over four times the uncertainty of the mean
multiplied by the corresponding Birge ratio $R_{\rm B}$. 

(iv) The large shift in the recommended value of $d_{220}$ from 2002 to 2006 is
due to the fact that in the 2002 adjustment only the NMIJ result for
$d_{220}({\rm {\scriptstyle NR3}})$ was included, while in the 2006 adjustment
this result (but updated by more recent NMIJ measurements) was included
together with the PTB result for $d_{220}({\rm {\scriptstyle W4.2a}})$ and the
new INRIM results for $d_{220}({\rm {\scriptstyle W4.2a}})$ and $d_{220}({\rm
{\scriptstyle MO^*}})$.  Moreover, the NMIJ value of $d_{220}$ inferred from
$d_{220}({\rm {\scriptstyle NR3}})$ strongly disagrees with the values of
$d_{220}$ inferred from the other three results. 

(v) The marginally significant shift in the recommended value of $g_{\rmssmu}$
from 2002 to 2006 is mainly due to the following: In the 2002 adjustment, the
principal hadronic contribution to the theoretical expression for $a_{\rmssmu}$
was based on both a calculation that included only ${\rm e}^+{\rm e}^-$
annihilation data and a calculation that used data from hadronic decays of the
$\rmtau$ in place of some of the ${\rm e}^+{\rm e}^-$ annihilation data.  In
the 2006 adjustment, the principal hadronic contribution was based on a
calculation that used only annihilation data because of various concerns that
subsequently arose about the reliability of incorporating the $\rmtau$ data in
the calculation; the calculation based on both ${\rm e}^+{\rm e}^-$
annihilation data and $\rmtau$ decay data was only used to estimate the
uncertainty of the hadronic contribution.  Because the results from the two
calculations are in significant disagreement, the uncertainty of
$a_{\rmssmu}$(th) is comparatively large: $u_{\rm r} = 1.8 \times 10^{-6}$. 

(vi) The reduction of the uncertainties of the magnetic moment ratios $\mu_{\rm
p}/\mu_{\rm B}$, $\mu_{\rm p}/\mu_{\rm N}$, $\mu_{\rm d}/\mu_{\rm N}$,
$\mu_{\rm e}/\mu_{\rm p}$, and $\mu_{\rm d}/\mu_{\rm p}$ are due to the new NMR
measurement of $\mu_{\rm p}$(HD)$/\mu_{\rm d}$(HD) and careful re-examination
of the calculation of the D-H shielding correction difference $\sigma_{\rm
dp}$.  Because the value of the product $(\mu_{\rm p}/\mu_{\rm e})(\mu_{\rm
e}/\mu_{\rm d})$ implied by the new measurement is highly consistent with the
same product implied by the individual measurements of $\mu_{\rm
e}$(H)$/\mu_{\rm p}$(H) and $\mu_{\rm d}$(D)$/\mu_{\rm e}$(D), the changes in
the values of the ratios are small. 

In summary, the most important differences between the 2006 and 2002
adjustments are that the 2006 adjustment had available new experimental and
theoretical results for $a_{\rm e}$, which provided a dramatically improved
value of $\alpha$, and a new result for $K_{\rm J}^2\,R_{\rm K}$, which
provided a significantly improved value of $h$.  These two advances from 2002
to 2006 have resulted in major reductions in the uncertainties of many of the
2006 recommended values compared with their 2002 counterparts. 

\subsection{Some implications of the 2006 CODATA recommended values and
adjustment for physics and metrology}

A number of conclusions that can be drawn from the 2006 adjustment concerning
metrology and the basic theories and experimental methods of physics are
presented here, where the focus is on those conclusions that are new or are
different from those drawn from the 2002 and 1998 adjustments. 
\\ \indent
\textit{Conventional electric units}. 
One can interpret the adoption of the conventional values $K_{\rm J-90} =
483\,597.9$~GHz/V and $R_{\rm K-90} = 25\,812.807~{\rm \Omega}$ for the
Josephson and von Klitzing constants as establishing conventional, practical
units of voltage and resistance, $V_{90}$ and ${\it \Omega}_{90}$, given by
$V_{90} = (K_{\rm J-90}/K_{\rm J})$ V and ${\it \Omega}_{90} = (R_{\rm
K}/R_{\rm K-90})~{\rm \Omega}$.  Other conventional electric units follow from
$V_{90}$ and ${\it \Omega}_{90}$, for example, $A_{90} = V_{90}/{\it
\Omega}_{90}$, $C_{90} = A_{90}$~s, $W_{90} = A_{90}V_{90}$, $F_{90} =
C_{90}/V_{90}$, and $H_{90} = {\it \Omega}_{90}$~s, which are the conventional,
practical units of current, charge, power, capacitance, and inductance,
respectively \cite{2001027}.  For the relations between $K_{\rm J}$ and $K_{\rm
J-90}$, and $R_{\rm K}$ and $R_{\rm K-90}$, the 2006 adjustment gives 
\begin{eqnarray}
K_{\rm J} &=& K_{\rm J-90} [1-1.9(2.5)\times10^{-8}] 
\\
R_{\rm K} &=& R_{\rm K-90} [1+2.159(68)\times10^{-8}] \ ,
\end{eqnarray}
which lead to
\begin{eqnarray}
V_{90} &=& [1 + 1.9(2.5) \times  10^{-8}]~{\rm V} 
\label{eq:c901}
\\
{\it \Omega}_{90} &=& [1 + 2.159(68) \times  10^{-8}]~{\rm \Omega} 
\label{eq:c902}
\\
A_{90} &=& [1 - 0.3(2.5) \times  10^{-8}]~{\rm A} 
\label{eq:c903}
\\
C_{90} &=& [1 - 0.3(2.5) \times  10^{-8}]~{\rm C} 
\label{eq:c904}
\\
W_{90} &=& [1 + 1.6(5.0) \times  10^{-8}]~{\rm W} 
\label{eq:c905}
\\
F_{90} &=& [1 - 2.159(68) \times  10^{-8}]~{\rm F} 
\label{eq:c906}
\\
H_{90} &=& [1 + 2.159(68) \times  10^{-8}]~{\rm H} \ .
\label{eq:c907}
\end{eqnarray}
Equations~(\ref{eq:c901}) and (\ref{eq:c902}) show that $V_{90}$ exceeds V and
${\it \Omega}_{90}$ exceeds $\Omega$ by $1.9(2.5) \times 10^{-8}$ and
$2.159(68) \times 10^{-8}$, respectively.  This means that measured voltages
and resistances traceable to the Josephson effect and $K_{\rm J-90}$ and the
quantum Hall effect and $R_{\rm K-90}$, respectively, are too small relative to
the SI by these same fractional amounts.  However, these differences are well
within the $40 \times 10^{-8}$ uncertainty assigned to $V_{90}/$V and the $10
\times 10^{-8}$ uncertainty assigned to ${\it \Omega}_{90}/\Omega$ by the
Consultative Committee for Electricity and Magnetism (CCEM) of the CIPM
\cite{2001223,1989052}. 
\\ \indent
\textit{Josephson and quantum Hall effects}.
The study in Sec.~\ref{sssec:epstests} provides no statistically significant
evidence that the fundamental Josephson and quantum Hall effect relations
$K_{\rm J} = 2e/h$ and $R_{\rm K} = h/e^2$ are not exact.  The theories of two
of the most important phenomena of condensed-matter physics are thereby further
supported. 
\\ \indent
\textit{Antiprotonic helium}.  
The good agreement between the value of $A_{\rm r}({\rm e})$ obtained from the
measured values and theoretical expressions for a number of transition
frequencies in antiprotonic $^4$He and $^3$He with three other values obtained
by entirely different methods indicates that these rather complex atoms are
reasonably well understood both experimentally and theoretically.
\\ \indent
\textit{Newtonian constant of gravitation}. 
Although the inconsistencies among the values of $G$ have been reduced somewhat
as a result of modifications to two of the eight results available in 2002, the
situation remains problematic; there is no evidence that the historic
difficulty of measuring $G$ has been overcome.
\\ \indent
\textit{Tests of QED}.  
The good agreement of the highly accurate values of $\alpha$ inferred from
$h/m({\rm ^{133}Cs})$ and $h/m({\rm ^{87}Rb})$, which are only weakly dependent
on QED theory, with the values of $\alpha$ inferred from $a_{\rm e}$, muonium
transition frequencies, and H and D transition frequencies, provide support for
the QED theory of $a_{\rm e}$ as well as the bound-state QED theory of muonium
and H and D.  In particular, the weighted mean of the two values of $\alpha$
inferred from $h/m({\rm ^{133}Cs})$ and $h/m({\rm ^{87}Rb})$, $\alpha^{-1} =
137.035\,999\,34(69)$ $[5.0 \times 10^{-9}]$, and the weighted mean of the two
values of $\alpha$ inferred from the two experimental values of $a_{\rm e}$,
$\alpha^{-1} = 137.035\,999\,680(94)$ $[6.9 \times 10^{-10}]$, differ by only
0.5$u_{\rm diff}$, with $u_{\rm diff} = 5.1 \times 10^{-9}$. This is a truly
impressive confirmation of QED theory.
\\ \indent
\textit{Physics beyond the Standard Model}. 
If the principal hadronic contribution to $a_{\rmssmu}$(th) obtained from the
${\rm e}^+{\rm e}^-$ annihilation-data plus $\rmtau$ hadronic-decay-data
calculation (see previous section) is completely ignored, and the value based
on the annihilation-data-only calculation with its uncertainty of $45 \times
10^{-11}$ is used in $a_{\rmssmu}$(th), then the value of $\alpha$ inferred
from the BNL experimentally determined value of $a_{\rmssmu}$(exp),
$\alpha^{-1} = 137.035\,670(91)$ $[6.6 \times 10^{-7}]$, differs from the
$h/m({\rm ^{133}Cs})$-$h/m({\rm ^{87}Rb})$ mean value of $\alpha$ by $3.6u_{\rm
diff}$. Although such a large discrepancy may suggest ``New Physics,'' the
consensus is that such a view is premature \cite{pc06md}.
\\ \indent
\textit{Electrical and silicon crystal-related measurements}.  
The previously discussed inconsistencies involving the watt-balance
determinations of $K_{\rm J}^{2}R_{\rm K}$, the mercury electrometer and
voltage balance measurements of $K_{\rm J}$, the XROI determinations of the
\{220\} lattice spacing of various silicon crystals, the measurement of
$h/m_{\rm n}d_{220}({\scriptstyle{\rm W04}})$, and the measurement of $V_{\rm
m}$(Si) hint at possible problems with one or more of these these rather
complex experiments.  This suggests that some of the many different measurement
techniques required for their execution may not be as well understood as is
currently believed. 
\\ \indent
\textit{Redefinition of the kilogram}. 
There has been considerable discussion of late about the possibility of the
24th General Conference on Weights and Measures (CGPM), which convenes in 2011,
redefining the kilogram, ampere, kelvin, and mole by linking these SI base
units to fixed values of $h$, $e$, $k$, and $N_{\rm A}$, respectively
\cite{2006320, 2006021}, in much the same way that the current definition of
the meter is linked to a fixed value of $c$ \cite{2006107}.  Before such a
definition of the kilogram can be accepted, $h$ should be known with a $u_{\rm
r}$ of a few parts in $10^{-8}$.  It is therefore noteworthy that the 2006
CODATA recommended value of $h$ has $u_{\rm r} = 5.0 \times10^{-8}$ and the
most accurate measured value of $h$ (the 2007 NIST watt-balance result) has
$u_{\rm r} = 3.6 \times10^{-8}$. 

\subsection{Outlook and suggestions for future work}

Because there is little redundancy among some of the key input data, the 2006
CODATA set of recommended values, like its 2002 and 1998 predecessors, does not
rest on as solid a foundation as one might wish.  The constants $\alpha$, $h$,
and $R$ play a critical role in determining many other constants, yet the
recommended value of each is determined by a severely limited number of input
data.  Moreover, some input data for the same quantity have uncertainties of
considerably different magnitudes and hence these data contribute to the final
adjustment with considerably different weights. 

The input datum that primarily determines $\alpha$ is the 2006 experimental
result for $a_{\rm e}$ from Harvard University with $u_{\rm r} = 6.5 \times
10^{-10}$; the uncertainty $u_{\rm r} = 37 \times 10^{-10}$ of the next most
accurate experimental result for $a_{\rm e}$, that reported by the University
of Washington in 1987, is 5.7 times larger. Furthermore, there is only a single
value of the eighth-order coefficient $A_1^{(8)}$, that due to Kinoshita and
Nio; it plays a critical role in the theoretical expression for $a_{\rm e}$
from which ${\alpha}$ is obtained and requires lengthy QED calculations. 

The 2007 NIST watt-balance result for $K_{\rm J}^2\,R_{\rm K}$ with $u_{\rm r}
= 3.6 \times 10^{-8}$ is the primary input datum that determines $h$, since the
uncertainty of the next most accurate value of $K_{\rm J}^2\,R_{\rm K}$, the
NIST 1998 result, is 2.4 times larger.  Further, the 2005 consensus value of
$V_{\rm m}$(Si) disagrees with all three high accuracy measurements of $K_{\rm
J}^2\,R_{\rm K}$ currently available. 

For $R$, the key input datum is the 1998 NIST value based on speed-of-sound
measurements in argon using a spherical acoustic resonator with $u_{\rm r} =
1.7 \times 10^{-6}$. The uncertainty of the next most accurate value, the 1979
NPL result, also obtained from speed of sound measurements in argon but using
an acoustic interferometer, is 4.7 times larger. 

Lack of redundancy is, of course, not the only difficulty with the 2006
adjustment. An equally important but not fully independent issue is the several
inconsistencies involving some of the electrical and silicon crystal-related
input data as already discussed, including the recently reported preliminary
result for $K_{\rm J}^2\,R_{\rm K}$ from the NPL watt balance given in
Sec~\ref{sssec:kj2rknpl}. There is also the issue of the recently corrected
(but still tentative) value for the coefficient $A_{1}^{(8)}$ in the
theoretical expression for $a_{\rm e}$ given in Sec~\ref{sec:mmagf}, which
would directly effect the recommended value of $\alpha$.

With these problems in mind, some of which impact the possible redefinition of
the kilogram, ampere, kelvin, and mole in terms of exact values of $h$, $e$,
$k$, and $N_{\rm A}$ in 2011, we offer the following ``wish list'' for new
work. If these needs, some of which appeared in our similar 2002 list, are
successfully met, the key issues facing the precision measurement-fundamental
constants and fundamental metrology fields should be resolved. As a
consequence, our knowledge of the values of the constants, together with the
International System of Units (SI), would be significantly advanced. 

(i) A watt-balance determination of $K_{\rm J}^2\,R_{\rm K}$ from a laboratory
other than NIST or NPL with a $u_{\rm r}$ fully competitive with $u_{\rm r} =
3.6 \times 10^{-8}$, the uncertainty of the most accurate value currently
available from NIST. 

(ii) A timely completion of the current international effort to determine
$N_{\rm A}$ with a $u_{\rm r}$ of a few parts in $10^{8}$ using highly enriched
silicon crystals with $x(^{28}\rm Si)> 0.999\, 85$ \cite{2006075}. This will
require major advances in determining the \{220\} lattice spacing, density, and
molar mass of silicon. 

(iii) A determination of $R$ (or Boltzmann constant $k = R/N_{\rm A}$) with a
$u_{\rm r}$ fully competitive with $u_{\rm r} = 1.7 \times 10^{-6}$, the
uncertainty of the most accurate value of $R$ currently available, preferably
using a method other than measuring the velocity of sound in argon. 

(iv) An independent calculation of the eighth order coefficient $A_1^{(8)}$ in
the QED theoretical expression for $a_{\rm e}$. 

(v) A determination of $\alpha$ that is only weakly dependent on QED theory
with a value of $u_{\rm r}$ fully competitive with $u_{\rm r} = 7.0
\times10^{-10}$, the uncertainty of the most accurate value currently available
as obtained from $a_{\rm e}$(exp) and $a_{\rm e}$(th). 

(vi) A determination of the Newtonian constant of gravitation $G$ with a
$u_{\rm r}$ fully competitive with $u_{\rm r} = 1.4 \times 10^{-5}$, the
uncertainty of the most accurate value of $G$ currently available. 

(vii) A measurement of a transition frequency in hydrogen or deuterium, other
than the already well-known hydrogen $1\rm S_{1/2}-2\rm S_{1/2}$ frequency,
with an uncertainty within an order of magnitude of the current uncertainty of
that frequency, $u_{\rm r} = 1.4 \times 10^{-14}$, thereby providing an
improved value of the Rydberg constant $R_\infty$. 

(viii) Improved theory of the principal hadronic contribution to the
theoretical expression for the muon magnetic moment anomaly $a_{\rmssmu}$(th)
and improvements in the experimental data underlying the calculation of this
contribution so that the origin of the current disagreement between
$a_{\rmssmu}$(th) and $a_{\rmssmu}$(exp) can be better understood. 

(ix) Although there is no experimental or theoretical evidence that the
relations $K_{\rm J}= 2e/h$ and $R_{\rm K} = h/e^{2}$ are not exact, improved
calculable-capacitor measurements of $R_{\rm K}$ and low-field measurements of
the gyromagnetic ratios of the shielded proton and shielded helion, which could
provide further tests of the exactness of these relations, would not be
unwelcome, nor would high accuracy results ($u_{\rm r} \approx 10^{-8}$) from
experiments to close the ``quantum electrical triangle'' \cite{2005284,
2007259}. 

It will be most interesting to see what portion, if any, of this very ambitious
program of work is completed by the 31 December 2010 closing date of the next
CODATA adjustment of the values of the constants.  Indeed, the progress made,
especially in meeting needs (i)-(iii), may very likely determine whether the
24th CGPM, which convenes in October 2011, will approve new definitions of the
kilogram, ampere, kelvin, and mole as discussed in the previous section. If
such new definitions are adopted, $h$, $e$, $k$, and $N_{\rm A}$ as well as a
number of other fundamental constants, for example, $K_{\rm J}$, $R_{\rm K}$
(assuming $K_{\rm J}= 2e/h$ and $R_{\rm K} = h/e^{2}$), $R$, and $\sigma$,
would be exactly known, and many others would have significantly reduced
uncertainties. The result would be a significant advance in our knowledge of
the values of the constants.

\section{Acknowledgments}

We gratefully acknowledge the help of our many colleagues who provided us
results prior to formal publication and for promptly and patiently answering
our many questions about their work. 

\bibliography{refs,final,errata}
\def\s#1{\hbox to #1pt{}}
\def\b{\hbox to 10 pt{}}
\begin{table*}[h]
\caption{An abbreviated list of the CODATA recommended values of the
fundamental constants of physics and chemistry based on the 2006 adjustment.}
\label{tab:abbr}
\begin{tabular}{lllll}
\toprule
& & & & Relative std. \\
\s{35}Quantity & \s{-10}Symbol & \s{15}Numerical value & \s{2}Unit 
& uncert. $u_{\rm r}$ \\
\colrule
speed of light in vacuum & $ c,c_0 $ & 299\,792\,458 & m~s$^{-1}$ & (exact)
\vbox to 12 pt {} \\
magnetic constant & $\mu_0$ & $ 4\rmpi\times10^{-7}$ & N~A$^{-2}$ & \\
& & $=12.566\,370\,614...\times10^{-7}$ & N~A$^{-2}$ & (exact) \\
electric constant 1/$\mu_0c^{2}$ & $\epsilon_0$ &
$8.854\,187\,817...\times 10^{-12}$ & F~m$^{-1}$ & (exact) \\
Newtonian constant & & & & \\
\, of gravitation & $ G $ & $ 6.674\,28(67)\times 10^{-11}$ & m$^{3}$~kg$^{-1}$~s$^{-2}$ & $ 1.0\times 10^{-4}$ \\
& & & & \\
Planck constant & $ h $ & $ 6.626\,068\,96(33)\times 10^{-34}$ & J~s & $ 5.0\times 10^{-8}$ \\
\b $h/2\rmpi$ & $\hbar$ & $ 1.054\,571\,628(53)\times 10^{-34}$ & J~s & $ 5.0\times 10^{-8}$ \\
elementary charge & $ e $ & $ 1.602\,176\,487(40)\times 10^{-19}$ & C & $ 2.5\times 10^{-8}$ \\
magnetic flux quantum $h$/2$e$ & ${\it \Phi}_0$ & $ 2.067\,833\,667(52)\times 10^{-15}$ & Wb & $ 2.5\times 10^{-8}$ \\
conductance quantum $2e^2\!/h$ & $G_0$ & $ 7.748\,091\,7004(53)\times 10^{-5}$ & S & $ 6.8\times 10^{-10}$ \\
& & & & \\
electron mass & $ m_{\rm e}$ & $ 9.109\,382\,15(45)\times 10^{-31}$ & kg & $ 5.0\times 10^{-8}$ \\
proton mass & $ m_{\rm p}$ & $ 1.672\,621\,637(83)\times 10^{-27}$ & kg & $ 5.0\times 10^{-8}$ \\
proton-electron mass ratio & $m_{\rm p}$/$m_{\rm e}$ & $ 1836.152\,672\,47(80)$ & & $ 4.3\times 10^{-10}$ \\
fine-structure constant $e^2\!/4\rmpi\epsilon_0 \hbar c$ & $\alpha$ & $ 7.297\,352\,5376(50)\times 10^{-3}$ & & $ 6.8\times 10^{-10}$ \\
\b inverse fine-structure constant & $\alpha^{-1}$ & $ 137.035\,999\,679(94)$ & & $ 6.8\times 10^{-10}$ \\
& & & & \\
Rydberg constant $\alpha^2m_{\rm e}c/2h$ & $ R_\infty$ & $ 10\,973\,731.568\,527(73)$ & m$^{-1}$ & $ 6.6\times 10^{-12}$ \\
Avogadro constant & $N_{\rm A},L$ & $ 6.022\,141\,79(30)\times 10^{23}$ & mol$^{-1}$ & $ 5.0\times 10^{-8}$ \\
Faraday constant $N_{\rm A}e$ & $ F $ & $ 96\,485.3399(24)$ & C~mol$^{-1}$ & $ 2.5\times 10^{-8}$ \\
molar gas constant & $ R $ & $ 8.314\,472(15)$ & J~mol$^{-1}$~K$^{-1}$ & $ 1.7\times 10^{-6}$ \\
Boltzmann constant $R$/$N_{\rm A}$ & $k$ & $ 1.380\,6504(24)\times 10^{-23}$ & J~K$^{-1}$ & $ 1.7\times 10^{-6}$ \\
Stefan-Boltzmann constant & & & & \\
\, ($\rmpi^2$/60)$k^4\!/\hbar^3c^2$ & $\sigma$ & $ 5.670\,400(40)\times 10^{-8}$ & W~m$^{-2}$~K$^{-4}$ & $ 7.0\times 10^{-6}$ \\
\multicolumn {5} {c} { \vbox to 12 pt {}
Non-SI units accepted for use with the SI} \\
electron volt: ($e$/{\rm C}) {\rm J} & eV & $ 1.602\,176\,487(40)\times 10^{-19}$ & J & $ 2.5\times 10^{-8}$ \\
(unified) atomic mass unit & & & & \\
\, 1 u $=m_{\rm u}= {1\over12}m(^{12}$C) & u & $ 1.660\,538\,782(83)\times 10^{-27}$ & kg & $ 5.0\times 10^{-8}$ \\
\, \, $=10^{-3}$ kg mol$^{-1}\!/N_{\rm A}$ & & & & \\
\botrule
\end{tabular}
\end{table*}

\clearpage
\def\b{\hbox to 12pt{}}
\def\s#1{\hbox to #1pt{}}
\renewcommand{\thefootnote}{$\alph{footnote}$}
\setcounter{footnote}{0}
\shortcites{2006110}
\begin{longtable*}{lllll}
\caption{The CODATA recommended values of the fundamental constants of physics
and chemistry based on the 2006 adjustment.\label{tab:constants}} \\
\toprule
& \T & & & Relative std. \\
\s{35}Quantity \B & \s{-15} Symbol & \s{15} Numerical value & \s{5}Unit & uncert. $u_{\rm r}$ \\
\colrule
\endfirsthead

\caption{{\it (Continued).}} \\
\colrule
& \T & & & Relative std. \\
\s{35}Quantity \B & \s{-15} Symbol & \s{15} Numerical value & \s{5}Unit & uncert. $u_{\rm r}$ \\
\colrule
\vspace{-5pt}
\endhead
\colrule
\endfoot
\endlastfoot
\multicolumn {5} {c} { \vbox to 12 pt {}
UNIVERSAL} \\
speed of light in vacuum & $ c,c_0 $ & $299\,792\,$458 & m~s$^{-1}$ & (exact) \\
magnetic constant & $\mu_0$ & 4$\rmpi\times10^{-7}$ & N~A$^{-2}$ & \\
& & $=12.566\,370\,614...\times10^{-7}$ & N~A$^{-2}$ & (exact) \\
electric constant 1/$\mu_0c^2$ & $\epsilon_0$ & $8.854\,187\,817...\times10^{-12}$ & F~m$^{-1}$ & (exact)\\
characteristic impedance & & & & \\
\, of vacuum $\sqrt{\mu_0/\epsilon_0} = \mu_0c$ & $Z_0$ & $376.730\,313\,461...$ & ${\rm \Omega}$ & (exact)\\
&&&&\\
Newtonian constant & & & & \\
\, of gravitation & $ G $ & $ 6.674\,28(67)\times 10^{-11}$ & m$^3$~kg$^{-1}$~s$^{-2}$ & $ 1.0\times 10^{-4}$ \\
& $G/\hbar c $ & $ 6.708\,81(67)\times 10^{-39}$ & $ ({\rm GeV}/c^2)^{-2}$ & $ 1.0\times 10^{-4}$ \\
Planck constant & $ h $ & $ 6.626\,068\,96(33)\times 10^{-34}$ & J~s & $ 5.0\times 10^{-8}$ \\
\b\b in eV s & & $ 4.135\,667\,33(10)\times 10^{-15}$ & eV~s & $ 2.5\times 10^{-8}$ \\
\b $h/2\rmpi$ & $\hbar$ & $ 1.054\,571\,628(53)\times 10^{-34}$ & J~s & $ 5.0\times 10^{-8}$ \\
\b\b in eV s & & $ 6.582\,118\,99(16)\times 10^{-16}$ & eV~s & $ 2.5\times 10^{-8}$ \\
\b $\hbar c$ in MeV fm & & $ 197.326\,9631(49)$ & MeV~fm & $ 2.5\times 10^{-8}$ \\
&&&&\\
Planck mass~$(\hbar c/G)^{1/2}$ & $m_{\rm P}$ & $ 2.176\,44(11)\times 10^{-8}$ & kg & $ 5.0\times 10^{-5}$ \\
\b energy equivalent in GeV & $m_{\rm P}c^2$ & $ 1.220\,892(61)\times 10^{19}$ & GeV & $ 5.0\times 10^{-5}$ \\
Planck temperature~$(\hbar c^5/G)^{1/2}/k$ & $T_{\rm P}$ & $ 1.416\,785(71)\times 10^{32}$ & K & $ 5.0\times 10^{-5}$ \\
Planck length~$\hbar/m_{\rm P}c=(\hbar G/c^3)^{1/2}$ & $l_{\rm P}$ & $ 1.616\,252(81)\times 10^{-35}$ & m & $ 5.0\times 10^{-5}$ \\
Planck time $l_{\rm P}/c=(\hbar G/c^5)^{1/2}$ & $t_{\rm P}$ & $ 5.391\,24(27)\times 10^{-44}$ & s & $ 5.0\times 10^{-5}$ \\
\multicolumn {5} {c} { \vbox to 12 pt {}
ELECTROMAGNETIC} \\
elementary charge & $e$ & $ 1.602\,176\,487(40)\times 10^{-19}$ & C & $ 2.5\times 10^{-8}$ \\
& $e/h$ & $ 2.417\,989\,454(60)\times 10^{14}$ & A~J$^{-1}$ & $ 2.5\times 10^{-8}$ \\
& & & & \\
magnetic flux quantum $h/2e$ & ${\it \Phi}_0$ & $ 2.067\,833\,667(52)\times 10^{-15}$ & Wb & $ 2.5\times 10^{-8}$ \\
conductance quantum $2e^2\!/h$ & $G_0$ & $ 7.748\,091\,7004(53)\times 10^{-5}$ & S & $ 6.8\times 10^{-10}$ \\
\b inverse of conductance quantum & $G_0^{-1}$ & $ 12\,906.403\,7787(88)$ & ${\rm \Omega}$ & $ 6.8\times 10^{-10}$ \\
Josephson constant\footnote{See Table~\ref{tab:adopted} for the conventional value adopted 
internationally for realizing representations of the volt using the Josephson effect.
}
2$e/h$ & $K_{\rm J}$ & $ 483\,597.891(12)\times 10^{9}$ & Hz~V$^{-1}$ & $ 2.5\times 10^{-8}$ \\
von Klitzing constant\footnote{See Table~\ref{tab:adopted}
for the conventional value adopted internationally
for realizing representations of the ohm using the quantum Hall effect.
}
& & & & \\
\, $h/e^2=\mu_0c/2\alpha$ & $R_{\rm K}$ & $ 25\,812.807\,557(18)$ & ${\rm \Omega}$ & $ 6.8\times 10^{-10}$ \\
& & & & \\
Bohr magneton $e\hbar/2m_{\rm e}$ & $\mu_{\rm B}$ & $ 927.400\,915(23)\times 10^{-26}$ & J~T$^{-1}$ & $ 2.5\times 10^{-8}$ \\
\b in eV T$^{-1}$ & & $ 5.788\,381\,7555(79)\times 10^{-5}$ & eV~T$^{-1}$ & $ 1.4\times 10^{-9}$ \\
\b & $\mu_{\rm B}/h$ & $ 13.996\,246\,04(35)\times 10^{9}$ & Hz~T$^{-1}$ & $ 2.5\times 10^{-8}$ \\
\b & $\mu_{\rm B}/hc$ & $ 46.686\,4515(12)$ & m$^{-1}~$T$^{-1}$ & $ 2.5\times 10^{-8}$ \\
\b & $\mu_{\rm B}/k$ & $ 0.671\,7131(12)$ & K~T$^{-1}$ & $ 1.7\times 10^{-6}$ \\
& & & & \\
nuclear magneton $e\hbar/2m_{\rm p}$ & $\mu_{\rm N}$ & $ 5.050\,783\,24(13)\times 10^{-27}$ & J~T$^{-1}$ & $ 2.5\times 10^{-8}$ \\
\b in eV T$^{-1}$ & & $ 3.152\,451\,2326(45)\times 10^{-8}$ & eV~T$^{-1}$ & $ 1.4\times 10^{-9}$ \\
\b & $\mu_{\rm N}/h$ & $ 7.622\,593\,84(19)$ & MHz~T$^{-1}$ & $ 2.5\times 10^{-8}$ \\
\b & $\mu_{\rm N}/hc$ & $ 2.542\,623\,616(64)\times 10^{-2}$ & m$^{-1}~$T$^{-1}$ & $ 2.5\times 10^{-8}$ \\
\b & $\mu_{\rm N}/k$ & $ 3.658\,2637(64)\times 10^{-4}$ & K~T$^{-1}$ & $ 1.7\times 10^{-6}$ \\
\multicolumn {5} {c} { \vbox to 12 pt {}
ATOMIC AND NUCLEAR} \\
\multicolumn {5} {c} {General} \\
fine-structure constant $e^2\!/4\rmpi\epsilon_0\hbar c$ & $\alpha$ & $ 7.297\,352\,5376(50)\times 10^{-3}$ & & $ 6.8\times 10^{-10}$ \\
\b inverse fine-structure constant & $\alpha^{-1}$ & $ 137.035\,999\,679(94)$ & & $ 6.8\times 10^{-10}$ \\
&&&&\\
Rydberg constant $\alpha^{2}m_{\rm e}c/2h$ & $R_\infty$ & $ 10\,973\,731.568\,527(73)$ & m$^{-1}$ & $ 6.6\times 10^{-12}$ \\
\b & $R_\infty c$ & $ 3.289\,841\,960\,361(22)\times 10^{15}$ & Hz & $ 6.6\times 10^{-12}$ \\
\b & $R_\infty hc$ & $ 2.179\,871\,97(11)\times 10^{-18}$ & J & $ 5.0\times 10^{-8}$ \\
\b $R_\infty hc$ in eV & & $ 13.605\,691\,93(34)$ & eV & $ 2.5\times 10^{-8}$ \\
&&&&\\
Bohr radius $\alpha/4\rmpi R_\infty=4\rmpi\epsilon_0\hbar^2\!/m_{\rm e}e^2$ & $a_{\rm 0}$ & $ 0.529\,177\,208\,59(36)\times 10^{-10}$ & m & $ 6.8\times 10^{-10}$ \\
Hartree energy $e^2\!/4\rmpi\epsilon_{\rm 0}a_{\rm 0}=2R_\infty hc$ & & & & \\
\, $=\alpha^2m_{\rm e}c^2$ & $E_{\rm h}$ & $ 4.359\,743\,94(22)\times 10^{-18}$ & J & $ 5.0\times 10^{-8}$ \\
\b in eV & & $ 27.211\,383\,86(68)$ & eV & $ 2.5\times 10^{-8}$ \\
quantum of circulation & $h/2m_{\rm e}$ & $ 3.636\,947\,5199(50)\times 10^{-4}$ & m$^2~$s$^{-1}$ & $ 1.4\times 10^{-9}$ \\
& $h/m_{\rm e}$ & $ 7.273\,895\,040(10)\times 10^{-4}$ & m$^2~$s$^{-1}$ & $ 1.4\times 10^{-9}$ \\
\multicolumn {5} {c} { \vbox to 12 pt {}
Electroweak} \\
Fermi coupling constant\footnote{Value recommended by the Particle Data Group \cite{2006110}.} & $G_{\rm F}/(\hbar c)^3$ & $ 1.166\,37(1)\times 10^{-5}$ & GeV$^{-2}$ & $ 8.6\times 10^{-6}$ \\
weak mixing angle\footnote{Based on the ratio of the masses of the W and Z bosons $m_{\rm W}/m_{\rm Z}$ recommended by the Particle Data Group \cite{2006110}.
The value for ${\rm sin}^2{\theta}_{\rm W}$
they recommend, which
is based on a particular variant of the modified minimal subtraction
$({\scriptstyle {\rm \overline{MS}}})$ scheme, is
${\rm sin}^2\hat{\theta}_{\rm W}(M_{\rm Z}) = 0.231\,22(15)$.}
$\theta_{\rm W}$ (on-shell scheme) & & & & \\
\, $\sin^2\theta_{\rm W} = s^2_{\rm W} \equiv 1-(m_{\rm W}/m_{\rm Z})^2$ & $\sin^2\theta_{\rm W}$ & $ 0.222\,55(56)$ & & $ 2.5\times 10^{-3}$ \\
\multicolumn {5} {c} {\vbox to 12 pt {}
Electron, e$^-$} \\
electron mass & $m_{\rm e}$ & $ 9.109\,382\,15(45)\times 10^{-31}$ & kg & $ 5.0\times 10^{-8}$ \\
\b\b in u, $m_{\rm e}=A_{\rm r}({\rm e})$~u~(electron &&&&\\
\b\b\, relative atomic mass times~u) & & $ 5.485\,799\,0943(23)\times 10^{-4}$ & u & $ 4.2\times 10^{-10}$ \\
\b energy equivalent & $m_{\rm e}c^2$ & $ 8.187\,104\,38(41)\times 10^{-14}$ & J & $ 5.0\times 10^{-8}$ \\
\b\b in MeV & & $ 0.510\,998\,910(13)$ & MeV & $ 2.5\times 10^{-8}$ \\
&&&&\\
electron-muon mass ratio & $m_{\rm e}/m_{\rmssmu}$ & $ 4.836\,331\,71(12)\times 10^{-3}$ & & $ 2.5\times 10^{-8}$ \\
electron-tau mass ratio & $m_{\rm e}/m_{\rmsstau}$ & $ 2.875\,64(47)\times 10^{-4}$ & & $ 1.6\times 10^{-4}$ \\
electron-proton mass ratio & $m_{\rm e}/m_{\rm p}$ & $ 5.446\,170\,2177(24)\times 10^{-4}$ & & $ 4.3\times 10^{-10}$ \\
electron-neutron mass ratio & $m_{\rm e}/m_{\rm n}$ & $ 5.438\,673\,4459(33)\times 10^{-4}$ & & $ 6.0\times 10^{-10}$ \\
electron-deuteron mass ratio & $m_{\rm e}/m_{\rm d}$ & $ 2.724\,437\,1093(12)\times 10^{-4}$ & & $ 4.3\times 10^{-10}$ \\
electron to alpha particle mass ratio & $m_{\rm e}/m_{\rmssalpha}$ & $ 1.370\,933\,555\,70(58)\times 10^{-4}$ & & $ 4.2\times 10^{-10}$ \\
&&&&\\
electron charge to mass quotient & $-e/m_{\rm e}$ & $ -1.758\,820\,150(44)\times 10^{11}$ & C~kg$^{-1}$ & $ 2.5\times 10^{-8}$ \\
electron molar mass $N_{\rm A}m_{\rm e}$& $M({\rm e}),M_{\rm e}$ & $ 5.485\,799\,0943(23)\times 10^{-7}$ & kg mol$^{-1}$ & $ 4.2\times 10^{-10}$ \\
Compton wavelength $h/m_{\rm e}c$ & $\lambda_{\rm C}$ & $ 2.426\,310\,2175(33)\times 10^{-12}$ & m & $ 1.4\times 10^{-9}$ \\
\b $\lambda_{\rm C}/2\rmpi=\alpha a_{\rm 0}=\alpha^2\!/4\rmpi R_\infty$ & $\lbar_{\rm C}$ & $ 386.159\,264\,59(53)\times 10^{-15}$ & m & $ 1.4\times 10^{-9}$ \\
 classical electron radius $\alpha^2a_{\rm 0}$ & $r_{\rm e}$ & $ 2.817\,940\,2894(58)\times 10^{-15}$ & m & $ 2.1\times 10^{-9}$ \\
Thomson cross section (8$\rmpi/3)r^2_{\rm e}$ & $\sigma_{\rm e}$ & $ 0.665\,245\,8558(27)\times 10^{-28}$ & m$^2$ & $ 4.1\times 10^{-9}$ \\
&&&&\\
electron magnetic moment & $\mu_{\rm e}$ & $ -928.476\,377(23)\times 10^{-26}$ & J~T$^{-1}$ & $ 2.5\times 10^{-8}$ \\
\b to Bohr magneton ratio & $\mu_{\rm e}/\mu_{\rm B}$ & $ -1.001\,159\,652\,181\,11(74)$ & & $ 7.4\times 10^{-13}$ \\
\b to nuclear magneton ratio & $\mu_{\rm e}/\mu_{\rm N}$ & $ -1838.281\,970\,92(80)$ & & $ 4.3\times 10^{-10}$ \\
electron magnetic moment & & & & \\
\, anomaly $|\mu_{\rm e}|/\mu_{\rm B}-1$ & $a_{\rm e}$ & $ 1.159\,652\,181\,11(74)\times 10^{-3}$ & & $ 6.4\times 10^{-10}$ \\
electron $g$-factor $-2(1+a_{\rm e})$ & $g_{\rm e}$ & $ -2.002\,319\,304\,3622(15)$ & & $ 7.4\times 10^{-13}$ \\
&&&&\\
electron-muon & & & & \\
\, magnetic moment ratio & $\mu_{\rm e}/\mu_{\rmssmu}$ & $ 206.766\,9877(52)$ & & $ 2.5\times 10^{-8}$ \\
electron-proton & & & & \\
\, magnetic moment ratio & $\mu_{\rm e}/\mu_{\rm p}$ & $ -658.210\,6848(54)$ & & $ 8.1\times 10^{-9}$ \\
electron to shielded proton & & & & \\
\, magnetic moment ratio & $\mu_{\rm e}/\mu^\prime_{\rm p}$ & $ -658.227\,5971(72)$ & & $ 1.1\times 10^{-8}$ \\
\,\, (H$_2$O, sphere, 25 $^\circ$C)& & & & \\
&&&&\\
electron-neutron & & & & \\
\, magnetic moment ratio & $\mu_{\rm e}/\mu_{\rm n}$ & $ 960.920\,50(23)$ & & $ 2.4\times 10^{-7}$ \\
electron-deuteron & & & & \\
\, magnetic moment ratio & $\mu_{\rm e}/\mu_{\rm d}$ & $ -2143.923\,498(18)$ & & $ 8.4\times 10^{-9}$ \\
electron to shielded helion
& & & & \\
\, magnetic moment ratio & $\mu_{\rm e}/\mu^\prime_{\rm h}$ & $ 864.058\,257(10)$ & & $ 1.2\times 10^{-8}$ \\
\,\, (gas, sphere, 25 $^\circ$C)& & & & \\
electron gyromagnetic ratio $2|\mu_{\rm e}|/\hbar$ & $\gamma_{\rm e}$ & $ 1.760\,859\,770(44)\times 10^{11}$ & s$^{-1}~$T$^{-1}$ & $ 2.5\times 10^{-8}$ \\
& $\gamma_{\rm e}/2\rmpi$ & $ 28\,024.953\,64(70)$ & MHz~T$^{-1}$ & $ 2.5\times 10^{-8}$ \\
\multicolumn {5} {c} { \vbox to 12 pt {}
Muon, ${\rmmu}^-$} \\
muon mass & $m_{\rmssmu}$ & $ 1.883\,531\,30(11)\times 10^{-28}$ & kg & $ 5.6\times 10^{-8}$ \\
\b\b in u, $m_{\rmssmu}=A_{\rm r}({\rmmu})$~u~(muon &&&&\\
\b\b\, relative atomic mass times~u) & & $ 0.113\,428\,9256(29)$ & u & $ 2.5\times 10^{-8}$ \\
\b energy equivalent & $m_{\rmssmu}c^2$ & $ 1.692\,833\,510(95)\times 10^{-11}$ & J & $ 5.6\times 10^{-8}$ \\
\b\b in MeV & & $ 105.658\,3668(38)$ & MeV & $ 3.6\times 10^{-8}$ \\
&&&&\\
muon-electron mass ratio & $m_{\rmssmu}/m_{\rm e}$ & $ 206.768\,2823(52)$ & & $ 2.5\times 10^{-8}$ \\
muon-tau mass ratio & $m_{\rmssmu}/m_{\rmsstau}$ & $ 5.945\,92(97)\times 10^{-2}$ & & $ 1.6\times 10^{-4}$ \\
muon-proton mass ratio & $m_{\rmssmu}/m_{\rm p}$ & $ 0.112\,609\,5261(29)$ & & $ 2.5\times 10^{-8}$ \\
muon-neutron mass ratio & $m_{\rmssmu}/m_{\rm n}$ & $ 0.112\,454\,5167(29)$ & & $ 2.5\times 10^{-8}$ \\
muon molar mass $N_{\rm A}m_{\rmssmu}$& $M({\rmmu}),M_{\rmssmu}$ & $ 0.113\,428\,9256(29)\times 10^{-3}$ & kg mol$^{-1}$ & $ 2.5\times 10^{-8}$ \\
&&&&\\
muon Compton wavelength $h/m_{\rmssmu}c$ & $\lambda_{{\rm C},{\rmssmu}}$ & $ 11.734\,441\,04(30)\times 10^{-15}$ & m & $ 2.5\times 10^{-8}$ \\
\b $\lambda_{{\rm C},{\rmssmu}}/2\rmpi$ & $\lbar_{{\rm C},{\rmssmu}}$ & $ 1.867\,594\,295(47)\times 10^{-15}$ & m & $ 2.5\times 10^{-8}$ \\
muon magnetic moment & $\mu_{\rmssmu}$ & $ -4.490\,447\,86(16)\times 10^{-26}$ & J~T$^{-1}$ & $ 3.6\times 10^{-8}$ \\
\b to Bohr magneton ratio & $\mu_{\rmssmu}/\mu_{\rm B}$ & $ -4.841\,970\,49(12)\times 10^{-3}$ & & $ 2.5\times 10^{-8}$ \\
\b to nuclear magneton ratio & $\mu_{\rmssmu}/\mu_{\rm N}$ & $ -8.890\,597\,05(23)$ & & $ 2.5\times 10^{-8}$ \\
&&&&\\
muon magnetic moment anomaly & & & & \\
\, $|\mu_{\rmssmu}|/(e\hbar/2m_{\rmssmu})-1$ & $a_{\rmssmu}$ & $ 1.165\,920\,69(60)\times 10^{-3}$ & & $ 5.2\times 10^{-7}$ \\
muon $g$-factor $-2(1+a_{\rmssmu}$) & $g_{\rmssmu}$ & $ -2.002\,331\,8414(12)$ & & $ 6.0\times 10^{-10}$ \\
muon-proton & & & & \\
\, magnetic moment ratio & $\mu_{\rmssmu}/\mu_{\rm p}$ & $ -3.183\,345\,137(85)$ & & $ 2.7\times 10^{-8}$ \\
\multicolumn {5} {c} { \vbox to 12 pt {}
Tau, ${\rmtau}^-$} \\
tau mass\footnote{This and all other values involving $m_{\rmsstau}$ are based on
the value of $m_{\rmsstau}c^2$ in MeV recommended by the
Particle Data Group \cite{2006110},
but with a standard uncertainty of
$0.29$ MeV rather than the quoted uncertainty of $-0.26$ MeV, $+0.29$ MeV.
}
& $m_{\rmsstau}$ & $ 3.167\,77(52)\times 10^{-27}$ & kg & $ 1.6\times 10^{-4}$ \\
\b\b in u, $m_{\rmsstau}=A_{\rm r}({\rmsstau})$~u~(tau &&&&\\
\b\b\, relative atomic mass times~u) & & $ 1.907\,68(31)$ & u & $ 1.6\times 10^{-4}$ \\
\b energy equivalent & $m_{\rmsstau}c^2$ & $ 2.847\,05(46)\times 10^{-10}$ & J & $ 1.6\times 10^{-4}$ \\
\b\b in MeV & & $ 1776.99(29)$ & MeV & $ 1.6\times 10^{-4}$ \\
&&&&\\
tau-electron mass ratio & $m_{\rmsstau}/m_{\rm e}$ & $ 3477.48(57)$ & & $ 1.6\times 10^{-4}$ \\
tau-muon mass ratio & $m_{\rmsstau}/m_{\rmssmu}$ & $ 16.8183(27)$ & & $ 1.6\times 10^{-4}$ \\
tau-proton mass ratio & $m_{\rmsstau}/m_{\rm p}$ & $ 1.893\,90(31)$ & & $ 1.6\times 10^{-4}$ \\
tau-neutron mass ratio & $m_{\rmsstau}/m_{\rm n}$ & $ 1.891\,29(31)$ & & $ 1.6\times 10^{-4}$ \\
tau molar mass $N_{\rm A}m_{\rmsstau}$& $M({\rmtau}),M_{\rmsstau}$ & $ 1.907\,68(31)\times 10^{-3}$ & kg mol$^{-1}$ & $ 1.6\times 10^{-4}$ \\
&&&&\\
tau Compton wavelength $h/m_{\rmsstau}c$ & $\lambda_{{\rm C},{\rmsstau}}$ & $ 0.697\,72(11)\times 10^{-15}$ & m & $ 1.6\times 10^{-4}$ \\
\b $\lambda_{{\rm C},{\rmsstau}}/2\rmpi$ & $\lbar_{{\rm C},{\rmsstau}}$ & $ 0.111\,046(18)\times 10^{-15}$ & m & $ 1.6\times 10^{-4}$ \\
\multicolumn {5} {c} {\vbox to 12 pt {} Proton, p} \\
proton mass & $m_{\rm p}$ & $ 1.672\,621\,637(83)\times 10^{-27}$ & kg & $ 5.0\times 10^{-8}$ \\
\b\b in u, $m_{\rm p}=A_{\rm r}({\rm p})$~u~(proton &&&&\\
\b\b\, relative atomic mass times~u) & & $ 1.007\,276\,466\,77(10)$ & u & $ 1.0\times 10^{-10}$ \\
\b energy equivalent & $m_{\rm p}c^2$ & $ 1.503\,277\,359(75)\times 10^{-10}$ & J & $ 5.0\times 10^{-8}$ \\
\b\b in MeV & & $ 938.272\,013(23)$ & MeV & $ 2.5\times 10^{-8}$ \\
&&&&\\
proton-electron mass ratio & $m_{\rm p}/m_{\rm e}$ & $ 1836.152\,672\,47(80)$ & & $ 4.3\times 10^{-10}$ \\
proton-muon mass ratio & $m_{\rm p}/m_{\rmssmu}$ & $ 8.880\,243\,39(23)$ & & $ 2.5\times 10^{-8}$ \\
proton-tau mass ratio & $m_{\rm p}/m_{\rmsstau}$ & $ 0.528\,012(86)$ & & $ 1.6\times 10^{-4}$ \\
proton-neutron mass ratio & $m_{\rm p}/m_{\rm n}$ & $ 0.998\,623\,478\,24(46)$ & & $ 4.6\times 10^{-10}$ \\
proton charge to mass quotient & $e/m_{\rm p}$ & $ 9.578\,833\,92(24)\times 10^{7}$ & C kg$^{-1}$ & $ 2.5\times 10^{-8}$ \\
proton molar mass $N_{\rm A}m_{\rm p}$& $M$(p), $M_{\rm p}$ & $ 1.007\,276\,466\,77(10)\times 10^{-3}$ & kg mol$^{-1}$ & $ 1.0\times 10^{-10}$ \\
&&&&\\
proton Compton wavelength $h/m_{\rm p}c$ & $\lambda_{\rm C,p}$ & $ 1.321\,409\,8446(19)\times 10^{-15}$ & m & $ 1.4\times 10^{-9}$ \\
\b $\lambda_{\rm C,p}/2\rmpi$ & $\lbar_{\rm C,p}$ & $ 0.210\,308\,908\,61(30)\times 10^{-15}$ & m & $ 1.4\times 10^{-9}$ \\
proton rms charge radius & $R_{\rm p}$ & $ 0.8768(69)\times 10^{-15}$ & m & $ 7.8\times 10^{-3}$ \\
proton magnetic moment & $\mu_{\rm p}$ & $ 1.410\,606\,662(37)\times 10^{-26}$ & J~T$^{-1}$ & $ 2.6\times 10^{-8}$ \\
\b to Bohr magneton ratio & $\mu_{\rm p}/\mu_{\rm B}$ & $ 1.521\,032\,209(12)\times 10^{-3}$ & & $ 8.1\times 10^{-9}$ \\
\b to nuclear magneton ratio & $\mu_{\rm p}/\mu_{\rm N}$ & $ 2.792\,847\,356(23)$ & & $ 8.2\times 10^{-9}$ \\
&&&&\\
proton $g$-factor $2\mu_{\rm p}/\mu_{\rm N}$ & $g_{\rm p}$ & $ 5.585\,694\,713(46)$ & & $ 8.2\times 10^{-9}$ \\
proton-neutron & & & & \\
\, magnetic moment ratio & $\mu_{\rm p}/\mu_{\rm n}$ & $ -1.459\,898\,06(34)$ & & $ 2.4\times 10^{-7}$ \\
shielded proton magnetic moment & $\mu^\prime_{\rm p}$ & $ 1.410\,570\,419(38)\times 10^{-26}$ & J~T$^{-1}$ & $ 2.7\times 10^{-8}$ \\
\, (H$_{2}$O, sphere, 25 $^\circ$C) & & & & \\
\b to Bohr magneton ratio & $\mu^\prime_{\rm p}/\mu_{\rm B}$ & $ 1.520\,993\,128(17)\times 10^{-3}$ & & $ 1.1\times 10^{-8}$ \\
\b to nuclear magneton ratio & $\mu^\prime_{\rm p}/\mu_{\rm N}$ & $ 2.792\,775\,598(30)$ & & $ 1.1\times 10^{-8}$ \\
proton magnetic shielding & & & & \\
\, correction $1-\mu^\prime_{\rm p}/\mu_{\rm p}$ & $\sigma^\prime_{\rm p}$ & $ 25.694(14)\times 10^{-6}$ & &  $ 5.3\times 10^{-4}$ \\
\,\, (H$_{2}$O, sphere, 25 $^\circ$C) & & & & \\
&&&&\\
proton gyromagnetic ratio $2\mu_{\rm p}/\hbar$ & $\gamma_{\rm p}$ & $ 2.675\,222\,099(70)\times 10^{8}$ & s$^{-1}~$T$^{-1}$ & $ 2.6\times 10^{-8}$ \\
& $\gamma_{\rm p}/2\rmpi$ & $ 42.577\,4821(11)$ & MHz~T$^{-1}$ & $ 2.6\times 10^{-8}$ \\
shielded proton gyromagnetic & & & & \\
\, ratio $2\mu^\prime_{\rm p}/\hbar$ & $\gamma^\prime_{\rm p}$ & $ 2.675\,153\,362(73)\times 10^{8}$ & s$^{-1}~$T$^{-1}$ & $ 2.7\times 10^{-8}$ \\
\,\, (H$_{2}$O, sphere, 25 $^\circ$C)& & & & \\
& $\gamma^\prime_{\rm p}/2\rmpi$ & $ 42.576\,3881(12)$ & MHz~T$^{-1}$ & $ 2.7\times 10^{-8}$ \\
\multicolumn {5} {c} {\vbox to 12 pt {} Neutron, n} \\
neutron mass & $m_{\rm n}$ & $ 1.674\,927\,211(84)\times 10^{-27}$ & kg & $ 5.0\times 10^{-8}$ \\
\b\b in u, $m_{\rm n}=A_{\rm r}({\rm n})$~u~(neutron &&&&\\
\b\b\, relative atomic mass times~u) & & $ 1.008\,664\,915\,97(43)$ & u & $ 4.3\times 10^{-10}$ \\
\b energy equivalent & $m_{\rm n}c^2$ & $ 1.505\,349\,505(75)\times 10^{-10}$ & J & $ 5.0\times 10^{-8}$ \\
\b\b in MeV & & $ 939.565\,346(23)$ & MeV & $ 2.5\times 10^{-8}$ \\
&&&&\\
neutron-electron mass ratio & $m_{\rm n}/m_{\rm e}$ & $ 1838.683\,6605(11)$ & & $ 6.0\times 10^{-10}$ \\
neutron-muon mass ratio & $m_{\rm n}/m_{\rmssmu}$ & $ 8.892\,484\,09(23)$ & & $ 2.5\times 10^{-8}$ \\
neutron-tau mass ratio & $m_{\rm n}/m_{\rmsstau}$ & $ 0.528\,740(86)$ & & $ 1.6\times 10^{-4}$ \\
neutron-proton mass ratio & $m_{\rm n}/m_{\rm p}$ & $ 1.001\,378\,419\,18(46)$ & & $ 4.6\times 10^{-10}$ \\
neutron molar mass $N_{\rm A}m_{\rm n}$ & $M({\rm n}),M_{\rm n}$ & $ 1.008\,664\,915\,97(43)\times 10^{-3}$ & kg mol$^{-1}$ & $ 4.3\times 10^{-10}$ \\
&&&&\\
neutron Compton wavelength $h/m_{\rm n}c$ & $\lambda_{\rm C,n}$ & $ 1.319\,590\,8951(20)\times 10^{-15}$ & m & $ 1.5\times 10^{-9}$ \\
\b $\lambda_{\rm C,n}/2\rmpi$ & $\lbar_{\rm C,n}$ & $ 0.210\,019\,413\,82(31)\times 10^{-15}$ & m & $ 1.5\times 10^{-9}$ \\
neutron magnetic moment & $\mu_{\rm n}$ & $ -0.966\,236\,41(23)\times 10^{-26}$ & J~T$^{-1}$ & $ 2.4\times 10^{-7}$ \\
\b to Bohr magneton ratio & $\mu_{\rm n}/\mu_{\rm B}$ & $ -1.041\,875\,63(25)\times 10^{-3}$ & & $ 2.4\times 10^{-7}$ \\
\b to nuclear magneton ratio & $\mu_{\rm n}/\mu_{\rm N}$ & $ -1.913\,042\,73(45)$ & & $ 2.4\times 10^{-7}$ \\
&&&&\\
neutron $g$-factor $2\mu_{\rm n}/\mu_{\rm N}$ & $g_{\rm n}$ & $ -3.826\,085\,45(90)$ & & $ 2.4\times 10^{-7}$ \\
neutron-electron & & & & \\
\, magnetic moment ratio & $\mu_{\rm n}/\mu_{\rm e}$ & $ 1.040\,668\,82(25)\times 10^{-3}$ & & $ 2.4\times 10^{-7}$ \\
neutron-proton & & & & \\
\, magnetic moment ratio & $\mu_{\rm n}/\mu_{\rm p}$ & $ -0.684\,979\,34(16)$ & & $ 2.4\times 10^{-7}$ \\
neutron to shielded proton & & & & \\
\, magnetic moment ratio & $\mu_{\rm n}/\mu_{\rm p}^\prime$ & $ -0.684\,996\,94(16)$ & & $ 2.4\times 10^{-7}$ \\
\,\, (H$_2$O, sphere, 25 $^\circ$C) & & & & \\
neutron gyromagnetic ratio $2|\mu_{\rm n}|/\hbar$ & $\gamma_{\rm n}$ & $ 1.832\,471\,85(43)\times 10^{8}$ & s$^{-1}~$T$^{-1}$ & $ 2.4\times 10^{-7}$ \\
& $\gamma_{\rm n}/2\rmpi$ & $ 29.164\,6954(69)$ & MHz~T$^{-1}$ & $ 2.4\times 10^{-7}$ \\
\multicolumn {5} {c} {\vbox to 12 pt {} Deuteron, d} \\
deuteron mass & $m_{\rm d}$ & $ 3.343\,583\,20(17)\times 10^{-27}$ & kg & $ 5.0\times 10^{-8}$ \\
\b\b in u, $m_{\rm d}=A_{\rm r}({\rm d})$~u~(deuteron &&&&\\
\b\b\, relative atomic mass times~u) & & $ 2.013\,553\,212\,724(78)$ & u & $ 3.9\times 10^{-11}$ \\
\b energy equivalent & $m_{\rm d}c^2$ & $ 3.005\,062\,72(15)\times 10^{-10}$ & J & $ 5.0\times 10^{-8}$ \\
\b\b in MeV & & $ 1875.612\,793(47)$ & MeV & $ 2.5\times 10^{-8}$ \\
&&&&\\
deuteron-electron mass ratio & $m_{\rm d}/m_{\rm e}$ & $ 3670.482\,9654(16)$ & & $ 4.3\times 10^{-10}$ \\
deuteron-proton mass ratio & $m_{\rm d}/m_{\rm p}$ & $ 1.999\,007\,501\,08(22)$ & & $ 1.1\times 10^{-10}$ \\
deuteron molar mass $N_{\rm A}m_{\rm d}$& $M({\rm d}),M_{\rm d}$ & $ 2.013\,553\,212\,724(78)\times 10^{-3}$ & kg mol$^{-1}$ & $ 3.9\times 10^{-11}$ \\
&&&&\\
deuteron rms charge radius & $R_{\rm d}$ & $ 2.1402(28)\times 10^{-15}$ & m & $ 1.3\times 10^{-3}$ \\
deuteron magnetic moment & $\mu_{\rm d}$ & $ 0.433\,073\,465(11)\times 10^{-26}$ & J~T$^{-1}$ & $ 2.6\times 10^{-8}$ \\
\b to Bohr magneton ratio & $\mu_{\rm d}/\mu_{\rm B}$ & $ 0.466\,975\,4556(39)\times 10^{-3}$ & & $ 8.4\times 10^{-9}$ \\
\b to nuclear magneton ratio & $\mu_{\rm d}/\mu_{\rm N}$ & $ 0.857\,438\,2308(72)$ & & $ 8.4\times 10^{-9}$ \\
&&&&\\
deuteron $g$-factor $\mu_{\rm d}/\mu_{\rm N}$ & $g_{\rm d}$ & $ 0.857\,438\,2308(72)$ & & $ 8.4\times 10^{-9}$ \\
deuteron-electron & & & & \\
\, magnetic moment ratio & $\mu_{\rm d}/\mu_{\rm e}$ & $ -4.664\,345\,537(39)\times 10^{-4}$ & & $ 8.4\times 10^{-9}$ \\
deuteron-proton & & & & \\
\, magnetic moment ratio & $\mu_{\rm d}/\mu_{\rm p}$ & $ 0.307\,012\,2070(24)$ & & $ 7.7\times 10^{-9}$ \\
deuteron-neutron & & & & \\
\, magnetic moment ratio & $\mu_{\rm d}/\mu_{\rm n}$ & $ -0.448\,206\,52(11)$ & & $ 2.4\times 10^{-7}$ \\
\multicolumn {5} {c} {\vbox to 12 pt {} Triton, t} \\
triton mass & $m_{\rm t}$ & $ 5.007\,355\,88(25)\times 10^{-27}$ & kg & $ 5.0\times 10^{-8}$ \\
\b\b in u, $m_{\rm t}=A_{\rm r}({\rm t})$~u~(triton &&&&\\
\b\b\, relative atomic mass times~u) & & $ 3.015\,500\,7134(25)$ & u & $ 8.3\times 10^{-10}$ \\
\b energy equivalent & $m_{\rm t}c^2$ & $ 4.500\,387\,03(22)\times 10^{-10}$ & J & $ 5.0\times 10^{-8}$ \\
\b\b in MeV & & $ 2808.920\,906(70)$ & MeV & $ 2.5\times 10^{-8}$ \\
&&&&\\
triton-electron mass ratio & $m_{\rm t}/m_{\rm e}$ & $ 5496.921\,5269(51)$ & & $ 9.3\times 10^{-10}$ \\
triton-proton mass ratio & $m_{\rm t}/m_{\rm p}$ & $ 2.993\,717\,0309(25)$ & & $ 8.4\times 10^{-10}$ \\
triton molar mass $N_{\rm A}m_{\rm t}$& $M({\rm t}),M_{\rm t}$ & $ 3.015\,500\,7134(25)\times 10^{-3}$ & kg mol$^{-1}$ & $ 8.3\times 10^{-10}$ \\
&&&&\\
triton magnetic moment & $\mu_{\rm t}$ & $ 1.504\,609\,361(42)\times 10^{-26}$ & J~T$^{-1}$ & $ 2.8\times 10^{-8}$ \\
\b to Bohr magneton ratio & $\mu_{\rm t}/\mu_{\rm B}$ & $ 1.622\,393\,657(21)\times 10^{-3}$ & & $ 1.3\times 10^{-8}$ \\
\b to nuclear magneton ratio & $\mu_{\rm t}/\mu_{\rm N}$ & $ 2.978\,962\,448(38)$ & & $ 1.3\times 10^{-8}$ \\
&&&&\\
triton $g$-factor $2\mu_{\rm t}/\mu_{\rm N}$ & $g_{\rm t}$ & $ 5.957\,924\,896(76)$ & & $ 1.3\times 10^{-8}$ \\
triton-electron & & & & \\
\, magnetic moment ratio & $\mu_{\rm t}/\mu_{\rm e}$ & $ -1.620\,514\,423(21)\times 10^{-3}$ & & $ 1.3\times 10^{-8}$ \\
triton-proton & & & & \\
\, magnetic moment ratio & $\mu_{\rm t}/\mu_{\rm p}$ & $ 1.066\,639\,908(10)$ & & $ 9.8\times 10^{-9}$ \\
triton-neutron & & & & \\
\, magnetic moment ratio & $\mu_{\rm t}/\mu_{\rm n}$ & $ -1.557\,185\,53(37)$ & & $ 2.4\times 10^{-7}$ \\
\multicolumn {5} {c} {\vbox to 12 pt {} Helion, h} \\
helion mass\footnotemark[5] & $m_{\rm h}$ & $ 5.006\,411\,92(25)\times 10^{-27}$ & kg & $ 5.0\times 10^{-8}$ \\
\b\b in u, $m_{\rm h}=A_{\rm r}({\rm h})$~u~(helion &&&&\\
\b\b\, relative atomic mass times~u) & & $ 3.014\,932\,2473(26)$ & u & $ 8.6\times 10^{-10}$ \\
\b energy equivalent & $m_{\rm h}c^2$ & $ 4.499\,538\,64(22)\times 10^{-10}$ & J & $ 5.0\times 10^{-8}$ \\
\b\b in MeV & & $ 2808.391\,383(70)$ & MeV & $ 2.5\times 10^{-8}$ \\
&&&&\\
helion-electron mass ratio & $m_{\rm h}/m_{\rm e}$ & $ 5495.885\,2765(52)$ & & $ 9.5\times 10^{-10}$ \\
helion-proton mass ratio & $m_{\rm h}/m_{\rm p}$ & $ 2.993\,152\,6713(26)$ & & $ 8.7\times 10^{-10}$ \\
helion molar mass $N_{\rm A}m_{\rm h}$& $M({\rm h}),M_{\rm h}$ & $ 3.014\,932\,2473(26)\times 10^{-3}$ & kg mol$^{-1}$ & $ 8.6\times 10^{-10}$ \\
shielded helion magnetic moment & $\mu^\prime_{\rm h}$ & $ -1.074\,552\,982(30)\times 10^{-26}$ & J~T$^{-1}$ & $ 2.8\times 10^{-8}$ \\
\, (gas, sphere, 25 $^\circ$C) & & & & \\
\b to Bohr magneton ratio & $\mu^\prime_{\rm h}/\mu_{\rm B}$ & $ -1.158\,671\,471(14)\times 10^{-3}$ & & $ 1.2\times 10^{-8}$ \\
\b to nuclear magneton ratio & $\mu^\prime_{\rm h}/\mu_{\rm N}$ & $ -2.127\,497\,718(25)$ & & $ 1.2\times 10^{-8}$ \\
shielded helion to proton & & & & \\
\b magnetic moment ratio & $\mu^\prime_{\rm h}/\mu_{\rm p}$ & $ -0.761\,766\,558(11)$ & & $ 1.4\times 10^{-8}$ \\
\, (gas, sphere, 25 $^\circ$C) & & & & \\
&&&&\\
shielded helion to shielded proton & & & & \\
\b magnetic moment ratio & $\mu^\prime_{\rm h}/\mu^\prime_{\rm p}$ & $ -0.761\,786\,1313(33)$ & & $ 4.3\times 10^{-9}$ \\
\, (gas/H$_2$O, spheres, 25 $^\circ$C) & & & & \\
shielded helion gyromagnetic & & & & \\
\, ratio $2|\mu^\prime_{\rm h}|/\hbar$ & $\gamma^\prime_{\rm h}$ & $ 2.037\,894\,730(56)\times 10^{8}$ & s$^{-1}$ T$^{-1}$ & $ 2.8\times 10^{-8}$ \\
\,\, (gas, sphere, 25 $^\circ$C)& & & & \\
& $\gamma^\prime_{\rm h}/2\rmpi$ & $ 32.434\,101\,98(90)$ & MHz~T$^{-1}$ & $ 2.8\times 10^{-8}$ \\
\multicolumn {5} {c} {\vbox to 12 pt {} Alpha particle, ${\rmalpha}$} \\
alpha particle mass & $m_{\rmssalpha}$ & $ 6.644\,656\,20(33)\times 10^{-27}$ & kg & $ 5.0\times 10^{-8}$ \\
\b\b in u, $m_{\rmssalpha}=A_{\rm r}({\rmalpha})$~u~(alpha particle &&&&\\
\b\b\, relative atomic mass times~u) & & $ 4.001\,506\,179\,127(62)$ & u & $ 1.5\times 10^{-11}$ \\
\b energy equivalent & $m_{\rmssalpha}c^2$ & $ 5.971\,919\,17(30)\times 10^{-10}$ & J & $ 5.0\times 10^{-8}$ \\
\b\b in MeV & & $ 3727.379\,109(93)$ & MeV & $ 2.5\times 10^{-8}$ \\
&&&&\\
alpha particle to electron mass ratio & $m_{\rmssalpha}/m_{\rm e}$ & $ 7294.299\,5365(31)$ & & $ 4.2\times 10^{-10}$ \\
alpha particle to proton mass ratio & $m_{\rmssalpha}/m_{\rm p}$ & $ 3.972\,599\,689\,51(41)$ & & $ 1.0\times 10^{-10}$ \\
alpha particle molar mass $N_{\rm A}m_{\rmssalpha}$& $M({\rmalpha}),M_{\rmssalpha}$ & $ 4.001\,506\,179\,127(62)\times 10^{-3}$ & kg mol$^{-1}$ & $ 1.5\times 10^{-11}$ \\
\multicolumn {5} {c} {\vbox to 12 pt {} PHYSICOCHEMICAL} \\
Avogadro constant & $N_{\rm A},L$ & $ 6.022\,141\,79(30)\times 10^{23}$ & mol$^{-1}$ & $ 5.0\times 10^{-8}$ \\
atomic mass constant & & & & \\
\, $m_{\rm u}=\frac {1}{12}m(^{12}{\rm C})= 1$ u & $m_{\rm u}$ & $ 1.660\,538\,782(83)\times 10^{-27}$ & kg & $ 5.0\times 10^{-8}$ \\
\vbox to 10pt{}\, \, $=10^{-3}$ kg mol$^{-1}\!/N_{\rm A}$ & & & & \\
\b energy equivalent & $m_{\rm u}c^2$ & $ 1.492\,417\,830(74)\times 10^{-10}$ & J & $ 5.0\times 10^{-8}$ \\
\b\b in MeV & & $ 931.494\,028(23)$ & MeV & $ 2.5\times 10^{-8}$ \\
Faraday constant\footnote{The numerical value of $F$ to be used in coulometric chemical measurements
is $ 96\,485.3401(48)$~[$ 5.0\times 10^{-8}$] when the relevant current is measured
in terms of representations of the volt and ohm based on the Josephson
and quantum Hall effects and the internationally adopted conventional
values of the Josephson and von Klitzing constants $K_{\rm J-90}$ and
$R_{\rm K-90}$ given in Table~\ref{tab:adopted}.
}
$N_{\rm A}e$ & $F$ & $ 96\,485.3399(24)$ & C~mol$^{-1}$ & $ 2.5\times 10^{-8}$ \\
&&&&\\
molar Planck constant & $N_{\rm A}h$ & $ 3.990\,312\,6821(57)\times 10^{-10}$ & J~s~mol$^{-1}$ & $ 1.4\times 10^{-9}$ \\
& $N_{\rm A}hc$ & $ 0.119\,626\,564\,72(17)$ & J~m~mol$^{-1}$ & $ 1.4\times 10^{-9}$ \\
molar gas constant & $R$ & $ 8.314\,472(15)$ & J~mol$^{-1}~$K$^{-1}$ & $ 1.7\times 10^{-6}$ \\
Boltzmann constant $R/N_{\rm A}$ & $k$ & $ 1.380\,6504(24)\times 10^{-23}$ & J~K$^{-1}$ & $ 1.7\times 10^{-6}$ \\
\b\b in eV K$^{-1}$  & & $ 8.617\,343(15)\times 10^{-5}$ & eV~K$^{-1}$ & $ 1.7\times 10^{-6}$ \\
\b & $k/h$ & $ 2.083\,6644(36)\times 10^{10}$ & Hz~K$^{-1}$ & $ 1.7\times 10^{-6}$ \\
\b & $k/hc$ & $ 69.503\,56(12)$ & m$^{-1}~$K$^{-1}$ & $ 1.7\times 10^{-6}$ \\
&&&&\\
molar volume of ideal gas $RT/p$ & & & & \\
\b $T=273.15\ {\rm K},\,p=101.325\ {\rm kPa}$ & $V_{\rm m}$ & $ 22.413\,996(39)\times 10^{-3}$ & m$^{3}~$mol$^{-1}$ & $ 1.7\times 10^{-6}$ \\
\b Loschmidt constant $N_{\rm A}/V_{\rm m}$ & $n_0$ & $ 2.686\,7774(47)\times 10^{25}$ & m$^{-3}$ & $ 1.7\times 10^{-6}$ \\
\b $T=273.15\ {\rm K},\,p=100\ {\rm kPa}$ & $V_{\rm m}$ & $ 22.710\,981(40)\times 10^{-3}$ & m$^{3}~$mol$^{-1}$ & $ 1.7\times 10^{-6}$ \\
&&&&\\
Sackur-Tetrode constant & & & & \\
\, (absolute entropy constant)\footnote{The entropy of an ideal monoatomic gas of relative atomic mass
$A_{\rm r}$ is given by $S = S_0 +{3\over2} R\, \ln A_{\rm r}
-R\, \ln(p/p_0) +{5\over2}R\,\ln(T/{\rm K}).$
}
& & & & \\
\, $\frac {5} {2}+\ln[(2\rmpi m_{\rm u}kT_1/h^2)^{3/2}kT_1/p_0]$ & & & & \\
\b $T_1=1\ {\rm K},\,p_0\,=\,100\ {\rm kPa}$ & $S_0/R$ & $ -1.151\,7047(44)$ & & $ 3.8\times 10^{-6}$ \\
\b $T_1=1\ {\rm K},\,p_0\,=\,101.325\ {\rm kPa}$ & & $ -1.164\,8677(44)$ & & $ 3.8\times 10^{-6}$ \\
&&&&\\
Stefan-Boltzmann constant & & & & \\
\, ($\rmpi^2/60)k^4\!/\hbar^3c^2$ & $\sigma$ & $ 5.670\,400(40)\times 10^{-8}$ & W~m$^{-2}~$K$^{-4}$ & $ 7.0\times 10^{-6}$ \\
first radiation constant 2$\rmpi hc^2$ & $c_1$ & $ 3.741\,771\,18(19)\times 10^{-16}$ & W~m$^{2}$ & $ 5.0\times 10^{-8}$ \\
first radiation constant for spectral radiance 2$hc^2$ & $c_{\rm 1L}$ & $ 1.191\,042\,759(59)\times 10^{-16}$ & W~m$^{2}$~sr$^{-1}$ & $ 5.0\times 10^{-8}$ \\
second radiation constant $hc/k$ & $c_2$ & $ 1.438\,7752(25)\times 10^{-2}$ & m~K & $ 1.7\times 10^{-6}$ \\
Wien displacement law constants & & & & \\
\, $b=\lambda_{\rm max}T=c_2/4.965\,114\,231...$ & $b$ & $ 2.897\,7685(51)\times 10^{-3}$ & m~K & $ 1.7\times 10^{-6}$ \\
\, $b^\prime=\nu_{\rm max}/T=2.821\,439\,372...\,c/c_2 $ & $b^\prime$ & $ 5.878\,933(10)\times 10^{10}$ & Hz~K$^{-1}$ & $ 1.7\times 10^{-6}$ \\
\botrule
\end{longtable*}

\begin{table*}
\caption{The variances, covariances, and correlation coefficients of the values
of a selected group of constants based on the 2006 CODATA adjustment.  The
numbers in bold above the main diagonal are $10^{16}$ times the numerical
values of the relative covariances; the numbers in bold on the main diagonal
are $10^{16}$ times the numerical values of the relative variances; and the
numbers in italics below the main diagonal are the correlation
coefficients.$^a$}
\label{tab:varmatrix}
\def\sp{\hbox to 10.7mm {}}
\begin{tabular} {c@{\sp}r@{\sp}r@{\sp}r@{\sp}r@{\sp}r@{\sp}r@{\sp}r}
\toprule
 \B \T & $\alpha$~~~ & $h$~~~  & $e$~~~  & $m_{\rm e}$~~~ & $N_{\rm A}$~~~  & $m_{\rm e}/m_{\rm \mu}$  & $F$~~~ \\
\colrule
$\alpha$  \T & ${\bf  0.0047}$ & ${\bf  0.0002}$ & ${\bf  0.0024}$ & ${\bf  -0.0092}$ & ${\bf  0.0092}$ & ${\bf  -0.0092}$ & ${\bf  0.0116}$ \\
$h$ & ${\it  0.0005}$ & ${\bf  24.8614}$ & ${\bf  12.4308}$ & ${\bf  24.8611}$ & ${\bf  -24.8610}$ & ${\bf  -0.0003}$ & ${\bf  -12.4302}$ \\
$e$ & ${\it  0.0142}$ & ${\it  0.9999}$ & ${\bf  6.2166}$ & ${\bf  12.4259}$ & ${\bf  -12.4259}$ & ${\bf  -0.0048}$ & ${\bf  -6.2093}$ \\
$m_{\rm e}$ & ${\it  -0.0269}$ & ${\it  0.9996}$ & ${\it  0.9992}$ & ${\bf  24.8795}$ & ${\bf  -24.8794}$ & ${\bf  0.0180}$ & ${\bf  -12.4535}$ \\
$N_{\rm A}$ & ${\it  0.0269}$ & ${\it  -0.9996}$ & ${\it  -0.9991}$ & ${\it  -1.0000}$ & ${\bf  24.8811}$ & ${\bf  -0.0180}$ & ${\bf  12.4552}$ \\
$m_{\rm e}/m_{\rm \mu}$ & ${\it  -0.0528}$ & ${\it  0.0000}$ & ${\it  -0.0008}$ & ${\it  0.0014}$ & ${\it  -0.0014}$ & ${\bf  6.4296}$ & ${\bf  -0.0227}$ \\
$F$ & ${\it  0.0679}$ & ${\it  -0.9975}$ & ${\it  -0.9965}$ & ${\it  -0.9990}$ & ${\it  0.9991}$ & ${\it  -0.0036}$ & ${\bf  6.2459}$ \\
\botrule
\end{tabular}
\begin{minipage}{7 in}
\footnotetext[1]{The relative covariance is 
$u_{\rm r}(x_i,x_j) = u(x_i,x_j)/(x_ix_j)$, where $u(x_i,x_j)$ is the covariance of
$x_i$ and $x_j$; the relative variance is
$u_{\rm r}^2(x_i) = u_{\rm r}(x_i,x_i)$:
and the correlation coefficient is
$r(x_i,x_j) = u(x_i,x_j)/[u(x_i)u(x_j)]$.
}\end{minipage}
\end{table*}

\begin{table*}
\caption{Internationally adopted values of various quantities.}
\label{tab:adopted}
\def\sp{\hbox to 6.5mm {}}
\begin{tabular}{l@{\sp}l@{\sp}l@{\sp}l@{\sp}l@{\sp}l}
\toprule
& \T & & & Relative std. \\
\B ~~~~~~Quantity & Symbol~~~~~ & Numerical value~~~~~ & Unit & uncert. $u_{\rm r}$ \\
\colrule
relative atomic mass\footnotemark[1] of $^{12}$C & $A_{\rm r}(^{12}$C) & $12$ &  & ~~~(exact) \vbox to 12 pt {} \\
molar mass constant  & $M_{\rm u}$ & $1 \times 10 ^{-3}$ & kg mol$^{-1}$ & ~~~(exact) \\
molar mass of $^{12}$C & $M(^{12}$C) & $12 \times 10 ^{-3}$ & kg mol$^{-1}$ & ~~~(exact) \\
conventional value of Josephson constant\footnotemark[2]
& $K_{\rm J-90}$ & 483\,597.9 & GHz V$^{-1}$ & ~~~(exact) \\
conventional value of von Klitzing
constant\footnotemark[3] ~~~~~
& $R_{\rm K-90}$ & 25\,812.807 & ${\rm \Omega}$ & ~~~(exact) \\
standard atmosphere \B & & $101\,325$ & Pa & ~~~(exact) \\
\botrule
\end{tabular}
\begin{minipage}{7 in}
\footnotetext[1]{The relative atomic mass $A_{\rm r}(X)$ of particle $X$ with
 mass $m(X)$ is defined by $A_{\rm r}(X) = m(X) /m_{\rm u}$, where
$m_{\rm u} = m(^{12}{\rm C})/12 = M_{\rm u}/N_{\rm A} = 1~{\rm u}$ is the
atomic mass constant, $M_{\rm u}$ is the molar mass constant,
$N_{\rm A}$ is the Avogadro constant, and u is the unified
atomic mass unit.  Thus the mass of particle $X$ is $m(X) = A_{\rm r}(X)$~u
and the molar mass of $X$ is $M(X) = A_{\rm r}(X)M_{\rm u}$.}
\footnotetext[2]{This is the value adopted internationally
for realizing representations of the volt using the Josephson effect.}
\footnotetext[3]{This is the value adopted internationally
for realizing representations of the ohm using the quantum Hall effect.}
\end{minipage}
\end{table*}

\begin{table*}[!]
\caption{Values of some x-ray-related quantities based on the 2006 CODATA
adjustment of the values of the constants.}
\label{tab:xrayvalues}
\def\sp{\hbox to 8.4mm {}}
\begin{tabular}{l@{\sp}l@{\sp}l@{\sp}l@{\sp}l@{\sp}l}
\toprule
\T & & & & Relative std. \\
\B ~~~~~~Quantity & Symbol & ~~~~Numerical value & Unit & uncert. $u_{\rm r}$ \\
\colrule
Cu x unit: $\lambda({\rm CuK}{\rm \alpha}_{\rm 1}) / 1\,537.400 $ & ${\rm xu}({\rm CuK}{\rm \alpha}_{\rm 1})$ & $ 1.002\,076\,99(28)\times 10^{-13}$ & m & $ 2.8\times 10^{-7}$ 
\vbox to 12 pt {} \\
Mo x unit: $\lambda({\rm MoK}{\rm \alpha}_{\rm 1}) / 707.831 $ & ${\rm xu}({\rm MoK}{\rm \alpha}_{\rm 1})$ & $ 1.002\,099\,55(53)\times 10^{-13}$ & m & $ 5.3\times 10^{-7}$ \\
{\aa}ngstrom star$: \lambda({\rm WK}{\rm \alpha}_{\rm 1}) / 0.209\,010\,0 $ & \AA$^{\ast}$ & $ 1.000\,014\,98(90)\times 10^{-10}$ & m & $ 9.0\times 10^{-7}$ \\
\vbox to 12 pt {}
lattice parameter\footnotemark[1]
of Si & $a$ & $ 543.102\,064(14)\times 10^{-12}$ & m & $ 2.6\times 10^{-8}$ \\
\, (in vacuum, 22.5 $^\circ$C)&&&&\\
\{220\} lattice spacing of Si $a/\sqrt{8}$ & $d_{\rm 220}$ & $ 192.015\,5762(50)\times 10^{-12}$ & m & $ 2.6\times 10^{-8}$ \\
\, (in vacuum, 22.5 $^\circ$C)&&&&\\
molar volume of Si & & & & \\
\, $M({\rm Si})/\rho({\rm Si})=N_{\rm A}a^{3}\!/8$ & $V_{\rm m}$(Si) & $ 12.058\,8349(11)\times 10^{-6}$ & m$^{3}$ mol$^{-1}$ & $ 9.1\times 10^{-8}$ \\
\, \, (in vacuum, 22.5 $^\circ$C)&&&&\\
\botrule
\end{tabular}
\begin{minipage}{7 in}
\footnotetext[1]{This is the lattice parameter (unit cell edge length) 
of an ideal single crystal of naturally occurring
Si free of impurities and imperfections, and is deduced from
measurements on extremely pure and nearly perfect single crystals of Si
by correcting for the effects of impurities.}
\end{minipage}
\end{table*}

\thispagestyle{empty}
\begin{table*}[!]
\caption{The values in SI units of some non-SI units
based on the 2006 CODATA adjustment of the values of the constants.}
\label{tab:units}
\def\sp{\hbox to 24 pt{}}
\begin{tabular}{l@{\sp}l@{\sp}l@{\sp}l@{\sp}l@{\sp}l}
\toprule
\T & & & & Relative std. \\
\B \s{35}Quantity & \s{-3}Symbol & \s{17}Numerical value & \s{2}Unit & uncert. $u_{\rm r}$ \\
\colrule
\multicolumn {5} {c} { \vbox to 12 pt {} Non-SI units accepted for use with the SI} \\
\multicolumn {5} {c} {} \\
electron volt: ($e/{\rm C}$) {\rm J} & eV & $ 1.602\,176\,487(40)\times 10^{-19}$ & J & $ 2.5\times 10^{-8}$ \\
(unified) atomic mass unit: & & & & \\
\, 1 u $=m_{\rm u}= {1\over12}m(^{12}$C) & u & $ 1.660\,538\,782(83)\times 10^{-27}$ & kg & $ 5.0\times 10^{-8}$ \\
\, \, $=10^{-3}$ kg mol$^{-1}\!/N_{\rm A}$ & & & & \\
\multicolumn {5} {c} {} \\
\multicolumn {5} {c} {Natural units (n.u.)} \\
\multicolumn {5} {c} {} \\
n.u. of velocity: &&&&\\
\, speed of light in vacuum & $c,c_0$ & 299\,792\,458 & m s$^{-1}$ & (exact) \\
n.u. of action: &&&&\\
\, reduced Planck constant $(h/2\rmpi)$ & $\hbar$ & $ 1.054\,571\,628(53)\times 10^{-34}$ & J s & $ 5.0\times 10^{-8}$ \\
\b\b in eV s & & $ 6.582\,118\,99(16)\times 10^{-16}$ & eV s & $ 2.5\times 10^{-8}$ \\
\b\b in MeV fm & $\hbar c$ & $ 197.326\,9631(49)$ & MeV fm & $ 2.5\times 10^{-8}$ \\
n.u. of mass: &&&&\\
\, electron mass & $m_{\rm e}$ & $ 9.109\,382\,15(45)\times 10^{-31}$ & kg & $ 5.0\times 10^{-8}$ \\
n.u. of energy & $m_{\rm e}c^2$ & $ 8.187\,104\,38(41)\times 10^{-14}$ & J & $ 5.0\times 10^{-8}$ \\
\b\b in MeV & & $ 0.510\,998\,910(13)$ & MeV & $ 2.5\times 10^{-8}$ \\
&&&&\\
n.u. of momentum & $m_{\rm e}c$ & $ 2.730\,924\,06(14)\times 10^{-22}$ & kg m s$^{-1}$ & $ 5.0\times 10^{-8}$ \\
\b\b in MeV/$c$ & & $ 0.510\,998\,910(13)$ & MeV/$c$ & $ 2.5\times 10^{-8}$ \\
n.u. of length $(\hbar/m_{\rm e}c)$& $\lbar_{\rm C}$ & $ 386.159\,264\,59(53)\times 10^{-15}$ & m & $ 1.4\times 10^{-9}$ \\
n.u. of time & $\hbar/m_{\rm e}c^2$ & $ 1.288\,088\,6570(18)\times 10^{-21}$ & s & $ 1.4\times 10^{-9}$ \\
\multicolumn {5} {c} {} \\
\multicolumn {5} {c} {Atomic units (a.u.)} \\
\multicolumn {5} {c} {} \\
a.u. of charge: &&&&\\
\, elementary charge & $e$ & $ 1.602\,176\,487(40)\times 10^{-19}$ & C & $ 2.5\times 10^{-8}$ \\
a.u. of mass: &&&&\\
\, electron mass & $m_{\rm e}$ & $ 9.109\,382\,15(45)\times 10^{-31}$ & kg & $ 5.0\times 10^{-8}$ \\
a.u. of action: &&&&\\
\, reduced Planck constant $(h/2\rmpi)$& $\hbar$ & $ 1.054\,571\,628(53)\times 10^{-34}$ & J s & $ 5.0\times 10^{-8}$ \\
a.u. of length: & & & & \\
\, Bohr radius (bohr) $(\alpha/4\rmpi R_\infty)$& $a_0$ & $ 0.529\,177\,208\,59(36)\times 10^{-10}$ & m & $ 6.8\times 10^{-10}$ \\
a.u. of energy: & & & & \\
\, Hartree energy (hartree) & $E_{\rm h}$ & $ 4.359\,743\,94(22)\times 10^{-18}$ & J & $ 5.0\times 10^{-8}$ \\
\, ($e^2\!/4\rmpi\epsilon_0a_0=2R_\infty hc = \alpha^2m_{\rm e}c^2)$&&&& \\
&&&&\\
a.u. of time & $\hbar/E_{\rm h}$ & $ 2.418\,884\,326\,505(16)\times 10^{-17}$ & s & $ 6.6\times 10^{-12}$ \\
a.u. of force & $E_{\rm h}/a_0$ & $ 8.238\,722\,06(41)\times 10^{-8}$ & N & $ 5.0\times 10^{-8}$ \\
a.u. of velocity ($\alpha c$) & $a_0E_{\rm h}/\hbar$ & $ 2.187\,691\,2541(15)\times 10^{6}$ & m s$^{-1}$ & $ 6.8\times 10^{-10}$ \\
a.u. of momentum & $\hbar/a_0$ & $ 1.992\,851\,565(99)\times 10^{-24}$ & kg m s$^{-1}$ & $ 5.0\times 10^{-8}$ \\
a.u. of current & $eE_{\rm h}/\hbar$ & $ 6.623\,617\,63(17)\times 10^{-3}$ & A & $ 2.5\times 10^{-8}$ \\
a.u. of charge density & $e/a_0^3$ & $ 1.081\,202\,300(27)\times 10^{12}$ & C m$^{-3}$ & $ 2.5\times 10^{-8}$ \\
&&&&\\
a.u. of electric potential & $E_{\rm h}/e$ & $ 27.211\,383\,86(68)$ & V & $ 2.5\times 10^{-8}$ \\
a.u. of electric field & $E_{\rm h}/ea_0$ & $ 5.142\,206\,32(13)\times 10^{11}$ & V m$^{-1}$ & $ 2.5\times 10^{-8}$ \\
a.u. of electric field gradient & $E_{\rm h}/ea_0^2$ & $ 9.717\,361\,66(24)\times 10^{21}$ & V m$^{-2}$ & $ 2.5\times 10^{-8}$ \\
a.u. of electric dipole moment & $ea_0$ & $ 8.478\,352\,81(21)\times 10^{-30}$ & C m & $ 2.5\times 10^{-8}$ \\
a.u. of electric quadrupole moment & $ea_0^2$ & $ 4.486\,551\,07(11)\times 10^{-40}$ & C m$^2$ & $ 2.5\times 10^{-8}$ \\
&&&&\\
a.u. of electric polarizability & $e^2a_0^2/E_{\rm h}$ & $ 1.648\,777\,2536(34)\times 10^{-41}$ & C$^2$ m$^2$ J$^{-1}$ & $ 2.1\times 10^{-9}$ \\
a.u. of 1$^{\rm st}$ hyperpolarizability & $e^3a_0^3/E_{\rm h}^2$ & $ 3.206\,361\,533(81)\times 10^{-53}$ & C$^3$ m$^3$ J$^{-2}$ & $ 2.5\times 10^{-8}$ \\
a.u. of 2$^{\rm nd}$ hyperpolarizability & $e^4a_0^4/E_{\rm h}^3$ & $ 6.235\,380\,95(31)\times 10^{-65}$ & C$^4$ m$^4$ J$^{-3}$ & $ 5.0\times 10^{-8}$ \\
a.u. of magnetic flux density & $\hbar/ea_0^2$ & $ 2.350\,517\,382(59)\times 10^{5}$ & T & $ 2.5\times 10^{-8}$ \\
a.u. of magnetic & & & & \\
\, dipole moment ($2\mu_{\rm B}$) & $\hbar e/m_{\rm e}$ & $ 1.854\,801\,830(46)\times 10^{-23}$ & J T$^{-1}$ & $ 2.5\times 10^{-8}$ \\
a.u. of magnetizability & $e^2a_0^2/m_{\rm e}$ & $ 7.891\,036\,433(27)\times 10^{-29}$ & J T$^{-2}$ & $ 3.4\times 10^{-9}$ \\
a.u. of permittivity $(10^7/c^2)$ & $e^2/a_0E_{\rm h}$ & $ 1.112\,650\,056\ldots\times 10^{-10}$ & F m$^{-1}$ & (exact) \\
\botrule
\end{tabular}
\end{table*}

\begin{table*}[!]
\caption{The values of some energy equivalents derived from the relations
$E=mc^2 = hc/\lambda = h\nu = kT$, and based on the 2006 CODATA adjustment of the values of
the constants; 1~eV~$=(e/{\rm C})$~J,
1~u $= m_{\rm u} = \textstyle{1\over12}m(^{12}{\rm C}) 
= 10^{-3}$~kg~mol$^{-1}\!/N_{\rm A}$, and
$E_{\rm h} = 2R_{\rm \infty}hc = \alpha^2m_{\rm e}c^2$ is the 
Hartree energy (hartree).}
\label{tab:enconv1}
\begin{tabular} {lllll}
\toprule
\multicolumn{5} {c} {Relevant unit}\B\T \\
\colrule
    &   \s{30}J &  \s{30}kg &  \s{30}m$^{-1}$  &   \s{30}Hz  \vbox to 12 pt {} \B  \\
\colrule
        &                    &                    &               &            \\
1~J    & $(1\ {\rm J})=$    &  (1 J)/$c^2=$     &       (1 J)/$hc=$   &        (1 J)/$h=$              \\
 & 1 J   & $ 1.112\,650\,056\ldots\times 10^{-17}$ kg  & $ 5.034\,117\,47(25)\times 10^{24}$ m$^{-1}$ & $ 1.509\,190\,450(75)\times 10^{33}$ Hz \\
        &        &             &           &  \\
1~kg    &        (1 kg)$c^2=$   &   $(1 \ {\rm kg})=$       &        (1 kg)$c/h=$ &        (1 kg)$c^2/h=$          \\
 &  $ 8.987\,551\,787\ldots\times 10^{16}$ J & 1 kg   & $ 4.524\,439\,15(23)\times 10^{41}$ m$^{-1}$ & $ 1.356\,392\,733(68)\times 10^{50}$ Hz \\
 &           &            &         &   \\
1~m$^{-1}$   &  (1 m$^{-1})hc=$   &       (1 m$^{-1})h/c=$    &  $(1$ m$^{-1})=$    &   (1 m$^{-1})c=$  \\
 &  $ 1.986\,445\,501(99)\times 10^{-25}$ J  & $ 2.210\,218\,70(11)\times 10^{-42}$ kg  & 1 m$^{-1}$  & $ 299\,792\,458$ Hz \\
 &           &            &         &   \\
1~Hz   &  (1 Hz)$h=$  &  (1 Hz)$h/c^{2}=$   &        (1 Hz)/$c=$  &  $(1$ Hz$)=$  \\
 &  $ 6.626\,068\,96(33)\times 10^{-34}$ J  & $ 7.372\,496\,00(37)\times 10^{-51}$ kg & $ 3.335\,640\,951\ldots\times 10^{-9}$ m$^{-1}$ & 1 Hz \\
 &           &            &         &   \\
1~K  &  (1 K)$k=$  &   (1 K)$k/c^{2}=$  &    (1 K)$k/hc=$ &  (1 K)$k/h=$   \\
 &  $ 1.380\,6504(24)\times 10^{-23}$ J  & $ 1.536\,1807(27)\times 10^{-40}$ kg & $ 69.503\,56(12)$ m$^{-1}$ & $ 2.083\,6644(36)\times 10^{10}$ Hz \\
         &            &            &           &        \\
1~eV    &  (1 eV) =  &  $(1~{\rm eV})/c^{2}=$  &   $(1~{\rm eV})/hc=$   &     $(1~{\rm eV})/h=$   \\
 &  $ 1.602\,176\,487(40)\times 10^{-19}$ J  & $ 1.782\,661\,758(44)\times 10^{-36}$ kg & $ 8.065\,544\,65(20)\times 10^{5}$ m$^{-1}$ & $ 2.417\,989\,454(60)\times 10^{14}$ Hz \\
         &          &             &         &     \\
1~u   &   $(1~{\rm u})c^{2}=$   &  (1 u) =   &  $(1~{\rm u})c/h=$  &  $(1~{\rm u})c^{2}/h=$    \\
 &  $ 1.492\,417\,830(74)\times 10^{-10}$ J  & $ 1.660\,538\,782(83)\times 10^{-27}$ kg & $ 7.513\,006\,671(11)\times 10^{14}$ m$^{-1}$ & $ 2.252\,342\,7369(32)\times 10^{23}$ Hz \\
         &          &             &        &       \\
1~$E_{\rm h}$   &  $(1~E_{\rm h})=$  & $(1~E_{\rm h})/c^2=$ &  $(1~E_{\rm h})/hc=$  &  $(1~E_{\rm h})/h=$   \\
\B &  $ 4.359\,743\,94(22)\times 10^{-18}$ J  & $ 4.850\,869\,34(24)\times 10^{-35}$ kg & $ 2.194\,746\,313\,705(15)\times 10^{7}$ m$^{-1}$ & $ 6.579\,683\,920\,722(44)\times 10^{15}$ Hz \\
\botrule
\end{tabular}
\end{table*}

\begin{table*}[!]
\caption{The values of some energy equivalents derived from the relations
$E=mc^2 = hc/\lambda = h\nu = kT$, and based on the 2006 CODATA adjustment of the values of
the constants; 1~eV~$=(e/{\rm C})$~J,
1~u $= m_{\rm u} = \textstyle{1\over12}m(^{12}{\rm C})
= 10^{-3}$~kg~mol$^{-1}\!/N_{\rm A}$, and
$E_{\rm h} = 2R_{\rm \infty}hc = \alpha^2m_{\rm e}c^2$ is the
Hartree energy (hartree).}
\label{tab:enconv2}
\begin{tabular} {lllll}
\toprule
\multicolumn{5} {c} {Relevant unit} \B\T \\
\colrule
\B  &   \s{30}K &  \s{30}eV &  \s{30}u  &   \s{30}$E_{\rm h}$  \vbox to 12 pt {}   \\
\colrule
        &                    &                    &               &            \\
1~J     & (1 J)/$k=$   &  (1 J) =   &       (1 J)/$c^2$ = &   (1 J) =              \\
 &  $ 7.242\,963(13)\times 10^{22}$ K  & $ 6.241\,509\,65(16)\times 10^{18}$ eV  & $ 6.700\,536\,41(33)\times 10^{9}$ u & $ 2.293\,712\,69(11)\times 10^{17}$ $E_{\rm h}$ \\
        &        &             &           &  \\
1~kg    &        (1 kg)$c^2/k=$    &   (1 kg)$c^2$ =    &    (1 kg) =  &  (1 kg)$c^2=$   \\
 &  $ 6.509\,651(11)\times 10^{39}$ K  & $ 5.609\,589\,12(14)\times 10^{35}$ eV  & $ 6.022\,141\,79(30)\times 10^{26}$ u  & $ 2.061\,486\,16(10)\times 10^{34}$ $E_{\rm h}$ \\
 &           &            &         &   \\
1~m$^{-1}$   &  (1 m$^{-1})hc/k=$  &       (1 m$^{-1})hc=$  &  (1 m$^{-1})h/c$ =   &   (1 m$^{-1})hc=$ \\
 &  $ 1.438\,7752(25)\times 10^{-2}$ K  & $ 1.239\,841\,875(31)\times 10^{-6}$ eV  & $ 1.331\,025\,0394(19)\times 10^{-15}$ u & $ 4.556\,335\,252\,760(30)\times 10^{-8}$ $E_{\rm h}$ \\
         &           &            &            &      \\
1~Hz   &  (1 Hz)$h/k=$  &  (1 Hz)$h=$  &   (1 Hz)$h/c^2$ = & (1 Hz)$h=$  \\
 &  $ 4.799\,2374(84)\times 10^{-11}$ K  & $ 4.135\,667\,33(10)\times 10^{-15}$ eV  & $ 4.439\,821\,6294(64)\times 10^{-24}$ u & $ 1.519\,829\,846\,006(10)\times 10^{-16}$ $E_{\rm h}$ \\
           &             &          &           &        \\ 
1~K  & $(1$ K$)=$  &   (1 K)$k=$ &    (1 K)$k/c^2=$ &  (1 K)$k=$   \\
 & 1 K   &   $ 8.617\,343(15)\times 10^{-5}$ eV & $ 9.251\,098(16)\times 10^{-14}$ u & $ 3.166\,8153(55)\times 10^{-6}$ $E_{\rm h}$ \\
         &            &            &           &        \\
1~eV    &  (1 eV)/$k=$  &  $(1$ eV$)=$  &   $(1~{\rm eV})/c^2=$  &     $(1~{\rm eV})=$   \\
 &  $ 1.160\,4505(20)\times 10^{4}$ K  & 1 eV  & $ 1.073\,544\,188(27)\times 10^{-9}$ u & $ 3.674\,932\,540(92)\times 10^{-2}$ $E_{\rm h}$ \\
         &          &             &         &     \\
1~u   &   $(1~{\rm u})c^{2}/k=$   &  $(1~{\rm u})c^2=$ &  $(1$ u$)=$  &  $(1~{\rm u})c^2=$   \\
 &  $ 1.080\,9527(19)\times 10^{13}$ K  & $ 931.494\,028(23)\times 10^{6}$ eV  & 1 u & $ 3.423\,177\,7149(49)\times 10^{7}$ $E_{\rm h}$ \\
         &          &             &        &       \\
1~$E_{\rm h}$   &  $(1~E_{\rm h})/k=$  & $(1~E_{\rm h})=$ &  $(1~E_{\rm h})/c^2=$ & $(1~E_{\rm h})=$    \\
\B &  $ 3.157\,7465(55)\times 10^{5}$ K  & $ 27.211\,383\,86(68)$ eV  & $ 2.921\,262\,2986(42)\times 10^{-8}$ u  & $1~E_{\rm h}$  \\
\botrule
\end{tabular}
\end{table*}

\end{document}